# The Role of Acetylene in the Chemical Evolution of Carbon Complexity

*Evgeniy O. Pentsak*[1], *Maria S. Murga*[2], *Valentine P. Ananikov*[1,*]

[1] Zelinsky Institute of Organic Chemistry, Russian Academy of Sciences, Moscow, 119991, Russia.

[2] Institute of Astronomy, Russian Academy of Sciences, Pyatnitskaya str. 48, Moscow 119017, Russia.

ABSTRACT

Acetylene, among the multitude of organic molecules discovered in space, plays a distinct role in the genesis of organic matter. Characterized by its unique balance of stability and reactivity, acetylene is the simplest unsaturated organic molecule known to have a triple bond. In addition to its inherent chemical properties, acetylene is one of the most prevalent organic molecules found across the Universe, spanning from the icy surfaces of planets and satellites and the cold interstellar medium with low temperatures to hot circumstellar envelopes where temperatures surge to several thousand kelvins. These factors collectively position acetylene as a crucial building block in the molecular diversification of organic molecules and solids present in space. This review comprehensively discusses the formation and expansion of carbon skeletons involving acetylene, ranging from the formation of simple molecules to the origination of the first aromatic ring and ultimately to the formation of nanosized carbon particles. Mechanisms pertinent to both hot environments, such as circumstellar envelopes, and cold environments, including molecular clouds and planetary atmospheres, are explored. In addition, this review contemplates the role of acetylene in the synthesis of prebiotic molecules. A distinct focus is accorded to the recent advancements and future prospects of research into catalytic processes involving acetylene molecules, which is a significant instrument in driving the evolution of carbon complexity in the Universe. The insights garnered from this review underscore the significance of acetylene in astrochemistry and potentially contribute to our understanding of the chemical evolution of the Universe.

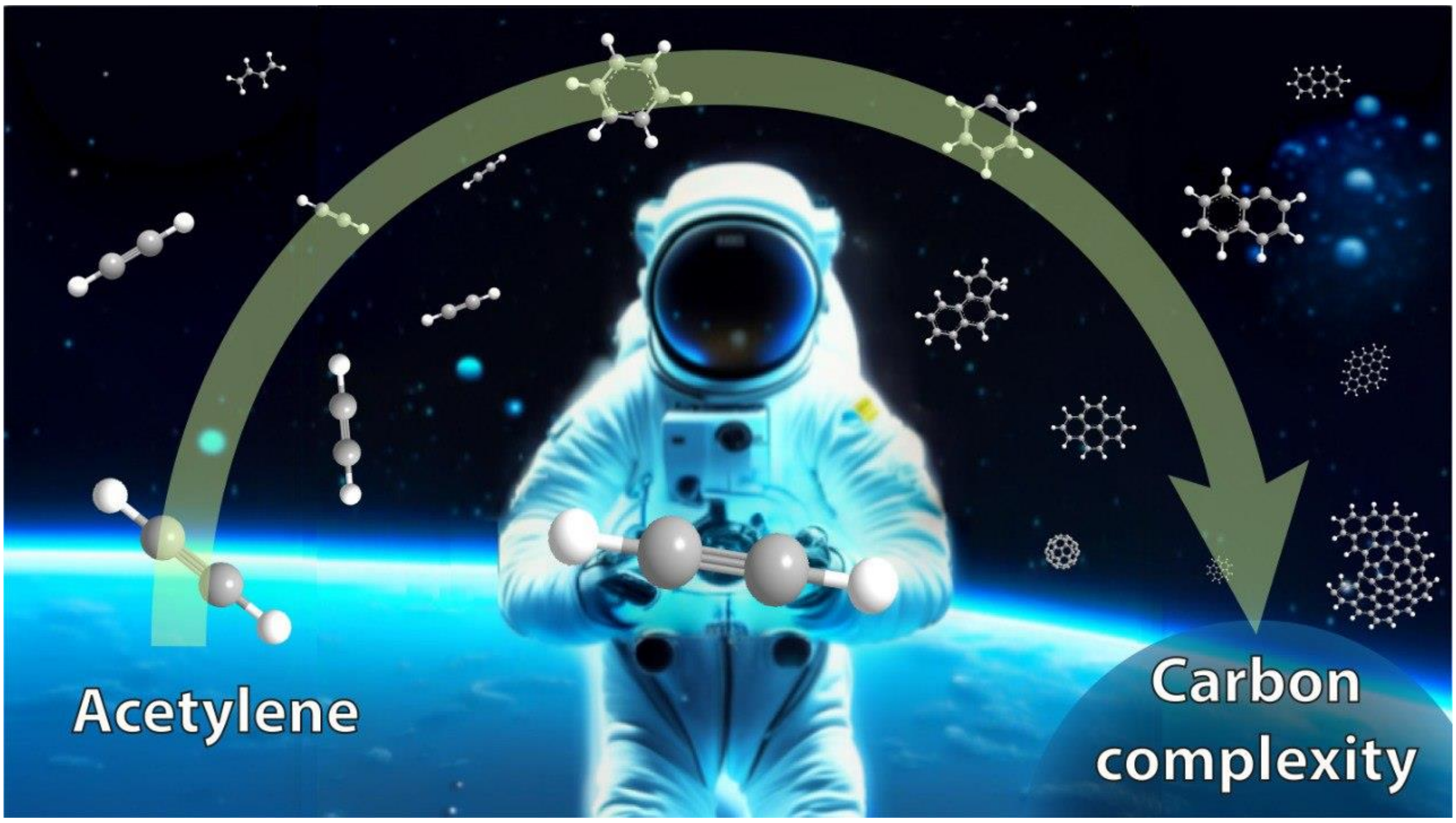

# Contents

# 1. Introduction

The past decade has witnessed substantial advancements in the chemistry of acetylene ($C_2H_2$), sparking novel insights into its role in the development of organic compounds and carbon structures associated with various celestial bodies. Rapid progress in areas such as carbon materials and carbocatalysis has led to the accumulation of a wealth of knowledge concerning acetylene chemistry.[1] This new understanding could inspire fresh breakthroughs in astrochemistry, particularly in relation to catalytic processes.

On Earth, acetylene is both a well-recognized and thoroughly studied molecule. Since its discovery by Davy in 1832[2] and later by Berthelot in 1860[3], acetylene has proven to be both interesting and valuable. Its primary source is industrial production through the pyrolysis of hydrocarbon raw materials.[4] Acetylene serves as a precursor for many industrial products, including vinyl chloride, vinyl acetate, acrylic acid, 1,4-butynediol, and carbon black. The production and industrial use of acetylene are among the global market indicators[4,5], and new potential applications of acetylene have recently been discovered.[6,7,8,9] For example, new methods for producing carbon materials such as carbon nanotubes and graphene are of particular interest.[10,11,12,13,14,15,16] The development of cycloaddition reactions involving acetylene has experienced a resurgence in recent years and is another exciting area.[17]

In the context of space, the situation around acetylene becomes even more intricate and fascinating. Chemical evolution from monatomic particles to complex carbon systems is a central issue in astrochemistry. The creation of complex aromatic systems is a thermodynamically favored process that pervades the Universe. The existence of numerous polyaromatic compounds, carbon structures such as fullerenes, and macroscopic carbon particles in space attests to this fact.[18,19,20,21,22,23,24,25] Acetylene is widely considered a primary element in the formation of these structures.

Acetylene is among the simplest organic molecules found abundantly in outer space. It is detectable and readily formed in many different environments: circumstellar envelopes — the hot and dense layers that are lost by asymptotic giant branch (AGB) stars and moving away from them — the atmospheres of planets within and likely beyond the Solar System — in the cold interstellar medium (ISM) at extremely low temperatures and others. Recent studies underscore its significance in the chemistry of exoplanets.[26,27]

The majority of related research has focused on the mechanisms of chemical reactions triggered by ultraviolet (UV) radiation and cosmic rays in the upper layers of planetary atmospheres, as well as in circumstellar and interstellar media. However, the evolution of organic matter on Earth-like planets or satellites may exhibit unique features. Recent discoveries in the chemistry of atmospheric and surface hydrocarbons on Titan and Enceladus confirm the crucial role of acetylene in the evolution of complex organic structures.[28,29,30,31,32]

Acetylene molecules serve as excellent precursors for polyaromatics[33,34] and carbon dust[35,36] (Figure 1). Although the hydrogen removal/acetylene addition (HACA) paradigm remains the prevailing model for explaining the formation of simple polycyclic aromatic hydrocarbons (PAHs) in space, alternative mechanisms warrant consideration.[37,38,39] Recent discoveries in chemistry have revealed a catalytic pathway for the formation of aromatic compounds from acetylene molecules. The trimerization of acetylene catalyzed by carbon molecules could unlock a wide variety of organic molecules (Figure 1).

Acetylene is hypothesized to play a significant role in prebiotic systems.[40,41] Recent studies have indicated its involvement in the metabolic processes of primitive life forms and suggested that it is a probable precursor for numerous prebiotic molecules.[42] According to the PAH world hypothesis, polyaromatic compounds contribute to the primary assembly of self-replicating informational polymers and likely have other essential roles in prebiotic environments.[43,44,45] Unraveling the mechanisms of self-sustaining the growth of polyaromatic molecules is a key element in resolving these questions.

Consequently, a wealth of data suggest that acetylene may form bedrock for numerous chemical processes in outer space, leading to the evolution of complex organic molecules and carbon macroparticles from simpler carbon particles (Figure 1). In essence, acetylene could be viewed as a chemical platform for catalyzing the emergence of more intricate carbon species. Our fundamental understanding of its "cosmic" significance is only now beginning to unfold.

Generally, acetylene participates in almost all types of chemical reactions that lead to molecular growth (bottom-up reactions): neutral-neutral reactions, ion-molecular reactions, surface reactions, and nucleation. Neutral-neutral reactions mostly have an activation barrier and dominate in hot and dense environments, such as stellar envelopes and photodissociation regions (PDRs); however, neutral-neutral reactions are not excluded at low temperatures in molecular clouds, as some reactions are barrierless or because the reactants are radicals. Ion-molecular reactions more easily proceed in diffuse ISM (DISM) and in molecular clouds and PDRs due to dipole interactions. The ionization of atoms and molecules is achieved by cosmic rays and UV radiation. Surface chemistry becomes relevant in molecular clouds and photodissociation regions, especially for denser and darker objects and prestellar cores, where many gaseous species are adsorbed on the dust grain surface. Some faraway objects of the Solar System, such as Kuiper's belt objects or comets, can undergo similar chemical reactions because the temperature is also low there. At low temperatures the dust grains are covered by an icy mantle, and interactions between the ice and surface species also occur. Moreover, chemical reactions occur inside the icy mantle. The major trigger of many reactions at low temperatures is cosmic rays (CRs), which, unlike UV photons, penetrate dense cloud cores. However, not only does surface chemistry take place in cold objects, but it is also relevant for radically different conditions typical of hot stellar envelopes where aromatic molecules can be formed on or from the dust surface. Under the same conditions, the nucleation process plays a crucial role in promoting the growth of macroparticles from molecules. Regardless of the type of reaction, acetylene is a starting point for many pathways, leading to cosmic complexity and variety.

Acetylene is also a key reactant of the opposite 'top-down' chemistry, where it is a fragment of dissociation of macromolecules or macroparticles induced by factors such as UV radiation, bombardment by CRs, and shock waves. Thus, acetylene is in some sense a connecting element of the carbon lifecycle. We schematically illustrate the role of acetylene in different types of chemical reactions in Figure 1.

To guide the reader throughout the paper, we provide a summary table (Table 1), which includes chemical reaction types and corresponding sections where these reactions are considered, astronomical objects and their typical physical conditions (number density $n_H$, temperature T, ionizing source, UV or CR radiation) where these reactions occur. We consider most space objects where acetylene is observed and/or plays an important role in chemical processing of the medium. We note that one type of reactions (not necessarily the same reaction) can correspond to objects with very distinct conditions, and vice versa; many types of reactions can proceed under the same conditions. Further subsections contain more specific tables where individual reactions or chains of reactions under corresponding conditions are presented, although some crossing between reactions may occur because the conditions are similar; thus, it is not always possible to strictly indicate the objects where specific reactions proceed. Some chains are short and fully presented, while long chains are shortened. If rate constants are provided in original papers or alternatively in chemical databases, they are presented in the tables of subsections. Rates are often dependent on temperature, pressure or other parameters; therefore, in these cases, we calculated the rates at fixed values of the parameters, which are averages for the objects listed in the 3rd column. In some cases, quantum yields or other values, instead of rate constants, are given in the original papers; thus, we do not present the kinetics. Finally, there are reactions whose rates are undetermined or not provided.

**Table 1**. Summary of acetylene chemistry in different astronomical media.

| **Type of C chemistry** | **Astronomical objects** | **$n_H$, $cm^{-3}$ (if gas phase)** | **T, K** | **$G_0$*/CR** | **Subsection** |
|---|---|---|---|---|---|
| UV photon-initiated chemistry | Atmospheres of Jovian planets and their satellites[46,47] | $10^{10}$-$10^{16}$ | 50-300 | up to $10^2$ | 2.1, 2.2, 2.7, 3.2, 4.2, 4.4, 5.4, 8.1, 8.2 |
| | Kuiper belt objects, comets[48] | | 50 | ~1 | |
| | PDRs[49] | $10^2$-$10^6$ | 20-1000 | up to $10^6$ | |
| | Protoplanetary nebulae [50] | $10^7$ | >200 | up to $10^6$ | |
| CR/UV photon initiated chemistry in ice | Molecular clouds, prestellar cores | $10^2$-$10^7$ | 10 | any $G_0$/ Galactic CR field[51,52,53] | 2.1, 2.2, 2.3, 2.6, 3.3, 4.4, 5.4, 8.2 |
| | Atmospheres of planets and their satellites | $10^{10}$-$10^{16}$ | 50-300 | | |
| | Kuiper belt objects, comets | | 50 | | |
| Neutral-neutral chemistry | Molecular and dust formation zone of carbon stellar envelopes[54] | $10^8$-$10^{15}$ | 500-2500 | not important | 2.5, 3.1, 3.2, 4.1, 4.3, 5.2.1, 5.2.2, 5.2.3, 5.2.4, 5.3, 5.4, 6.1, 7.2, 8.2 |
| | Molecular clouds, prestellar cores | $10^2$-$10^7$ | 10 | | |
| | Hot cores[55] | $\geq 10^7$ | 100-300 | | |
| Ion-molecular chemistry | PDRs | $10^2$-$10^6$ | 100-2000 | up to $10^6$ | 2.7, 3.1, 3.2, 4.2, 4.4, 8.2 |
| | Protoplanetary nebulae; | $10^7$ | > 200 | up to $10^6$ | |
| | Molecular clouds prestellar cores | $10^2$-$10^7$ $cm^{-3}$ | 10 | very low | |
| | Atmospheres of Jovian planets and their satellites | $10^{10}$-$10^{16}$ | 50-300 | up to $10^2$ | |
| | DISM[49] | 50-100 | 100 | 0.1-100 | |
| Nucleation | Dust formation zone of carbon stellar envelopes. | $10^8$-$10^{12}$ | 500-2000 | not important | 6.2 |
| Catalytic chemistry | Molecular clouds, prestellar cores | $10^2$-$10^7$ | 10 | Galactic CR field | 7.2, 7.3, 8.2 |
| | Dust formation zone of carbon stellar envelopes | $10^8$-$10^{12}$ | 500-2000 | not important | |
| | Planet's satellites | $10^{10}$-$10^{16}$ | 50-300 | up to $10^2$ / Galactic CR field | |
| | PDRs, DISM | $10^2$-$10^6$ | 100-2000 | up to $10^6$ | |

* - $G_0$ is a factor generally used to describe the level of the UV radiation field in space objects. The value of the factor is the ratio between integral intensity of the radiation field in the wavelength interval from 6 to 13.6 eV in an object and the same integral intensity in the solar neighborhood. The spectrum of the radiation field intensity in the Solar neighborhood was adopted from the work of Habing.[56] Strictly speaking, using this factor for estimation of the radiation field for the planets is incorrect due to the planets being within the HII region of the Sun while the factor is used for characterizing the conditions out of HII regions. However, we are not determined to give any precise estimations of the radiation field intensity, our goal is to show its approximate level to compare to other space objects in common units.

This review elucidates the most pertinent issues related to chemical processes involving acetylene molecules and their derivatives in outer space. The first section presents a brief introduction. The second section offers a concise overview of the mechanisms of acetylene formation and its prevalence in space. The third section discusses various chemical reactions not associated with aromatic compounds, where acetylene plays a pivotal role. The fourth and fifth sections delve into the formation of an aromatic ring and the expansion of polyaromatic hydrocarbons, respectively. The sixth section examines the creation of nanoparticles, including fullerenes and nanotubes. The seventh section is dedicated to the role of catalytic processes involving acetylene in the spatiotemporal evolution of organic compounds. The eighth section speculates on the possible role of acetylene in prebiotic processes. Finally, the ninth section concludes the review with some closing remarks.

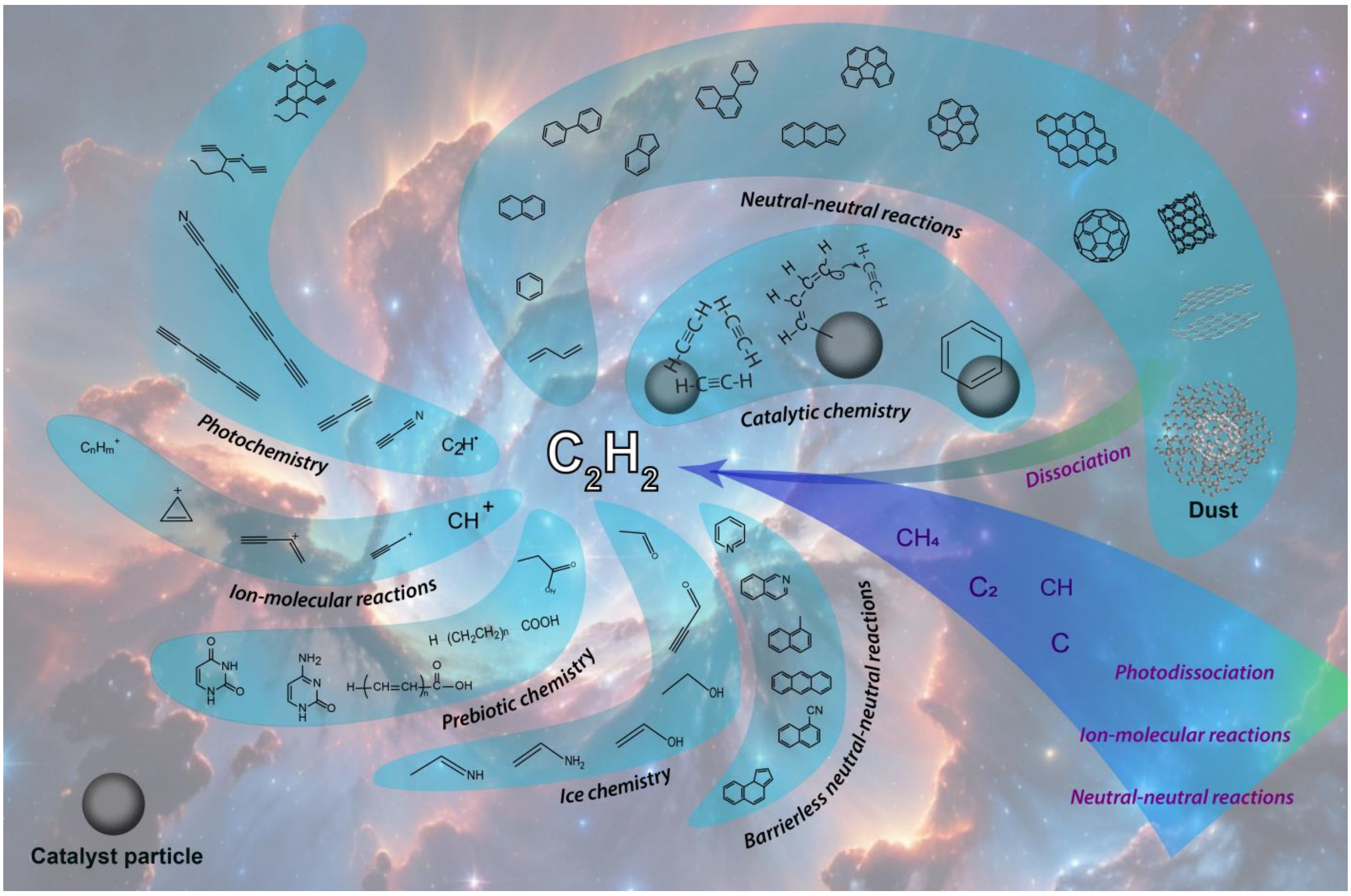

**Figure 1.** Evolution of carbon species in space with the participation of acetylene.

# 2. Acetylene in the Universe. Origin and occurrence

Carbon is mainly formed in thermonuclear fusion reaction at the stars of intermediate masses (from ~ 0.8M$_\odot$ to 8M$_\odot$).[57] Furthermore, carbon spreads into the ISM, driven by the stellar winds of these stars. Some fraction of carbon can be produced in massive stars and thrown into the ISM during supernova explosions.[58] Simple carbonaceous molecules, including acetylene and others (e.g., CO, $CO_2$, and CN), are formed from this carbon directly inside the parent stellar atmosphere[59], and the formation of even more molecules occurs further in the ISM. As a result, carbonaceous molecules are widespread throughout the Universe.[60,61]

Acetylene does not have a permanent dipole moment and therefore does not have the rotational spectrum observed in the radio wave range. Instead, acetylene can be detected by rovibrational transitions that fall into the mid-infrared (mid-IR) range. The rovibrational spectrum of acetylene was both theoretically calculated and obtained in the laboratory.[62,63] Some of the fundamental modes from this spectrum are observed in many space objects, namely, $\nu^3$ at approximately 3 µm, $\nu^4 + \nu^5$ at 7.5 µm and $\nu^5$ at 13.7 µm. Additionally, several weaker bands corresponding to these modes have been observed via high-resolution spectroscopy.[54] Acetylene also exhibits many absorption bands in the far-UV range between wavelengths of 0.15 and 0.19 µm, and these bands are also used for acetylene detection.[47,64] Over the past half century, acetylene has been discovered in a fairly large number of space objects, both inside and outside the Solar system, in the atmospheres of giant planets and their satellites, comets, and circumstellar and interstellar environments. We summarize the observational evidence of the presence of acetylene in space objects in Table S1 in the Supporting Information. Below, we provide detailed information on the detection of acetylene along with the proposed pathways of its formation.

## 2.1. The Jovian Planets

For the first time, acetylene in space was detected in the atmosphere of Jupiter in 1974 based on the analysis of IR spectra obtained at the Kitt Peak National Observatory.[65,66] The fundamental mode $\nu^5$ at nearly 13.7 µm was found. As a measure of abundance, the term 'mixing ratio' is used in studies of planetary atmospheres. This ratio represents the ratio between the specific gas and the dominant gas in the atmosphere. In the Jupiter atmosphere, the dominant gas is $H_2$ (similar to that in other Jovian planets), and the mixing ratio of acetylene was estimated to be $8 \cdot 10^{-5}$ relative to $H_2$. A little later, in 1980, acetylene was discovered on Saturn using the NASA Infrared Telescope Facility (IRTF) at Mauna Kea.[67,68,69] The acetylene mixing ratio was found to be approximately $2 \cdot 10^{-7}$.[70] The launch of Voyager 1 and Voyager 2[71] in 1977 made it possible to significantly extend the available data on the abundance of acetylene in the atmosphere of the planet in the Solar system. In 1981, the IR spectra obtained during the passage of Voyager 1 through the Saturn system were published and showed the presence of acetylene in atmospheres of Saturn and Titan.[70,72] In 1986, the mixing ratios for Jupiter and Saturn were estimated and found to be $1 \cdot 10^{-7}$ and $3 \cdot 10^{-7}$, respectively.[73] Acetylene was determined to be the third most abundant hydrocarbon in the atmosphere of these planets after $CH_4$ and $C_2H_6$, while the most abundant molecule was $H_2$.

In late 2000 and early 2001, the Cassini spacecraft passed near Jupiter, performing a gravity-assisted maneuver on its way to Saturn. The Cassini orbiter was equipped with an ultraviolet imaging spectroscope (UVIS) connected to a camera (ISS) and two infrared spectrometers: a composite infrared spectrometer (CIRS) and a visible and infrared mapping spectrometer (VIMS).[74] CIRS provides the highest resolution map of the acetylene distribution on Jupiter (Figure 2).[75] A greater intensity of emissions in the region of the poles can be associated with both higher temperature and a greater abundance of gas.[76] In addition, recent observations of polar regions have revealed local enhancements in the abundances of acetylene and its derivatives that correlate with the location of diffuse auroras.[77,78] The abundance of acetylene in the upper stratosphere derived from UVIS data is IR thermal emission from the middle stratosphere. In 2020, Melin et al. produced a vertical acetylene abundance profile above 1 mbar, which ensured the consistency of the data obtained by CIRS and UVIS. According to the new model, the abundance of acetylene was $1.21 \pm 0.07 \cdot 10^{-6}$ at 0.1 mbar.[64]

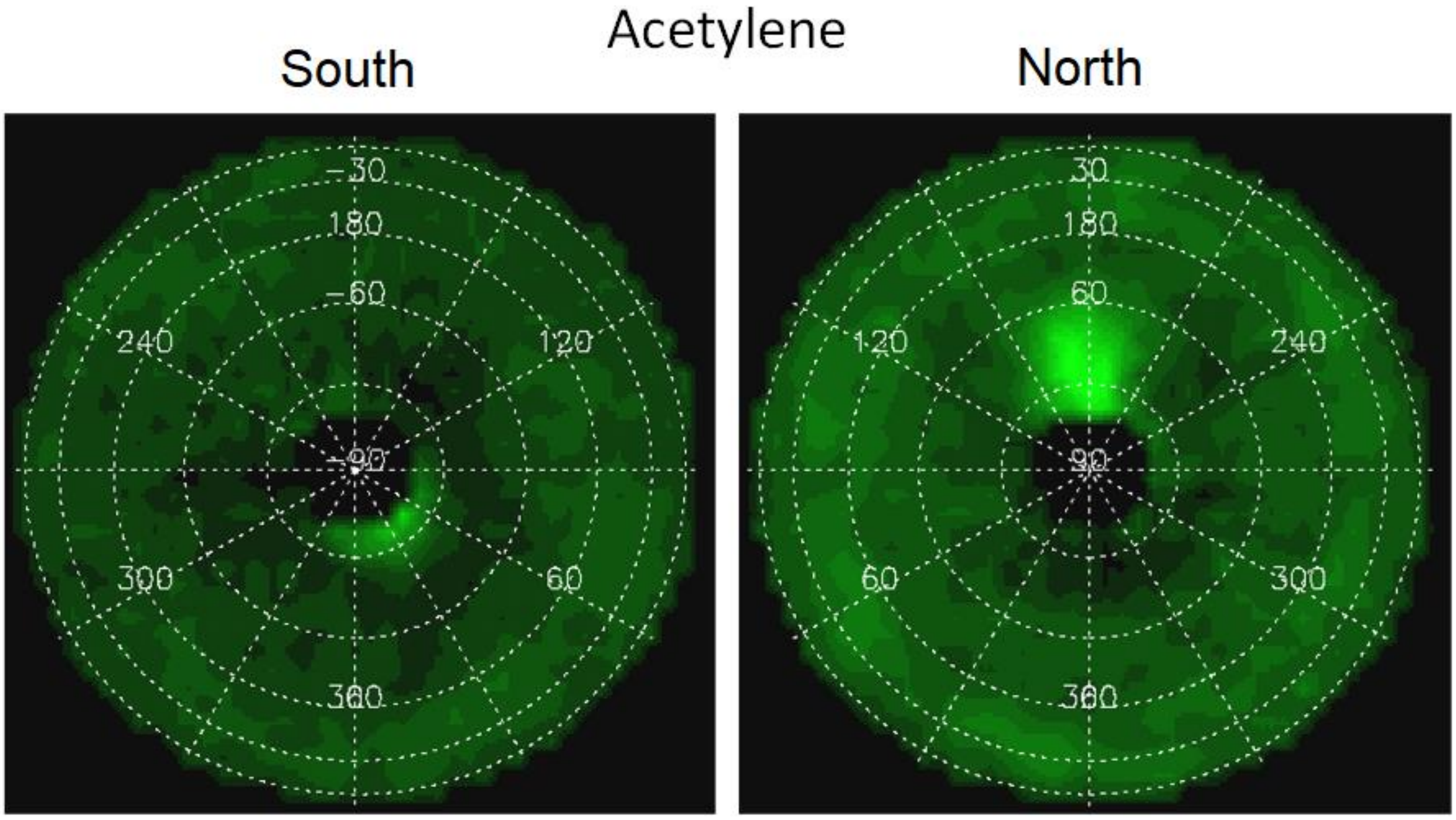

**Figure 2.** Distribution of acetylene at the north and south poles of Jupiter based on data obtained by the CIRS instrument during the Cassini-Huygens mission. Dec. 2000 – Jan. 2001. Reproduced with permission from ref 75. Copyright 2010 NASA/JPL/GSFC.

Acetylene was detected on other planets, the Uranus and Neptune. In 1986, Broadfoot et al.[47] measured the acetylene mixing ratio in Uranus as $2{\cdot}10^{-7}$ using data from a Voyager Ultraviolet Spectrometer. This discovery confirmed the findings of Encrenaz et al.[46] in 1986 based on International Ultraviolet Explorer (IUE) satellite observations. $C_2H_2$ is the dominant constituent above the 1 mbar level.[79] Moreover, in 1987, Orton et al.[80] reported $C_2H_2$ IR emission at 13.7 µm obtained with NASA IRTF, which is consistent with the maximum mixing ratio of $C_2H_2$ being equal to $9{\cdot}10^{-9}$. In the same work, Orton et al. reported recording a Neptune spectrum at approximately 13.7 µm, which is consistent with the emission of stratospheric acetylene with a maximum mixing ratio of $9 \cdot 10^{-7}$. Caldwell et al. in 1988, based on IUE data, independently reported estimates of the acetylene mixing ratio, equal to $4{\cdot}10^{-9}$, which is approximately 2 orders less than the estimates from the IR spectra.[81] In 1989, Voyager 2 reached the Neptune system, and its atmosphere was examined using an infrared spectrometer (IRIS).[82] Acetylene was detected with a mixing ratio of $10^{-7}$ to $10^{-6}$ in the atmosphere (approximately 1.5 mbar). For Neptune, Saturn and Jupiter, acetylene is the third most abundant hydrocarbon in the atmosphere after $CH_4$ and $C_2H_6$.[83]

Since the first detections, the contributions of UV radiation and thunderstorms to the origin of acetylene in the atmosphere of gas giants have been actively discussed.[84] To date, the formation of acetylene has been associated with the photodissociation of methane in the upper stratosphere by UV radiation and the production of radicals that recombine in hydrocarbons with two or more carbon atoms.[85,86,87] The photolysis of methane with the formation of acetylene occurs in the wavelength range of 120-190 nm.[88] According to Bar-Nun[84], incident solar radiation up to 140 nm is absorbed by methane in the atmosphere down to ~$5 \cdot 10^{-4}$ bar; at 145 nm, approximately 40% of the radiation is absorbed at $10^{-2}$ bar.

It should be noted that the degree of branching of $CH_4$ during UV photodissociation is still debatable. Thus, Romani et al.[89], based on the results of numerical modeling of methane photochemistry in the Neptune stratosphere, showed the photolytic formation of excited methylene $^1CH_2$, ground-state methylene $^3CH_2$, and CH radicals but not the formation of methyl radical $CH_3$ as a result of $CH_4$ photolysis. In Romani's scheme, the formation of acetylene occurs through the addition of a second methane molecule to the CH radical (the product of photolysis of $CH_4$) to form an ethylene molecule and subsequent photolysis. Another pathway to acetylene is the photolysis of ethane. In turn, ethane is formed as a result of the recombination of methyl radicals formed from methylene or methylidene radicals (Scheme 1A).[90]

Moreover, experimental studies have shown the possibility of the formation of a methyl radical as a result of the photolysis of methane when the sample is irradiated at a vacuum UV (VUV) wavelength of approximately 120 nm.[91,92] Mordaunt et al.[93] also pointed to the formation of $CH_3$ (and CH) fragments rather than methylene radicals from the photodissociation of $CH_4$ at 121,6 nm and the importance of this pathway for the chemistry of the atmospheres of the outer planets. As a result, the formation of methyl groups occurs due to the photolysis of methane, subsequent recombination of methyl radicals to form ethane or a methyl radical with a methylene radical to form ethylene, and subsequent photolysis of ethane or ethylene may constitute alternative pathways for the formation of acetylene (Scheme 1B).[94]

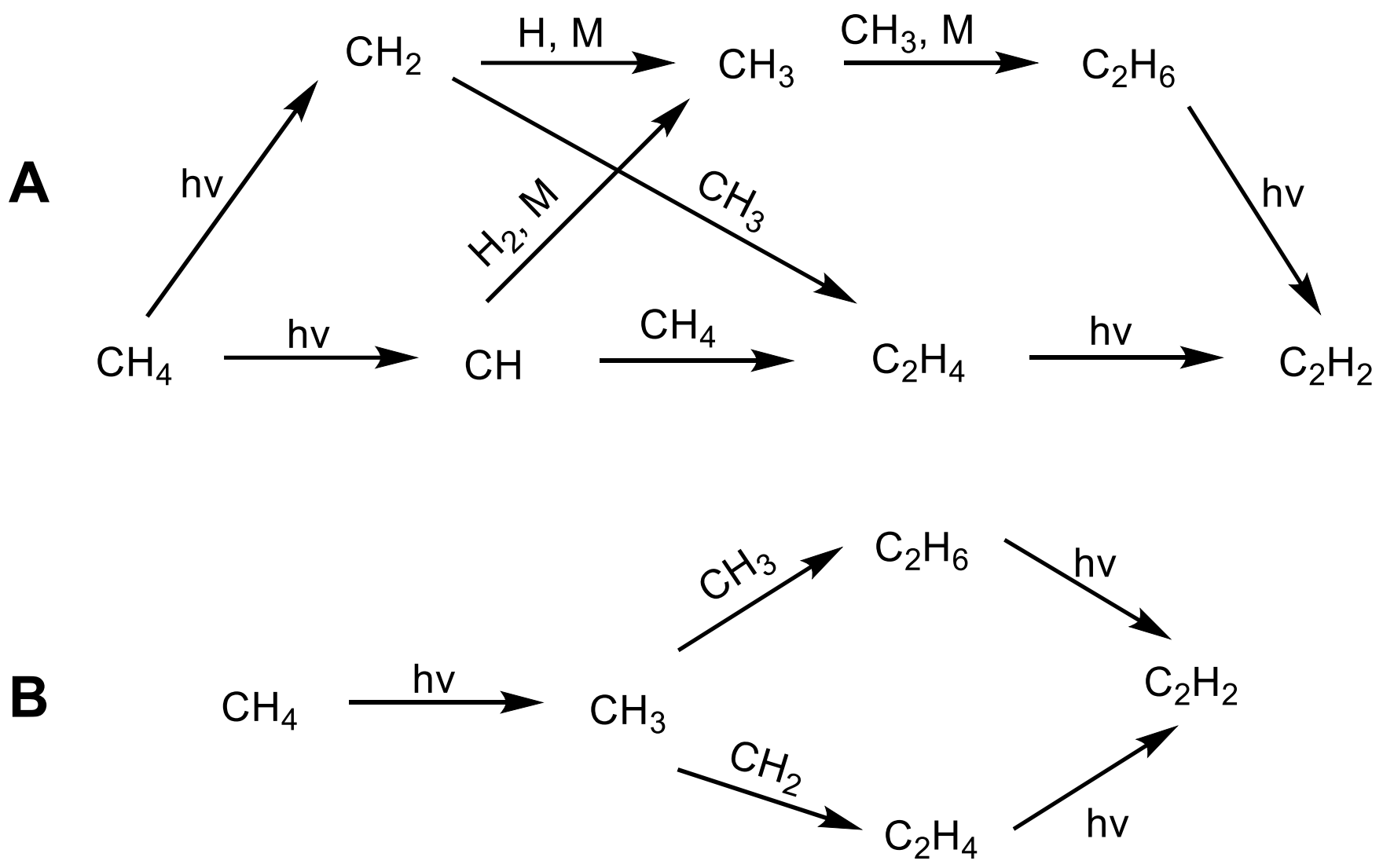

**Scheme 1.** Acetylene formation reactions during methane photolysis. Pathways based on Romani model adapted from the work 89 (A). Pathways through methyl radical based on the work 94 (B).

Thunderstorms in the clouds of Jupiter and Saturn are assumed to be an additional source of acetylene.[84,95,96] Thunderstorms create extreme conditions by releasing energy from lightning strikes. The electric discharge of lightning causes intense heating of the gas along the current flow up to 30000 K,[97] which causes thermal dissociation of atmospheric species. In addition, the shock wave resulting from the sudden thermal expansion of the gas has an additional effect. These factors promote the dissociation of molecules into radicals. In the case of Jovian atmospheres, $CH_x$ radicals can combine to form the corresponding compounds. Bar-Nun's calculation[84] showed that the lightning energy flux on Jupiter is 7.5 $erg \cdot cm^{-2} \cdot sec^{-1}$, and the efficiency of converting heat into lightning energy on Jupiter is $5 \cdot 10^{-4}$. Furthermore, based on experimental inventions[98], Bar-Nun noted that a thunderstorm column 5 km in length and 10 cm in diameter at a pressure of 7 bars converted 2.5 moles of methane (34%) to acetylene. As a result, Bar-Nun concluded that 245 strokes per $km^2$ per year will result in $7.8 \cdot 10^{-16}$ $mol \cdot cm^{-2} \cdot sec^{-1}$ acetylene, which is comparable in scale to the photolytic acetylene formation pathway. However, based on high-spectral-resolution Cassini/CIRS data, Hurley et al.[99] reported that there was no correlation between the content of upper tropospheric acetylene and thunderstorms 2000 km in size on Saturn.  The main reactions described in this section are summarized in Table 2.

**Table 2.** Reactions of acetylene formation in Jovian planets, their kinetics and conditions where the reactions are relevant.

| Reactions | Kinetics | Objects & conditions |
|---|---|---|
| 1.<br>a. $CH_4 + h\nu \rightarrow CH + H + H_2$<br>b. $CH + CH_4 \rightarrow C_2H_4 + H$<br>c. $C_2H_4 + h\nu \rightarrow C_2H_2 + 2H$[89] | $k_{1a} = 2.4 \cdot 10^{-10}$ s$^{-1}$<br>($G_0$=1, $A_V$=0) (KIDA[100])<br>$k_{1b} = 2.3 \cdot 10^{-10}$ s$^{-1}$ (T = 100 K)<br>$k_{1c} = 3 \cdot 10^{-9}$ s$^{-1}$ ($G_0$ = 1, $A_V$=0) | Atmospheres of Jovian planets. Conditions:<br>$n_H = 10^{10}$-$10^{16}$ cm$^{-3}$<br>T = 50-300 K<br>$G_0$ = up to $10^2$ |
| 2.<br>a. $CH_4 + h\nu \rightarrow CH_3 + H$[86]<br>b. $CH_3 + CH_3 \rightarrow C_2H_6$<br>c. $C_2H_6 + h\nu \rightarrow C_2H_4 + H_2$<br>d. $C_2H_4 + h\nu \rightarrow C_2H_2 + 2H$ | $k_{2a} = 2 \cdot 10^{-8}$ s$^{-1}$ (p = $10^{-8}$ mbar) – $3.6 \cdot 10^{-11}$ s$^{-1}$ (p = $10^{-3}$ mbar)[86]<br>$k_{2b} \approx 10^{-24}$ cm$^6$ s$^{-1}$ (T = 150 K, $n_H = 10^{12}$ cm$^{-3}$)<br>$k_{2c} = 2.5 \cdot 10^{-9}$ s$^{-1}$ (p = $10^{-8}$ mbar) – $1.1 \cdot 10^{-10}$ s$^{-1}$ (p = $10^{-3}$ mbar)<br>$k_{2d} = 1.1 \cdot 10^{-7}$ s$^{-1}$ (p = $10^{-8}$ mbar) - $6.6 \cdot 10^{-8}$ s$^{-1}$ (p = $10^{-3}$ mbar) | |

## 2.2. Satellites and Pluto

In addition to giant planets, acetylene has also been found on other objects in the Solar system. Saturn's moon Titan has been particularly well studied. Unlike giant planets, Titan's atmosphere consists mostly of nitrogen, but hydrocarbons are also present. Voyager 1 flew by Titan in 1980 and probed its atmosphere with IRIS, registering the $\nu^5$ band at 13.7 μm.[101,102] The acetylene mixing ratio was estimated to be $2 \cdot 10^{-6}$ relative to $N_2$ (for Titan's atmosphere) at an altitude of 130 km. Titan's atmosphere was later explored by the Cassini spacecraft.[103] Using CIRS mid-IR spectra, the mixing ratio in the stratosphere was found to be 3-4 $\cdot 10^{-6}$.[104,105] The Ion and Neutral Mass Spectrometer (INMS) data of the Cassini spacecraft showed that the acetylene mixing ratio in the thermosphere was $3.1 \pm 1.1 \cdot 10^{-4}$.[105,106,107] Based on these studies, acetylene was found to be the third most abundant hydrocarbon in atmosphere of Titan (after $CH_4$ and $C_2H_6$).[108] In addition, acetylene is the most abundant unsaturated hydrocarbon in Titan's atmosphere. Similarly, as in the case of the Jovian planets, acetylene is formed as a result of the photolysis of methane in the case of Titan.[109]

Using VIMS Cassini, acetylene was also found in solid form in three equatorial regions of Titan: Tui Regio, eastern Shangri-La, and Fensal–Aztlan/Quivira.[110] Additionally, with the use of a gas chromatography mass spectrometer (GCMS) of Huygens probe, a trace amount of $C_2H_2$ gas sublimated/evaporated from the surface was detected.[111] Another moon on which acetylene was detected in Saturn is Enceladus. Cassini's INMS recorded the presence of acetylene while passing at an altitude 168 km from the surface of the satellite in 2005.[32]

Finally, the recent New Horizons[112] mission in 2015 provided information on the composition of Pluto's atmosphere and allowed modeling of photochemical processes.[113] The atmosphere of Pluto is very rarefied and consists of gases, mainly nitrogen with an admixture of methane (approximately 0.25%). Moreover, the data from New Horizons's UV spectrograph Alice indicate the absorption signatures of the products of the photochemical transformations of methane to ethylene and acetylene.[114] Photochemical models predict that most of the precipitation on Pluto is formed by acetylene.[115]

Notably, there are near-Earth objects where acetylene was not detected, despite the presence and diversity of organic matter. For example, the Hayabusa2 mission examined the asteroid 162173 Ryugu and returned samples from its surface to Earth in 2020.[116,117] Analysis of soluble organics revealed the presence of PAHs, heterocyclic compounds, carboxylic acids, amino acids, amines, etc.[118] In addition to the absence of acetylene, the abundance of methane and ethane in the gas samples was low.[119] The reactions of acetylene formation in satellites are similar to those for the Jovian planets given in Table 2.

## 2.3. Comets

Comets are another important class of objects in the Solar system in which acetylene was discovered. The first comet in which acetylene was detected was comet C/1996 B2 (Hyakutake). The detection was performed in 1996 with the NASA IRTF.[120] The abundance of acetylene was estimated to be 0.3–0.9% relative to water, the dominant molecule in cometary ice.[121] Subsequently, acetylene has been detected on many other comets using a cryogenic echelle spectrometer (CSHELL) at NASA IRTF and via IR spectroscopy with NIRSPEC at the Keck Observatory on Mauna Kea[122] in amounts up to 0.5% relative to water. In addition, the presence of acetylene in the comets Halley and Churyumov-Gerasimenko was confirmed by the data received by the space missions Giotto[123] in 1999 and Rosetta[124] in 2015, respectively.

The Rosetta mission revealed a high acetylene abundance compared to that of other Jupiter family comets. Moreover, according to data from the Rosetta Orbiter Sensor for Ion and Neutral Analysis (ROSINA), which analyzed the chemical composition of cometary coma, the abundance of $C_2H_2$ in the summer hemisphere was 0.045% relative to that in water, while in the winter hemisphere, it reached 0.55%. Léna Le Roy et al.[124] noted that the abundance of ethane is also much greater than that of any other comet. One of the possible pathways for the formation of ethane in comets may involve the passage of acetylene, which is produced in the gas phase, condenses into icy grains, and hydrogenates into dense interstellar clouds.[125] The amount of acetylene in comets is comparable to its amount in interstellar ices; therefore, this acetylene has an interstellar origin. It might be formed through gas phase ion-molecular or neutral-neutral reactions and then adsorbs onto the icy mantle of dust grains.[125,120] We refer readers to sections 2.6 and 2.7, where such reactions are discussed.

## 2.4. Exoplanets

In the last decade, research on the chemical composition of discovered exoplanets has been actively conducted. Acetylene is considered an essential building block for prebiotic systems and may indicate the possibility of abiogenetic processes occurring on planets.[29,42] Therefore, acetylene is one of the molecules identified in the spectra of exoplanet atmospheres.[126] Tsiaras et al. (2016) suggested that acetylene may be present in the atmosphere of super-Earth 55 Cancri.[127] The prediction of the presence of acetylene in the atmosphere of exoplanets was based on data obtained by modeling and data from IR telescopes (see SI Table S1). However, the lack of spectral data over a broad range and at high temperature did not allow us to firmly establish the presence of acetylene. Despite the difficulties in detecting acetylene on exoplanets, the probability of the presence of acetylene molecules on super-Earth exoplanets is estimated to be high.[126] Tennyson and Yurchenko noted[26] that acetylene is probably the most important polyatomic molecule for exoplanet and cool star research, for which there is still no comprehensive hot line list. For a polyatomic molecule, the size of the acetylene and the transitions between rovibrational levels at exoplanet and cool star temperatures give rise to spectra typically spanning the IR region. The problem with acetylene is due to its linear geometry in the equilibrium structure. The rovibrational spectra of linear acetylene requires special consideration when calculating the rovibrational energies. Nevertheless, computational processing is feasible: an extensive effective Hamiltonian fit is available, and the calculations essentially require solving the Schrödinger equation of nuclear motion with some approximations.[128]

To facilitate the detection of important molecular species, such as acetylene, in atmospheres of exoplanets and cool stars, the ExoMol spectroscopic database was created.[26,27] Molecules suggested to be important for exoplanet spectroscopy are included in the ExoMol molecular line list. Spectral lines of molecules can indicate the type of exoplanet. Spectroscopy of hot gas giants is very different from that of hot rocky objects and cooler objects. The search for acetylene signals on exoplanets continues.[127,129,130,131,132] There is no information on formation routes; for some exoplanets, routes can be the same as for Solar system planets. Although there is no information on general formation routes for exoplanets, in some cases, the routes can be the same as those for Solar system planets.

## 2.5. Carbon-rich stars

Acetylene was first discovered outside the Solar system in 1976 in the IR spectrum of IRC +10216 with the 4-m Mayall telescope at the Kitt Peak observatory.[133] The detection was based on the band at ≈2.4 μm, which is a combination of $\nu^1$ and $\nu^5$ modes. The object IRC +10216 is a well-studied carbon AGB star that is embedded in a thick dust envelope. The detection of acetylene was important, as acetylene is thought to be a precursor of dust. This star was observed in the IR range repeatedly afterwards. For example, Fonfria et al.[54] performed a detailed analysis of the high-resolution spectra at 11-14 μm and identified many acetylene bands corresponding to the modes $\nu^2$, $\nu^4$, $\nu^5$ and their combinations, although many modes are not observed properly. The abundance of $C_2H_2$ in the envelope of IRC +10216 obtained from the observations is $3 \cdot 10^{-8} – 1 \cdot 10^{-7}$ relative to that of $H_2$.[134,135] Subsequently, with a great contribution from space IR missions, the Infrared Space Observatory (ISO)[136] and Spitzer Space Telescope[137], the presence of acetylene was also established by the IR spectra in other carbon stars[138,139] both in the Milky Way and other galaxies[140,141,142] and in the protoplanetary nebula CRL 618[143].

Acetylene is one of the most abundant molecules after CO in envelopes around C-rich stars and has a carbon-to-oxygen ratio (C/O) greater than 1.[140] Below (Scheme C and Table 3), the chain of reactions leading to the formation of $C_2H_2$ is presented, although other chains of neutral-neutral reactions are also possible.[144,145] First, $C_2$ is formed in the reaction between atomic C and radical CH (Scheme 2A). Then, an ethynyl radical ($C_2H$) is formed from the $C_2$ chain and a hydrogen molecule just after the shock wave front (Scheme 2B). Finally, acetylene appears as a result of an exchange reaction with its radical (Scheme 2C).

**A** $C + CH^{\cdot} \longrightarrow C_2 + H^{\cdot}$

**B** $C_2 + H_2 \longrightarrow C_2H^{\cdot} + H^{\cdot}$

**C** $C_2H^{\cdot} + H_2 \longrightarrow C_2H_2 + H^{\cdot}$

**Scheme 2.** Reactions of acetylene formation.

**Table 3.** Chain of reactions leading to acetylene formation in carbon-rich stellar envelopes.

| **Reactions** | **Kinetics** | **Objects & conditions** |
|---|---|---|
| a. $C + CH\cdot \rightarrow C_2 + H\cdot$<br>b. $C_2 + H_2 \rightarrow C_2H^{\cdot} + H\cdot$<br>c. $C_2H\cdot + H_2 \rightarrow C_2H_2 + H\cdot$[144] | $k_a = 2.4 \cdot 10^{-10}$ cm³ s⁻¹ (KIDA)<br>$k_b = 8 \cdot 10^{-11}$ $cm^3\ s^{-1}$ (T = 2000 K) (KIDA)<br>$k_c = 10^{-10}$ $cm^3\ s^{-1}$ (UMIST)[146] | Stellar envelopes<br>$n_H = 10^8$-$10^{15}$ $cm^{-3}$<br>T = 500-2500 K |

## 2.6. Cold interstellar medium

In cold environments, acetylene can be locked in icy mantles on dust grain surfaces. Astrochemical models predict that the abundance of acetylene is approximately $10^{-3}$ relative to that of water in ice.[147,148] The detection of acetylene in solid form is difficult because acetylene has weak features that are blended with much stronger features of $H_2O$.[149] At the same time, acetylene was indicated in the gas phase by its vibrational modes in a number of star-forming regions and young stellar objects.[150,151] The observed abundances are relatively high, e.g., a few of $10^{-8}$ relative to $H_2$ in the star-forming region Cepheus A East in the work of Sonnentrucker et al.[150] or ~$10^{-7}$ relative to $H_2$ in young stellar objects in the work of Lahius et al.[151] Gas-phase chemical reactions provide lower abundances of acetylene – from $10^{-10}$ to $10^{-8}$ relative to $H_2$.[152] Therefore, it is assumed that the observed gas-phase acetylene is a result of sublimation of interstellar ice. Sublimation can occur via thermal or nonthermal desorption. In the case of young stellar objects where the temperature exceeds 300 K[151], thermal desorption is efficient, while in the case of colder objects, desorption can be a result of sputtering caused by shocks.[150]

In the gas phase, acetylene can be formed through neutral-neutral and ion-molecular reactions. For example, acetylene can be generated through electron recombination of $C_2H_3^+$, which in turn results in a reaction of methane and $C^+$.[153] Acetylene cations can be generated directly through the same reaction between methane and $C^+$.[153] Alternative pathways of $C_2H_2$ formation occur on dust ice surfaces. Based on experiments with hydrocarbons in water ice, Hudson & Moore noted that there is no evidence for $C_2H_2$ formation in ice and suggested that $C_2H_2$ is part of the natal ice composition.[154] However, the observed abundance is often greater than that predicted by reaction network models.[155,156] In 2023 in experimental work using the Reaction Apparatus for Surface Characteristic Analysis at Low Temperature (RASCAL) accompanied by temperature-programmed desorption (TPD) and a combination of photostimulated desorption and resonance-enhanced multiphoton ionization (PSD-REMPI) methods. Tsuge et. al. [157] reported that C atoms are weakly bound to the surface of ice and can diffuse on the surface of amorphous solid water, which is the major component of the icy mantle of cosmic dust in cold regions, at temperatures above approximately 30 K. It was determined that the diffusion reaction of C atoms is activated at approximately 22 K on interstellar ice. The authors were able to demonstrate the formation of $C_2$ molecules. TPD measurements showed an increase in the $C_2H_6$ signal (m/z 28) in the temperature range from 70 to 90 K for the water ice sample with carbon atoms deposited at 10 K followed by exposure to H atoms. The formation of stable methane ($CH_4$) or ethane ($C_2H_6$) molecules through successive hydrogenation reactions of species C and $C_2$ can be detected by the TPD method. Importantly, the reactions involving the sequential addition of hydrogen atoms to $C_2$ and $C_2H$ species to form acetylene $C_2H_2$ are barrier-free. It was also previously shown that the hydrogenation of acetylene to ethylene and ethane can successfully occur in water ice at 10 and 20 K even in the environment of a dark molecular cloud.[158] Moreover, the authors[157] noted that the contribution of the interaction of two $CH_3$ species to the formation of $C_2H_6$ should be insignificant in the TPD process. Under the experimental conditions during hydrogenation, $CH_3$ radicals are immediately converted to methane $CH_4$ due to the barrier-free reaction of $CH_3$ and H; thus, they do not remain on the surface of water ice.

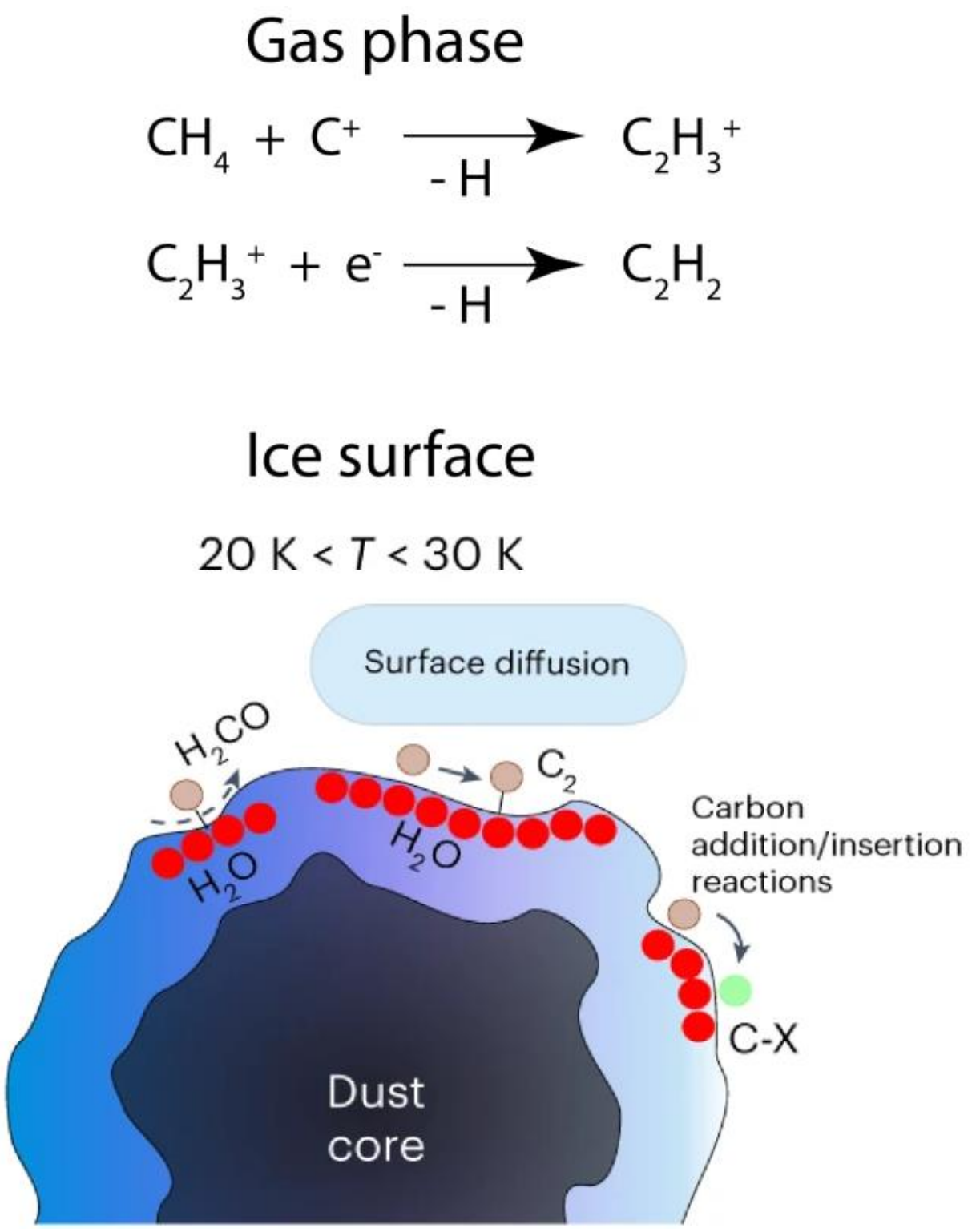

**Figure 3.** Chemical processes in the gas phase and on the grains. At $20\,K < T < 30\,K$ physisorbed C atoms can diffuse, promoting insertion reactions with other C atoms with formation $C_2$ species. Reproduced with permission from ref 157. Copyright 2023 Springer Nature.

Laboratory studies[159,160] of the irradiation of methane ice at 10 K energetic electrons as an imitation of processes caused by cosmic rays have shown that the recombination of the resulting methyl radicals, as well as radiolysis with the elimination of atomic hydrogen, can lead to the formation of acetylene. This process can explain the synthesis of acetylene in interstellar ice and comets and in aerosol particles and organic haze layers of hydrocarbon-rich atmospheres of planets and their satellites, such as Titan, which was mentioned above. This pathway is illustrated in Figure 4.

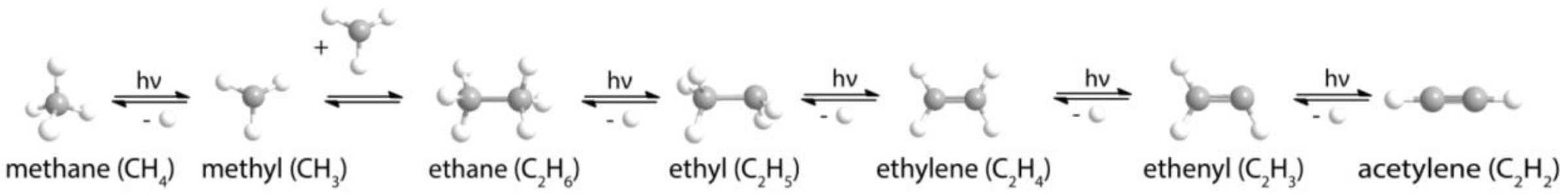

**Figure 4.** Reaction pathway for the CR- and photoinitiated formation of acetylene from methane.

In addition, Feldman et al. showed[161] that the radiolysis of ethylene and ethane at 5 K in noble gas matrices under the influence of X-ray radiation at 20 keV leads to the formation of $C_2H_2$ molecules as a result of progressive dehydrogenation. The widespread occurrence of acetylene in harsh space environments with high levels of radiation is due not only to the ease of its formation but also to the high stability of the acetylene molecule and its radiation resistance in the series of $C_2$ hydrocarbons (ethane < ethylene < acetylene). We summarize the reactions considered in this section in Table 4.

**Table 4.** The reactions of acetylene formation in cold interstellar medium.

| **Reactions** | **Kinetics** | **Objects & conditions** |
|---|---|---|
| 1. Ion-molecular gas phase reactions[162]<br>a. $CH_4 + C^+ \rightarrow C_2H_3^+ + H$<br>b. $C_2H_3^+ + e^- \rightarrow C_2H_2 + H$<br>2. $CH_4 + C^+ \rightarrow C_2H_2^+ + H_2$ | $k_{1a} = 10^{-9}$ cm$^3$ s$^{-1}$ (UMIST)<br>$k_{1b} = 3 \cdot 10^{-7}$ cm$^3$ s$^{-1}$ (T = 100 K) (UMIST)<br>$k_2 = 3.89 \cdot 10^{-10}$ cm$^3$ s$^{-1}$ (UMIST) | DISM, molecular clouds.<br>Conditions:<br>T = 10 - 500 K<br>$n_H = 10^{-1} - 10^7$ cm$^{-3}$<br>$G_0$ = from very low to $10^2$<br>CR field: Galactic |
| 3. Reactions in methane ice[163]<br>a. $2CH_4 \rightarrow C_2H_6^{\cdot} + 2H/H_2$<br>b. $C_2H_6^{\cdot} \rightarrow C_2H_4 + 2H/H_2$<br>c. $C_2H_4^{\cdot} \rightarrow C_2H_2 + 2H/H_2$ | $k_{3a} = 6.14 \cdot 10^{-5}$ s$^{-1}$<br>$k_{3b} = 3.18 \cdot 10^{-3}$ s$^{-1}$<br>$k_{3c} = 1.22 \cdot 10^{-3}$ s$^{-1}$ | Ice on dust grains in molecular clouds, in comets, aerosol particles and organic haze layers of planets and their satellites.<br>Conditions:<br>T = 5-100 K<br>CR field: Galactic<br>$G_0$ from very low to ~$10^2$ |
| 4. Diffuse chemistry on ice surface[157]<br>a. $C + C \rightarrow C_2$<br>b. $C_2 + H \rightarrow C_2H$<br>c. $C_2H + H \rightarrow C_2H_2$ | n.d. | |

## 2.7 Photo-dissociation regions and diffuse interstellar medium

Acetylene is not detected in PDRs or DISM; moreover, it is highly sensitive to UV radiation.[38] Nevertheless, some amount of acetylene may be present in the PDRs and participate in the reactions leading to the different small hydrocarbons $C_mH_n$, which were detected.[195,193] Several pathways for the formation of acetylene and acetylene cations at low temperatures were shown above in Section 2.6. As shown above, acetylene can be formed through ion-molecular reactions, namely, through electron recombination of $C_2H_3^+$. At temperatures less than ~300 K, $C_2H_3^+$ can be generated as a result of the reactions 1,2 shown in Table 4.[153] At higher temperatures, the reactions with excited $H_2$ can be activated as shown in Scheme 3[145,164] and Table 5.

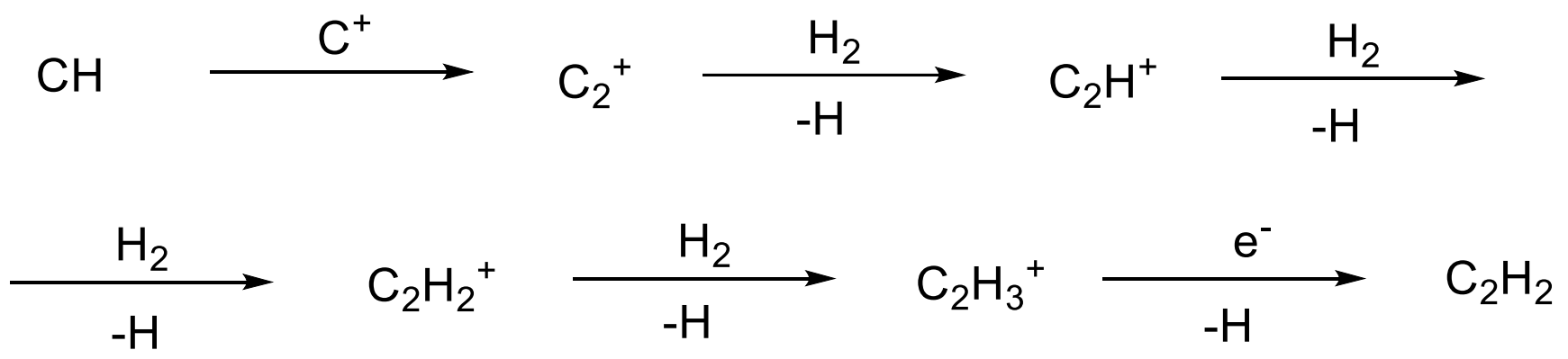

**Scheme 3.** Possible routes of acetylene formation in PDRs and DISM.

On the other hand, the acetylene molecule can be obtained from a top-down mechanism. This molecule is the main carbonaceous dissociation channel for PAHs[165,166,167]. This channel becomes even more important if PAHs are hydrogenated[168,169] while hydrogenated PAHs are likely present in the ISM.[170] In PDRs, PAHs absorb UV photons with energies up to 13.6 eV. This energy efficiently converts to PAH internal energy,[171] which becomes higher than the dissociation energy of approximately 4-5~eV.[172] Although the dissociation process competes with ionization and IR relaxation, its probability was estimated to be high enough that small PAHs would be destroyed in the ISM.[165] The minimum size of PAHs that survive in the ISM varies from 15 to 50 depending on the calculation method and parameters.[165,173,174] This size is greater under harsh UV radiation from PDRs; hereby, small PAHs are destroyed much quicker than in the diffuse ISM, i.e., this origin of acetylene is time limited. [175] Larger PAHs can live longer than smaller PAHs, but their dissociation probability is lower; hence, large PAHs should not be considered significant contributors to acetylene. The destruction of PAH clusters likely compensates for the abundance of small PAHs, as they seem to be destroyed in PDRs[176]. However, calculations of the impact of PAH cluster destruction have not yet been performed.

The grains from hydrogenated amorphous carbon (HAC) are larger than those from PAHs, and they are not planar. However, irradiation may also lead to the release of acetylene, as shown by Alata et al. (2015). Moreover, they lose other small hydrocarbons ($C_2H_4$, $C_3H_4$, $C_3H_6$, $C_4H_4$, etc.[177,178]), whole PAHs and even fullerenes.[179] However, these experimental results should be used with caution because the hydrogenation level of HAC grains in the ISM should be much lower than that in laboratory samples.[180,181] Enrichment of the ISM by small hydrocarbons ceases quickly without reverse hydrogenation because hydrogen loss is a much more favorable dissociation channel, while dehydrogenated HAC grains become stable during UV irradiation.[182] Table 5 summarizes the reactions of acetylene formation considered in this section.

**Table 5.** The reactions of the acetylene formation in PDRs and DISM.

| **Reactions** | **Kinetics** | **Objects & conditions** |
|---|---|---|
| 1. Ion-molecular reactions with excited $H_2$[145, 183]<br>a. $C^+ + CH \rightarrow C_2^+ + H$<br>b. $C_2^+ + H_2 \rightarrow C_2H^+ + H$<br>c. $C_2H^+ + H_2 \rightarrow C_2H_2^+ + H$<br>d. $H_2 + C_2H_2^+ \rightarrow C_2H_3^+ + H$<br>e. $C_2H_3^+ + e^- \rightarrow C_2H_2 + H$ | $k_{1a} = 3.8 \cdot 10^{-10}$ cm$^3$ s$^{-1}$ (UMIST)<br>$k_{1b} = 1.1 \cdot 10^{-9}$ cm$^3$ s$^{-1}$ (UMIST)<br>$k_{1c} = 1.1 \cdot 10^{-9}$ cm$^3$ s$^{-1}$ (UMIST)<br>$k_{1d} = 1.1 \cdot 10^{-11}$ cm$^3$ s$^{-1}$ (UMIST)<br>$k_{1e} = 3 \cdot 10^{-7}$ cm$^3$ s$^{-1}$ (T=100 K) (UMIST) | The reactions may take place in many objects (PDRs, DISM, molecular clouds, protoplanetary disks) where ions can be produced by UV radiation or by CRs.<br>Conditions:<br>T = 10 - 500 K<br>$n_H = 10^{-1} - 10^7$ cm$^{-3}$<br>$G_0$ = from very low to $10^5$<br>CR field: Galactic |
| 2. Ion-molecular reactions<br>a. $C^+ + H_2 \rightarrow CH_2 + h\nu$<br>b. $CH_2^+ + H_2 \rightarrow CH_3^+ + H$<br>c. $CH_3^+ + e \rightarrow CH_2 + H$<br>d. $CH_2 + C^+ \rightarrow C_2H^+ + H$<br>e. $C_2H^+ + H \rightarrow C_2H_2^+ + H$[184] | $k_{2a} = 3 \cdot 10^{-16}$ cm$^3$ s$^{-1}$ (T = 100 K) (KIDA)<br>$k_{2b} = 1.2 \cdot 10^{-9}$ cm$^3$ s$^{-1}$ (KIDA)<br>$k_{2c} = 5 \cdot 10^{-7}$ cm$^3$ s$^{-1}$ (KIDA)<br>$k_{2d} = 10^{-9}$ cm$^3$ s$^{-1}$ (100 K) (KIDA)<br>$k_{2e} = 1.1 \cdot 10^{-9}$ cm$^3$ s$^{-1}$ (KIDA) | |
| 3. Top-down chemistry:<br>PAH/HAC $\rightarrow$ n($C_2H_2$) | Rates are sensitive to a parent particle (type, size, charge, C/H) and $G_0$, $n_H$, T[175,185]<br>For a $PAH^0$ with $N_C = 50$ and $G_0 = 1$: $k = 10^{-15}$ s$^{-1}$ | |

## 2.8 Young stellar objects

$C_2H_2$ was also detected in young stellar objects and on protoplanetary disks via rovibrational modes.[151, 186,187] The abundance of $C_2H_2$ was found to be $10^{-8}$-$10^{-6}$ relative to $H_2$, and this value increases as the temperature increases, which is consistent with the hypothesis that at low temperature, $C_2H_2$ is locked in ice and appears in gas due to ice sublimation when a young star heats the surrounding gas.[151]

The gas phase abundance is normally consistent with the estimates of its abundance in ice.[151] Recently, $C_2H_2$ in protoplanetary disks was observed with the James Webb Telescope.[188] While the ratio between the abundances of $C_2H_2$ and $CO_2$ is on the order of unity in three objects, it is $10^4$ in disk J 160532.[189] This high abundance cannot be explained by evaporation from ice. Van Dischoek et al.[190] suggested that the enhanced abundance is a result of thermal and/or photodestruction of carbonaceous grains, whether they are PAHs or HACs, similar to what occurs in PDRs (Section 2.7).

The above examples demonstrate the presence of acetylene in a variety of conditions throughout the Universe, both in objects inside the Solar System, in the atmosphere and on the surface of planets and satellites, and beyond, in stellar and interstellar objects. Acetylene molecules can be found in different states and forms, from the cold ice of comets to the hot gases in the circumstellar space. The variety of states and conditions determines the variety of chemical processes. The possibility of acetylene formation along the bottom-up and top-down pathways provides greater dynamism of organic chemical environments and the possibility of the formation of new structures when conditions change.

The wide distribution of acetylene makes it a universal molecule, the activity of which is capable of providing a wide network of chemical transformations of both aliphatic organic molecules and large aromatic particles, “pure” hydrocarbon structures and compounds with oxygen, nitrogen, etc. Thus, acetylene can determine and influence the chemistry of various space environments, promoting chemical evolution in space.

# 3. Acetylene in the network of chemical reactions in space

Acetylene plays an important role in the chemistry of the ISM and planetary systems, including the Solar system. It is a key reactant in the formation of a great number of molecules. The widely used astrochemistry network of chemical reactions, KIDA (Kinetic Database for Astrochemistry),[191] includes approximately 9 thousand reactions, approximately 5 hundreds of which are reactions with acetylene. This molecule participates in many different reactions, including general reactions such as charge exchange, absorption of photons or cosmic rays, and specific reactions involving the direct formation of organic molecules from small hydrocarbons to complex large molecules as a parent molecule or a transitional product. While some products of the reactions with acetylene are observed, e.g., small hydrocarbons, and consequently, chemical models can be checked, the detection of complex organic molecules (COMs) is more complicated, and the progression of the corresponding reactions remains questionable. In this section, we consider the reactions with acetylene resulting in nonaromatic products, and the next sections will focus on the network of reactions related to aromatic molecules and dust.

## 3.1. Acetylene as a precursor of small hydrocarbons

Small hydrocarbons ($C_2H$, $C_4H$, c-$C_3H_2$, l-$C_3H^+$, etc.) are observed in various space objects.[192,193,194] Several pathways lead to their appearance, but the pathways mainly start from acetylene. Acetylene can react with CH via a neutral-neutral reaction leading to $C_3H_2$ (we omit c- and l-) (Scheme 4A) or participate in a series of ion-molecular reactions that end in $C_3H_2$ (Scheme 4B). Other examples, namely, the formation of $C_4H_3^+$, $C_4H_2^+$, etc., are given in Table 6.

However, the presented chemical routes do not provide enough hydrocarbons, which are found in some PDRs.[193,195,196] To explain the high abundance of these hydrocarbons, reactions with excited hydrogen molecules have been suggested.[183] Hydrogen molecules can be excited if the temperature exceeds a few hundred Kelvin. In particular, a chain of such reactions starting from acetylene can be described as shown in Scheme 4.

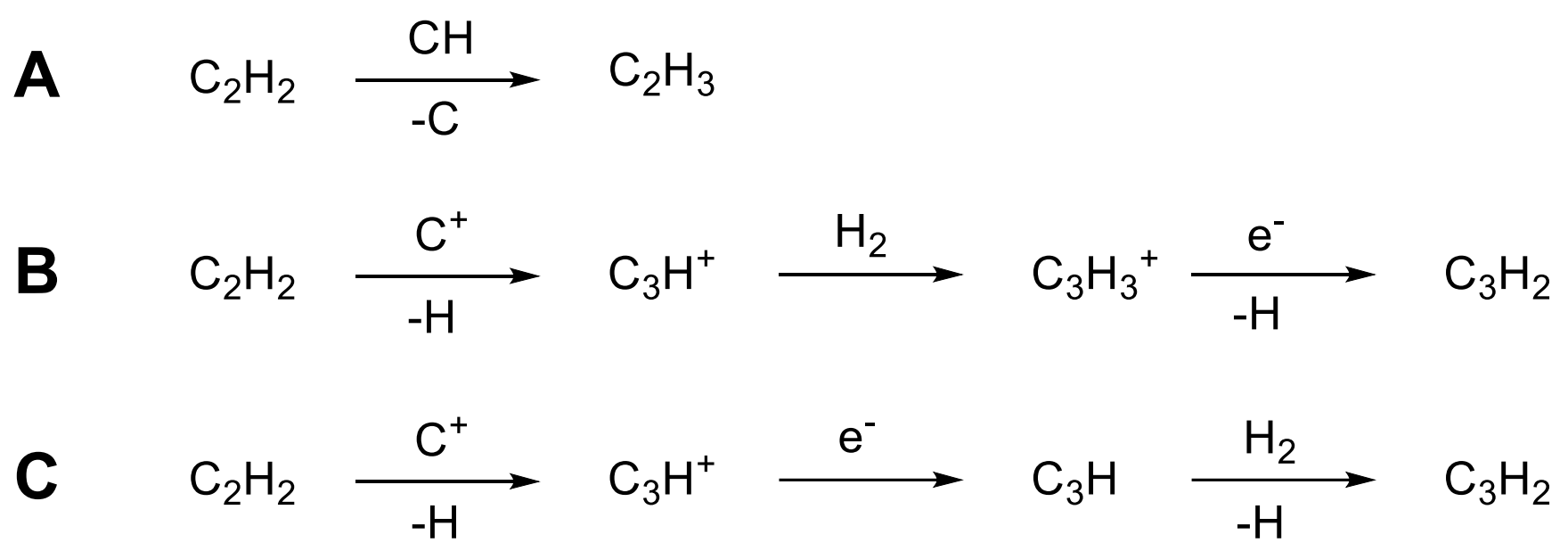

**Scheme 4.** Formation of $C_3H_2$ molecule from acetylene.

It is worth mentioning that these reactions take place only in hot PDRs, such as the Orion Bar PDR, but not in all PDRs with the increased abundance of small hydrocarbons. For instance, the PDR Horsehead nebula is relatively cold, and these reactions do not proceed there. Dissociative reactions induced by CRs and photons can introduce some $C_3H_2$ but still cannot explain the observed abundance of $C_3H_2$ and other small hydrocarbons. This underestimation is mainly due to the small abundance of the parent acetylene. As discussed above in Section 2.7, acetylene can be a result of PAH dissociation or other carbonaceous solid particles that are present in such PDRs[178,193]; however, Murga et al.[175,182] showed that this channel also cannot help achieve the observed abundances. Thus, alternative mechanisms that can provide additional acetylene or reactions that convert acetylene or related molecules to small hydrocarbons are needed.

**Table 6.** The reactions that start from acetylene and lead to formation of small hydrocarbons.

| Reactions | Kinetics | Objects & conditions |
|---|---|---|
| Neutral-neutral reactions:<br>1. $C_2H_2 + CH \rightarrow C_2H_3 + H$<br>2. $C_2H + C_2H_2 \rightarrow C_4H_2 + H$ | $k_1 = 3.39 \cdot 10^{-10}$ cm$^3$ s$^{-1}$ (UMIST)<br>$k_2 = 1.2 \cdot 10^{-10}$ cm$^3$ s$^{-1}$ (T = 100 K) (UMIST) | The reactions may take place in many objects (PDR's, diffuse ISM, molecular clouds, protoplanetary disks) where ions can be produced by UV radiation or by CRs.<br>Conditions:<br>T = 10 - 500 K<br>$n_H = 10^{-1} - 10^7$ cm$^{-3}$<br>$G_0$ = from very low to $10^5$<br>CR field: Galactic |
| Ion-molecular reactions[145,183]<br>3.<br>a. $C_2H_2 + C^+ \rightarrow C_3H^+ + H$<br>b. $C_3H^+ + H_2 \rightarrow$ l/c-$C_3H_3^+ + h\nu$<br>c. c-$C_3H_3^+ + e^- \rightarrow$ c-$C_3H_2 + H$ | $k_{3a} = 1.78 \cdot 10^{-9}$ cm$^3$ s$^{-1}$ (UMIST)<br>$k_{3b} = 5.5 \cdot 10^{-12}$ cm$^3$ s$^{-1}$ (T=100 K) (KIDA)<br>$k_{3c} = 7 \cdot 10^{-7}$ cm$^3$ s$^{-1}$ (T = 100 K) (KIDA) | |
| 4.<br>a. $C_2H_2 + C^+ \rightarrow C_3H^+ + H$<br>b. $C_3H^+ + e^- \rightarrow C_3 + H$<br>c. $C_3 + H_2 \rightarrow$ l/c-$C_3H + H$<br>d. l/c-$C_3H + H_2 \rightarrow$ l/c-$C_3H_2 + H$ | $k_{4a} = 1.78 \cdot 10^{-9}$ cm$^3$ s$^{-1}$ (UMIST)<br>$k_{4b} = 2.5 \cdot 10^{-7}$ cm$^3$ s$^{-1}$ (T = 100 K) (UMIST)<br>$k_{4c} = 1.2 \cdot 10^{-5}$ cm$^3$ s$^{-1}$ (T = 100 K) (KIDA)<br>$k_{4d} = 1.5 \cdot 10^{-7}$ cm$^3$ s$^{-1}$ (T = 100 K) (KIDA) | |
| 5. $C_2H_2 + C_2H_2^+ \rightarrow C_4H_2^+ + H_2$<br>6. $C_2H_2 + C_2H_2^+ \rightarrow C_4H_3^+ + H$[197] | $k_5 = 5.2 \cdot 10^{-10}$ cm$^3$ s$^{-1}$ (KIDA)<br>$k_6 = 8.8 \cdot 10^{-10}$ cm$^3$ s$^{-1}$ (KIDA) | |

## 3.2. Acetylene as a precursor of polyynes

Acetylene is a parent molecule for different polyynes, cyanopolyynes and methyl-substituted cyanopolyynes. The study of circumstellar envelopes, molecular clouds, hot cores and other objects indicates the presence of polyyne ($HC_nH$) and cyanopolyyne ($HC_nN$) chains.[198,199,200,201,202]

Cernicharo proposed a model[203] in which polyynes, cyanopolyynes, and methyl-polyynes can be formed in PDRs through a network of reactions between $C_n + C_mH_2$, $C_nH + C_mH$, $C_nN + H_2$, $C_nN + C_mH_2$ and others, many of which include acetylene or its derivatives. These reactions could reproduce the observed abundance of polyynes, although Cernicaro emphasized the necessity of both theoretically and experimental checking the parameters of the reactions. The author also suggested that in protoplanetary nebulae such as CRL 618, the large abundance of small cumulenes ($H_2C_nH_2$) indicates that the growth of small carbon grains with $C/H \gg 1$ could be dominated by reactions between polyynes and cumulenes.

Reactions such as reactions between the cyan radical CN and acetylene $C_2H_2$ and its derivative $HC_nH$ have been widely used in astrochemical models of molecular clouds and hot cores to explain $HC_nN$.[204,205] It is assumed that the rate constants increase as the temperature decreases. The experiments carried out by Balucani et al. confirmed that some of these reactions (e.g., $C_2H_2$ + CN) may proceed under conditions of circumstellar envelopes in molecular clouds and hot cores.[206]

In turn, Agundez et al.[200] presented a model of the formation of acetylenic chains for the stellar outflow of the C-rich star IRC +10216. This model also includes neutral-neutral reactions between cyan radicals and unsaturated hydrocarbons but is slightly different. According to this model, $C_2H_2$ and prussic acid (HCN) are formed inside a stellar envelope, while the formation of acetylenic chains occurs further outside the envelope but from the parent molecules $C_2H_2$ and HCN. First, $C_2H_2$ and HCN are dissociated by UV photons from the outer radiation field and leave $C_2H$ and CN radicals. The reactions between $C_2H$ and $C_2$, $C_2H$, and $C_4H$ result in $C_4H$, $HC_4H$, and $HC_6H$, respectively. The subsequent photoinduced H loss eventually gives $C_4H$ and $C_6H$. The reactions between $C_2H_2$ and its derivatives $HC_4H$, $HC_6H$ and CN lead to $HC_3N$, $HC_5N$ and $HC_7N$, while the subsequent photoinduced H loss from $HC_3N$ and $HC_5N$ provides $C_3N$ and $C_5N$. The scheme is illustrated in Figure 5.

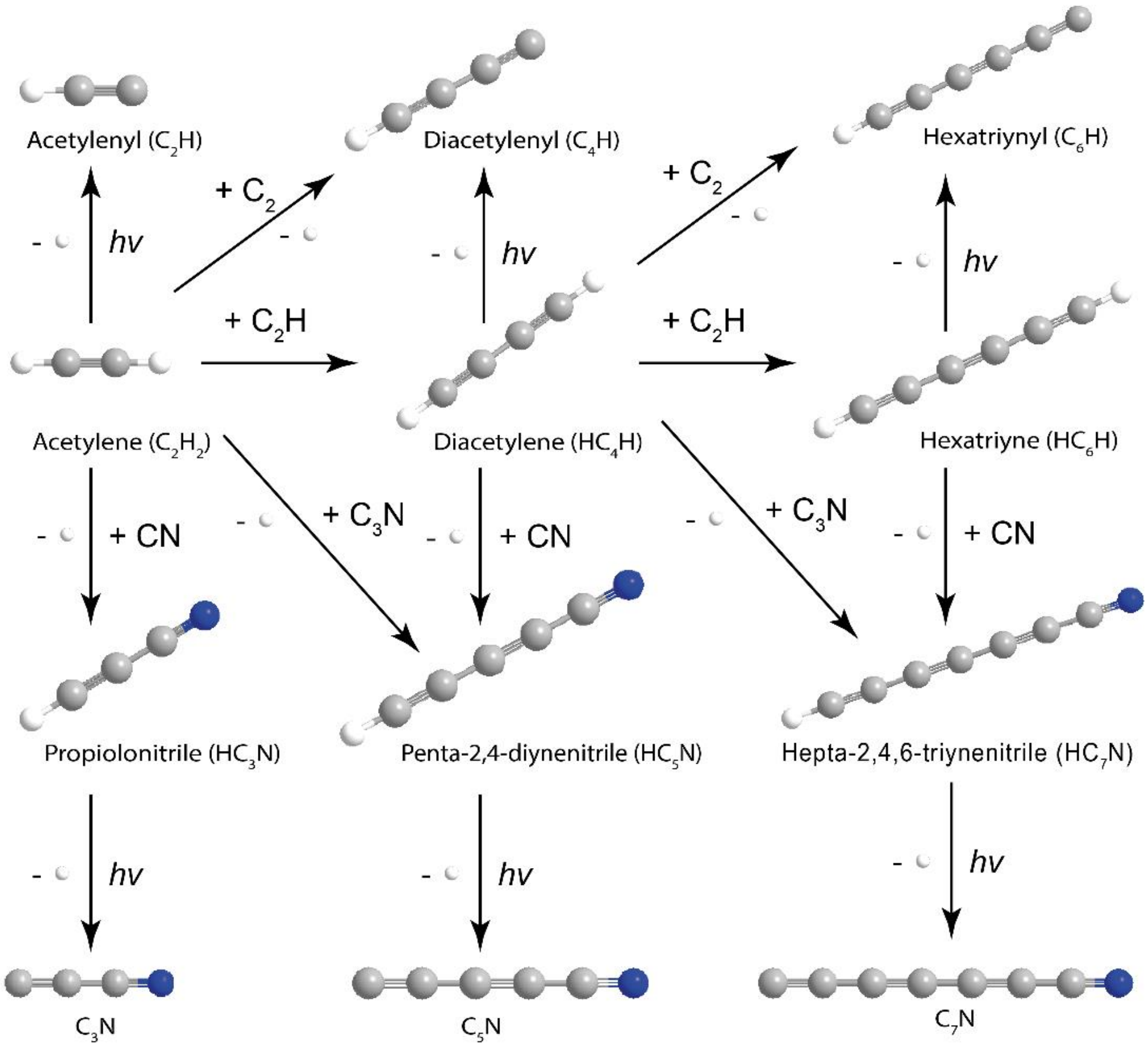

**Figure 5**. Pathways from acetylene to polyynes and cyanopolyynes.

Santoro et al.[207] experimentally reproduced the conditions of the circumstellar envelopes and studied the chemical evolution of the $C_2/C_2H_2$ gas mixture. Among the products, polyynes were found, confirming the pathways described above. They emphasized the enormous role of acetylene in the evolution of chemical composition.

In addition to reactions such as CN + $C_2H_2$, which can proceed at low temperature according to Balucani et al.[206], there are other barrierless reactions. Gu et al.[208] reported that the reaction between ethynyl radical and diacetylene occurs without a barrier, forming triacetylene. Other experiments at low temperature showed that methyltriacetylene ($CH_3(C\equiv C)_3H$) can be generated by a barrierless reaction between diacetylene (HCCCCH) and a 1-propynyl radical ($CH_3CC$).[209] This molecule has been detected at low temperature in the molecular cloud TMC-1[210]; therefore, these reactions most likely take place in such objects and may explain the presence of different polyynes.

Cuylle et al. studied[211] the chemistry of pure $C_2H_2$ ice under vacuum ultraviolet (VUV) irradiation. Using a combination of infrared and ultraviolet-visible (UV-VIS) spectroscopy, the authors showed the predominance of polymerization and polyyne formation. In this case, the growth of the polyyne chains reached that of $HC_{20}H$ and larger polyyne-like molecules.

The formation of polyyne chains in acetylene ice upon irradiation was also observed in the experiment by G. Compagnini et al.[212] A total of 2 μm-thick acetylene $C_2H_2$ was deposited at 16 K onto a KBr support. Ion irradiation was carried out with an $H^+$ beam with an energy of 200 keV. Irradiation of the sample caused the formation of linear carbon chains, the signals of which appeared at 2100, 2170 and 2190 $cm^{-1}$. The small line widths of these emergent structures indicated that the chains had a distinct length and that they were likely isolated from the rest of the ice matrix. The authors suggested that the chains in question probably contained 8–12 carbon atoms, citing the position of the peak. The results obtained support the hypothesis that polyynes could be synthesized in comets under the influence of irradiation by CRs and released in gas evaporating from the comet nuclei.

In addition to the neutral carbon chains mentioned above, anions of chains such as $C_4H^-$, $C_6H^-$, and $C_8H^-$ were detected in circumstellar envelopes, dark clouds and other objects.[213,214,215] As electron attachment to $C_nH$ is estimated to be inefficient[215], it was suggested[216] that such anionic carbon chains are formed through ion-molecular reactions with acetylene (Scheme 5). The corresponding experiments were carried out, and it was concluded that such reactions could be important for the formation of carbon chain anions.

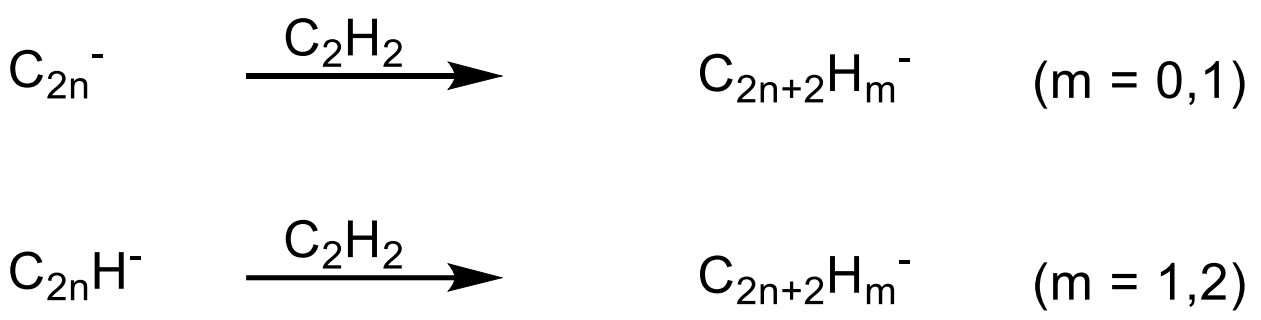

**Scheme 5.** Formation of $C_{2n+2}H_m^-$ anions through ion-molecular reactions with acetylene.

Polyynes and cyanopolyynes can be important components of Titan's haze (see Section 5.4). Kaiser and Mebel, in studies with crossed molecular beams, showed that isoelectronic ethynyl ($C_2H(X_2\Sigma^+)$) and cyano ($CN(X_2\Sigma^+)$) radicals can form on Titan as a result of the photolysis of acetylene ($C_2H_2$) and hydrogen cyanide (HCN) from solar UV photons.[217] The data obtained showed that these species, when reacting with unsaturated hydrocarbons (such as acetylene, methylacetylene, and diacetylene), can contribute to the complexation of Titan's haze layers, forming molecules of polyynes and cyanopolyynes.[217] As shown below in Sections 5.2.4, 5.4 and 8.1, polyynes can participate in PAH formation and in the formation of organonitrogen compounds, which can play an important role in the formation of the topography of Titan and other objects of the Solar system and could play a role in the development of prebiotic chemistry. In Table 7, we summarize the reactions considered in this section.

**Table 7.** The reactions that start from acetylene and lead to formation of polyynes and cyanopolyynes.

| **Reactions** | **Kinetics** | **Objects & conditions** |
|---|---|---|
| Neutral-neutral reactions:<br>1. $C_2H_2 + CN \rightarrow HC_3N + H$ | $k_1 = 2.38 \cdot 10^{-10}$ cm$^3$ s$^{-1}$ (T=10 K)<br>(KIDA)<br>$k_1 = 1.42 \cdot 10^{-10}$ cm$^3$ s$^{-1}$ (T = 1000 K) (KIDA) | Stellar envelopes. Conditions:<br>$n_H = 10^8$-$10^{15}$ cm$^{-3}$<br>T = 500-2500 K<br>Molecular clouds, prestellar cores. Conditions:<br>$n_H = 10^2$-$10^7$ cm$^{-3}$<br>T = 10 K<br>$G_0$ = very low<br>Galactic CR field |
| 2.<br>a. $C_2H_2 + h\nu \rightarrow C_2H + H$<br>b. $C_2H + C_2H_2 \rightarrow C_4H_2 + H$<br>c. $C_4H_2 + h\nu \rightarrow C_4H + H$<br>d. $C_4H + C_2H_2 \rightarrow C_6H_2 + H$<br>e. $C_6H_2 + h\nu \rightarrow C_6H + H$[200] | $k_{2a} = 4.3 \cdot 10^{-9}$ $cm^3$ $s^{-1}$<br>$k_{2b} = 1.3 \cdot 10^{-10}$ $cm^3$ $s^{-1}$<br>$k_{2c} = 1.1 \cdot 10^{-8}$ $cm^3$ $s^{-1}$<br>$k_{2d} = 3 \cdot 10^{-10}$ $cm^3$ $s^{-1}$<br>$k_{2e} = 1.7 \cdot 10^{-8}$ $cm^3$ $s^{-1}$<br>(all rates are given for T = 100 K, $G_0 = 1$)[200] | PDRs, DISM, molecular clouds, protoplanetary disks.<br>Conditions:<br>T = 10 - 500 K<br>$n_H = 10^{-1} - 10^7$ $cm^{-3}$<br>$G_0$ = from very low to $10^5$<br>Galactic CR field<br>Titan atmosphere, early Earth[217]<br>Conditions:<br>$n_H = 10^{10}$-$10^{16}$ $cm^{-3}$<br>T = 50-300 K<br>$G_0$: up to 10<br>Galactic CR field |
| 3.<br>a. $HCN + h\nu \rightarrow CN + H$<br>b. $CN + C_2H_2 \rightarrow HC_3N + H$<br>c. $HC_3N + h\nu \rightarrow C_3N + H$<br>d. $C_3N + C_2H_2 \rightarrow HC_5N + H$<br>e. $HC_5N + h\nu \rightarrow C_5N + H$ | $k_{3a} = 1.9 \cdot 10^{-9}$ $cm^3$ $s^{-1}$<br>$k_{3b} = 4.1 \cdot 10^{-10}$ $cm^3$ $s^{-1}$<br>$k_{3c} = 4.4 \cdot 10^{-9}$ $cm^3$ $s^{-1}$<br>$k_{3d} = 3.7 \cdot 10^{-10}$ $cm^3$ $s^{-1}$<br>$k_{3e} = 8.7 \cdot 10^{-9}$ $cm^3$ $s^{-1}$<br>(all rates are given for T = 100 K, $G_0 = 1$)[200] | |
| 4.<br>$H(C{=}C)_2H + CH_3CC \rightarrow CH_3(CC)_3H$[209] | n.d. | Molecular clouds, prestellar cores.<br>Conditions:<br>$n_H = 10^2$-$10^7$ $cm^{-3}$<br>T = 10 K<br>$G_0$ is very low<br>Galactic CR field |
| 5.<br>$C_{2n}^- + C_2H_2 \rightarrow C_{2n+2}H_m^-$ (m = 0,1)<br>$C_{2n}H^- + C_2H_2 \rightarrow C_{2n+2}H_m^-$ (m = 1,2)[216] | n.d. | Stellar envelopes. Conditions:<br>$n_H = 10^8$-$10^{15}$ $cm^{-3}$<br>T = 500-2500 K<br>Molecular clouds, prestellar cores.<br>Conditions:<br>$n_H = 10^2$-$10^7$ $cm^{-3}$<br>T = 10 K<br>$G_0$ is very low<br>Galactic CR field |

## 3.3. Acetylene as a precursor of complex organic molecules

Many COMs are observed in molecular clouds, prestellar cores and protostellar objects.[218] Among the detected COMs, there are many ethyl- and vinyl-bearing molecules, such as $C_2H_5OH$, $C_2H_3OH$, $C_2H_5CN$, $C_2H_3CCCH$, and $C_2H_3C_3N$.[194] Some and perhaps most COMs can form at low temperatures on the ice surface of dust grains or even inside the icy mantle[219]. The interstellar ice mainly consists of $H_2O$, CO, $CO_2$, $NH_3$, $CH_4$, and $CH_3OH$.[220,221] Many different species, including molecules, can be adsorbed on dust grains; these species may lie on or wander the surface and occasionally react with each other or with ice species.[222,223,224] Energy sources (UV photons, CRs or exothermic reactions) stimulate these reactions. The formed COMs may desorb to the gas phase if enough energy is available, and then, they are observed in radio waves.

Acetylene can be present among species on or inside ice. It can be generated from $CH_4$, as indicated above in Section 2.6, or trapped from the gas phase. The chemistry of acetylene in interstellar and cometary ices has become the subject of intensive experimental research in recent decades. Experimental studies[160] have shown that acetylene under cold ISM conditions can be the ancestor of a whole family of derivatives, such as diacetylene, vinylacetylene, methylacetylene, and propene, as shown in Figure 6. Specifically, in the experiments, the reactions of acetylene conversion to derivatives were initiated by highly energetic electrons that imitate secondary electrons produced by primary CRs. Furthermore, these derivatives can be the basis for the formation of more complex molecules and carbon nanoparticles (Sections 4, 5, 6).

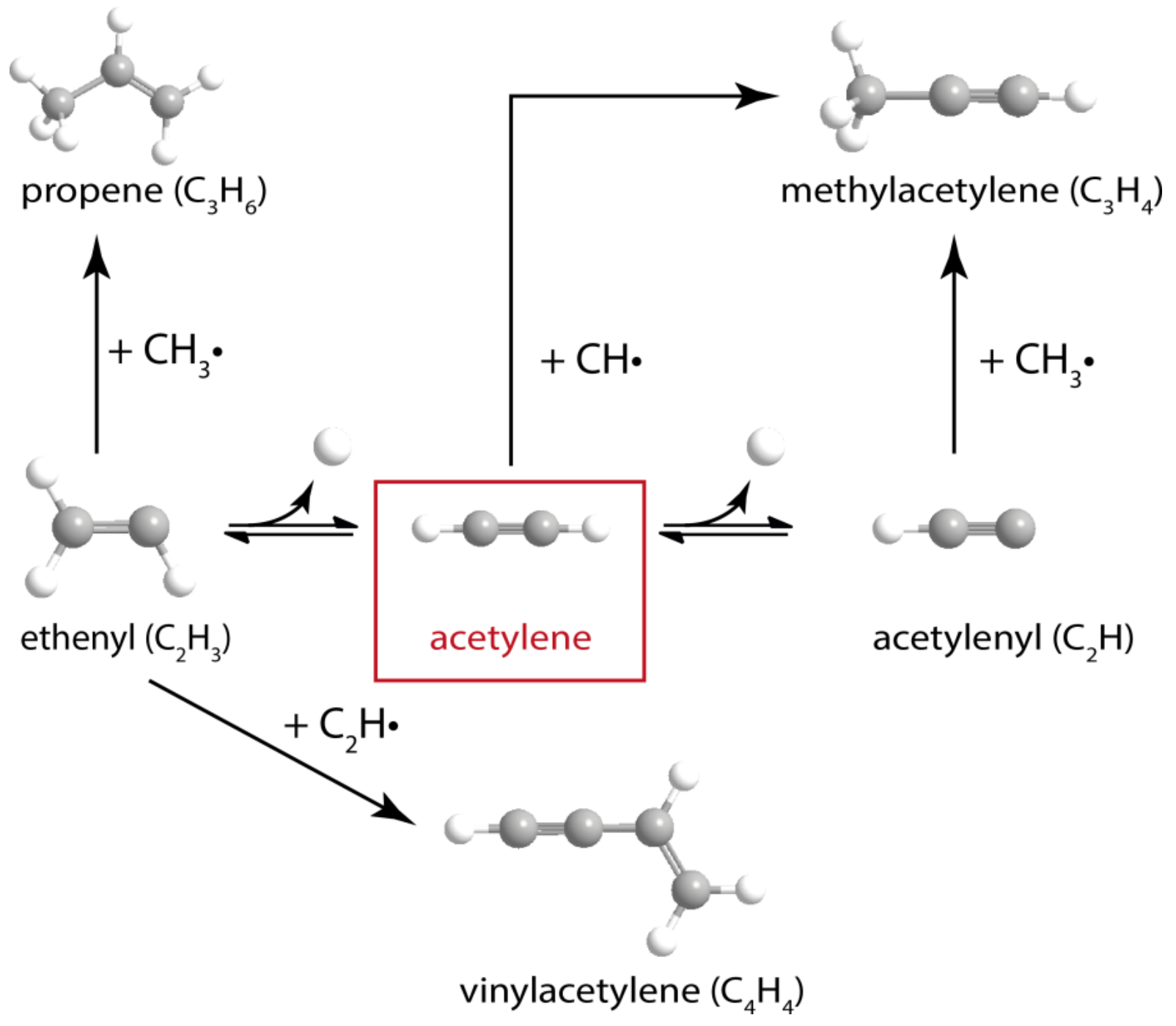

**Figure 6.** Experimentally determined[160] dominant reaction pathways for the formation of $C_2$-hydrocarbons as well as $C_3H_4$, $C_3H_6$ and $C_4H_4$ isomers from methane ices.

The transformations of acetylene in ice of various compositions under cryogenic conditions under the influence of radiation were studied by Feldman. In one of his works[161], he showed that the decomposition of $C_2H_2$ in a xenon ice matrix leads to the formation of $C_2$ species. The clear induction period during which $C_2$ accumulated indicates that the process proceeded through the decomposition of stabilized $C_2H$ radicals. Conjugated hydrocarbons were detected in acetylene ice in this work, from which the authors concluded that condensation predominates over decomposition during radiation treatment of acetylene ice. This finding is consistent with the findings of other works mentioned above and below.[211]

Reactions between acetylene and other species on/in ice may lead to the formation of COMs. One of the COMs, ketene ($CH_2CO$), was experimentally produced from acetylene-containing ice in a number of works. Specifically, ketene was formed as a result of photolysis and radiolysis of ice mixtures.[224,,225] Several low-temperature routes involving acetylene and oxygen were investigated experimentally.[226,227] Notably, these experiments showed that ketene is formed in an ice mixture of $C_2H_2$ and $CO_2$ at a temperature of 10 K. First, $CO_2$ dissociates to CO and O, and then O reacts with $C_2H_2$ to form ketene. These reactions can explain the observed presence of ketenes in molecular clouds.[228]

The interaction of acetylene with a water molecule in an ice matrix at a temperature of 5 K under radiation conditions was experimentally studied by Feldman et al.[229] The authors showed that the formation of a complex with water facilitates the radiation decay of the acetylene molecule. FTIR was used to detect ketene, which is a complex with hydrogen ($CH_2CO$–$H_2$), ketenyl radicals (HCCO), methane ($CH_4$) and carbon monoxide (CO). Such processes can occur in interstellar and cometary ices.

Irradiation of $H_2O/CO_2$ ices with acetylene also leads to the formation of other (semi)saturated hydrocarbons, such as $C_2H_4$, $C_2H_6$ and other COMs (vinyl alcohol [$CH_2CHOH$], acetaldehyde [$CH_3CHO$], ketene [$CH_2CO$], and ethanol [$CH_3CH_2OH$]).[154,230,231,232] Bergner et al.[225,226,233] suggested that ethylene and ethane are formed by subsequent hydrogenation of acetylene (Scheme 6). The hydrogenation of acetylene to ethane was proven to be efficient under conditions of amorphous solid water in the environment of a dark molecular cloud.[158]

**A** $C_2H_2 + H^{\bullet} \longrightarrow C_2H_3^{\bullet}$

**B** $C_2H_3 + H^{\bullet} \longrightarrow C_2H_4$

**C** $C_2H_4 + H^{\bullet} \longrightarrow C_2H_5^{\bullet}$

**D** $C_2H_5 + H^{\bullet} \longrightarrow C_2H_6$

**Scheme 6.** Ethylene, ethane formation by subsequent hydrogenation of acetylene in the ices.

Charnley[234] demonstrated how the combination of hydrogenation and oxidation in $H_2O$/ $C_2H_2$ ice leads to the formation of vinyl alcohol, acetaldehyde, and ethanol (Scheme 7).

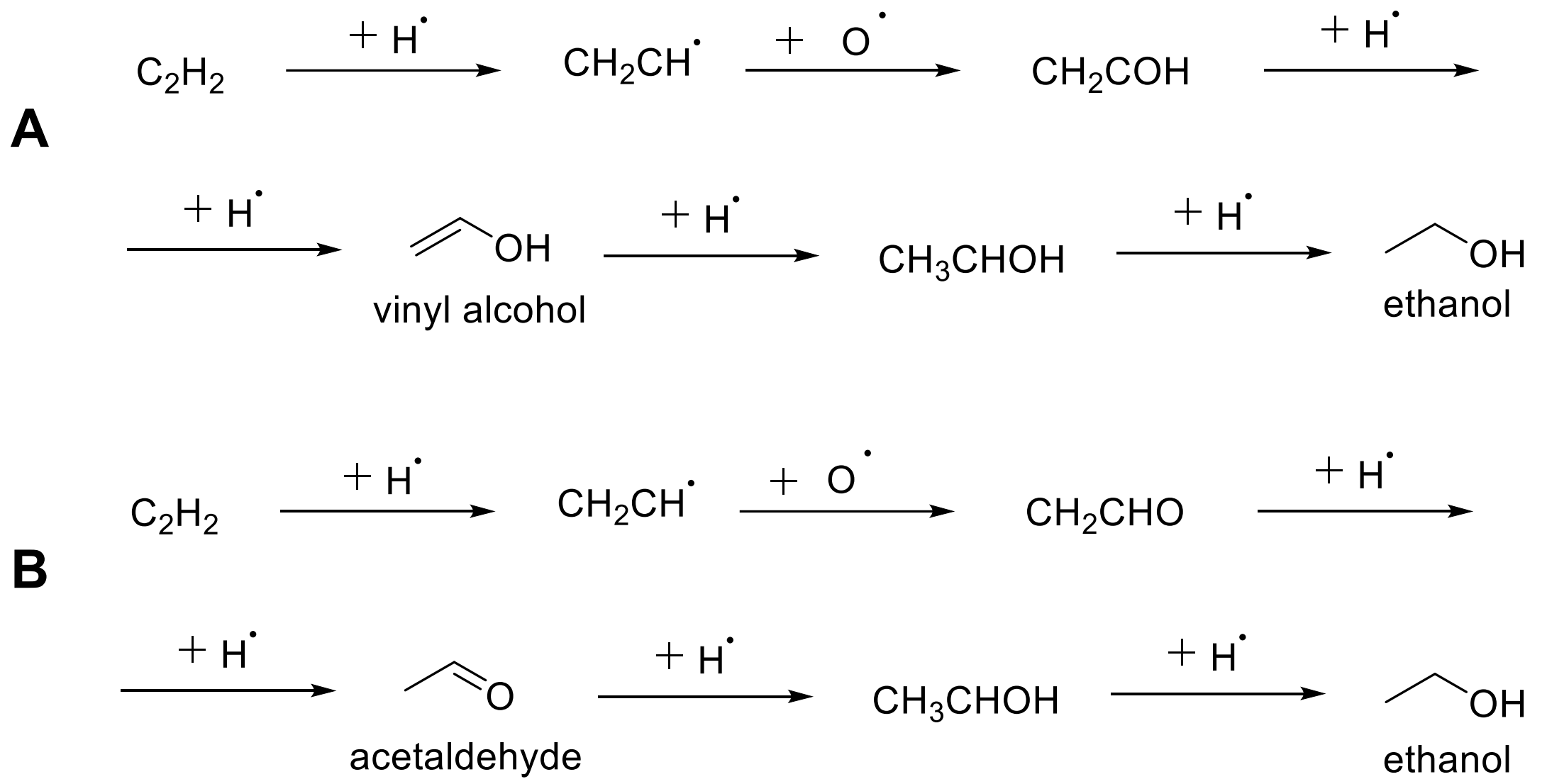

**Scheme 7.** Formation of formation of vinyl alcohol, acetaldehyde, and ethanol by combination of hydrogenation and oxidation of acetylene in $H_2O/C_2H_2$ ice.

Hudson et al. suggested that ethanol is formed in an ice mixture of $C_2H_2$ and $H_2O$ via the following sequence: A) radiolysis of water; B-D) hydrogenation of $C_2H_2$ to $C_2H_5$; and E) addition of an OH radical to $C_2H_5$ with $C_2H_5OH$ formation (Scheme 8).

**A** $H_2O \longrightarrow H^{\bullet} + OH^{\bullet}$ by $p^+$ radiolysis

**B** $HC{\equiv}CH + H \longrightarrow H_2C{=}CH^{\bullet}$

**C** $H_2C{=}CH^{\bullet} + H^{\bullet} \longrightarrow H_2C{=}CH_2$

**D** $H_2C{=}CH_2 + H^{\bullet} \longrightarrow H_3C{-}\dot{C}H_2$

**E** $H_3C{-}\dot{C}H_2 + OH^{\bullet} \longrightarrow C_2H_5OH$

**Scheme 8.** Ethanol formation in the ices.

According to Chuang et al. 2021, the produced COMs (acetaldehyde, ketene, ethanol) can be explained by reactions that start from the formation of vinyl alcohol in the reactions between $C_2H_2$ and H and/or OH radicals. Further transformation to acetaldehyde, ketene, or ethanol occurs either by hydrogenation or energy absorption. These COMs were also produced via a nonenergetic route in the experiment where $C_2H_2$, $O_2$ and H were codeposited on the surface and were not irradiated.[235]

Using DFT calculations, Perrerro et al.[236] showed that ethanol can be formed by the reaction of $C_2H$ (an acetylene derivative) with one $H_2O$ molecule. It was assumed in the model that the reaction proceeded in clusters of water ice. Such a process can be barrierless due to the possibility of proton transfer on the boundary molecules of ice water.

Zhou et al. studied[237] the transformation of the ice mixture $CO/C_2H_2$ at 10 K under high-energy electron irradiation in an ultrahigh vacuum. As a result, two isomers of the $C_3H_2O$ molecule could be detected. Propynal (HCCCHO) was synthesized from a carbon monoxide-acetylene complex via the [HCO... CCH] radical pair. In turn, the addition of triplet carbon monoxide to acetylene in the ground state (or vice versa) occurred with the formation of a cyclic structure (c-$C_3H_2O$). The authors noted that the heterogeneous formation of both $C_3H_2O$ isomers was orders of magnitude greater than that of gaseous chemistry. The production rates were 5 O $10^6$ molecules $cm^{-2}s^{-1}$ for propynal species (3 O $10^4$ molecules $cm^{-2}s^{-1}$ in gaseous chemistry) and 5 O $10^6$ molecules $cm^{-2}s^{-1}$ for c-$C_3H_2O$. Thus, the process of formation of $C_3H_2O$ isomers from acetylene can be initiated by energetic electrons produced by CRs penetrating interstellar ice grains in cold molecular clouds.

Wu et al. studied[238] acetylene in an environment of solid nitrogen ($N_2$) at a temperature of 10 K. Ice was irradiated by VUV photons and electrons. In the first case, the formation of $C_2H$, CN radicals and $C_2N_2$ isomers was observed. In the second case, the formation of $N_3$ and $C_2H$ radicals, as well as various nitriles, was observed. Photolysis of $C_2H_2/N_2$ ice samples produced 15 times more $C_2H$ radicals than did radiolysis.

The transformation of acetylene in ammonia ice was also studied experimentally. Namely, ice mixtures of $NH_3$:$C_2H_6$, $NH_3$:$C_2H_4$ and $NH_3$:$C_2H_2$ were irradiated by VUV photons at temperatures ranging from 10–50 K. As a result, the formation of imine ($CH_3CH$=NH) and acetonitrile ($CH_3CN$) was detected. The highest yields of imine (35%) and acetonitrile (5.2%) were achieved by irradiating the mixtures with $C_2H_2$ at 10 K.[239] An important discovery was made by Zhang et al.[240], who experimentally demonstrated the formation of vinylamine ($C_2H_3NH_2$) in the ice mixture $NH_3$:$C_2H_2$, which was exposed to energetic electrons. Quantum chemical calculations have shown that vinylamine ($C_2H_3NH_2$) can be formed in two different ways: through barrierless radical-radical recombination of amino ($NH_2$) and vinyl ($C_2H_3$) radicals and through a one-step concerted pathway.

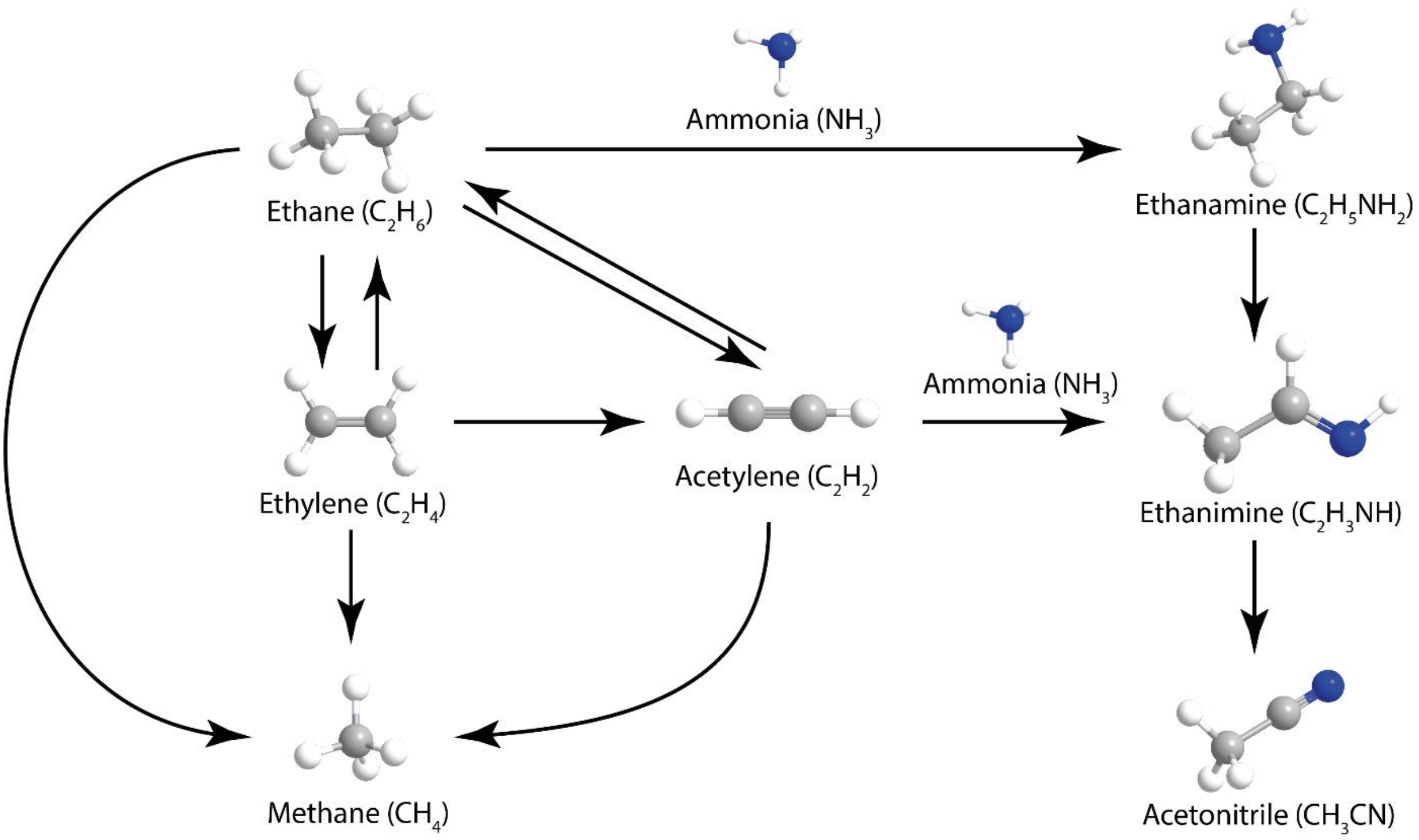

**Figure 7.** Proposed reaction network for the irradiation of $C_2H_6$, $C_2H_4$, $C_2H_2$ and ammonia ice mixtures.

Recently, Molpeceres and Rivilla[241] computationally (DFT) studied the surface reactions between acetylene, ethylene and ethane and radicals (H, OH, $NH_2$, $CH_3$). The reactions with the radicals H and OH were found to be the most important for COM formation, namely, for $C_2H_3$, $C_2H_5$, $C_2H_2OH$ and $C_2H_4OH$, while the reactions with $NH_2$ and $CH_3$ had a minor role. These formed radicals can further participate in reactions leading to more complex molecules, including those detected in space, although certain pathways have still not been established.

Another example of the likely interaction of acetylene with ammonia is Jupiter's Great Red Spot (GRS).[242] In laboratory simulations,[243] a material was obtained by reacting acetylene and ammonia under VUV irradiation, corresponding to solar radiation on Jupiter within and above the upper troposphere. HR-MS and IR spectroscopy confirmed the presence of aliphatic azine, azo, and diazo compounds in the resulting chromophore material. The spectral properties of the resulting chromophore compound were in excellent agreement with the 0.35–1.05 μm spectra of the core of the GRS obtained by VIMS during the Cassini–Huygens mission. In the high troposphere, upwelling $NH_3$ molecules can photodissociate and react with downwelling $C_2H_2$ molecules. The resulting compounds can coat ammonia grains at the top of the GRS clouds. Tholins, which may be products of the reaction of ammonia and acetylene, may also be responsible for the orange-red color of objects on a number of planets and moons in the Solar system (see Section 5.4). More details about the nature of the tholins and the chemical mechanism of action of acetylene with ammonia in the formation of important nitrogen-containing compounds are discussed in Section 8.1. In Table 8, we present the main reactions discussed in the section, although kinetics information is not yet available.

**Table 8.** The reactions of formation of COMs with acetylene as a key reactant.

| **Reactions** | **Objects & conditions** |
|---|---|
| **$CO_2/C_2H_2$ ice**<br>1. a $CO_2 \rightarrow CO + O$<br>b. $C_2H_2 + O \rightarrow CH_2CO$ [226·227·234] | Molecular clouds, prestellar cores.<br>Conditions:<br>T = 10 K<br>$G_0$ = very low<br>Galactic CR field |
| **$H_2O/C_2H_2$ ice**<br>2. a. $H_2O \rightarrow H\cdot + OH\cdot$<br>b. $C_2H_2 + H\cdot \rightarrow C_2H_3$<br>c. $C_2H_3 + H\cdot \rightarrow C_2H_4$<br>d. $C_2H_4 + H\cdot \rightarrow C_2H_5$<br>e. $C_2H_5 + H\cdot \rightarrow C_2H_6$<br>e. $C_2H_5 + OH\cdot \rightarrow C_2H_5OH$[230] | |
| 3. a. $C_2H_2$ + CR (or hν) $C_2H + H$<br>b. $C_2H + H_2O \rightarrow HCCHOH$<br>c. $HCCHOH + H\cdot \rightarrow CH_3CHOH$<br>d. $CH_3CHOH + H\cdot \rightarrow C_2H_5OH$[236] | |
| 4. a. $C_2H_2 + OH\cdot$ (from $H_2O$)<br>$CH_2CHOH$ (vinyl alcohol)<br>b. $CH_2CHOH \rightarrow CH_3CHO$ (acetaldehyde)<br>c. $CH_2CHOH$ + CR/ hν $\rightarrow CH_2CO$ (ketene)<br>d. $CH_2CHOH + 2H\cdot \rightarrow CH_3CH_2OH$ (ethanol)[225] | |
| 5. a. $C_2H_2 + H\cdot \rightarrow CH_2CH$<br>b. $CH_2CH + O \rightarrow CH_2COH$<br>c. $CH_2COH + H\cdot \rightarrow CH_2CHOH$<br>d. $CH_2CHOH + H\cdot \rightarrow CH_3CHOH$<br>e. $CH_3CHOH + H\cdot \rightarrow C_2H_5OH$[234] | |
| 6. a. $C_2H_2 + H\cdot \rightarrow CH_2CH$<br>b. $CH_2CH + O \rightarrow CH_2CHO$<br>c. $CH_2CHO + H\cdot \rightarrow CH_3CHO$<br>d. $CH_3CHO + H\cdot \rightarrow CH_3CHOH$<br>e. $CH_3CHOH + H\cdot \rightarrow C_2H_5OH$[234] | |
| 7. a $C_2H_2 + H\cdot \rightarrow C_2H_3$<br>b. $C_2H_2 + OH\cdot \rightarrow C_2H_2OH$<br>c. $C_2H_4 + H\cdot \rightarrow C_2H_5$<br>d. $C_2H_4 + OH\cdot \rightarrow C_2H_4OH$[241] | |

## 4. Benzene formation as the first step in PAH evolution

The formation of an aromatic ring is a turning point in chemical evolution from acetylene. This is the first step toward complex PAH molecules. The initial molecules are benzene $C_6H_6$ and the phenyl radical $C_6H_5$, which further act as elementary building blocks for all other high-molecular-weight polycyclic aromatic systems and carbon nanoparticles as well as soot particles and carbon dust. Unlike $C_2H_2$, benzene was found only in a few objects, in the protoplanetary nebulae CRL 618[143] and SMP LMC 11[244] and in the protoplanetary disk J160532.[188] Additionally, benzonitrile was detected in the TMC-1 molecular cloud.[245] Nevertheless, benzene is present in space objects and plays an important role in chemical reaction networks.

The formation of the initial aromatic ring is considered a limiting step that determines the rate of formation of larger polycyclic aromatic systems and particles.[20,246] This process has even been named a “bottleneck” in macromolecular growth and dust nucleation.[38] Moreover, it is important to note that the transition of hydrocarbons from an aliphatic structure to a closed aromatic structure ensures that the system achieves the minimum energy. That is, the process of formation of the benzene ring is thermodynamically determined. Despite its importance, the mechanism of formation of the first aromatic system from aliphatic precursors in various regions of space is still the subject of discussion.

Traditionally, acetylene was considered to be the main precursor of aromatic molecules. Chemical mechanisms for the formation of an aromatic system in outer space were proposed in the 1980s, when the mid-IR interstellar emission bands were interpreted as related to PAH molecules.[247] The conditions differ significantly for different space objects; therefore, there are a variety of mechanisms that are determined by the specific physical conditions and chemical composition. Nevertheless, the possible formation of aromatic molecules can be related to two distinct environments: a cold and dense environment, which includes molecular clouds, planets and moons; and a hot/warm environment, which includes stellar envelopes and their remnants, protoplanetary and planetary nebulae. Furthermore, the main pathways are presented. The mechanisms proposed in astrophysics traditionally do not imply the participation of transition metals and almost do not consider catalytic processes. These mechanisms will be discussed in a separate section.

## 4.1. Circumstellar envelopes and flame burning models

As one of the main sources of cosmic carbon, C-rich AGB stars have attracted the attention of researchers as the basis for models of aromatic ring formation. Several mechanisms for the formation of the benzene ring have been proposed for astrochemical models from the chemistry of flame combustion.[248,249,250,251] In 1984, Cole et al. suggested that the main mechanism for the formation of aromatic compounds in a flame is the reaction of 1,3-butadienyl with acetylene (Scheme 9A).[252] Recently, Frenklach et al. described the main cyclization pathways in a low-pressure acetylene flame soot simulation study. According to their mechanism, the formation of a phenyl radical occurs upon the interaction of 1-buten-3-ynyl with an acetylene molecule through the cyclization of linear $C_6H_5$ (Scheme 9B).[253,254] In 1992, Miller and Melius first proposed a mechanism in which ring closure occurs after the reactions of two propargyl radicals as the main pathway for the formation of the benzene ring (Scheme 9C).[255]

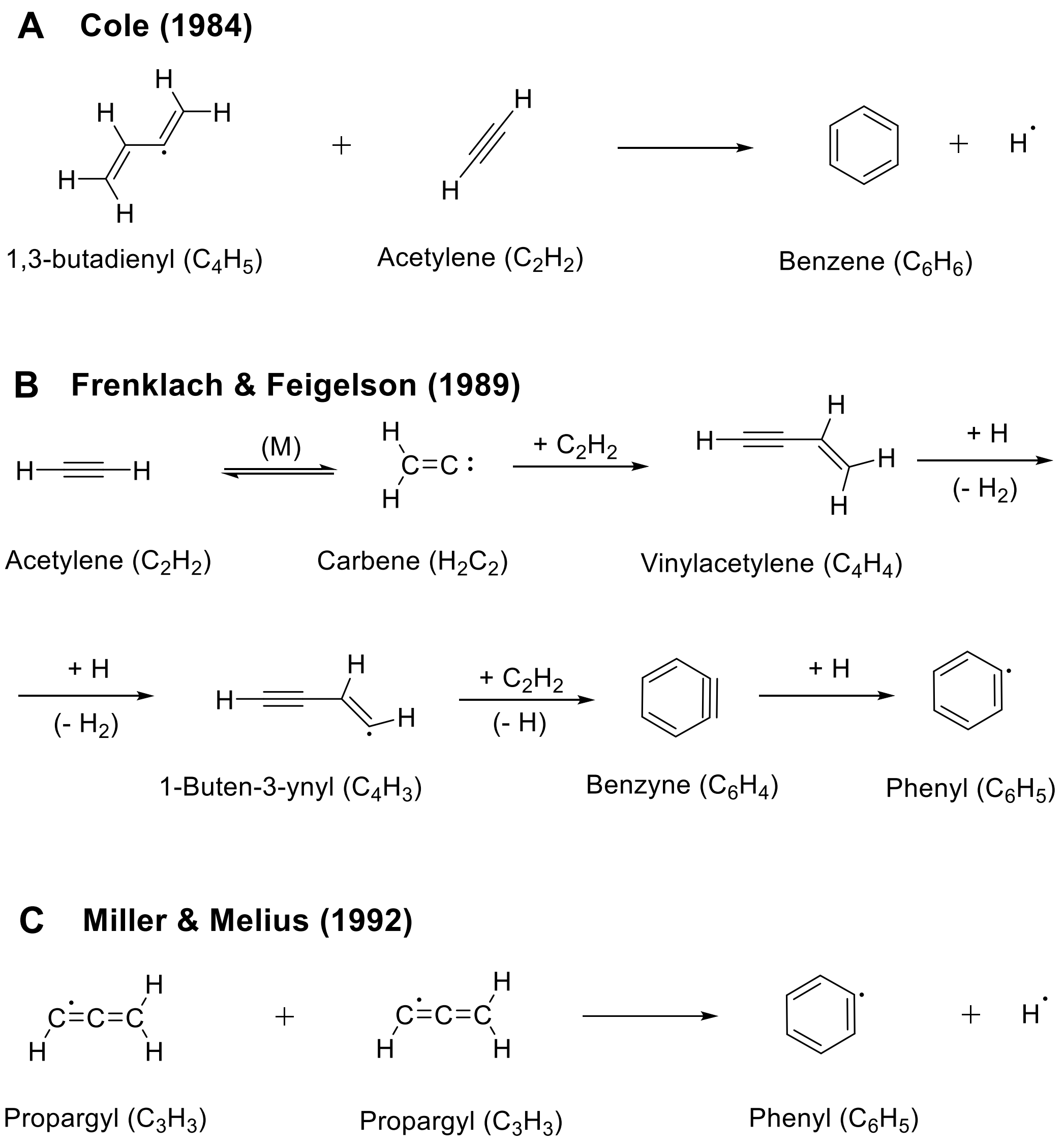

**Scheme 9.** Various schemes for the formation of an aromatic ring at hot temperature.

Subsequently, Cherchneff noted that the propargyl cyclization pathway is dominant, while Cole's and Frenklach's mechanisms are minor; moreover, these pathways will be realized if the reactants are present in the circumstellar gas.[246] The formation of benzene as a result of barrierless self-recombination of the propargyl radical under conditions simulating combustion and in a circumstellar environment using a high-temperature chemical reactor was confirmed in a study by Zhao et al. 2021.[256]

Regardless of the underlying mechanism, acetylene is the main precursor that initiates this reaction. The formation of the propargyl radical occurs as a result of the reaction of acetylene ($C_2H_2$) and methylene ($CH_2$). Other agents originate from vinylacetylene ($C_4H_4$) by the addition of atomic hydrogen to form 1,3-butadienyl ($C_4H_5$) or its elimination to form 1-buten-3-ynyl ($C_4H_3$). In this case, the precursor of vinylacetylene ($C_4H_4$) is also acetylene.

Other possible mechanisms for the formation of the first aromatic ring have been described.[257] For example, in addition to the recombination of propargyl radicals, the reaction between vinylacetylene ($C_4H_4$) and vinyl radical ($C_2H_3$) promoted the formation of benzene.[258,259] Another interesting route is the formation of benzene via cyclopentadienyl (c-$C_5H_5$), which is generated during the reaction of acetylene with propargyl.[260] Lin et al. showed that the reaction of methyl and cyclopentadienyl radicals is a potential source of benzene at high temperatures.[261,262] This pathway combines the advantages of a highly stable propargyl radical and acetylene as a widely available starting building block.

The main reactions of benzene formation are given in Table 9. It should be noted that the mechanisms presented above, based on the analysis of combustion, control aromatic chemistry at temperatures of approximately 1000 K and are not applicable to cold interstellar environments (~ 10 K)[263] or to atmospheres of planets and satellites (~ 90 K temperature of the Titan surface, for example[264]). In addition, these reaction systems do not consider the effects of UV radiation or ion-molecular chemistry.

**Table 9.** The reactions of benzene formation which start from acetylene or its derivatives.

| **Reactions** | **Kinetics** | **Objects & conditions** |
|---|---|---|
| 1.<br>$C_4H_5 + C_2H_2 \rightarrow C_6H_6 + H$[252] | $k_1 = 5.3 \cdot 10^{-13}$ cm$^3$ s$^{-1}$ (T = 1000 K)[252] | Dust formation zone in stellar envelopes.<br>Conditions:<br>$n_H = 10^8$-$10^{12}$ cm$^{-3}$<br>T = 500-2000 K |
| 2.<br>a. $C_2H_2 + M \rightarrow H_2C_2$<br>b. $H_2C_2 + C_2H_2 \rightarrow C_4H_4$<br>c. $C_4H_4 + H \rightarrow C_4H_3\cdot + H_2$<br>d. $C_4H_3\cdot + C_2H_2 \rightarrow C_6H_6$[254] | $k_{2a} = 2.6 \cdot 10^{-16}$ cm$^3$ s$^{-1}$<br>$k_{2b} = 1.9 \cdot 10^{-12}$ cm$^3$ s$^{-1}$<br>$k_{2c} = 1.4 \cdot 10^{-12}$ cm$^3$ s$^{-1}$<br>$k_{2d} = 2.5 \cdot 10^{-14}$ cm$^3$ s$^{-1}$<br>(all rates are given for T = 1000 K)[254] | |
| 3.<br>$C_3H_3\cdot + C_3H_3\cdot \rightarrow C_6H_5 + H\cdot$ | $k_3 = 5.65 \cdot 10^{-11}$ cm$^3$ s$^{-1}$ [265] | |

## 4.2. Benzene formation in protoplanetary nebulae

The presence of benzene as well as di- and triacetylene chains was recorded in protoplanetary nebulae[266]; therefore, it was proposed that these chains can be formed in such environments. These objects represent the next and short stage after an AGB star's death. In the literature, they are also called post-AGB stars. The fusion process in these stars stops, but the surrounding shell still has a high temperature (several hundred kelvins) and is relatively dense (~ $10^7$ cm$^{-3}$). Additionally, the UV radiation intensity from a star gradually increases and may exceed the mean interstellar intensity $G_0$ by 5 orders of magnitude. This factor can play an important role in the activation of ion-molecular chemical reactions. To the best of the authors' knowledge, to date, a chain of ion-molecular reactions leading to benzene formation has been suggested only in the work of Woods et al.[50] They presented a gas-phase chemical model of the protoplanetary nebula CRL 618 and showed that the efficient production of benzene is associated with a high flux of ionizing radiation. Woods et al.[50] reported that $HCO^+$ species can transfer a proton to the $C_2H_2$ parent molecule, triggering a cascade of ionic-molecular transformations. The successive addition of acetylene molecules to cations leads first to a phenyl cation and then to a benzene molecule through the addition of hydrogen and a dissociative reaction (Figure 8). However, this chain of reactions has not been studied either experimentally or theoretically and therefore requires additional approval.

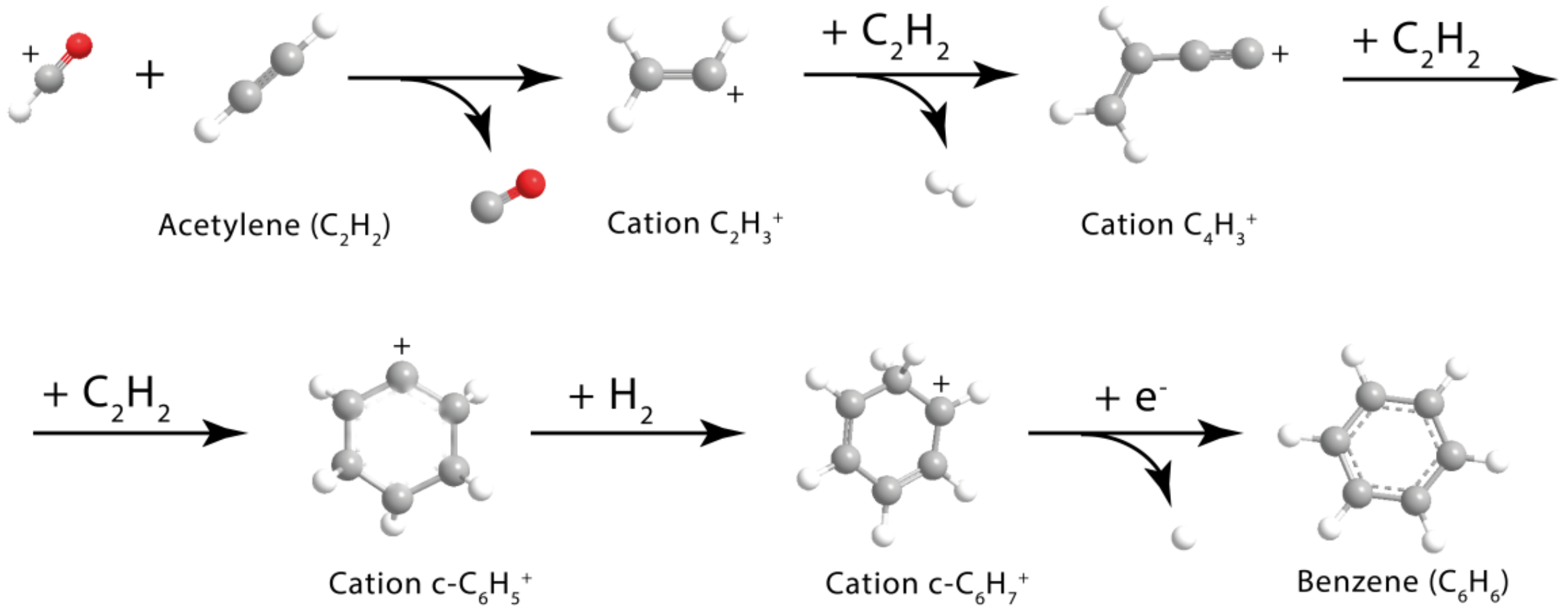

**Figure 8.** Proposed synthesis of benzene in protoplanetary nebula CRL 618.[50] Gray color is C atoms, white is hydrogen, and red is oxygen.

**Table 10.** The chain of ion-molecular reactions which lead to formation of benzene.

| **Reactions** | **Kinetics** | **Objects & conditions** |
|---|---|---|
| a. $HCO^+ + C_2H_2 \rightarrow C_2H_3^+ + CO$<br>b. $C_2H_3^+ + C_2H_2 \rightarrow C_4H_3^+ + H_2$<br>c. $C_4H_3^+ + C_2H_2 \rightarrow cC_6H_5^+$<br>d. $cC_6H_5^+ + H_2 \rightarrow cC_6H_7^+$<br>e. $cC_6H_7^+ + e \rightarrow C_6H_6 + H$[50] | $k_a = 1.4 \cdot 10^{-9}$ cm$^3$ s$^{-1}$ (KIDA)<br>$k_b = 7.2 \cdot 10^{-10}$ cm$^3$ s$^{-1}$ (KIDA)<br>$k_c = 1.4 \cdot 10^{-9}$ cm$^3$ s$^{-1}$ (T=250 K) (KIDA)<br>$k_d = 6.0 \cdot 10^{-11}$ cm$^3$ s$^{-1}$ (T=250 K) (KIDA)<br>$k_e = 8.7 \cdot 10^{-7}$ cm$^3$ s$^{-1}$ (KIDA) | Protoplanetary nebulae.<br>Conditions:<br>T = 250 K<br>$n_H = 10^7$ cm$^{-3}$<br>$G_0 = 2 \cdot 10^5$,<br>but suitable for similar PDR conditions |

## 4.3. Benzene formation in cold interstellar medium

Jones et al. analyzed the above-described mechanisms based on neutral–neutral reactions from combustion chemistry and ion–molecular mechanisms suitable for AGB and post-AGB stars, respectively, and concluded that these reactions cannot lead to the formation of benzene under conditions of low temperatures of 10 K and pressure in cold molecular clouds and prestellar cores, where the density of matter is $10^2$-$10^7$ cm$^{-3}$.[21,267] As a result, Jones et al., based on experiments with a crossed molecular beam in combination with statistical calculations and DFT calculations of the electronic structure, concluded that benzene can be synthesized through a barrierless exoergic reaction of an ethynyl radical and 1,3-butadiene under single collision conditions (Scheme 10).[21] The authors proposed the following sequence of reactions: 1) ethane is formed as a result of the addition of a hydrogen atom to acetylene or reactions between fragments of the methyl radical ($CH_3$); 2) the fast reaction of methylidyne radicals (CH) with ethane ($C_2H_6$) leads to the formation of propene ($C_3H_6$) and atomic hydrogen; 3) then, propene reacts rapidly with another methylidyne radical, forming 1,3-butadiene ($C_4H_6$); and 4) in the last step, the ethynyl radical and 1,3-butadiene give benzene and hydrogen. Additionally, the ethynyl radical can react with 1,3-butadiene to form an acyclic isomer.

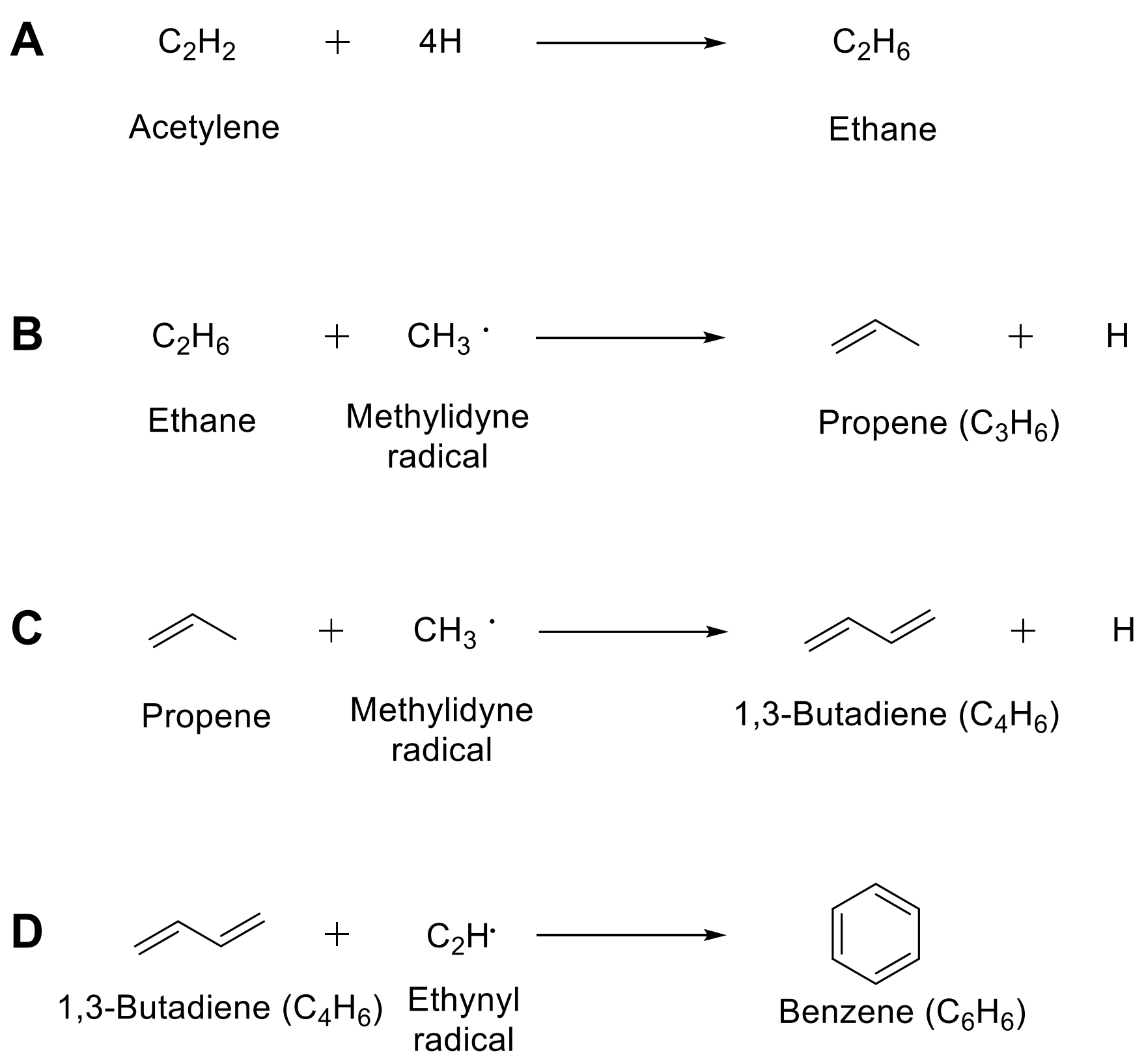

**Scheme 10.** Proposed mechanism of benzene formation through a barrierless exoergic reaction of an ethynyl radical and 1,3-butadiene.

Ethane is unlikely to form under gas chemistry conditions. Most likely, its synthesis occurs on dust particles inside the icy mantle from acetylene. Such reactions involving the formation of ethane were described earlier in experiments simulating the conditions of interstellar ice: acetylene hydrogenation of methyl radical recombination.[159,268] Additionally, high ethane abundance has been detected for the Oort cloud comets by IR observations.[121] When formed in ice[159,268], ethane can be released into the gas phase as a result of the thermal desorption of the icy mantle. However, this process occurs only at temperatures higher than ~60 K[159], and it is relevant for hot core objects or comets approaching perihelion. In the case of cold clouds, an icy mantle can be sublimated via nonthermal desorption due to ion bombardment and grain–grain collisions during the propagation of shock waves.[269,270]

Extensive experiments on barrierless reactions involving the formation of benzene or related molecules have been conducted. Independent experimental evidence for a one-step radiation-induced formation of benzene from isolated trimeric acetylene was presented in a recent study by Lukianova and Feldman.[271] Using Fourier IR spectroscopy, the authors showed that the decomposition of the acetylene trimer and the accumulation of benzene occur in a frozen inert medium at a very low temperature (6 K). Thus, the $(C_2H_2)_3$ trimer is an ideal precursor for the formation of benzene molecules due to the geometric correspondence between the trimer structure and the shape of the benzene molecule. It is assumed that the formation of benzene in cryogenic films may be related to the cationic mechanism known in the gas phase. This mechanism implies ionization of the trimer followed by almost barrier-free transformation into a stable benzene radical cation. The described mechanism of radiation-induced benzene production without any significant contribution from molecular diffusion could be considered an interesting route for the formation of aromatic rings under cold space conditions. However, the likelihood of the $C_2H_2$ trimer existing in space conditions is relatively low.

In addition to benzene, benzyl radicals and *ortho*-benzynes can form.[272,273] These compounds may further form benzene in subsequent reactions. These reactions have also been investigated with quantum mechanical calculations.[274,275,276] Thus, benzene has been established to be formed not only by fusion but also at very low temperatures.

**Table 11.** The reactions of benzene formation in cold interstellar medium.

| **Reactions** | **Kinetics** | **Objects & conditions** |
|---|---|---|
| 1.<br>a. $C_2H_2$ + 4H → $C_2H_6$ (surface reaction)<br>b. $C_2H_6$ + CH·→ $C_3H_6$<br>c. $C_3H_6$ + CH· → $C_4H_6$<br>d. $C_4H_6$ + $C_2H$· → $C_6H_6$[21] | $k_a$ = -<br>$k_b$ = 5.8 · $10^{-13}$ $cm^3$ $s^{-1}$ (KIDA)<br>$k_c$ = 3.6 · $10^{-10}$ $cm^3$ $s^{-1}$ (KIDA)<br>$k_d$ = 3.0 · $10^{-10}$ $cm^3$ $s^{-1}$ [21] | Molecular clouds, prestellar cores.<br>Conditions:<br>T = 10 K<br>$G_0$ = very low<br>Galactic CR field |
| 2. Trimerization<br>3$C_2H_2$ + CR → $C_6H_6$[271] | n.d. | |

## 4.4. Benzene formation in planets and moons

Benzene has been observed in the atmospheres of Jupiter, Saturn and Titan by ISO and interplanetary spacecrafts such as Voyager and Cassini.[277] The paths from small hydrocarbons to benzene in the atmospheres of the giant planets and their satellites are assumed to be similar to the paths considered. Wong et al. proposed a model for predicting the formation of benzene and PAHs in the atmosphere of Jupiter.[278] Wong et al. concluded that ion–molecular reactions play the most important role, so the model considers a system of ion–molecular processes and neutral chemistry (Scheme 11). The model included 288 reactions of ion-neutral exchange and 79 reactions of electron-ion recombination. The system of transformations begins with the destruction of methane by energetic particles, in addition to the reactions of acetylene. Neutral chemical reactions contribute only approximately a few percent of the total reactions. Among them are the most important $C_3H_3$ recombinations and the addition of $C_2H_2$ to $C_4H_3$. The described chemical processes, as the authors suggest, may also be relevant for other giant planets, such as Saturn, and extrasolar giant revolves close to their main orbit and receive high doses of far-UV radiation.

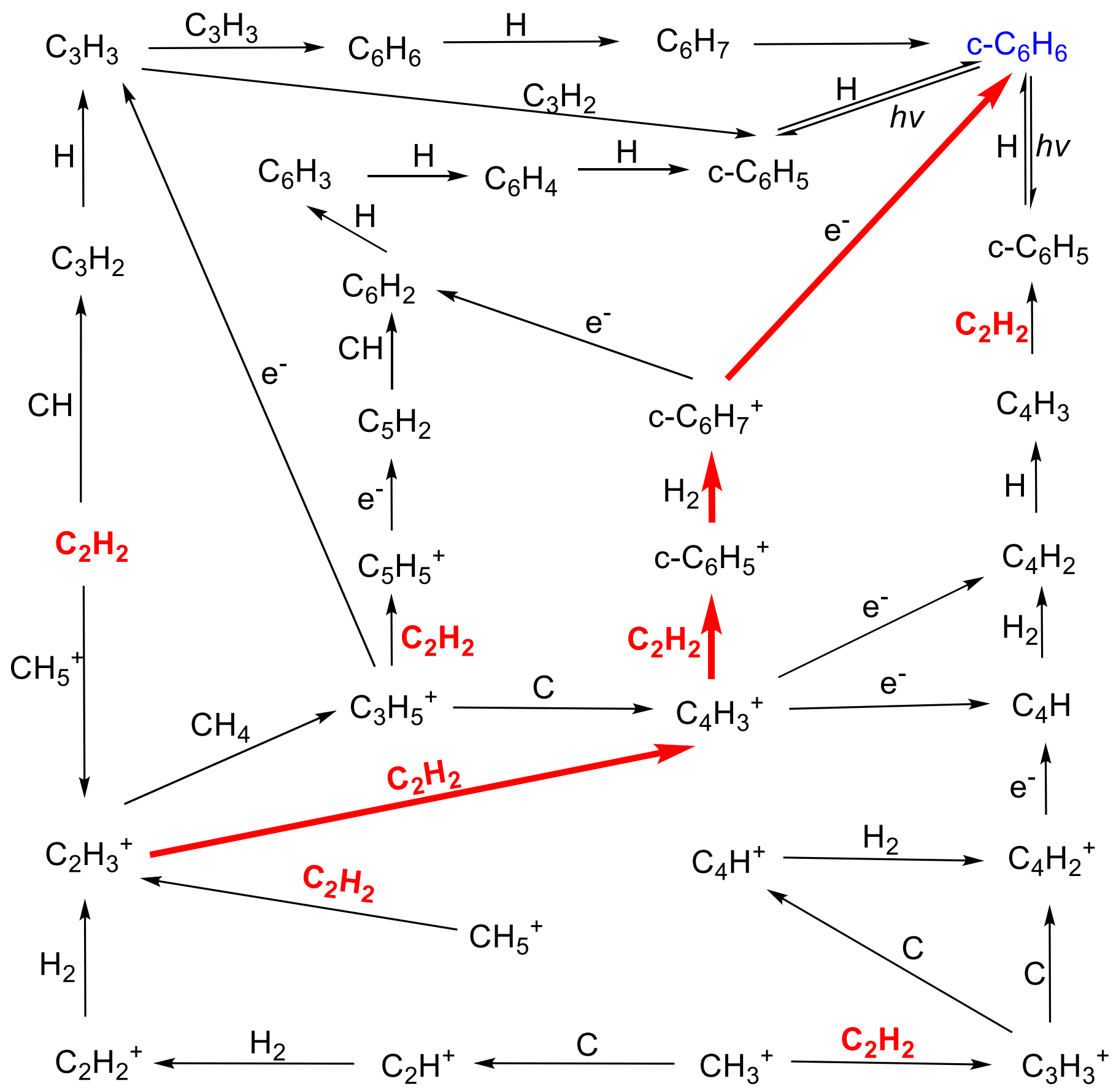

**Scheme 11.** The most important reaction pathways for the formation of benzene in Jupiter according to the model of Wong et al.[278] One of the main chains of reactions designated by red arrows.

Wilson et al. (2003)[279], Lebonnois (2005)[280], and Vardavas (2008)[109,281] explored possible ways for benzene to form in Titan's atmosphere (Figure 9). These pathways include 1) recombination of two propargyl radicals, 2) reaction between acetylene and butadiene radical and 3) subsequent reactions of diacetylene with hydrogen and acetylene. Thus, in these models for the conditions of planetary atmospheres and their satellites, the same reactions play a key role as in the models of benzene formation at temperatures of 1000 K and above.

The photochemical model proposed by Vardavas et al.[109] suggested that three-body reactions are the main route for benzene production. However, the predictions of this model give more than a twofold underestimation of the benzene abundance in the thermosphere of Titan. In addition, given the planet's rarefied thermosphere, three-body reactions are unlikely. In addition, the abundance of benzene in the thermosphere of Titan is greater than the abundance of benzene in the stratosphere.[282,283]

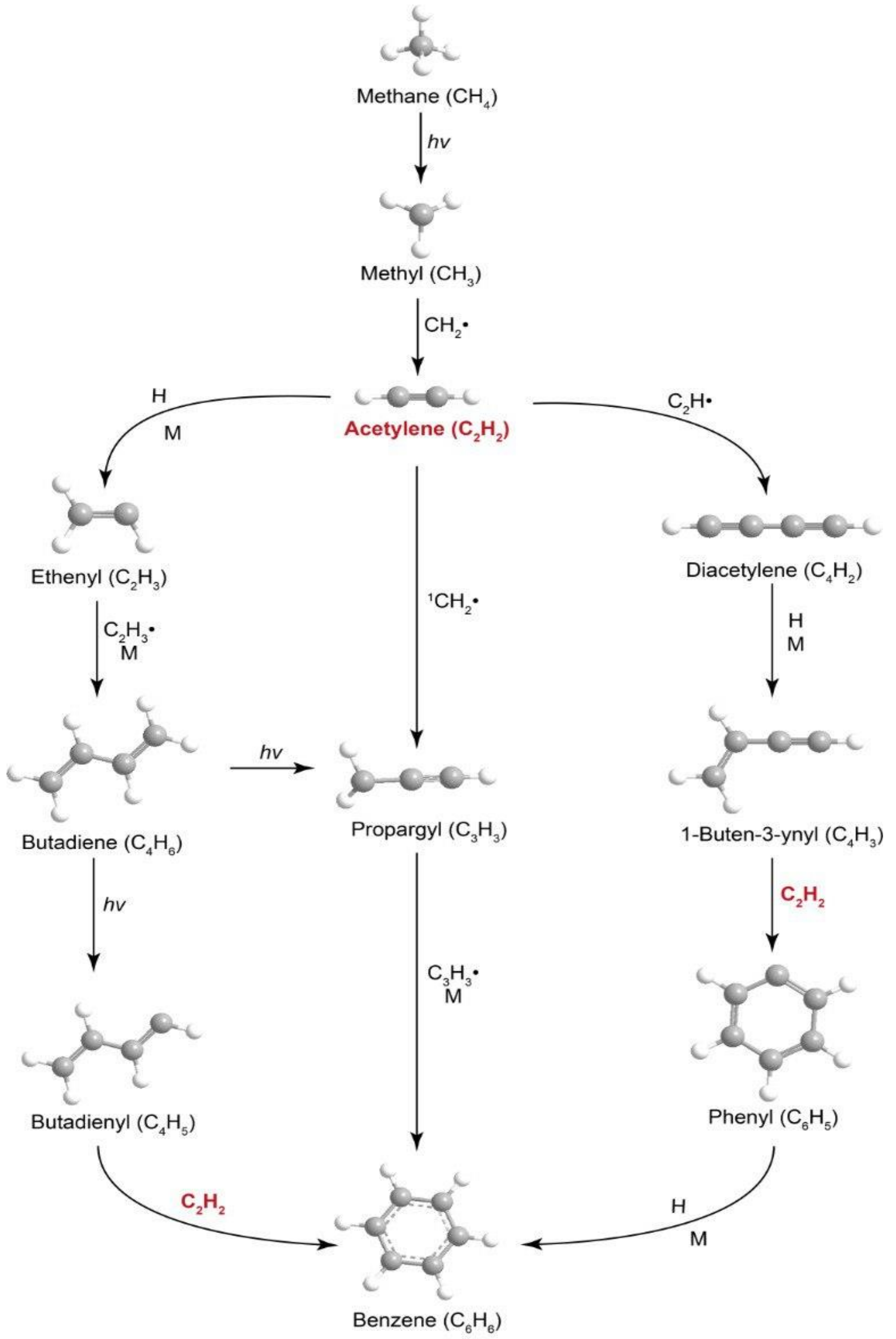

**Figure 9.** Schematic diagram of the mechanisms of formation of benzene according to the model of Wilson et al.[279] M is a third neutral molecule required for energy balance in the reactions.

Based on the analysis of altitude data, Vuitton et al.[284] concluded that benzene is synthesized precisely in the thermosphere and is not the result of diffusion processes from the stratosphere, where the mole fraction of benzene is lower (Scheme 12).[104] Vuitton et al. noted that the rate of neutral-neutral reactions is not enough to form a sufficient amount of benzene; therefore, a model based on the chemistry of ion-molecular reactions was proposed for the upper atmosphere (thermosphere and ionosphere) of Titan. Acetylene plays an important role in this model and is responsible for the formation of benzene precursors such as $C_4H_5^+$ and $C_6H_5^+$.[285]

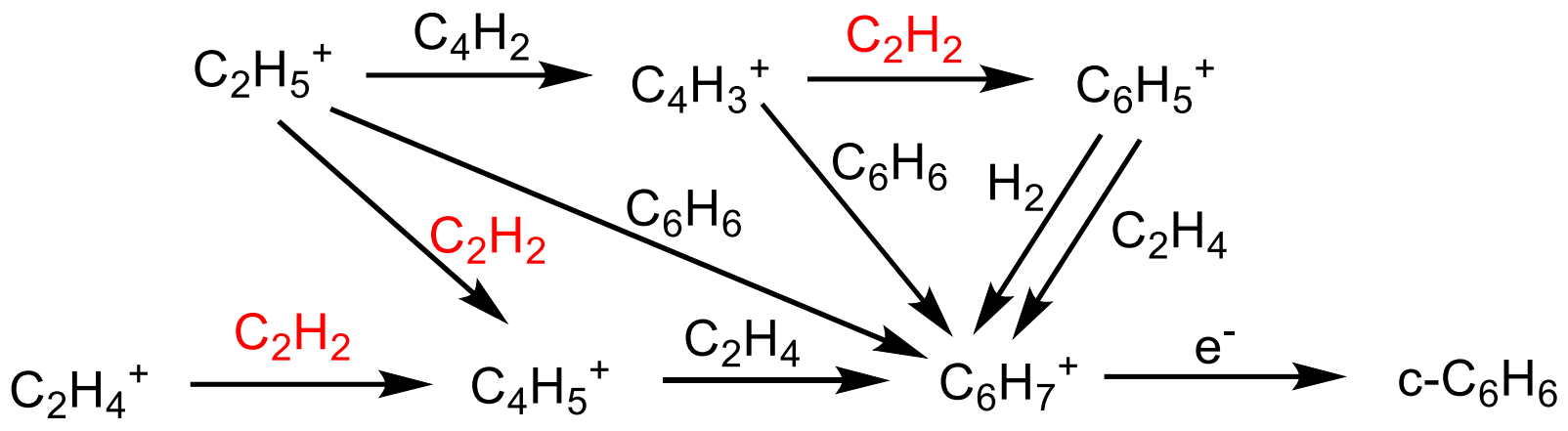

**Scheme 12.** The main ion-chemical reactions of the formation of benzene in the atmosphere of Titan.[284]

Among other substances, benzene can be formed directly from acetylene on the surface of Titan. Experiments on the irradiation of acetylene ices at low temperature with energetic electrons have shown that CR-mediated production of benzene directly from three acetylene molecules on the surface of Titan may be the dominant source of benzene.[286]

The two singlet ground-state acetylene molecules do not react with each other since the entrance barrier to the reaction is too large and equals to 140 kJ·mol$^{-1}$. However, the calculations showed that the acetylene molecule in the first excited triplet state $C_2H_2$ ($a^3B_2$) can indeed easily (the barrier value is only 4 kJ mol$^{-1}$) react with the acetylene molecule in the ground state to form the trans-$C_4H_4$ structure [1] (Figure 10). The resulting molecule can undergo trans-cis isomerization and form [2]. In turn, the triplet structure [2] can react with the second acetylene molecule in the ground state to form a structure [3]. The reaction of cis-CHCHCHCH [2] and $C_2H_2$ ($X^1\Sigma_g^+$) can be possible even at low temperatures if [2] retains enough energy gained from the initial interaction of two acetylene molecules. Then, the intermediate triplet can easily rearrange into a structure [4], followed by ring closure to triplet benzene [5].

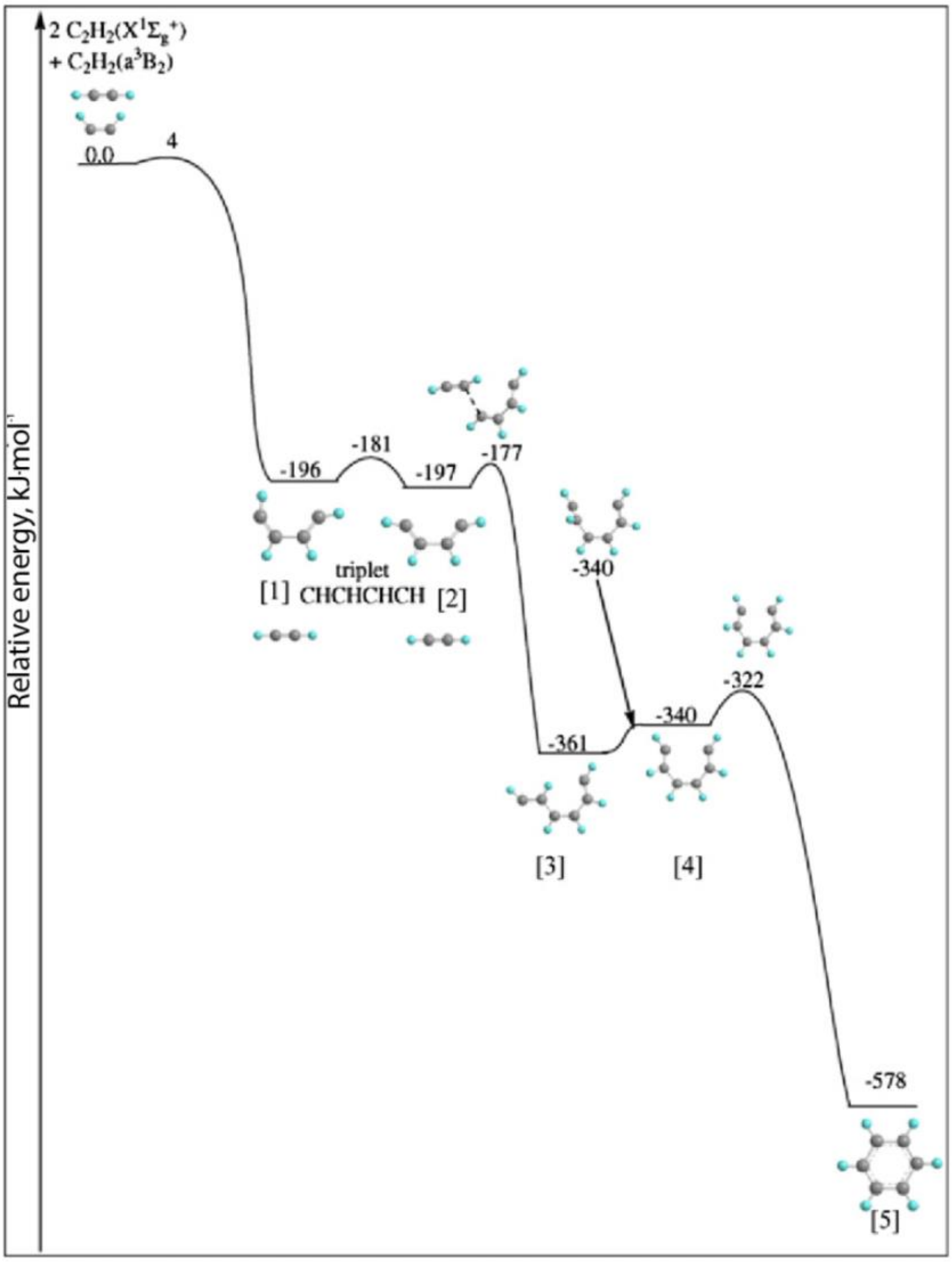

**Figure 10.** Part of the potential energy surface of the $C_6H_6$, triplet involved in the formation of the benzene triplet through reactions with excited acetylene. Reproduced with permission from ref 284. Copyright 2010 The American Astronomical Society.

In their experimental study, Kleimeier et al.[287] showed that the photolysis of solid acetylene by low-energy UV photons can lead to the formation of benzene. The authors suppose that this mechanism is the main mechanism for Pluto's atmosphere and other outer planets where acetylene is very abundant according to New Horizons's observations.

The reactions discussed in this section are given in Table 12. We emphasize that over the past few decades, ideas about the mechanisms of synthesis of the first aromatic ring have expanded from models borrowed from combustion chemistry for high-temperature environments to experiments with crossed molecular beams for barrierless reactions under single-collision conditions in cold interstellar environments. Moreover, for almost all the proposed mechanisms, acetylene, as a key building block, takes part, if not directly, in the mechanism of formation of the aromatic ring and then as a precursor of the initial reactants. In this regard, acetylene molecules are exemplary due to their structure, properties, and wide distribution in various cosmic environments, which facilitates their participation in a wide variety of reactions.

**Table 12.** Summary of considered reactions, kinetics (if available) and conditions where the reactions are relevant

| **Reactions** | **Kinetics** | **Objects & conditions** |
|---|---|---|
| Neutral-neutral reactions<br>1. $C_3H_3 + C_3H_3 \rightarrow C_6H_5 + H$ | $k_1 = 4.98 \cdot 10^{-12}$ cm$^3$ s$^{-1}$ [278] | Planetary atmospheres.<br>Conditions:<br>$n_H = 10^{10}$-$10^{16}$ cm$^{-3}$<br>T = 50-300 K<br>$G_0$ = up to $10^2$<br>Galactic CR field |
| 2. $C_3H_3 + C_3H_3 + M \rightarrow C_6H_6$ | $k_2 = 2.1 \cdot 10^{-23}$ cm$^6$ s$^{-1}$<br>(T = 150 K, $n_H = 10^{12}$ cm$^{-3}$)[280] | |
| 3. $C_2H_2 + C_4H_3 + M \rightarrow C_6H_5$ | $k_3 = 2.1 \cdot 10^{-29}$ cm$^6$ s$^{-1}$ (150 K)[278] | |
| 4.<br>a. $C_2H_2 + h_v \rightarrow C_2H + H$<br>b. $C_2H_2 + C_2H \rightarrow C_4H_2 + H$<br>c. $C_4H_2 + H + M \rightarrow C_4H_3$<br>d. $C_4H_3 + C_2H2 \rightarrow C_6H_4 + H$<br>e. $C_6H_4 + H + M \rightarrow C_6H_5$<br>f. $C_6H_5 + H + M \rightarrow C_6H_6$ | $k_{4d} = 6.3 \cdot 10^{-20}$ cm$^3$ s$^{-1}$ (150 K)<br>$k_{4f} = 1.7 \cdot 10^{-22}$ cm$^6$ s$^{-1}$ (150 K)[336] | |
| 5. $C_4H_5 + C_2H_2 \rightarrow C_6H_6 + H$<br>(Titan) | $k_5 = 2.9 \cdot 10^{-19}$ cm$^3$ s$^{-1}$ (150 K)[278] | |
| Ion-molecular reactions<br>6.<br>a. $C_2H_3^+ + C_2H_2 \rightarrow C_4H_3^+ + H_2$<br>b. $C_4H_3^+ + C_2H_2 \rightarrow cC_6H_5^+$<br>c. $cC_6H_5^+ + H_2 \rightarrow cC_6H_7^+$<br>d. $cC_6H_7^+ + e \rightarrow C_6H_6 + H$ | $k_{6a} = 2.16 \cdot 10^{-10}$ cm$^3$ s$^{-1}$<br>$k_{6b} = 2.2 \cdot 10^{-10}$ cm$^3$ s$^{-1}$<br>$k_{6c} = 6.0 \cdot 10^{-11}$ cm$^3$ s$^{-1}$<br>$k_{6d} = 7.1 \cdot 10^{-7}$ cm$^3$ s$^{-1}$ [278] | |
| 7.<br>a. $C_2H_4^+ + C_2H_2 \rightarrow C_4H_5^+ + H$<br>b. $C_4H_5^+ + C_2H_4 \rightarrow C_6H_7^+$<br>c. $C_6H_7^+ + e \rightarrow C_6H_6 + H$[284] | $k_{7a} = 1.9 \cdot 10^{-10}$ cm$^3$ s$^{-1}$<br>$k_{7b} = 7.4 \cdot 10^{-11}$ cm$^3$ s$^{-1}$<br>$k_{7c} = 5.9 \cdot 10^{-6}$ cm$^3$ s$^{-1}$ (150 K)[284] | |
| Trimerization:<br>8. $3C_2H_2 + CR \rightarrow C_6H_6$ | $k_8 = 3$-$5 \cdot 10^{-3}$ mol eV$^{-1}$ [286] | Titan's surface.<br>Conditions:<br>T~100 K,<br>Galactic CR field |

# 5. Mechanisms for the Formation of Polyaromatic Compounds in Space

## 5.1. PAHs in space and their role in carbon evolution

The formation of the first aromatic ring is the first step in the growth of a large number of polycyclic aromatic systems. PAHs are responsible for unidentified infrared bands (UIB) in the range from 3 to 20 µm.[288,289,290] The main bands are positioned at 3.3, 6.2, 7.7, 8.6, and 11.2 µm. These bands appear in most galactic and extragalactic sources[291]. The peak wavelengths and relative strengths of the bands can vary from object to object; nevertheless, the spectra are similar and indicate the ubiquitous presence of PAHs in the ISM. PAHs, their hydrogenated, dehydrogenated[292], protonated[293], alkylated[294,295,] ionized[296] derivatives, fullerenes and other related carbonaceous molecules are considered promising candidates for so-called diffuse interstellar bands, which are discrete absorption features superimposed on the interstellar extinction curve in the range from the blue region of visible light (400 nm) to near IR (1200 nm).[297,298] The extinction curve in the UV range has a prominent bump at 2175 Å, which is most likely related to electronic transitions in PAHs or related molecules. Moreover, the significant extinction in the UV range and its increase at wavelengths down to 0.1 μm indicate that a large number of small species are comparable to PAH sizes. Other observational features, such as extended red emission and anomalous microwave emission, can also be mentioned as signatures of PAH presence in the ISM, although decoding of these signatures is complicated.[299] Despite a great number of observations, specific interstellar PAHs have been scarcely identified. The most precise identification of interstellar PAHs to date involved the discovery of cyanonaphtalene[300] as well as indene[301]. PAHs have been identified in several dozen carbonaceous chondrites, such as Murchison, Allende, and Orgueil[18,302,303,304,305,306,307,308], and their supposed structures are illustrated in Figure 11.

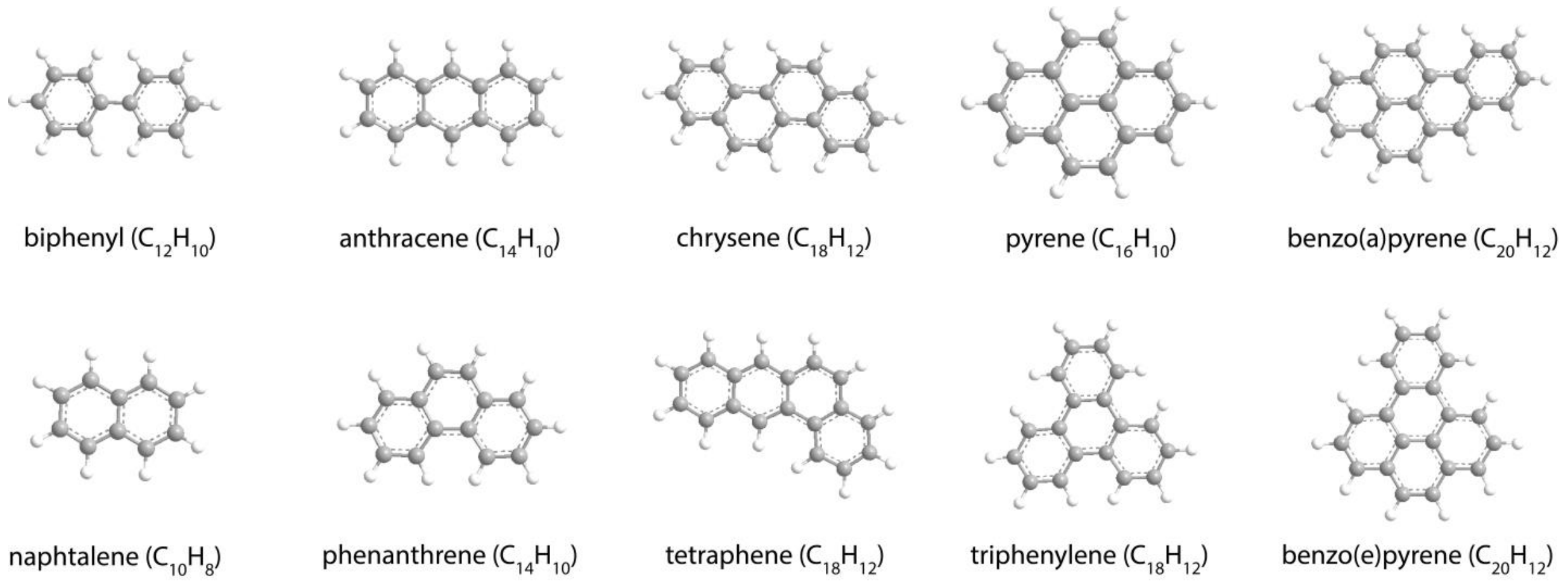

**Figure 11.** Possible structures of aromatic molecules found in carbonaceous meteorites.

PAHs and carbon dust are found throughout the cosmos: in circumstellar media; in planetary and reflection nebulae[309]; in molecular clouds; in interstellar, interplanetary and intergalactic media[310]; and finally in the atmosphere of planets and in meteorites. According to different estimates, up to 20% of carbon can be locked in the form of PAHs, the size of which can exceed 50 carbon atoms.[60,291,311,312,313,314] These molecules play an important role in the ISM, determining its thermal and ionization balance.[170] PAHs are intermediates between low-molecular-weight hydrocarbons and carbon particles and provide nucleation sites for the formation of carbon dust particles.[315] Tielens noted[49] that PAH molecules influence key processes in the evolution and characteristics of interstellar media and galaxies through interactions with gases, photons and energetic ions; provide active surfaces for chemical processes; participate in the regulation of opacity; and influence the heating and cooling of neutral atomic and molecular gases.

PAHs can participate in complex networks of transformations that occur in stellar ejecta and even play the role of catalysts, thereby affecting the overall chemical composition of star formation and planet formation regions.[291,316] The most impressive example of using PAHs as catalysts is the formation of hydrogen molecules.[317] In the atmospheres of planets and satellites, PAHs are responsible for the formation of atmospheric organic haze.[318,319] In addition, the PAH world hypothesis suggests that PAHs could play a role in astrobiological evolution as energy transduction elements, container elements and templating genetic components.[44,320,321] Therefore, the mechanisms of PAH formation are of fundamental importance for understanding molecular evolution in the universe, the origin of carbon chemistry and prebiotic systems.

In addition to the case of benzene, first, astrochemical models of PAH formation were built from models of combustion flames.[38] Considering the temperature in the circumstellar envelopes (~500-2500 K), such borrowing can be considered justified. However, current theoretical studies point out inconsistencies between the stellar yield of PAHs and the estimated abundance of interstellar PAHs. PAHs can be rapidly destroyed by photolysis, interstellar shocks and cosmic ray bombardment.[165,322,323] The destruction of PAHs occurs through either the loss of the $C_2H_2/C_2$-fragment from the periphery of the molecule with the preservation of the aromatic structure or direct ejection of atoms from a skeleton in the case of elastic collisions.[324,325] This destruction is not compensated by the rate of formation by the mechanisms proposed for AGB stars. For the lifetime of PAHs in the ISM, there are estimated to be $10^8$ years, which is noticeably shorter than the $2 \cdot 10^9$ years required for the injection of PAHs into the ISM from AGB stars.[322,323,324,326] Therefore, PAHs should not exist either in the ISM or in meteorites if we consider injection from C-rich AGB stars only. Therefore, the ubiquitous presence of PAHs in the ISM represents a key paradox.[290,39] On the one hand, this paradox motivates the study of mechanisms in hot stellar envelopes in more detail. On the other hand, it provides reasons to search for alternative mechanisms of PAH formation. Among these alternative mechanisms, a "top-down" scenario can be mentioned. Hirashita proposed a scenario in which PAHs are produced as a result of fragmentation of graphite or amorphous grains in grain–grain collisions at high velocities induced by interstellar shocks.[327] The fingerprint of dust in supernova remnants also hints at the possible formation of PAHs in such highly turbulent media.[328] In turn, Sandstrom et al.[329] showed that the abundance of PAHs is strongly correlated with CO emission, i.e., cold dense clouds, which indicates that PAHs can be generated in such environments. Moreover, the detection of cyanonaphthalene in molecular cloud TMC-1 indicates that the formation must occur in a straightforward manner in this cloud, as such small PAHs are unlikely to survive during a trip from parent stellar envelopes to molecular clouds.[165,325] To date, several mechanisms that can proceed at low temperatures have been suggested, and C-rich AGB stars are far from being the only source of PAHs.

Thus, the understanding of the formation of PAHs and the role of acetylene in the growth of hydrocarbon molecules has progressed substantially over the past couple of decades from high-temperature mechanisms that can only work in circumstellar envelopes to barrierless mechanisms that work in cold media and cosmic-ray-initiated reactions in acetylene ices on the surface of satellites and planets. Such significant advances in understanding the mechanisms of PAH synthesis are due to experiments involving cross-molecular beams under single-collision conditions, pyrolytic microreactor experiments, and high-level electronic structure calculations.

In a 2021 review, Kaiser and Hansen[33] mentioned five elementary stages that are responsible for the formation of PAHs in a wide range of extreme space conditions, from cold molecular clouds (10 K) and hydrocarbon-rich atmospheres of planets and their moons (50–300 K) to high-temperature circumstellar envelopes of C-rich AGB stars (~1000 K): 1) Hydrogen Abstraction–Acetylene Addition (HACA); 2) Hydrogen Abstraction–Vinylacetylene Addition (HAVA); 3) Phenyl Addition–Dehydrocyclization (PAC); 4) Radical-Radical Reactions (RRR); and 5) Methylidyne Addition-Cyclization-Aromatization (MACA). Recently, He et al.[330] supplemented the list with the Propargyl Addition–BenzAnnulation (PABA) mechanism. In fact, several alternative mechanisms have been proposed in the literature. For instance, PAHs are suggested to be formed through polyyne polymerization (Section 5.2.4.). In addition to bottom-up mechanisms, PAHs are thought to be generated as a result of the fragmentation of carbon grains exposed to interstellar shocks, as noted above.[323,327,331] The mechanisms of PAH formation discussed in this review are listed in Table S2 in the Supporting Information.

This variety of pathways makes PAH systems quite complex.[332] Some of the suggested mechanisms can work simultaneously, compete with each other and provide aromatic systems of different sizes and structures, which leads to complicated and varying observational signatures. The acetylene molecule plays an important role in most of the discussed mechanisms, either directly in the formation of complex aromatic systems or as a precursor of key reagents.

In this section, we aim to discuss the mechanisms of PAH formation related to acetylene chemistry, while we omit discussions on other possible PAH formation pathways here. We refer interested readers to a number of specialized reviews in which the full spectrum of mechanisms is described.[33,34,35,36,291,326] As in the previous chapter, we divided the mechanisms according to the conditions under which they work; namely, we considered the mechanisms for hot environments applicable to stellar envelopes and protoplanetary nebulae, for cold dense clouds and for Solar system objects. Additionally, we provide information on the reactions leading to the formation of PAHs with five-membered rings.

## 5.2. Mechanisms of PAH formation in hot environments

### 5.2.1. HACA mechanism

The most popular mechanism for the growth of an aromatic system larger than benzene is the HACA mechanism, which was explicitly proposed by Bockhorn et al.[333] Independently, in 1989, Frenklach & Feigelson investigated the mechanism of PAH formation using a chemokinetic approach.[254] They specified a similar sequence of chemical reactions for the formation of soot in a hydrocarbon flame and gave the mechanism the name "Hydrogen Abstraction–Acetylene ($C_2H_2$) Addition" (HACA), which reflects that the sequence includes repeating abstraction of hydrogen followed by the addition of acetylene.[334] On this basis, Frenklach & Feigelson created a model for astrophysical conditions, taking into account the lower pressure and density.

Frenklach noted[335] that the relationship between "thermodynamic resistance" and reaction reversibility and the kinetic driving force is a defining feature of the "HACA model". In a recent review, Frenklach and Mebel emphasized[35] that the kinetic-thermodynamic interaction indeed determines the HACA mechanism but not necessarily the nature of the carbon growth forms. The value of HACA can be extended to cover other forms of carbon growth as long as the kinetic-thermodynamic basis is maintained. In addition, aromatic site activation (the first step in the HACA mechanism) can proceed not only through abstraction but also through hydrogen addition. Thus, Frenklach and Mebel propose considering HACA more generally as an abbreviation for H-Activated-Carbon-Addition, characterized by an underlying kinetic-thermodynamic relationship, where "H-Activated" implies "H-Abstraction", "H-Addition" or even "H-Migration".

However, researchers usually consider the HACA mechanism to be a specific case of aromatic ring formation. In the simplest case, the HACA mechanism starts from benzene ($C_6H_6$) and leads first to the phenyl radical ($C_6H_5$) (Scheme 13A). Then, an acetylene molecule is added to form a styrenyl radical ($C_8H_7$) (Scheme 13B). In the next stage, acetylene is added to the $C_8H_7$ species (Scheme 13C). In the last stage, cyclization occurs via the elimination of hydrogen to form naphthalene ($C_{10}H_8$), the simplest PAH molecule (Scheme 13D). This process occurs at elevated temperatures up to 1000 K.[30,336]

**A** $C_6H_6 + H^{\bullet} \longrightarrow C_6H_5^{\bullet} + H_2$

**B** $C_6H_5^{\bullet} + C_2H_2 \longrightarrow C_8H_7^{\bullet}$

**C** $C_8H_7^{\bullet} + C_2H_2 \longrightarrow C_{10}H_9^{\bullet}$

**D** $C_{10}H_9^{\bullet} \longrightarrow C_{10}H_8 + H^{\bullet}$

**Scheme 13.** HACA mechanism.

Notably, the reaction described in Scheme 13 can proceed in two different ways.[33,337,338] The acetylene molecule can directly react with the styrenyl radical ($C_8H_7$) and form the 4-phenylbuta-1,3-dien-1-yl radical ($C_{10}H_9$). In this case, mutual transformations of the corresponding geometric isomers can occur. As a result, cyclization of the 4-phenyl-(Z)-buta-1,3-dien-1-yl radical ($C_{10}H_9$) occurs with the formation of the 1,8-dihydronaphthalen-1-yl radical ($C_{10}H_9$), from which hydrogen abstraction leads to the formation of naphthalene. This pathway is designated as the Bittner–Howard reaction sequence (Scheme 14).

The second pathway is associated with the isomerization of the styrenyl radical ($C_8H_7$) to the *ortho*-vinylphenyl radical ($C_8H_7$). Acetylene is attached to this radical, which results in the formation of *ortho*-vinylstyrenyl radicals ($C_{10}H_9$). Its cyclization leads to the formation of a 1,2-dihydronaphthalen-1-yl radical ($C_{10}H_9$). Similarly, the elimination of hydrogen leads to the formation of a molecule of naphthalene ($C_{10}H_8$). This pathway is designated as the Frenklach reaction sequence (Scheme 14). Other options for the Frenklach route have also been proposed, involving the formation of an *ortho*-vinylphenyl radical ($C_8H_7$) via the direct reaction of a phenyl radical ($C_6H_5$) and $C_2H_2$, isomerization of a styrenyl radical ($C_8H_7$), or a pathway involving H-addition styrene formation.[337]

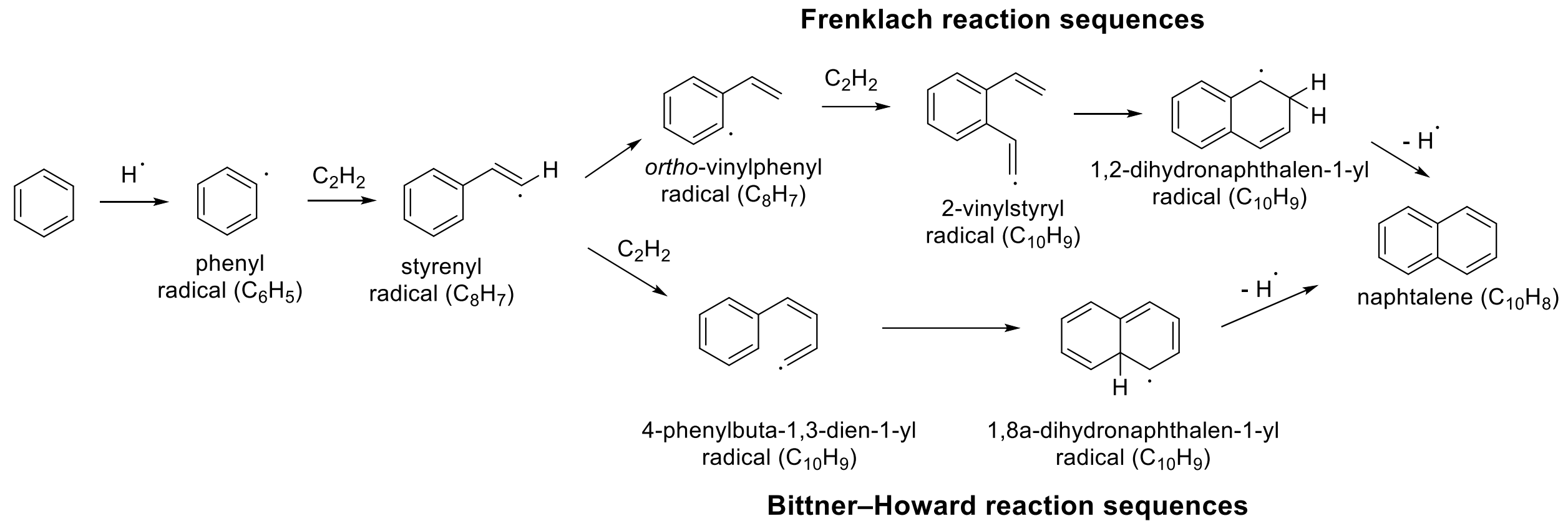

**Scheme 14**. Possible branches of the HACA mechanism after formation of styrenyl radical.

Initially, the mechanism was only hypothesized, but in 2014, it was experimentally confirmed by Parker et al.[37] who provided strong evidence that the naphthalene molecule together with phenylacetylene can be formed in a pyrolytic reactor from a phenyl radical and acetylene precursors via the HACA pathway. Photoionization mass spectrometry detection of naphthalene in combination with phenylacetylene successfully demonstrated reaction pathways consistent with the HACA mechanism. In 2016, Kaiser et al.[338] gave additional experimental proof of the easy formation of naphthalene in a simulated combustion environment by the HACA mechanism involving styrenyl $C_8H_7$ and *ortho*-vinylphenyl $C_8H_7$ radicals, which are key transitional species of the HACA mechanism.

Frenklach and Feigelson[254], who adapted the HACA mechanism to stellar envelopes, concluded that PAH formation should occur within the temperature range between 900 and 1100 K and that large densities and slower stellar winds are required to provide the observed PAH abundance. Their original chains were slightly different from those in Scheme 13 and are presented in Table 13 including the reaction rates. Cherchneff[339] developed the dust condensation model by adding shock-induced chemistry, and it was concluded that in this case, the observed carbonaceous dust mass can be explained by the HACA mechanism only, although subsequent dust destruction in the ISM has not been taken into account.

Another mechanism, the RRR, is supposed to work at high temperatures. For example, this process involves the recombination of two cyclopentadienyl radicals.[340] However, this pathway has not been studied in detail as much as HACA has. Therefore, the study of possible pathways is still ongoing.

**Table 13.** The chain of reactions from benzene to naphthalene by the HACA mechanism.

| **Reactions** | **Kinetics** | **Objects & conditions** |
|---|---|---|
| a. $C_6H_6 + H\cdot \rightarrow C_6H_5\cdot + H_2$<br>b. $C_6H_5\cdot + C_2H_2 \rightarrow C_8H_7\cdot$<br>c. $C_8H_7\cdot + C_2H_2 \rightarrow C_{10}H_8 + H$ | $k_a = 1.5 \cdot 10^{-12}$ cm$^3$ s$^{-1}$ (T = 1000 K)<br>$k_b = 1.7 \cdot 10^{-11}$ cm$^3$ s$^{-1}$<br>$k_c = 5 \cdot 10^{-10}$ cm$^3$ s$^{-1}$[254] | Stellar envelopes (outer edge).<br>Conditions:<br>$n_H = 10^8$-$10^{12}$ cm$^{-3}$<br>T = 500-1500 K |

## 5.2.2. Mechanism of formation of three- and four-ring PAHs in the circumstellar media

A number of experiments have shown that the formation of PAHs with more than two rings via the HACA mechanism is problematic. Kislov et al. performed DFT calculations followed by kinetic analysis of various reaction pathways and showed[341] that the HACA mechanism essentially ends with the formation of naphthalene ($C_{10}H_8$) and acenaphthalene ($C_{12}H_8$). More recently, Parker et al.[342] in 2015 confirmed these conclusions experimentally. Thus, once naphthalene is formed via the HACA mechanism, further acetylene addition leads to the formation of a five-membered ring instead of a six-membered ring, and acenaphthylene is formed instead of the expected anthracene or phenanthrene.[343] Therefore, the original HACA mechanism does not lead to the growth of six-membered planar PAHs.

Along with the HACA mechanism, it was suggested that aryl-aryl combination reactions can occur during PAH growth.[344] In 1994, Sarofim et.al noted that these reactions are underestimated in PAH and soot formation theories[345]. Biaryls themselves are highly reactive intermediates. They can undergo intramolecular cyclodehydrogenation, and then, upon the addition of acetylene, they turn into highly condensed PAHs.[346] Boehm et.al.[347] in 1998 investigated the reaction mechanism of PAH growth and showed that aromatic ring–ring condensation is more efficient than the HACA route for the production of high molecular weight aromatic compounds with a short reaction time, while the contribution of HACA increased over longer periods of time. In 2002, Dong and Hüttinger[348] reviewed PAH growth reactions and reported that aryl-aryl combination and acetylene addition are the most favorable reactions for PAH growth. The authors also named the model the "particle-filler model", where aromatic hydrocarbons act as molecular species and acetylene is a molecular filler. A general scheme of PAH growth according to this model is illustrated in Figure 12. Depending on the side where acetylene is attached, the final structure of the formed PAHs can be different. For example, Dong and Hüttinger demonstrated a pathway in which naphthalene ($C_{10}H_8$) and naphthalenyl radical ($C_{10}H_7$) combine to form binaphthalenyl ($C_{20}H_{14}$), after which the sequence of dehydrogenation and acetylene addition leads to the formation of PAH $C_{30}H_{14}$ (Figure 12).

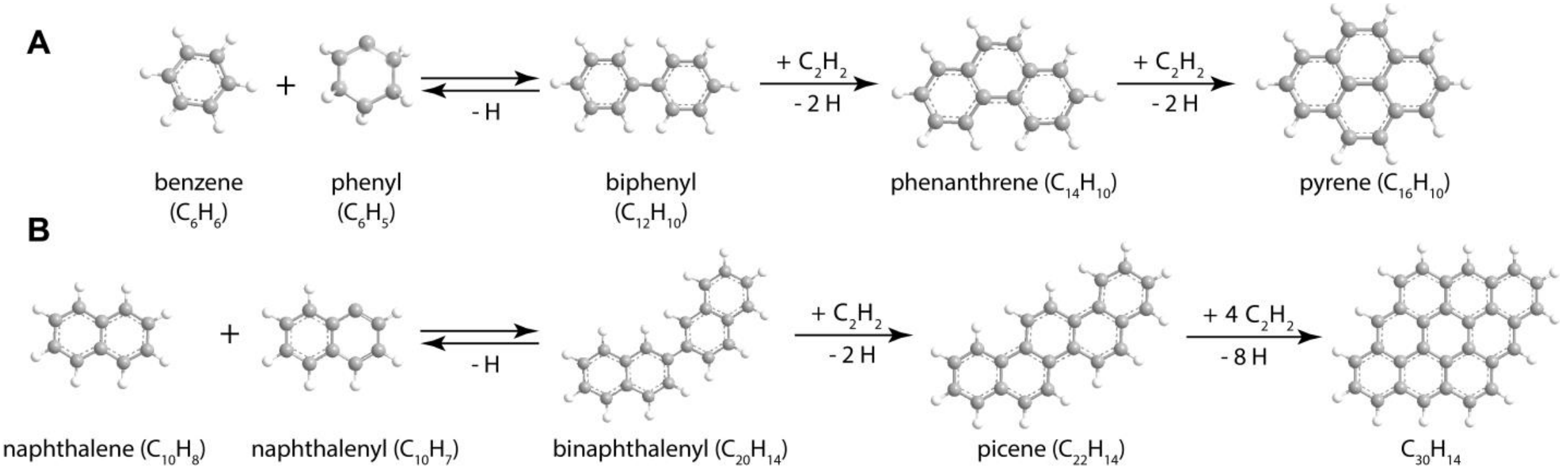

**Figure 12.** Examples of the «particle-filler model» pathway through aryl condensations and acetylene additions at 4-C bays: for formation of phenanthrene, pyrene (A) and the fully condensed planar PAH (B).[348]

Later, experiments with crossed molecular beams showed that biphenyl can be formed via a single collision event during the reaction of phenyl radicals with benzene.[349] In 2017, Yang et al.[39] presented experimental evidence for the formation of phenanthrene (a PAH with three aromatic rings) through the interaction of an *o*-biphenylyl radical and acetylene, as confirmed by DFT calculations. A diagram of the potential energy corresponding to this route is presented in Figure 13. The proposed pathway is initiated by abstraction of a hydrogen atom from the ortho position of the biphenyl molecule, which in turn can be formed through the elemental reaction of benzene with a phenyl radical.[349] Then, the radical center of *o*-biphenylyl was added to the acetylene molecule (the reaction barrier was only 10 kJ $mol^{-1}$). The ring closure of the $C_{14}H_{11}$ collision complex i1 occurs through the passage of a transition state located 20 kJ $mol^{-1}$ upstream of the complex, resulting in the formation of i2. Notably, the entry barrier can be easily overcome under high-temperature circumstellar conditions. The tricyclic intermediate $C_{14}H_{11}$ (i2) eventually removes atomic hydrogen and aromatizes through a tight transition state to form phenanthrene (p1) and atomic hydrogen. The authors noted that phenanthrene is the preferred product in the temperature range of 500–2500 K and is the main reaction product of *o*-biphenylyl with acetylene under typical conditions of circumstellar envelopes such as the C-rich star IRC +10216, where temperatures up to 2000 K are maintained near the central star. Generally, this mechanism can be considered a variation of the HACA mechanism, as it includes the same two steps: hydrogen abstraction and acetylene addition.

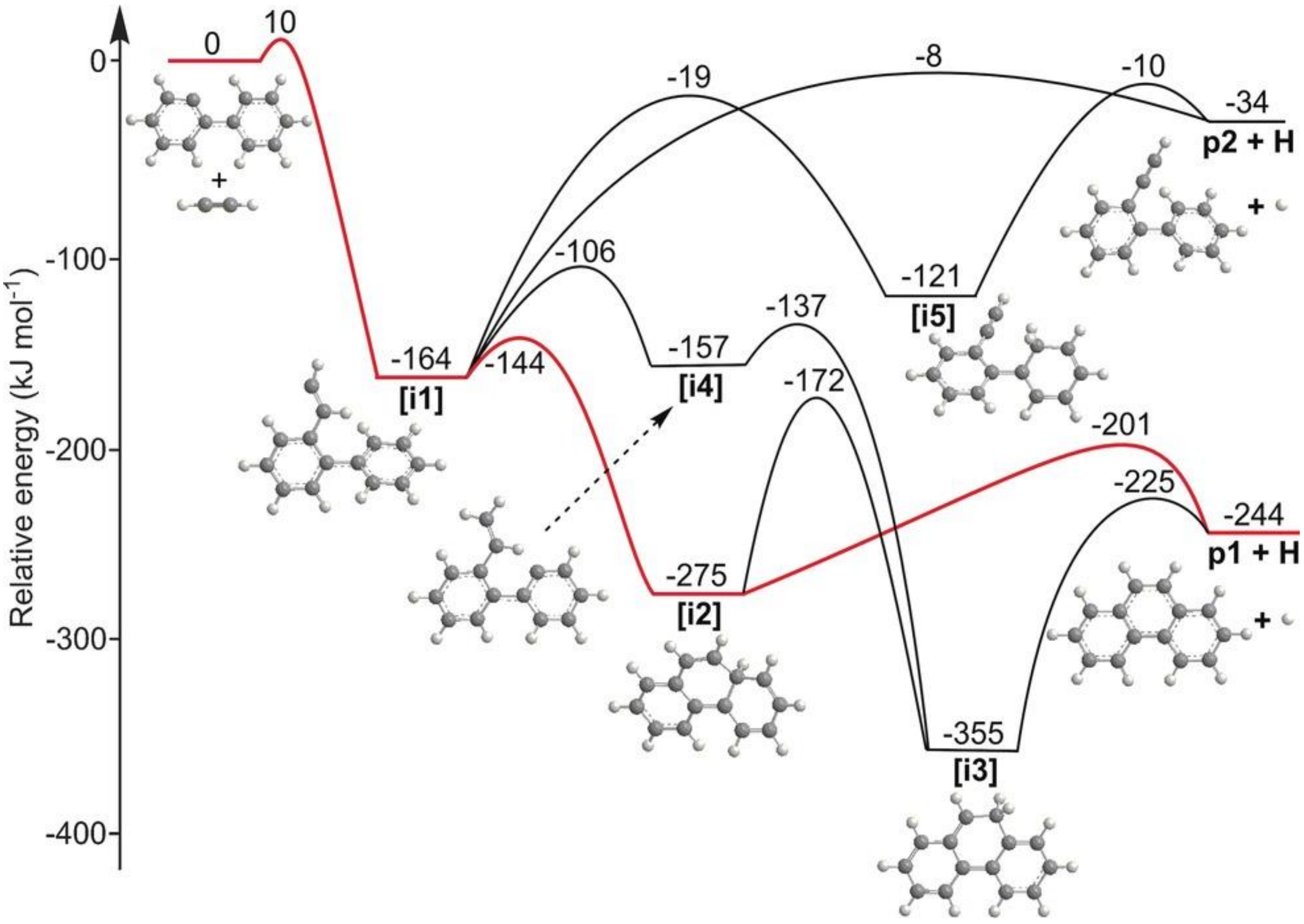

**Figure 13.** Proposed reaction mechanism based on stationary points on the $C_{14}H_{10}$ potential energy surface for *o*-biphenynyl reaction with acetylene. Red line indicates the favorable pathway leading to phenanthrene. Reproduced with permission from ref 39. Copyright 2017 John Wiley and Sons.

In 2018, Mebel et al. published a study[18] demonstrating the easy formation of the four-ring PAH pyrene ($C_{16}H_{10}$) in the reaction of the 4-phenanthrenyl radical ($C_{14}H_9$) with acetylene under prevailing conditions in C-rich circumstellar envelopes. According to the mechanism discussed in a previous study, phenanthrene ($C_{14}H_{10}$) molecules can undergo abstraction of a hydrogen atom from the 4-carbon position with the formation of a 4-phenanthrenyl radical.

The proposed route to pyrene through the bimolecular reaction of the 4-phenanthrenyl radical with acetylene occurs in many respects similar to the mechanism of phenanthrene formation discussed above. This process is illustrated in Figure 14. The reaction is initiated by the addition of a 4-phenanthrenyl radical species to an acetylene molecule through an entry barrier of 20 kJ mol$^{-1}$ to form intermediate i1. Then, the $C_{16}H_{11}$ collision complex ring closes, which leads to intermediate i2 through the entrance barrier at 26 kJ·mol$^{-1}$. In the last step, intermediate i2 undergoes removal of a hydrogen atom and aromatization to form pyrene as the end product p1. The authors confirmed their conclusions with convincing experimental evidence of the synthesis of a molecule with the molecular formula $C_{16}H_{10}$ in the 4-phenanthrenyl/acetylene system in a high-temperature chemical reactor.

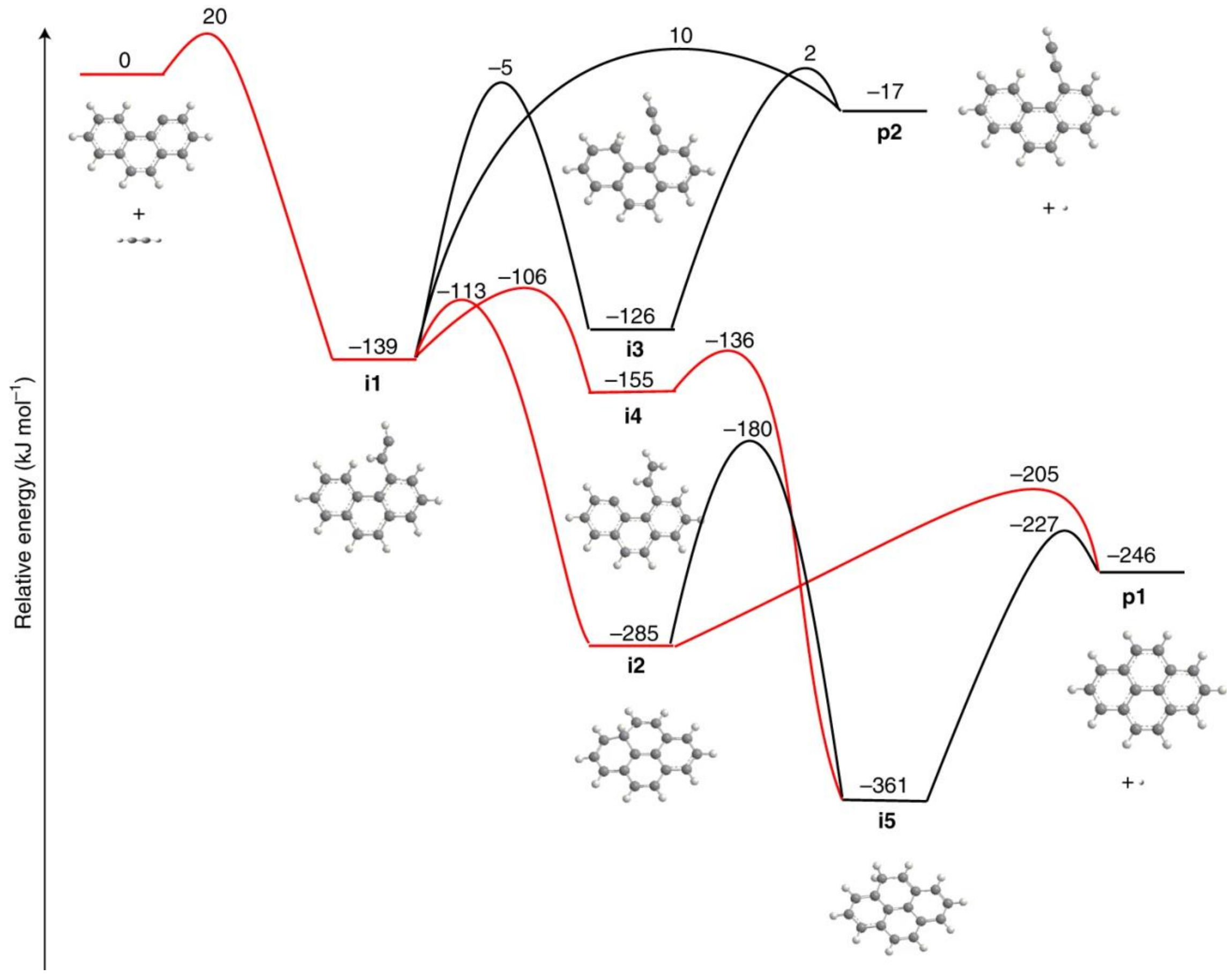

**Figure 14.** Proposed reaction mechanism based on stationary points on the $C_{16}H_{10}$ potential energy surface for 4-phenanthrenyl radical reaction with acetylene. Red line indicates the favorable pathway leading to pyrene. Reproduced with permission from ref 18. Copyright 2018 Springer Nature.

The pyrene molecule is a key intermediate in the mass growth of PAHs, ultimately leading to the formation of two-dimensional carbon nanostructures such as graphene. The formation of pyrene dimers is the basis for the formation of soot particles.[350] However, further experimental measurements of the kinetics of pyrene dimerization coupled with theoretical results showed that a mechanism involving further chemical growth of PAHs to sizes well beyond those of pyrene is necessary.[351] Mebel et al. proposed a model for the growth of PAHs.[14] The authors noted that the systematic expansion of the ring leading to more complex PAHs is the result of the cooperation of two mechanisms: HACA and HAVA. The HACA mechanism alone cannot explain the formation of PAHs such as benzo[*e*]pyrene ($C_{20}H_{12}$). The abstraction of hydrogen and the addition of acetylene to the pyrene molecule can lead to the formation of the $C_{18}H_{11}$ radical, which is expected to undergo cyclization with subsequent loss of atomic hydrogen to cyclopent[*cd*]pyrene ($C_{18}H_{10}$). The addition of a second acetylene molecule followed by cyclization and loss of hydrogen to $C_{20}H_{12}$ is unlikely according to electronic structure calculations[34,341] and experimental data for the naphthyl/acetylene system.[342] Therefore, alternative mechanisms to HACA should complement the formation of extensive polyaromatic systems. Starting with pyrene, the HAVA mechanism provides radial third-order perimeter growth, while the HACA mechanism responds to closing to bays to form five-membered rings. Sequential implementation of the HACA and HAVA mechanisms allows the synthesis of nonplanar carbon nanostructures containing corannulene subunits (see Section 5.2.3).

In recent studies, Mebel et al. showed that, along with the HACA and HAVA mechanisms, species such as allene ($C_3H_4$) and methylacetylene ($C_3H_4$) can also play a role in ring closure during the growth of condensed polyaromatic molecules.[352] Experiments in a high-temperature chemical microreactor combined with electronic structure calculations have shown that naphthalene-based isomers, such as 2-(prop-2-yn-1-yl)naphthalene ($C_{13}H_{10}$), 2-(prop-1-yn-1-yl)naphthalene ($C_{13}H_{10}$), and 2-(propa-1,2-dien-1-yl)naphthalene ($C_{13}H_{10}$), can be obtained from the aforementioned $C_3H_4$ species with the 2-naphthyl isomer. Moreover, it was possible to record the formation of 3H-benz[*e*]indene ($C_{13}H_{10}$) and 1H-benz[*f*]indene ($C_{13}H_{10}$), which each contain a five-membered ring.[352] Later, it was shown that the formation of 1H-phenalene can occur by a similar mechanism. The authors noted that the processes of graphene nanoflakes can start from 1H-phenalene via peripheral expansion of the 1H-phenalene species to the triangular condensed molecules 1,5-dihydrodibenzo[*cd,mn*]pyrene ($C_{22}H_{14}$) and then 6,11-dihydro-1H-tribenzo[*bc,hi,no*]coronene ($C_{33}H_{18}$).[353] Despite the fact that acetylene molecules are not directly involved in this mechanism, the molecules allene ($C_3H_4$) and methylacetylene ($C_3H_4$) are derivatives of acetylene, and as shown above, their formation can proceed by the interaction of acetylene with a methylene species.[160] This highlights the role of acetylene as a versatile building block for a variety of carbon molecules.

One additional mechanism, the PAC, should also be considered. In the case of PAC, a phenyl radical attaches to biphenyl followed by H abstraction. After subsequent dehydrogenation, cyclization and aromatization, the biphenyl-phenyl system is converted to triphenylene, i.e., four-ring PAHs. Zhao et al. demonstrated the formation of triphenylene through the PAC mechanism. This mechanism may occur in high-temperature and high-pressure regions of space, such as the circumstellar envelopes of C-rich AGB stars, e.g., IRC+10216. [354]

Notably, in 2021, Green et al. studied the reaction of naphthalene radicals with acetylene ($C_2H_2$). Using VUV photoionization time-of-flight mass spectrometry at 15 to 50 torr and 500 to 800 K, they recorded the formation of $C_{14}H_{10}$ species, which are probably phenanthrene and anthracene. This study provides experimental evidence that the HACA mechanism is a pathway for the formation of PAHs larger than naphthalene.[355]

**Table 14.** Summary of considered reactions, kinetics (if available) and conditions where the reactions are relevant

| Reactions | Kinetics | Objects & conditions: |
|---|---|---|
| 1.<br>a. $C_6H_5 + C_6H_6 \rightarrow C_{12}H_{10} + H$<br>b. $C_{12}H_{10}$ (biphenyl) + $C_2H_2 \rightarrow$ $C_{14}H_{10}$ (phenanthrene) + 2H<br>c. $C_{14}H_{10} + C_2H_2 \rightarrow C_{16}H_{10}$ (pyrene) + 2H | $k_{1a} = 3.1 \cdot 10^{-13}$ cm$^3$ s$^{-1}$ (T = 1000 K)[347]<br>$k_{1b} = 4.0 \cdot 10^{-13}$ - $2.6 \cdot 10^{-12}$ cm$^3$ s$^{-1}$ (p = 0.04-100 atm., T = 1000 K)[39]<br>$k_{1c} = 1.0 \cdot 10^{-12}$ cm$^3$ s$^{-1}$[18] | Stellar envelopes (outer edge).<br>Conditions:<br>$n_H = 10^8$-$10^{12}$ cm$^{-3}$<br>T = 500-1500 K |
| 2. $C_{10}H_7\cdot + C_3H_4 \rightarrow C_{13}H_{10} + H\cdot$<br>$C_{10}H_7\cdot + C_3H_4 \rightarrow C_{12}H_8 + CH_3\cdot$[353] | n.d. | |
| 3. $C_{12}H_{10} + C_6H_5\cdot \rightarrow C_{18}H_{12}$ (triphenylene) + H + $H_2$[354] | n.d. | |

### 5.2.3. Mechanism of formation of PAHs with five-membered rings

The formation of PAHs with five-membered rings is an important step in the formation of detected in space nonplanar carbon molecules, such as fullerene. Several alternative mechanisms for the formation of PAHs with five-membered rings have also been presented in the last decade. Parker et al.[356] provided experimental evidence that indene can be formed by the bimolecular gas-phase reaction of a benzyl radical ($C_7H_7$) with acetylene at 600 K. This mechanism was theoretically predicted as early as 1981 by Bittner and Howard.[357] This process includes the addition of $C_2H_2$ to the $CH_2$ moiety of the benzyl radical to form a doublet $C_9H_9$ intermediate. Then, the doublet $C_9H_9$ intermediate isomerizes through facile ring closure to the *ortho*-carbon of the phenyl ring to form bicyclic intermediate i2, which can be converted to an indene after H loss from the *ortho*-carbon of the phenyl ring (Figure 15).

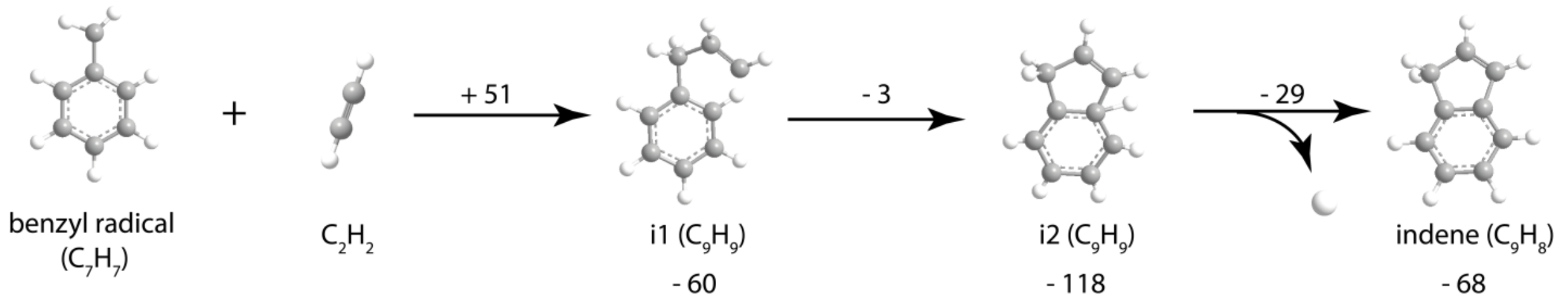

**Figure 15.** Proposed mechanism of indene formation via the reaction of the benzyl radical with acetylene. Energies are given in kJ mol$^{-1}$. Reproduced with permission from ref 354. Copyright 2015 John Wiley and Sons.

The growth of the aromatic system can further proceed through the mechanisms described above involving acetylene or its derivatives. Thus, 5- and 6-indenyl radicals ($C_9H_7^{\cdot}$) can interact with vinylacetylene ($C_4H_4$). The products of this interaction are tricyclic PAHs, 3H-cyclopenta[*a*]naphthalene ($C_{13}H_{10}$), 1H-cyclopenta[*b*]naphthalene ($C_{13}H_{10}$), and 1H-cyclopenta[*a*]naphthalene ($C_{13}H_{10}$)[360] (Figure 16). In addition, it has recently been shown that the reaction of a benzyl radical with a phenyl ($C_6H_5$) radical is a pathway for the formation of another important PAH containing a five-membered ring, 9H-fluorene.[358] Li et al. showed that the interaction of acetylene with methylene-substituted aromatic compounds carrying a hydrogen atom at the *ortho*-position of the ring can provide a universal pathway for accessing five-membered aromatic compounds ($C_{13}H_{10}$, as indicated above) at elevated temperatures.[359] These reactions were carried out in a high-temperature chemical microreactor at 1300 K. In addition to the experiments, corresponding calculations of the energies and molecular parameters of various intermediates and transition states were performed. These calculations have shown that tricyclic PAHs can be formed via a barrierless HAVA mechanism in cold molecular clouds at temperatures of approximately 10 K.[360]

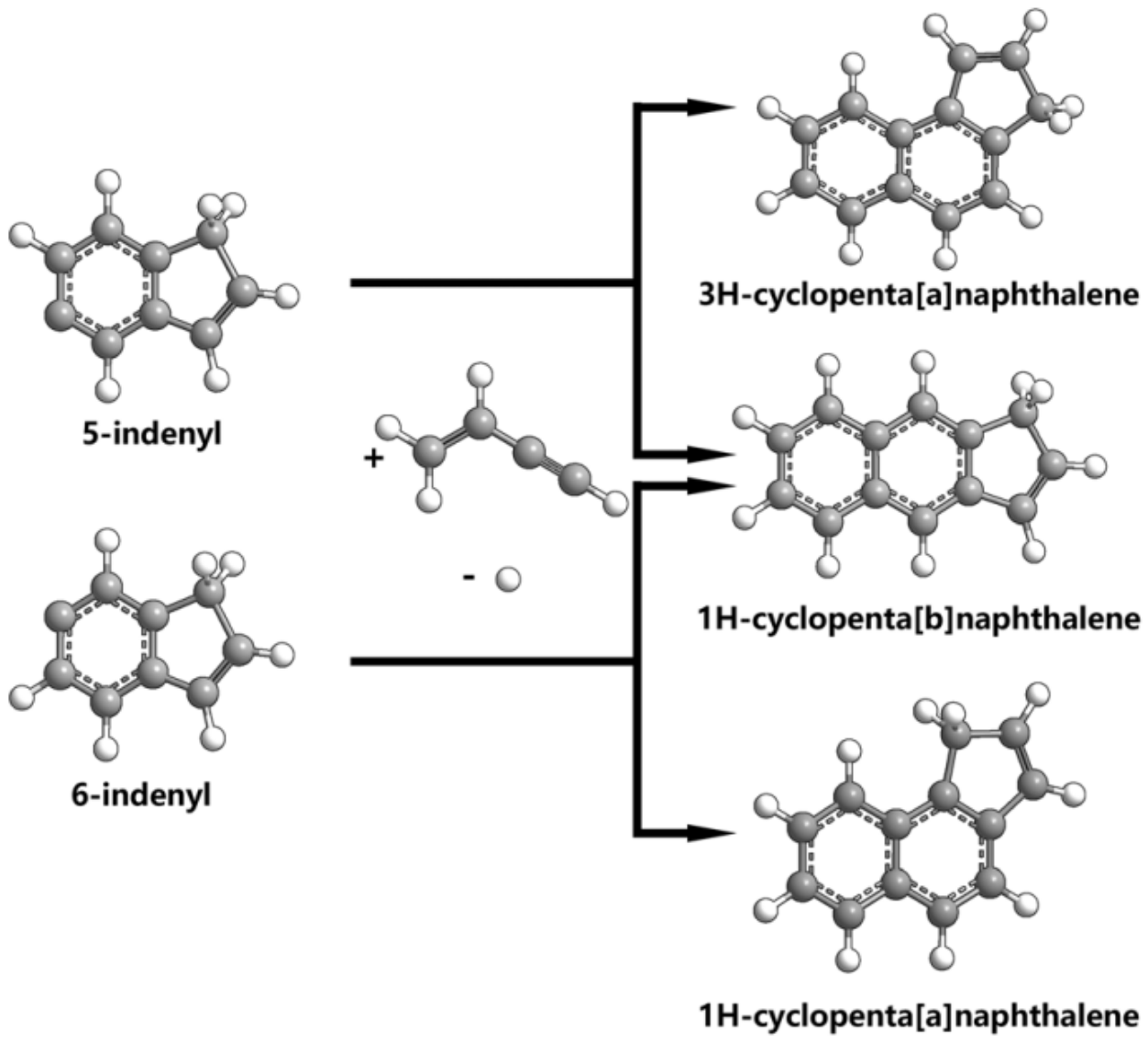

**Figure 16.** Proposed mechanism formation of 3H-cyclopenta[*a*]naphthalene, 1H-cyclopenta[*b*]naphthalene, and 1H-cyclopenta[*a*]naphthalene ($C_{13}H_{10}$) by reaction of 5-/6-indenyl radicals ($C_9H_7$) with vinylacetylene ($C_4H_4$). Reproduced with permission from ref 358. Copyright 2020 Royal Society of Chemistry.

An interesting experimental study of the formation of PAHs with five-membered rings under high-temperature conditions was carried out by Necula and Scott.[361] Using flash vacuum pyrolysis, the authors demonstrated the formation of large PAH molecules at a temperature of approximately 830 K and pressures in the range of $10^{-5}$-$10^{-4}$ atm through the $C_2$-accretion pathway. The key C−C bond-forming step involved the trapping of aryl radicals by acetylene ($C_2H_2$). The main reported product of the reaction of acetylene and 1-naphthyl radical was acenaphthylene ($C_{12}H_8$). Minor amounts of naphthalene ($C_{10}H_8$) and 2-ethynylnaphthalene ($C_{12}H_8$), and traces of pyracylene ($C_{14}H_8$) were also detected. A similar set of products was obtained when 2-naphthyl radicals were used. The proposed mechanism is presented in Scheme 15.

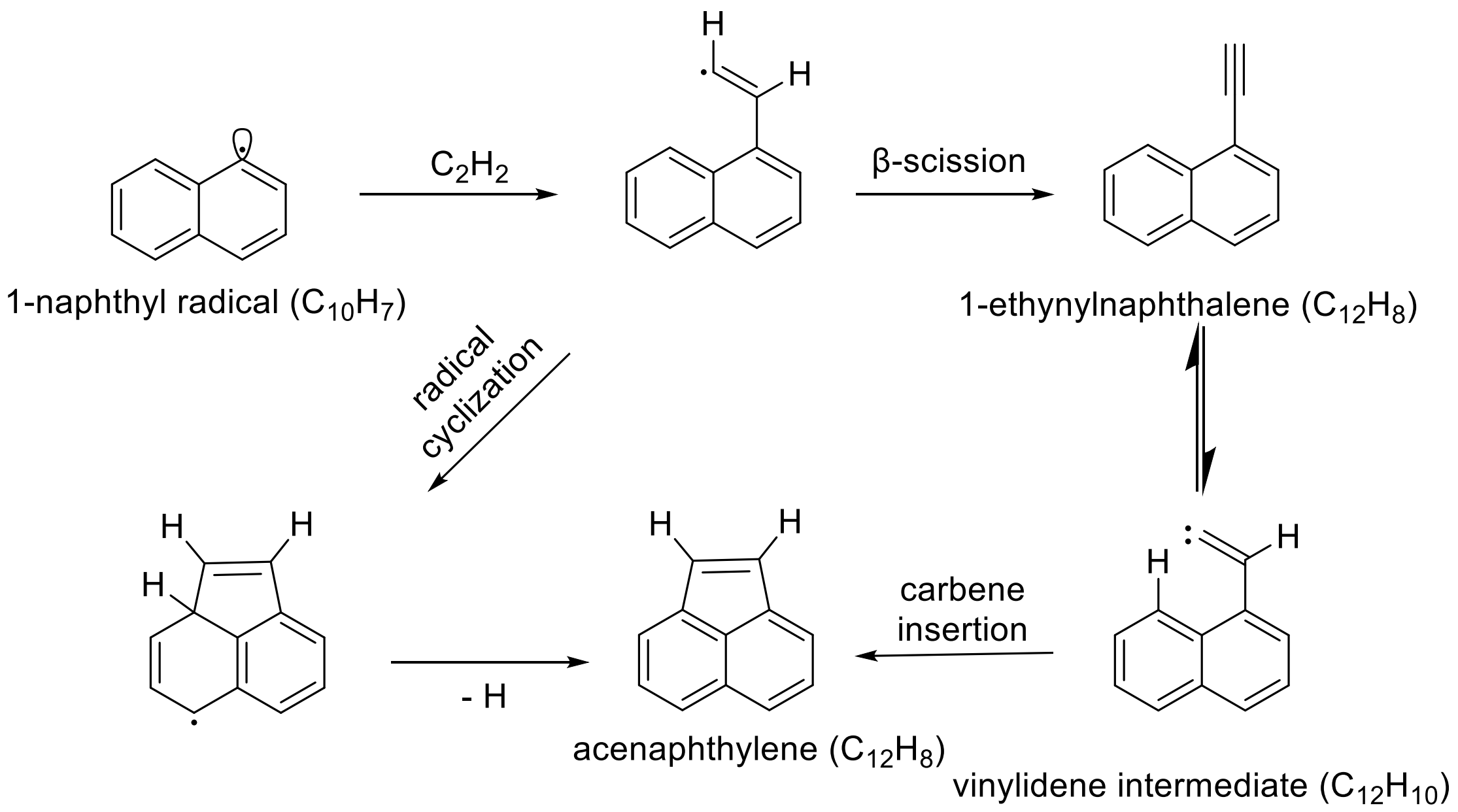

**Scheme 15.** Proposed reaction mechanism acenaphthylene formation by trapping aryl radicals with acetylene. [359]

One of the important molecules with a five-membered ring is corannulene ($C_{20}H_{10}$), which is a building block in the pathway of fullerene formation. A combination of molecular beam experiments and theoretical calculations demonstrated that corannulene can be synthesized in the gas phase through reactions of 7-fluoranthenyl ($C_{16}H_9$) and benzo[*ghi*]fluoranthenen-5-yl ($C_{18}H_9$) radicals with acetylene (Figure 17). This process is initiated by H abstraction from the 7- and 5-positions of fluoranthene ($C_{16}H_{10}$) and benzo[*ghi*]fluoranthenene ($C_{18}H_{10}$) and has barriers that can be easily overcome in high-temperature environments of C-rich circumstellar envelopes of AGB stars or under harsh UV radiation fields intrinsic to planetary nebulae.[362] Notably, fluoranthene can also be formed via the PAC mechanism when the phenyl radical reacts with naphthalene, similar to the reaction between the phenyl radical and biphenyl described above.

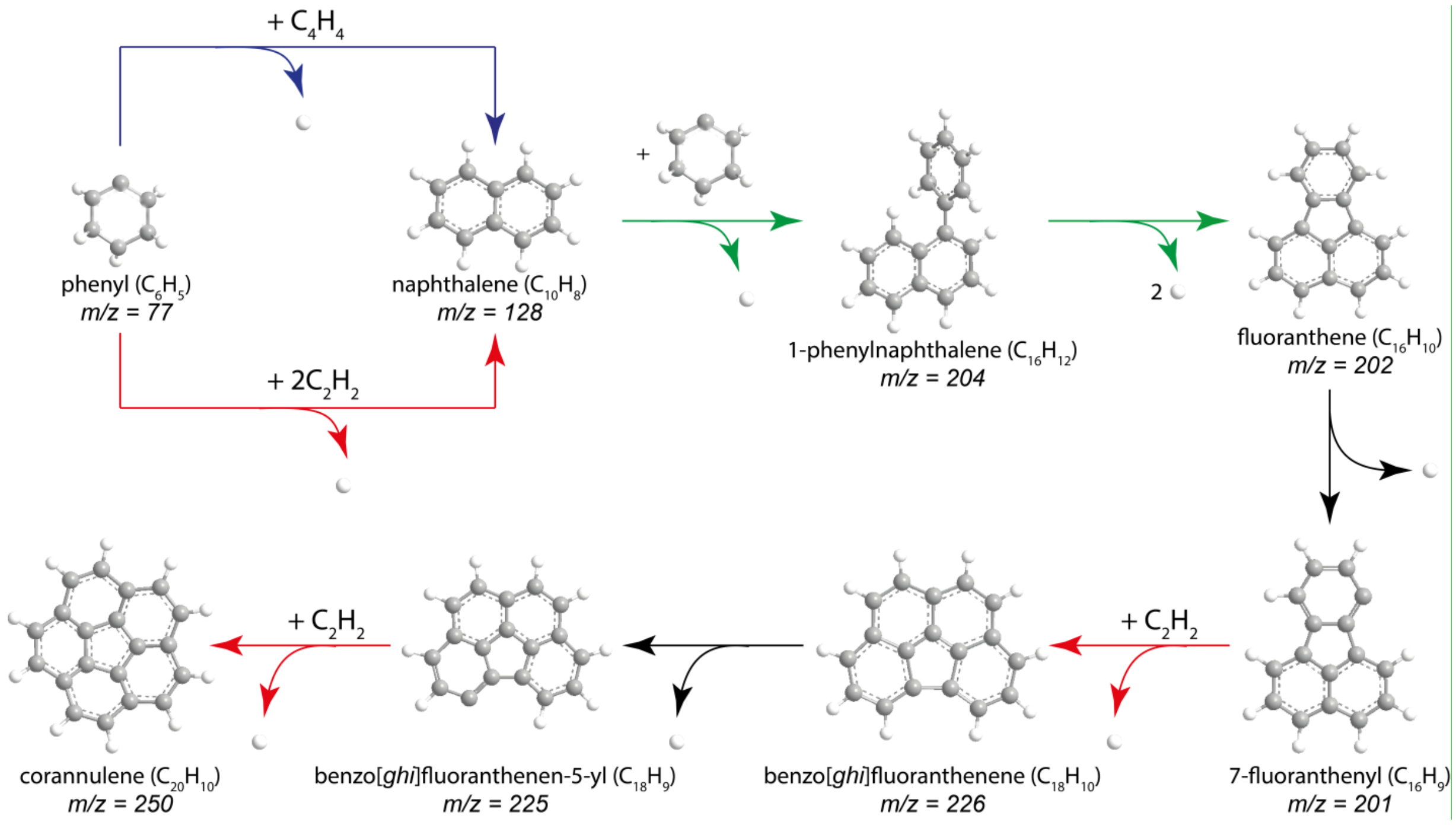

**Figure 17.** Scheme of the formation of corannulene through the following sequence: formation of fluoranthene from the phenyl radical involving steps of HACA (red), HAVA (blue), and PAC (green) mechanisms; the formation of benzo[*ghi*]fluoranthene and corannulene by the HACA mechanism through the reaction of the 7-fluoranthenyl radical with acetylene.[362]

**Table 15.** The reactions which lead to formation of PAHs with five-membered rings.

| **Reactions** | **Kinetics** | **Objects & conditions** |
|---|---|---|
| 1. $C_7H_7 + C_2H_2 \rightarrow C_9H_8 + H$[356] | n.d. | Stellar envelopes (outer edge).<br>Conditions:<br>$n_H = 10^8$-$10^{12}$ cm$^{-3}$<br>T = 500-1500 K |
| 2. $C_9H_7 + C_4H_4 \rightarrow C_{13}H_{10} + H$[360] | n.d. | Molecular clouds, prestellar cores.<br>Conditions:<br>$n_H = 10^2$-$10^7$ cm$^{-3}$<br>T = 10 K<br>$G_0$ = very low<br>Galactic CR field |
| 3. $C_7H_7 + C_6H_5 \rightarrow C_{13}H_{10} + H_2$[330] | $k_3 = 3 \cdot 10^{-11}$ cm$^3$ s$^{-1}$<br>(T = 1500 K, p = 0.01 atm)[330] | Stellar envelopes (outer edge) |
| 4.<br>a. $C_{16}H_9 + C_2H_2 \rightarrow C_{18}H_{10} + H$<br>b. $C_{18}H_{10} \rightarrow C_{18}H_9 + H$<br>c. $C_{18}H_9 + C_2H_2 \rightarrow C_{20}H_{10} + H$[362] | $k_{4a} \approx 1.0 \cdot 10^{-11}$ cm$^3$ s$^{-1}$<br>$k_{4c} \approx 3.0 \cdot 10^{-12}$ cm$^3$ s$^{-1}$<br>(both at 1500 K, p = 0.04-100 atm.)[362] | Stellar envelopes (outer edge) |

## 5.2.4. PAH formation via polyyne polymerization

In the 1990s, Krestinin et al. proposed an alternative mechanism for PAH and soot formation.[363,364,365,366] They considered polyyne polymerization as an alternative route for generating high-molecular-weight carbon species from acetylene. This hypothesis is consistent with the observations of small polyynes along with PAHs in the flame soot formation zone.[252] In addition, polyynes and small (transient) cyclic compounds can be formed in discharge plasma sources and are used to create a chemically reactive environment that simulates astrochemical conditions.[367] As mentioned above (Section 3.2), carbon-chain polyynic radicals ($C_nH$), polyynes ($HC_nH$), cyanopolyynes and methylpolyynes are quite abundant in the C-rich star IRC +10216 and protoplanetary nebula CRL 618 and are formed from acetylene.[143,266,368]

Bonn, Homann and Wagner suggested[369] that polyynes play a key role in the formation of soot. Proponents of the polyyne hypothesis note that the thermodynamic stability of acetylene (and the entire family of $HC_{2n}H$ polyynes, n = 2, 3,…) increases at higher temperatures, while the stability of all other hydrocarbons decreases.[370] Krestinin et al. also indicated that polyynes grow rapidly by a one-step ethynyl radical addition mechanism, as opposed to aromatic ring formation. As a result, the concentration of polyynes will be quite high (Scheme 16).

$$HC_{2n}H \quad + \quad C_2H^{\bullet} \longrightarrow HC_{2n+2}H \quad + \quad H^{\bullet}$$

**Scheme 16.** Polyynes growth via one-step ethynyl radical addition mechanism.

These polyynes can form complexes at the surface radical sites of the seed particles, followed by branching during chain polymerization (Scheme 17). As a result of cyclization occurring in such a complex, two new additional radical centers are created. This "propagation" of radical sites during polyyne polymerization is a key assumption that introduces nucleation into the model. This finding is also consistent with the data of Homann and Wagner[369], who noted that the electron spin resonance signal from nascent soot particles under combustion conditions exceeds that from mature soot by more than 100 times, indicating the generation of large amounts of radicals during nucleation. Cherchneff et al. noted that such a mechanism of PAH growth may be relevant in environments such as the wind of the hydrogen-rich R Coronae Borealis star V854 Cen.[246]

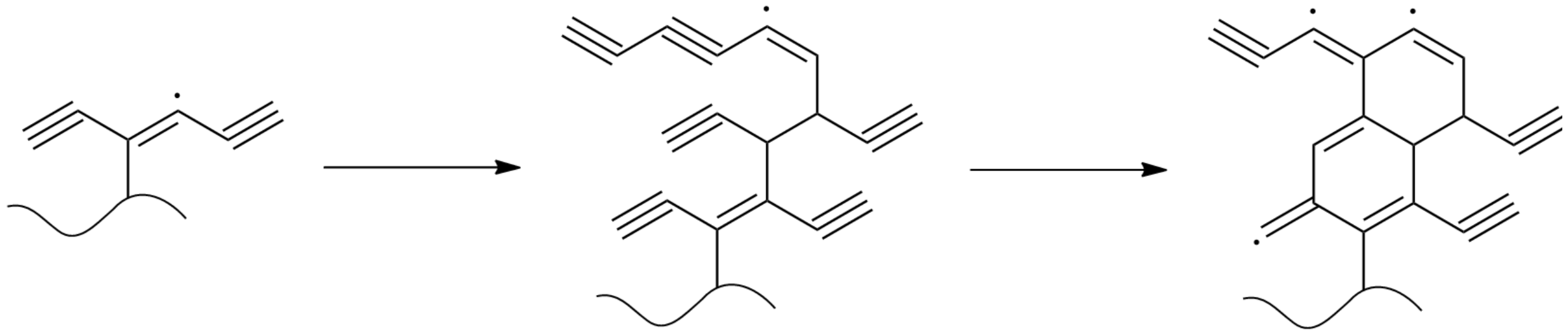

**Scheme 17.** Formation of a polyyne complex at the surface radical site and its transformation into a more stable structure with subsequent "propagation" of radical centers.[370]

**Table 16.** The reaction of polyyne growth via ethynyl radical addition mechanism.

| **Reactions** | **Kinetics** | **Objects & conditions** |
|---|---|---|
| $C_{2n}H_2 + C_2H \rightarrow C_{2n+2}H_2 + H$[363] | $k = 3.3 \cdot 10^{-11}$ cm$^3$ s$^{-1}$ | Stellar envelopes (outer edge).<br>Conditions:<br>$n_H = 10^8$-$10^{12}$ cm$^{-3}$<br>T = 500-1500 K |
| | | Protoplanetary nebulae.<br>Conditions:<br>T = 250 K<br>$n_H = 10^7$ cm$^{-3}$<br>$G_0 = 2 \cdot 10^5$ |

## 5.3. Mechanisms of PAH formation in cold interstellar medium

As mentioned above, the observations and theoretically estimated PAH lifetimes require a path of PAH formation in the cold ISM at a temperature of approximately 10 K. Due to such low temperatures and pressures, a vital criterion for potential reaction paths will be their barrierless nature and occurrence under single collision conditions. Reactions by the HACA mechanism involving acetylene have significant entry barriers ranging from 10 to 30 kJ mol$^{-1}$ and can occur only at high temperatures up to several thousand Kelvin.[371] However, in the last decade, it has been found that many reactions involving aromatic radicals leading to the formation of PAHs are essentially barrierless.[356] Such mechanisms (a single barrierless bimolecular collision) can play a key role in the formation of PAHs in extremely low-temperature environments, such as dense molecular clouds and atmospheres of planets and their satellites in the outer Solar system.

Mebel et al. proposed a number of barrierless mechanisms for the formation of PAHs in cold environments[372]. Among these growth mechanisms of the aromatic system, the most important are the ethynyl addition mechanism (EAM), HAVA, and the reaction between methylidyne and styrene, leading to the formation of an indene (MACA).[290] In all the mechanisms, acetylene can be a precursor of the chain initiator. We provide these reactions in Table 17. Apart from gas-phase reactions, PAHs can likely be formed in astrophysical ices, as will be shown below, together with mechanisms of formation in planetary and satellite atmospheres.

To determine the significance of these barrierless reactions, one needs to include them in astrochemical modeling. Such work was started by McGuire et al. 2022.[300] The authors considered the reactions between phenyl radicals and vinylacetylene and between phenyl radicals and 1,3-butadiene to form naphthalene, which further reacts with cyanide to form cyanonaphtalene. The authors concluded that the considered reactions do not provide the observed abundance of cyanonaphtalene; therefore, other pathways should be taken into account in future modeling.

**Table 17.** The reactions of PAH growth at low temperatures.

| **Reactions** | **Kinetics** | **Objects & conditions** |
|---|---|---|
| 1.<br>a. $C_6H_6 + C_2H \rightarrow C_6H_5C_2H + H$<br>b. $C_6H_5C_2H + C_2H \rightarrow C_{10}H_7$<br>or<br>c. $C_6H_5C_2H + C_2H \rightarrow C_6H_5(C_2H)_2$<br>d. $C_6H_5(C_2H)_2 + C_2H \rightarrow C_{12}H_8$[372] | $k_{1a,b,c,d}$ ~$10^{-10}$ $cm^3$ $s^{-1}$ [372] | Molecular clouds, prestellar cores.<br>Conditions:<br>$n_H = 10^2$-$10^7$ $cm^{-3}$<br>T = 10 K<br>$G_0$ = very low<br>Galactic CR field<br>+ Some reactions suit planetary atmospheres and satellites. |
| 2. $C_6H_5C_2H_3 + CH \rightarrow C_9H_8$[290] | $k_2$~$10^{-10}$ $cm^3$ $s^{-1}$[290] | The same as reactions #1. |
| 3. $C_6H_5 + C_4H_4 \rightarrow C_{10}H_8 + H$[34,373] | $k_3$~$10^{-14}$ $cm^3$ $s^{-1}$ [34] | Molecular clouds (not planets) |

## 5.4. Mechanisms of PAH formation for planets and satellites

In addition to circumstellar envelopes and the ISM, the atmospheres of planets and satellites are no less important for the production of PAHs. In particular, researchers have paid special attention to the chemistry of the atmosphere of Titan. PAHs are considered to be key components of Titan's haze and potential constituents of organic material on Titan's surface.[374,375] The atmosphere of Titan is a relatively low-temperature environment (surface temperature of 94 K)[376]; therefore, high-temperature mechanisms are difficult to use to explain the formation of PAHs under such conditions. However, the main explanations for the formation of PAHs in Titan's atmosphere have come from classical high-temperature mechanisms such as HACA.[377]

Due to the low temperature of Titan, any entry barrier effectively blocks the reaction. Recently, Mebel et al. proposed a new model for the growth of PAHs in the atmosphere of Titan based on the barrierless HAVA mechanism. The authors experimentally proved that tricyclic PAHs, such as anthracene ($C_{14}H_{10}$) and phenanthrene ($C_{14}H_{10}$), can be formed in the atmosphere through bimolecular gas-phase reactions of 1- and 2-naphthyl radicals ($C_{10}H_7$) with vinylacetylene ($CH_2$=CH–C≡CH).[378]

In addition to the atmosphere of planets, important chemical evolution processes can occur on the surface of planets. This is how the Cassini-Huygens mission discovered a remarkably rich world on Titan's surface, with complex landscapes and topography.[379] One of the amazing features of Titan's landscape is the large longitudinal dunes in the equatorial deserts.[380,381] Data from Cassini's VIMS indicate the presence of a significant amount of dark organics of undetermined chemical composition and origin.[382] It was assumed that the main components of these formations are aromatic hydrocarbons formed as a result of molecular weight growth initiated by solar photons through gas phase ion-molecular and neutral-neutral reactions and surface reactions on the satellite.[106,293,383],[384],[385] However, the discrepancy between the particle sizes of the material from the dark dunes at Shangri-La[386] (one of the three largest equatorial Titan's sand seas) and the particle size of atmospheric aerosols points to a special way of their formation.[380]

A comparison of data from the Cassini Synthetic Aperture Radar (SAR) on the location of Titan's dark dune and VIMS data on the location of acetylene ice allowed us to make an assumption about the link between acetylene ices and the formation of dune material.[28,110,110,387,388] At the same time, most of the energy sources are absorbed by Titan's thick atmosphere. However, energetic CRs are able to penetrate Titan's atmosphere and reach the surface.[389,390] The transfer of a part of the kinetic energy of the secondary electrons formed by CRs to the acetylene molecule can lead to electronically excited acetylene. As discussed in Section 4.5, the formation of benzene in ice can occur by the reaction of electronically excited acetylene with a neighboring singlet acetylene molecule. In this case, the barrier can be easily overcome with the help of vibrational energy in one of the reagents. Similar processes of aromatic ring growth with the participation of electronically excited species can lead to naphthalene and phenanthrene on the surface of Titan.

Recently, Kaiser et al. conducted an experiment in which ices of acetylene and deuterated acetylene ($C_2D_2$) irradiated energetic electrons at 5 K and pressures of a few $10^{-11}$ torr.[28] As a result, the authors recorded the formation of aromatic hydrocarbons, such as naphthalene and phenanthrene. An increase in the dose by a factor of approximately 10 to 27±2 eV per molecule promotes the synthesis of even more complex aromatic compounds, such as chrysene ($C_{18}H_{12}$), perylene ($C_{20}H_{12}$), pentacene ($C_{22}H_{14}$) and coronene ($C_{24}H_{12}$), which bear up to six benzene rings.[391]

X-ray-induced formation of naphthalene from an isolated styrene-acetylene complex in a solid krypton matrix at 6 K was demonstrated by Lukianova and Feldman in 2022.[392] The authors used $SF_6$ as an electron scavenger to determine the conversion mechanism. Using Fourier transform IR spectroscopy, they determined that the formation of a naphthalene radical cation occurred. Thus, the implementation of the cationic pathway is assumed: the cycle is closed due to the formation of an ionized complex, followed by the elimination of a hydrogen molecule (Scheme 18). Thus, it was shown that PAHs can form as a result of the impact of CRs on low-temperature acetylene ice on the surface of Titan.

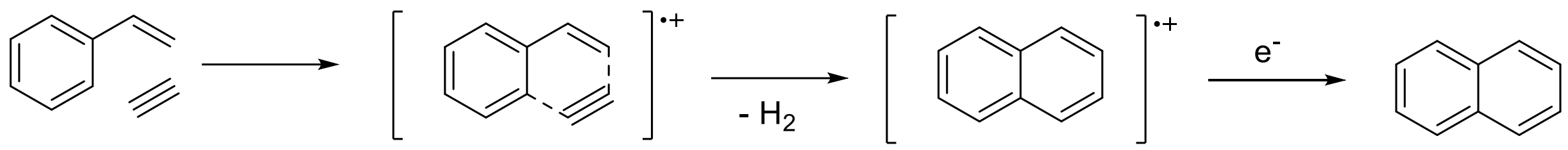

**Scheme 18.** Proposed mechanism of naphthalene formation from the styrene-acetylene complex under X-ray irradiation through radical cation.[392]

Similar processes, including the formation of aromatic molecules from acetylene at low temperatures, can occur not only on Titan but also on other objects of the Solar system, such as the dwarf planets Pluto[393], Makemake, Ceres[394,395], Orcus[396], Salacia[397], Saturn's satellites Hyperion, Iapetus and Phoebe[398,399,400], Jupiter's moons Ganymede and Callisto[388], satellites Pluto, Charon, Nikta, Hydra[401,402] and others, where regions with dark organic matter have been found. Recently, Kaiser et al. showed that the photochemistry of acetylene ice can lead to the formation of PAHs, including five ring molecules, such as dipropylperylene ($C_{26}H_{24}$). [403,287] In another experiment simulating the irradiation of methane and acetylene by galactic cosmic rays in the Kuiper belt in ultrahigh vacuum, Kaiser et al.[404] reported the initial formation of acetylene from methane. In this work, approximately 30 resulting organic molecules were identified, including mono- and tricyclic aromatic compounds (substituted benzenes, naphthalenes and biphenylenes, as well as phenanthrene, phenalene, acenaphthylene and 9H-fluorene). Methyl ($CH_3$), ethyl ($C_2H_5$), vinyl ($C_2H_3$) and allylic ($C_3H_5$) groups were determinated as substituents. The authors noted that these groups may be potential linkers that connect individual aromatic molecules to complex tholin-like structures. The authors also emphasized a stepwise process of molecular weight growth both in the series of mono-, bi- and tricyclic molecules through ring annulation and from pure aromatic molecules to mono- and dialkylated PAHs.

What makes this mechanism even more universal is that methane ices have been found on many of the outer bodies of the Solar system. [405] Methane, in turn, can be easily converted by ionizing radiation to acetylene.[159,404] Thus, acetylene becomes a bridge to a variety of complex organic compounds on the hydrocarbon-rich surfaces of planets and their satellites.

The reactions discussed in this section are given in Table 18. Important pathways of PAH formation based on experiments with crossed molecular beams and pyrolytic reactors combined with quantum chemical calculations are listed in Table S3 in the Supporting Information.

**Table 18.** The reactions of the PAH growth in planetary atmospheres and their satellites.

| Reactions | Kinetics | Objects & conditions |
|---|---|---|
| 1. $C_{10}H_7\cdot + C_4H_4 \rightarrow C_{14}H_{10} + H$ | $k_1$ ~ $10^{-17}$ $cm^3 s^{-1}$ (p = $10^8$ atm, T = 70-180 K)[378] | Planetary atmospheres and satellites. Conditions: $n_H = 10^{10}$-$10^{16}$ $cm^{-3}$ T = 50-300 K $G_0$ = up to $10^2$ Galactic CR field. |
| Ice reactions 2. a. $3C_2H_2 + CR \rightarrow C_6H_6$ b. $C_6H_6 + 2C_2H_2 + CR \rightarrow C_{10}H_8 + H_2$ c. $C_{10}H_8 + 2C_2H_2 + CR \rightarrow C_{14}H_{10} + H_2$[28,404] | $k_{2a}$ ~ 0.6/0.19 amu $eV^{-1}$ (10/40 K)[404] $k_{2b}$ ~ 0.055/0.08 amu $eV^{-1}$ (10/40 K)[404] | Specifically, Titan, Kuiper's belt objects. Conditions: Galactic CR field. T = 50-100 K $G_0$: up to 10 Galactic CR field. |
| 3. $C_6H_5C_2H_3 + C_2H_2 + CR \rightarrow C_{10}H_8$[392] | n.d. | Molecular clouds, prestellar cores. Conditions: $n_H = 10^2$-$10^7$ $cm^{-3}$ T: 10 K $G_0$ = very low Galactic CR field. + Planetary atmospheres and satellites. |

# 6. Formation of carbon nanoparticles and grains

## 6.1. Mechanism of formation of fullerenes and nanotubes

The formation of other important carbon macromolecules, such as fullerenes and nanotubes, has been studied. Fullerenes are detected in protoplanetary and planetary nebulae; therefore, they must be formed in these objects or, earlier, in stellar envelopes. These particles may form simultaneously or separately from PAHs. Traditionally, fullerenes are produced from carbon vapor in arc charge or, e.g., via laser ablation[406,407] but these methods scarcely work in stellar envelopes. Several mechanisms suitable for astrophysical conditions have been suggested, including a 'bottom-up' scenario[408] — closed network growth from C and $C_2$ gas — and a 'top-down' scenario[409,410,411,412,413] — formation due to photodissociation of PAHs or HACs.[409] Despite their diversity, the proposed formation scenarios are of interest and can be taken into account in astrochemical modeling.

Recently, Tuli et al.[414] suggested a new 'bottom-up' pathway to form a building block of fullerenes — a nanobowl $C_{40}H_{10}$. The nanobowl can be obtained through a chain of reactions starting from the corannulene molecule, whose formation was described above in Section 5.2.3. The chain of reactions includes 1) the reaction of the corannulene radical ($C_{20}H_9^{\cdot}$) with vynilacetylene ($C_4H_4$), which results in the formation of benzocorannulene ($C_{24}H_{12}$); 2) benzoannulation of benzocorannulene up to the formation of pentabenzocorranulene ($C_{40}H_{20}$); and 3) ring closure between five attached benzenes with H abstraction and formation of the nanobowl $C_{40}H_{10}$. Curvatures are caused by five-membered rings. The formed nanobowl serves as a building block for buckminsterfullerene $C_{60}$.

It is likely that along with fullerenes, nanotubes are present in space, although they have not yet been detected. Chen and Li studied the possibility of carbon nanotube formation in space via the HACA mechanism using DFT.[23] The model they proposed included, at the first stage, the formation of biphenyl by the addition of phenyl to benzene with the removal of hydrogen. The formation of triphenyls occurs via a process similar to that of the addition of a second phenyl ring, in addition to the removal of hydrogen. In the next stage, triphenyl isomerizes into a closed three-dimensional structure with the loss of hydrogen.

The next three stages represent the growth of a nanotube with the participation of acetylene according to the HACA mechanism (Figure 18). The authors studied the key transition states involved in forming a nanotube via a similar path. The most interesting point was the possibility of bending the triphenyl molecule to form a ring structure. The displacement vectors tended to close the structure. The authors also examined the internal reaction coordinates of the transition state to confirm the calculations. The calculated barrier for molecular bending was 4.59 eV, which is achievable in the circumstellar envelopes of AGB stars and planetary nebulae.

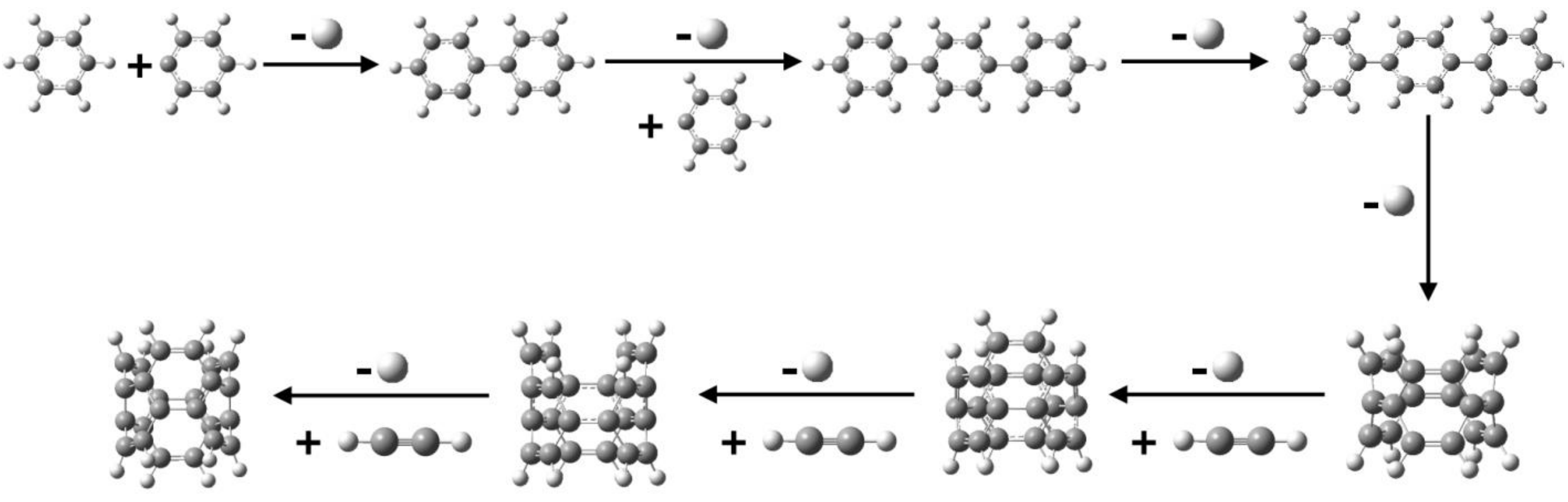

hnt-3-3-2 hnt-3-3-1

**Figure 18.** Scheme of molecular growth of a three-dimensional carbon nanostructure by the NACA mechanism. Reproduced with permission from ref . Copyright 2019 The European Southern Observatory (ESO).

**Table 19.** The reactions which lead to formation of a nanobowl and a nanotube.

| **Reactions** | **Kinetics** | **Objects & conditions** |
|---|---|---|
| 1.<br>a. $C_{20}H_{10} + H \rightarrow C_{20}H_9\cdot + H_2$<br>b. $C_{20}H_9\cdot + C_4H_4 \rightarrow C_{24}H_{12} + H$<br>c. 4 iterations of a and b:<br>$C_{24}H_{12} + 4x[+H] + 4x[C_4H_4] \rightarrow C_{40}H_{20} + 4x[H_2] + 4x[-H]$<br>d. $C_{40}H_{20} \rightarrow C_{40}H_{10}$ (nanobowl) + $5H_2$[414] | $k_{1b} \sim 3.4 \cdot 10^{-17}$ cm$^3$ s$^{-1}$ (T = 1200 K, p = 0.03 atm) [414]<br>$k_{1d} \sim 10^7$ s$^{-1}$<br>(T = 1200 K,<br>p = 0.03 atm)[414] | Stellar envelopes (outer edge).<br>Conditions:<br>$n_H = 10^8$-$10^{12}$ cm$^{-3}$<br>T = 500-1500 K |
| 2.<br>a. $C_6H_5 + C_6H_6 \rightarrow C_{12}H_{10}$ +H<br>b. $C_{12}H_{10} + C_6H_5 \rightarrow C_{18}H_{14}$ + H<br>c. $C_{18}H_{14} \rightarrow$ [isomerization] $\rightarrow C_{18}H_{12}$ (hnt 3-3-1) + 2H<br>d. $C_{18}H_{12} + 3C_2H_2 \rightarrow C_{24}H_{12}$ (hnt 3-3-2) (nanotube) + 6H[23] | n.d. | |

## 6.2. Soot formation

The formation of PAHs is only the first step in the formation of carbon dust grains. The study of dust formation has roots in the study of the practically significant process of carbon black formation.[415] Currently, it is generally accepted that the process of carbon black or soot formation during the pyrolysis of hydrocarbon feedstock includes dehydrogenation, pyrolysis, cracking, and aromatization. An important role is played by the intermediate formation of acetylene, which occurs in flames by pyrolysis of organic fuels at sufficiently high temperatures.[332,416,417] These processes eventually lead to the formation of a growth nucleus from PAHs, which are nanosized (1–2 nm) clusters of interconnected carbon atoms. All the underlying mechanisms are currently based on the key role of PAHs in this process (Figure 19).[418]

In recent work[35], Frenklach and Mebel critically reviewed soot nucleation mechanisms. They quantitatively estimated (by performing kinetic calculations coupled with thermodynamics and DFT calculations) a number of reactions of the dimerization of PAHs. The authors concluded that the most stable PAH dimer is the one in which two PAHs are connected by a doubly bonded bridge, which is termed the E-bridge. The connected rings on the edges are five-membered, and the bridges are aliphatic bonds. The authors emphasize that the dimerization mechanism is related to the HACA mechanism.

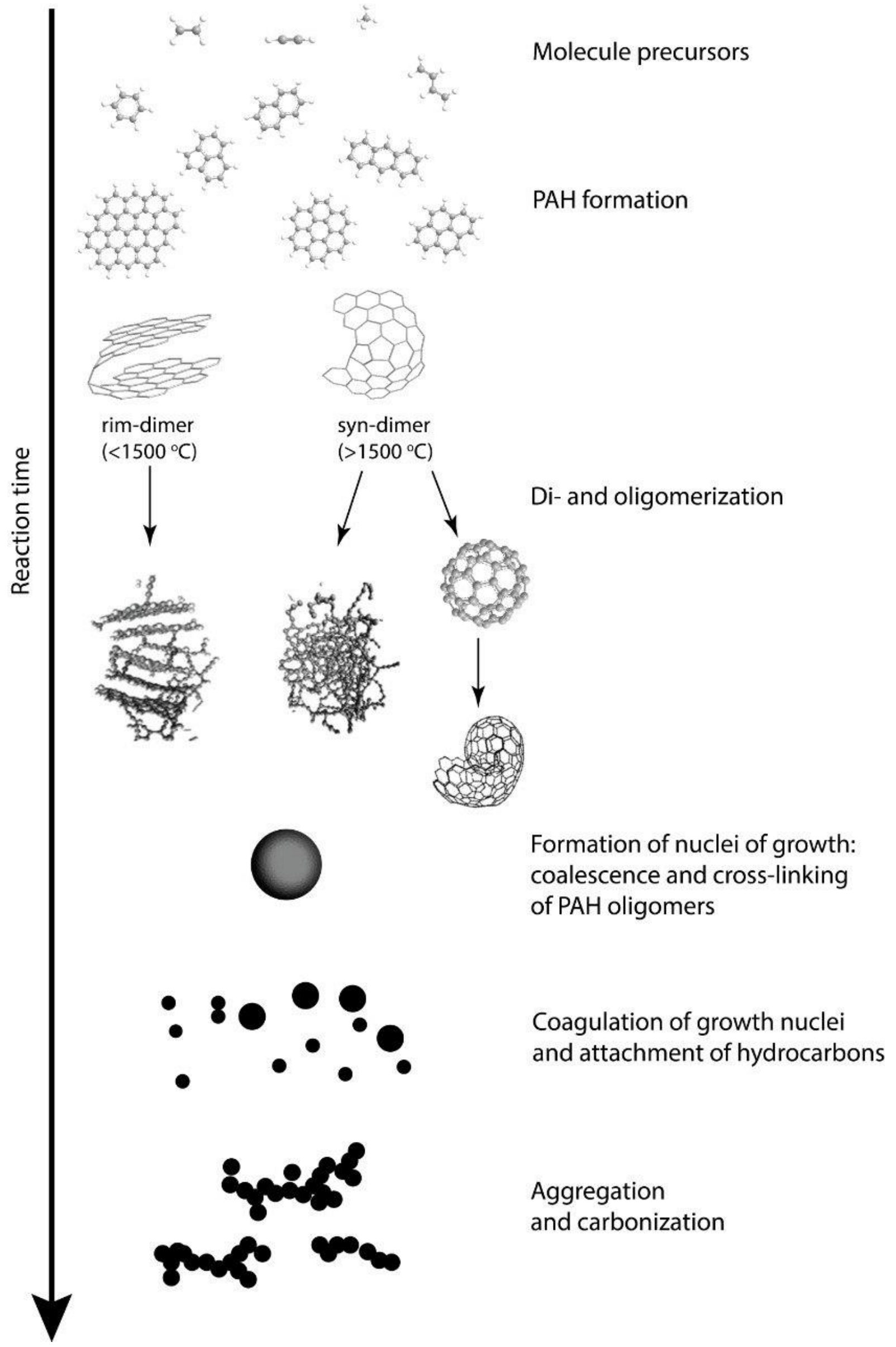

**Figure 19.** The main stages of the process of formation of soot particles with division into low-temperature and high-temperature regions with a conditional boundary of 1500 °C. Reproduced with permission from ref 416. Copyright 2022 Russian Chemical Reviews.

Polyacetylenic chains (polyynes) formed during combustion (see Sections 3.2 and 5.2.4) may play an important role in the formation of large carbon dust particles. Experiments using a special cosmic simulation chamber (COSmIC) facility showed that polyacetylenic chain radicals, as well as H and $C_2H_2$, can be formed during the opening of PAH cycles.[20] Molecular dynamics studies of ReaxFF by Mao et al.[419] showed that such radicals promote the growth of soot particles. Polyacetylene radicals can link large PAHs such as coronene, ovalene, and circumcoronene to form ‘carbon bridges’ on the edge (Figure 20, Table 20).

As a result, soot particles with stacked structures are formed. In addition, fullerene-like soot particles are formed from polyacetylenic chains at 2500 K.

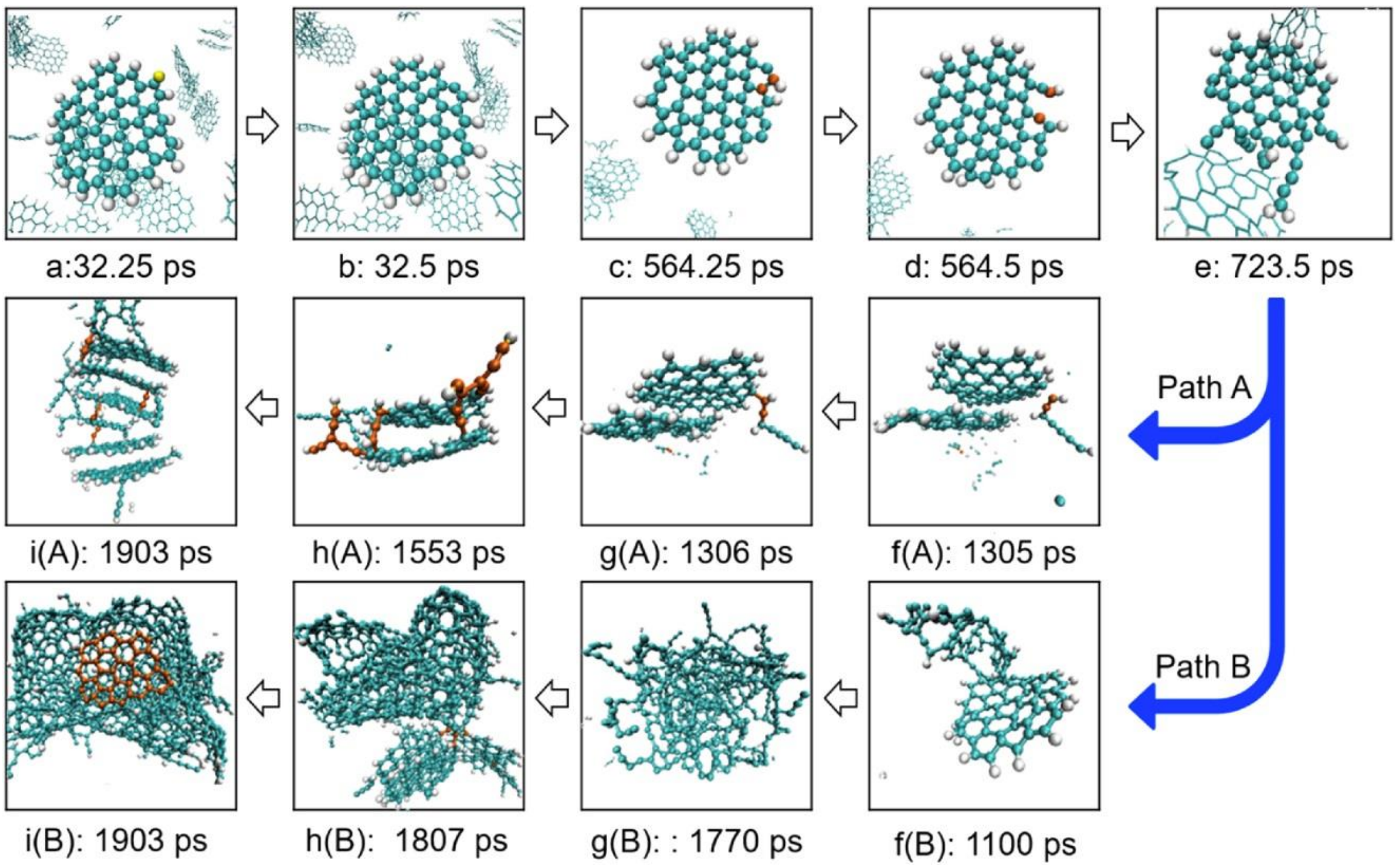

**Figure 20.** Key stages of PAH fragmentation and soot formation from circumrooten monomers at 2500 K. Reproduced with permission from ref 417. Copyright 2017 Elsevier.

Using the Stardust machine,[420] the final gas-phase molecules and solid nanoparticles obtained from the $C_2/C_2H_2$ gas mixture were investigated by Santoro et al.[207]. Among the gas-phase products, polyynes; different aromatic molecules, such as benzene, toluene, styrene, naphthalene, acenaphtalene, and biphenyl; and hydrogenated aliphatic $C_5$-, $C_9$-, and $C_{11}$-clusters were found. Compared with the results of previous experiments carried out in the absence of $C_2H_2$[421], the composition of the products was significantly different; therefore, they emphasized the key role of acetylene in the formation of dust analogs. The produced dust analogs were agglomerates with a diameter of several nanometers. The particles represent complex hydrocarbons consisting of sp, $sp^2$ and $sp^3$ bonds, i.e., they are a mix of aromatics connected by aliphatic or olefinic bonds. Therefore, all the above-mentioned molecules originated from acetylene, which includes small hydrocarbons, polyacetylenic chains, benzene and PAHs, and participate in the formation of carbon dust.

**Table 20.** Summary of considered reactions, kinetics (if available) and conditions where the reactions are relevant

| **Reactions** | **Kinetics** | **Objects & conditions** |
|---|---|---|
| PAH + PAH + $C_2H_2$ → $[PAH]_2$ (cluster)[419] | n.d. | Stellar envelopes (outer edge).<br>Conditions:<br>$n_H = 10^8$-$10^{12}$ cm$^{-3}$<br>T = 500-1500 K |

# 7. Catalytic reactions in space with acetylene

The abundance of molecules in cold ISM cannot be explained by considering only gas-phase reactions. The ISM, which is mostly gas, also contains approximately 1% dust particles ranging in size from a few tenths of nanometers to several microns.[60] These dust particles can catalyze many reactions because low-molecular-weight species can agglomerate on their surface. The interaction of gas species with grains can lead to the formation of COMs. Surface catalysis on solid interstellar particles can enable chemical pathways that cannot occur in the gas phase due to reaction barriers.[422,423] The importance of heterogeneous catalysis in astrochemistry has rapidly grown in recent years. In particular, rich grain-surface chemistry has been described in reviews by Tielens et al.[61] and Cuppen et al.[424] One exciting example of a catalytic reaction on the surface of a cosmic dust grain analog is the formation of cyclic prebiotic molecules such as pyrimidinone, uracil, cytosine, purine, urea, and dihydrouracil.[425,426] Simple compounds of elements such as carbon[427,428], silicon[429], titanium[430], iron[431] and many other transition metals[432], whose high catalytic activity has been well studied in organic synthesis reactions in recent decades, can act as catalytic centers of cosmic particles. Certainly, fascinating discoveries are coming in this astrochemical field in the future.

## 7.1. Catalytic reactions of acetylene in the laboratory

In our opinion, catalytic pathways for the formation of PAHs from acetylene might have a great potential. In the laboratory, the process of obtaining benzene and aromatic systems from acetylene and alkynes has become a convenient tool in organic chemistry since Reppe proposed the use of transition metals as cycloaddition catalysts in 1948.[433] These reactions make it possible to rapidly construct molecular scaffolds with various functional groups in a single step. For a long time, metal-catalyzed synthesis was limited to the trimerization of acetylene to benzene or led to complex mixtures of isomers in the case of substituted alkynes. For practical applications, product selectivity plays a key role. The creation of substituted benzenes is problematic due to the difficulty in controlling the chemoselectivity during metallacycle formation and regioselectivity upon the introduction of a third alkyne. Since then, the variety of proposed virtuous approaches for the synthesis of substituted benzene molecules using transition metal-catalyzed [2+2+2] cycloadditions has increased significantly.[434,435] Half a century later, selective trimerization of three different alkyne components was achieved with catalysts based on zirconium[436], titanium[437] and ruthenium[438]. Catalyst systems based on cobalt[439], rhodium[440], ruthenium[441], titanium[442] and nickel[443] help solve the problem of regioselectivity and produce 1,2,4-trisubstituted benzene rings. The use of a manganese catalyst made it possible to carry out the [2+2+2] cycloaddition of 1,3-dicarbonyls to terminal acetylenes.[444] The replacement of manganese with a rhenium catalyst led to the formation of pyrone adducts instead of benzenes.[445]

In addition to benzene derivatives, transition-metal-catalyzed cyclocotrimerizations are powerful tools for the synthesis of pyridines and their derivatives. Cyclo(co)trimerization of acetylene with nitriles is a convenient path for accessing a wide range of 2-substituted pyridines. These transformations occur using cobalt complexes as catalysts.[446]

The formation of bicyclic product as a result of [2 + 2 + 2]-cycloaddition reaction between acetylene and diynes was studied by More et.al.[447] This transformation is catalyzed by cyclopentadienylcobalt dicarbonyl. Notably, metal carbonyls may be widely available in space, where CO is one of the most abundant molecules. Moreover, metal carbonyls were found in interstellar dust clouds[448] and Jiange H5 chondrite[449].

Another important catalytic process is the preparation of triazoles in the click reactions of acetylene and azides in the presence of Cu(I).[450] In the presence of a Au(I) catalyst, it was recently discovered that acetylene can react as a dicarbene with alkenes, giving 1,1'-bi(cyclopropane).[451] Thus, acetylene is an excellent tool for the formation of a wide variety of cycles. These transformations were discussed in detail in a recent review devoted to cycloaddition reactions involving acetylene.[17] We summarize the catalytic reactions considered in this section in Table 21.

**Table 21.** Catalytic reactions known in industry but not adapted to astrochemical modeling.

| **Reactions** | **Catalytic metal** | **Objects & conditions** |
|---|---|---|
| 1. $2R_3$-$C_2H$ (1,3-dicarbonyl) + $C_2HOH$-$R_1R_2$(terminal acetylene) + Cat (Mn complex) → $C_6H_3OH$-$R_1R_2(R_3)_2$ → $C_6H_2$-$R_1R_2(R_3)_2$ (substituted benzene) + $H_2O$ 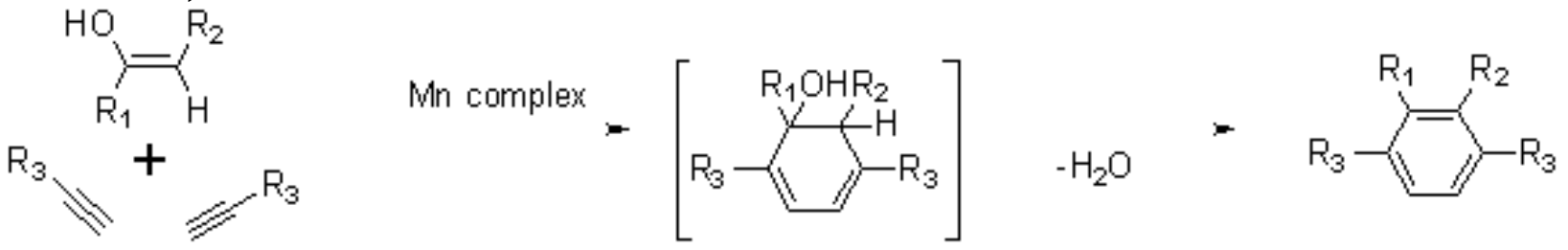  R1R2R3 - functional groups | Mn | not adapted to space objects |
| 2. $2C_2H_2$ + $C_5N(R)_6$ + Cat (Co complex) → $C_9N(R)_6$ (substituted pyridine) + $H_2$ 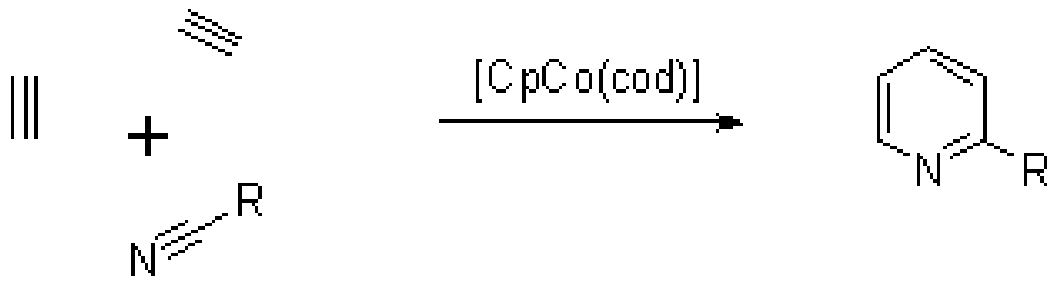  | Co | not adapted to space objects |
| 3. diyne + $C_2H_2$ + Cat ($CpCo(CO)_2$) → bicyclic product Cp – cyclopentadienyl 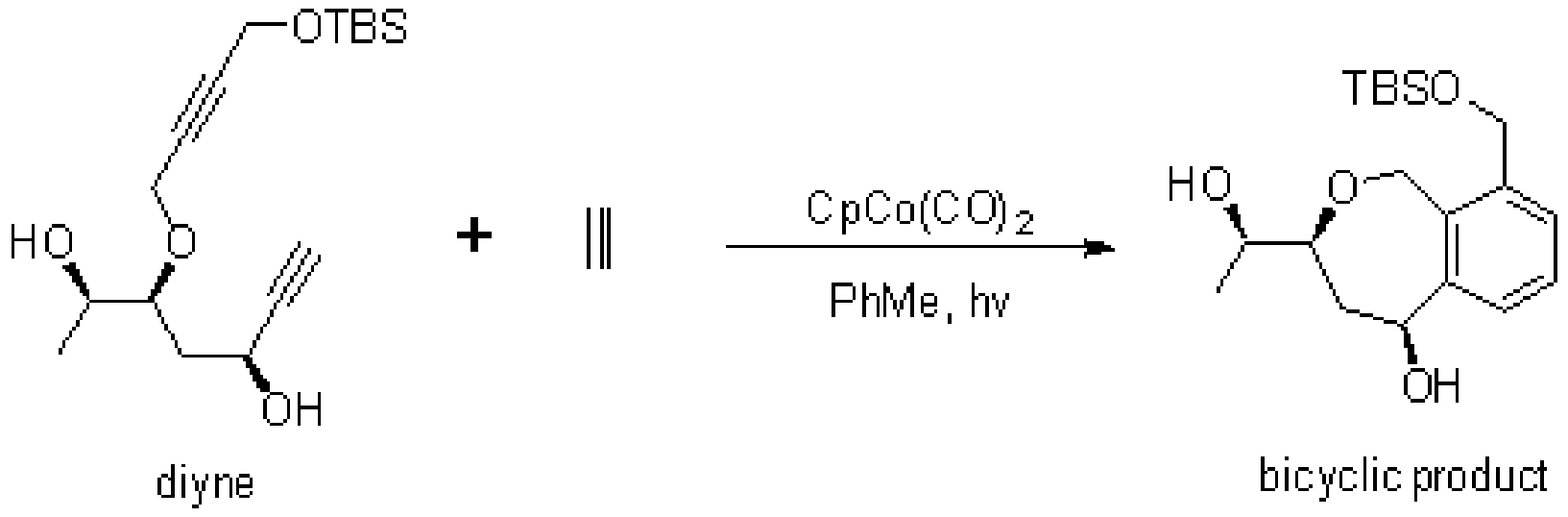  | Co | not adapted to space objects |
| 4. azide + $C_2H_2$ + Cat (CuI) → triazole 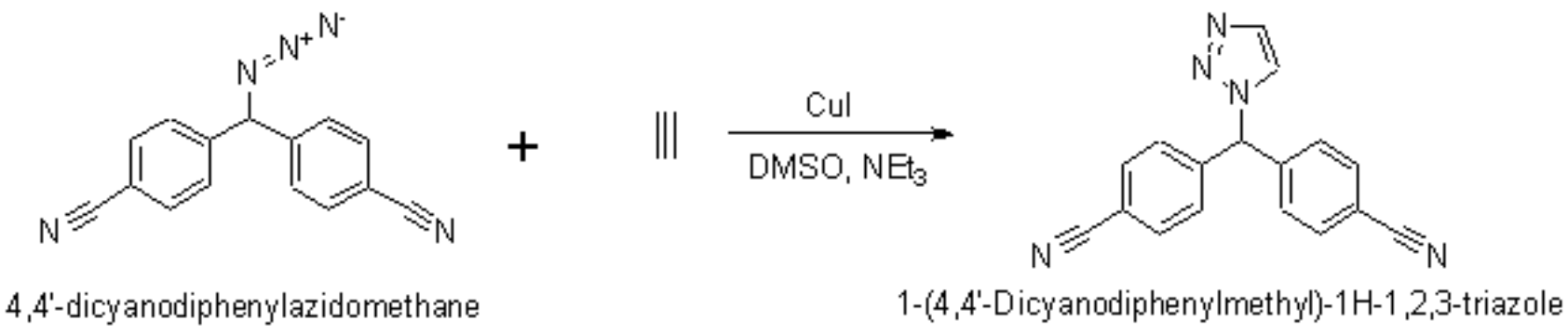  | Cu | not adapted to space objects |
| 5. alkene + $C_2H_2$ + Cat ($AuL^+$) → 1,1'-bi(cyclopropane) 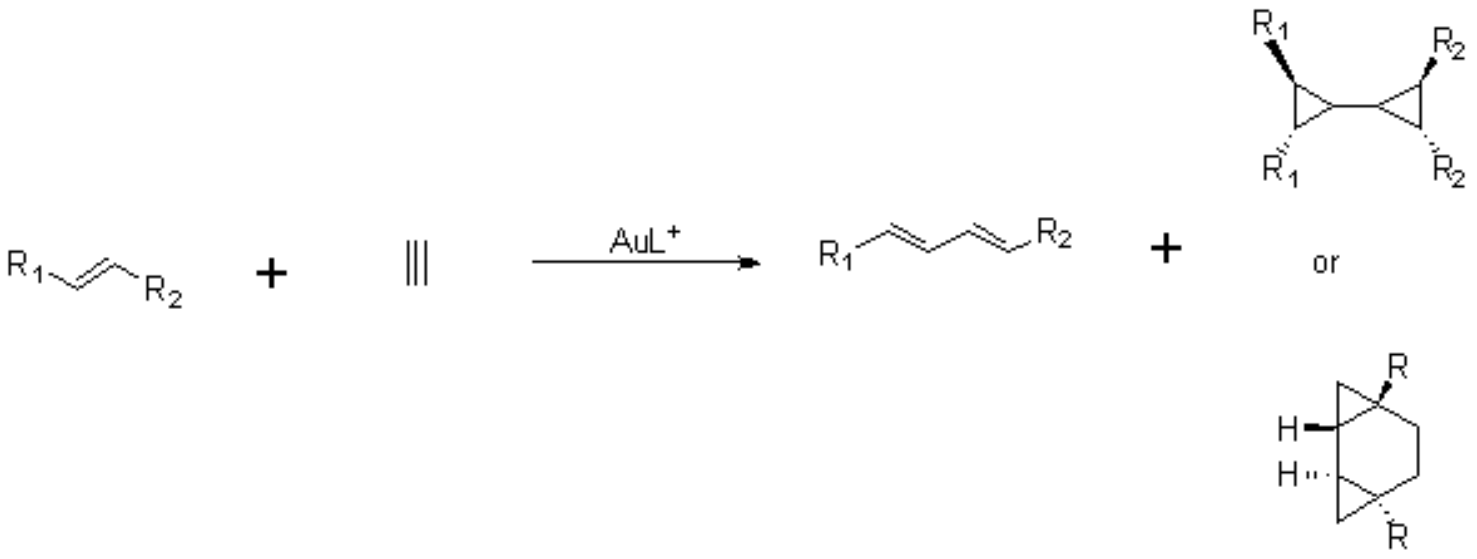  | Au | not adapted to space objects |

## 7.2. Catalytic reactions involving acetylene in space

The presence of catalytic species can provide a variety of pathways for the formation of aromatic systems. Metal complexes that are used in laboratories and industry are rare in outer space, although iron-nickel particles are present in meteorites and likely in cosmic dust.[452] The dust particles are either carbonaceous (HACs, graphites, and PAHs) or silicate. Silicate grains consist mostly of pyroxene ($MgSiO_3$) and olivine ($Mg_2SiO_4$).[197] Inclusions of aluminum, titanium and other metals were not excluded. Some carbides are also present in the ISM. Specifically, silicate carbide (SiC) is observed in many C-rich stellar envelopes[453], and titanium carbide is speculated to also be present in these envelopes.[454,455] Certainly, catalytic metals are present in the ISM in the gas phase as single atoms or ions, which are traced through their absorption/emission lines.[456] However, it is doubtful that these materials can act as catalysts in the gas phase, as three-body collisions are assumed to occur, which are quite rare under interstellar conditions.

In 2015, Zhao et al. demonstrated the catalytic conversion of acetylene to PAHs over SiC grains[457]. They observed the formation of aromatic molecules containing up to 20 carbon atoms and even a surface graphene-like coating in their experiments. The authors reported that acetylene activation occurs with the participation of dangling bonds on SiC surfaces, which are potential active sites for chemisorption and activation (Scheme 19). In the first step, acetylene adsorbs onto a silicon dangling bond to generate a silylacetenyl radical. Then, another acetylene molecule attaches to the formed radical to generate a 1-buten-3-ynyl radical, and the next acetylene molecule attaches to the diene. Finally, in the next step, the carbon radical attacks a neighboring Si dangling bond to form an octagonal ring. This ring has an unstable resonance structure with a positive and a negative charge at each end of the triene. The octatomic ring can undergo a rearrangement process to form a benzene ring to obtain the most stable structure. This process likely occurs in the dust formation zone because SiC grains are present there. Additionally, SiC was also found in presolar meteorites and comet materials returned to Earth by the Stardust mission[458]; therefore, the process may take place in other environments in addition to stellar envelopes.

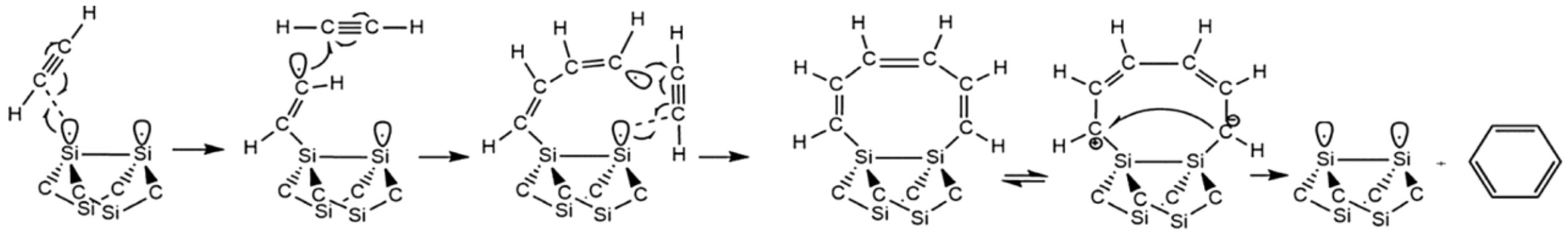

**Scheme 19.** The proposed mechanism for the catalytic formation of benzene from acetylene on a SiC particle. Reproduced with permission from ref 455. Copyright 2016 Royal Society of Chemistry.

The source of SiC can also be acetylene. Previously, Parker et al. experimentally showed that the silacyclopropenylidene (c-$SiC_2H_2$) molecule can be synthesized in the gas phase under single-collision conditions via the reaction of a silylidine radical (SiH) with acetylene. Later, Yang et al.[459] also investigated the reactions between silicon atoms and allene ($H_2CCCH_2$) and methylacetylene ($CH_3CCH$), and they obtained $SiC_3H_2$. In the above works, the authors noted that c-$SiC_2H_2$ and c-$SiC_3H_2$ can be photolyzed to carbon-silicon clusters such as silicon c-$SiC_3$ and c-$SiC_2$, respectively; thus, they can be precursors of SiC dust grains in the outflows of C-rich AGB stars.[460]

A recent study[30] showed that acetylene can be adsorbed on cosmic dust analogs in the temperature range found in Titan's atmosphere below 600 km. Compounds such as $Mg_2SiO_4$, $MgFeSiO_4$ and $Fe_2SiO_4$ can act as catalysts for the cyclotrimerization of acetylene to benzene. The authors showed that $C_6H_6$ production via $C_2H_2$ uptake on cosmic dust followed by cyclotrimerization and desorption is likely to be a competitive pathway between 80 and 120 km with gas-phase $C_6H_6$ production. Other recent studies have also shown that siliceous rocks containing siloxyl radicals can act as catalysts for complex organic matter synthesis.[461] The assumption about the formation of PAHs in interstellar clouds from acetylene due to a metal-catalyzed process was first put forward in 1992. The authors assumed that the formation of PAH organometallic compounds with Fe, Co, Ni, and other metals could be responsible for the observed metal transition depletion.[462]

Murakami et al. conducted a DFT study simulating acetylene trimerization in the ISM in water ice. The potential energy profiles of this reaction are given in Figure 21. The authors found that in the presence of acetylene and the $Fe^{+}(H_2O)_n$ cluster as a catalyst, benzene can be formed in one step, including one transition state structure of three acetylene molecules. Moreover, in the case of single-atom catalysis by $Fe^+$, benzene was formed in two stages, including the formation of one C–C σ-bond and two subsequent formations of C–C σ-bonds.[463] Subsequently, the authors expanded this study to include other transition metals: Sc, Ti, Mn, Co and Ni.[464] The number of electrons in the 3d orbitals of the transition metal cation significantly contributed to the catalytic efficiency of the acetylene cyclotrimerization process. The systems with $Mn^+$, $Co^+$ and $Ni^+$ had stable structures in which three acetylene molecules were coordinated to the cationic center of the transition metal. In this case, the energy levels of the transition complexes were higher than the energy levels of the initial reagents. In the case of systems with $Sc^+$ and $Ti^+$ cations, there is no significant barrier to the reaction path. That is, isolated acetylene and $(C_2H_2)_2$, together with $Sc^+$ and $Ti^+$ cations associated with the water cluster $(H_2O)_8$ and $NC^-$, can form a benzene-metal-cation complex without an entry barrier. Thus, according to calculations, $Sc^+$ and $Ti^+$ cations act as effective catalysts for the cyclotrimerization of acetylene. In Table 22, we summarize the catalytic reactions that were considered in astrochemistry.

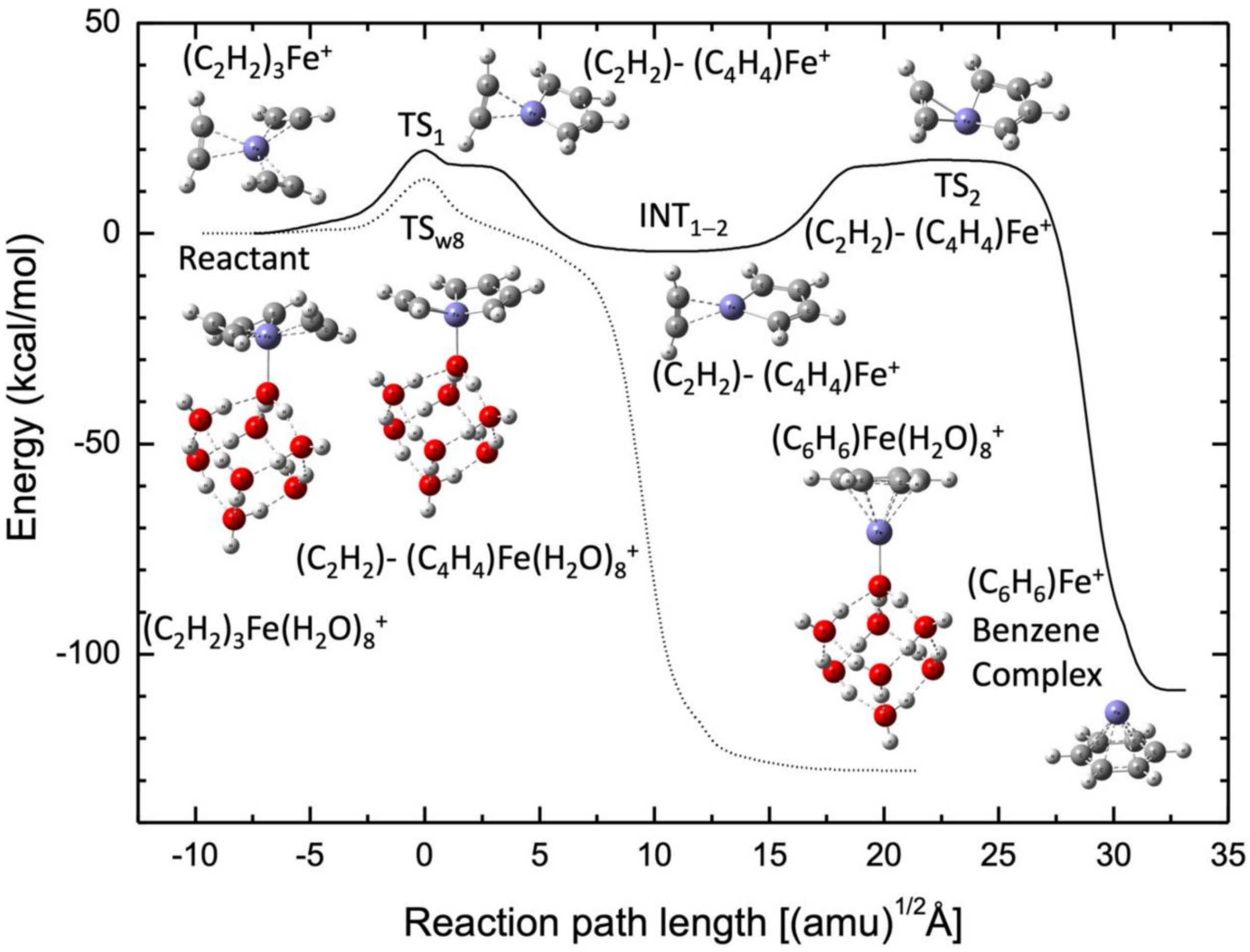

**Figure 21.** Potential energy profiles for the benzene formation by acetylene cyclotrimerization reactions catalyzed by $Fe^+$ and cluster $Fe^+(H_2O)_8$ obtained at the B3LYP(D3BJ)/def2-SVPP DFT level. Reproduced with permission from ref 461. Copyright 2022 MDPI.

**Table 22.** The catalytic reactions which were considered in astrochemistry.

| **Reactions** | **Kinetics** | **Objects & conditions** |
|---|---|---|
| 1. $SiC + 3C_2H_2 \rightarrow SiC + C_6H_6$[457] | n.d. | Stellar envelopes.<br>Conditions:<br>$n_H = 10^8\text{-}10^{15}$ cm$^{-3}$<br>T = 500-2500 K |
| 2.<br>a. $SiH + C_2H_2 \rightarrow c\text{-}SiC_2H_2$<br>b. $c\text{-}SiC_2H_2$ + energy (UV, shocks) → $c\text{-}SiC_2 + H_2$[460] | n.d. | a) Stellar envelopes.<br>Conditions:<br>$n_H = 10^8\text{-}10^{15}$ cm$^{-3}$<br>T = 500-2500 K<br><br>b) Stellar envelopes (outer edge).<br>Conditions:<br>$n_H = 10^8\text{-}10^{12}$ cm$^{-3}$<br>T = 500-1500 K.<br>+ PDRs, DISM. Conditions:<br>T = 10 - 500 K<br>$n_H = 10^{-1}$ - $10^5$ cm$^{-3}$<br>$G_0$ = 1 to $10^5$<br>Galactic CR field. |
| 3.<br>a. $Si + C_3H_4 \rightarrow c\text{-}SiC_3H_2 + H_2$<br>b. $c\text{-}SiC_3H_2 + h\nu \rightarrow c\text{-}SiC_3 + H_2$[459] | provided many pathways in the work of Yang [459] | The same as reactions #2 |
| 4. $3C_2H_2$ + M($Mg_2SiO_4$, $MgFeSiO_4$ or $Fe_2SiO_4$) → $C_6H_6$[30] | $k_4 = 1.2 \cdot 10^{-5}$ cm$^3$ s$^{-1}$ (T = 150 K)[30] | Titan's atmosphere.<br>Conditions:<br>$n_H$: $10^{10}\text{-}10^{16}$ cm$^{-3}$<br>T= 100-200 K<br>$G_0$: up to $10^2$<br>Galactic CR field. |
| 5. Ice reactions.<br>$3C_2H_2$ + M($Fe^+(H_2O)_n$) → $C_6H_6$ + M($Fe^+(H_2O)_n$)[463] | n.d. | Molecular clouds, prestellar cores, comets.<br>Conditions:<br>$n_H = 10^2\text{-}10^7$ cm$^{-3}$<br>T = 10 K<br>$G_0$ = very low<br>Galactic CR field. |

## 7.3. PAHs as a catalyst

The active development of carbocatalysis in the last decade has shown that important catalytic processes can proceed even without the participation of metals. Several decades ago, PAHs were noted to be appropriate for $H_2$ formation in PDRs.[465,466] Functionalized PAHs with oxygen-bearing groups can be a source of small oxygen-bearing molecules under conditions of UV irradiation in the ISM.[467,468] Recently, it has been shown that PAHs can also be catalysts for many other hydrogenation reactions in the ISM, resulting in the synthesis of molecules such as $H_2O$ and HCO.[469] Similarly, under certain conditions, PAHs can play a role in the synthesis of benzene from simple precursors. Recently, published DFT calculations and experimental data have confirmed the formation of benzene by trimerization of three acetylene molecules catalyzed by PAH carbene sites at ~900 K.[470] The authors considered the carbene sites on the zigzag edge of PAHs to be catalytic sites. The considered PAHs of various sizes and configurations from $C_6H_6$ to $C_{361}H_{51}$.

Quantumochemical calculations of the reaction mechanism made it possible to reconstruct the free energy profiles and visualize the spatial distributions of the spin density for all stages of benzene formation from three acetylene molecules on the carbene centers of PAH molecules (Figure 22). The stepwise addition of acetylene molecules at the carbene center was promoted by continuous spin migration onto the β-carbon of the added alkyne moiety. It has been shown that the trimerization process will dominate linear oligomerization after the addition of three acetylene molecules due to the free energy driving force. Due to the high strength of the C–C bond in the intermediate complex, the removal of benzene and regeneration of the carbocatalyst can be kinetically difficult. The authors considered two possible mechanisms for the elimination of benzene from a PAH. In the associative pathway, in which the benzene molecule is competitively replaced by an acetylene molecule, a higher activation energy was observed. The associative mechanism is characterized by the predominant localization of unpaired electrons at the carbene center. The dissociative pathway has a lower activation energy and is accompanied by the delocalization of unpaired electrons in a polyaromatic system. The authors also noted that the removal of a benzene molecule by the dissociative mechanism for monoradicals is characterized by a very low activation energy (1.0–3.0 kcal mol$^{-1}$) compared to that of carbenes (~26.0 kcal mol$^{-1}$). Therefore, polyaromatic monoradicals can be promising candidates for carbocatalysts in ISM, although thorough adaptation of this reaction to astrophysical conditions is needed. Interestingly, the catalytic cycle was controlled by reversible fluctuations in the spin density. These vibrations were responsible for the formation of benzene and the reactivation of active sites. This reaction is shown in Table 23, although the kinetics of this reaction, which can be used in astrochemistry, have yet to be determined.

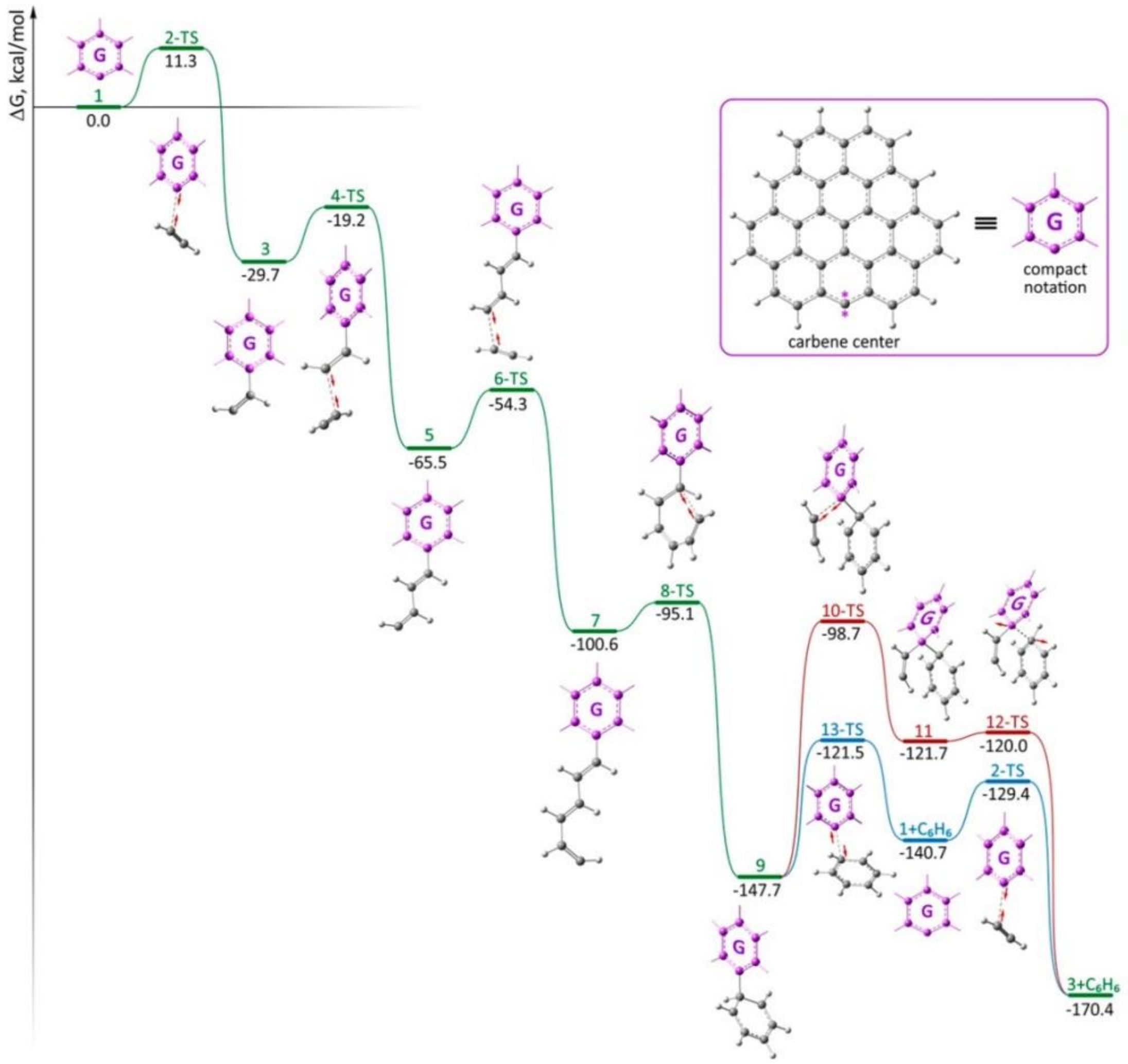

**Figure 22.** Free energy profile for the acetylene cyclotrimerization reaction catalyzed by $C_{37}H_{14}$. The associative pathway of the product elimination is shown in red; the dissociative pathway is shown in blue. The initial form of the $C_{37}H_{14}$ catalyst, location of the chemically reactive center, and designation of this $C_{37}H_{14}$ structure in the catalytic cycle representations are shown in the magenta frame. Reproduced with permission from ref 468. Copyright 2020 American Chemical Society.

Notably, carbenes are important intermediates in processes leading to carbon nanostructures in the ISM. Previously, carbene species were identified in the interstellar space.[471] Recently, Mebel et al., using crossed molecular beam data with electronic structure calculations and quasiclassical trajectory simulations, reported the fundamental reaction mechanisms leading to the formation of highly reactive carbene species under single-collision conditions relevant to extreme astrochemical conditions.[472] As a model, the authors used the reaction of methylidine (CH) with diacetylene (HCCCCH) under single collision conditions to form triplet pentadiynylidene (HCCCCCH) and singlet ethynylcyclopropenylidene (c-$C_5H_2$) carbenes. Considering the prevalence of carbenes in various cosmic media, their role in promoting the evolution of high-molecular-weight organic compounds can be underestimated. The mechanism of the catalytic trimerization of acetylene on the active sites of PAHs and carbon dust particles could be one of the possible ways to synthesize hydrocarbons on aerosol particles and organic haze layers in hydrocarbon-rich atmospheres on planets and their moons. The role of catalytic processes in the evolution of hydrocarbon molecules on the surface of interstellar dust particles, ice, or atmospheres of planets and satellites has yet to be assessed.

**Table 23.** The reaction of acetylene trimerization on PAHs.

| **Reactions** | **Kinetics** | **Objects & conditions** |
|---|---|---|
| $3C_2H_2$ + PAHs → $C_6H_6$ + PAHs[470] | n.d. | Stellar envelopes (outer edge).<br>Conditions:<br>$n_H = 10^8$-$10^{12}$ $cm^{-3}$<br>T = 500-1500 K |

# 8. Acetylene in prebiotic chemistry evolution

## 8.1. Acetylene in reactions producing prebiotic molecules

One important aspect that determines the significance of acetylene chemistry research in space is its possible role in prebiotic chemistry evolution. Acetylene could play an important role in the origin of life, and the presence of acetylene in planetary atmospheres may indicate the possible existence of an extraterrestrial anaerobic ecosystem.[42] Acetylene may serve as a signature for young Earth-like exoplanets with reducing atmospheres undergoing heavy bombardment.[126] Such conditions are considered to be similar to those that were intrinsic to the primordial Earth during the Hadean Eon (>4 Gyr ago), when chemical evolution and prebiotic accumulation took place. Theoretical models of the planetary atmosphere and surface that consider photochemical reactions predict steady-state surface $C_2H_2$ quantities of 0.4% for Earth-like young planets.[126]

HCN is considered one of the key precursors of prebiotic molecules. In turn, an important source of HCN is the photolysis of a mixture of $C_2H_2$ and $NH_3$ in planetary atmospheres at temperatures of 130-298 K.[473] Recent studies of the photolysis mechanisms of $NH_3/C_2H_2$ mixtures have shown that this process can be a source of many organonitrogen compounds that are important for prebiological organic chemistry.[41] Under similar conditions, the formation of so-called tholins[474], which are nitrogen-containing polymeric compounds that are formed on the surface of many planets and icy bodies of the Solar system, primarily Titan, can occur. [283,375] The possible role of tholins in the evolution of the prebiotic chemistry of planets, including Pluto[475], has been noted. [476] It is likely that tholin formation proceeds through acetylene chemistry analogous to that of cyanopolyynes (see Section 3.2) based on the hypothesized structures of the tholins. However, this pathway has yet to be determined.

In a number of recent experimental and theoretical works, Cable et al. demonstrated the formation of acetylene co-crystals with various organic molecules on the surface of Titan and their influence on the landscape of the satellite. It was experimentally shown that cocrystal $C_2H_2$-$NH_3$ can be formed on Titan at 90 K.[477] In addition, DFT calculations predicted the existence of three-component co-crystals with the composition 2c-$C_6H_6$:$C_2H_2$:HCN under the surface conditions of Titan.[478] Cable et al.[479] investigated the formation of co-crystals of acetylene ($C_2H_2$) and acetonitrile ($CH_3CN$) on the surface of Titan. The authors noted that similar molecular minerals can be widely distributed in the labyrinth terrains of Saturn's moon. Finally, the same authors[480] demonstrated the stability of a cocrystal of acetylene and pyridine in the temperature range from 90 K to 180 K. The authors suggested that such acetylene cocrystals could provide concentration, storage and transport of prebiotic and high-energy molecules for the functioning of hypothetical biological systems.

A number of papers have been devoted to formulating speculative hypotheses about the presence of life on Titan using acetylene as an energy source, citing the deficiency of $C_2H_2$ and $H_2$ on the satellite's surface as a possible biosignature.[487,481,482,483,484,485] The proposed models of acetylene metabolism on Titan included the following exothermic reactions (Scheme 20).

**A** $$C_2H_2 + 2H_2 \longrightarrow C_2H_6 \quad \Delta G_0 = -242\ \text{kJ}$$

**B** $$C_2H_2 + 3H_2 \longrightarrow 2CH_4 \quad \Delta G_0 = -375\ \text{kJ}$$

**Scheme 20.** Proposed models of acetylene metabolism on Titan.

Such metabolism could be stimulated by a strong thermodynamic benefit. However, a number of factors make the existence of life under such conditions unlikely. One of the factors is the low solubility of large molecules in liquid hydrocarbons, which creates problems for the functioning of biopolymers. This factor becomes even more critical given the extremely low temperatures.[486] In addition, calculations of the biomass value predicted for life on Titan based on similar biochemistry on Earth have shown that the existence of bacteria using acetylene as an energy source on Titan is unlikely.[483]

Finally, Abbas and Schulze-Makuch proposed three interesting, albeit obviously flawed, pathways for the formation of amino acids as essential prebiogenic components involving acetylene in hydrothermal vents or subsurface Titan.[487] Catalytic reactions play a key role in the first model, which is illustrated in Scheme 21. The first stage was the trimerization of acetylene with the participation of metal catalysts or clay minerals and zeolites. The formation of substituted benzene could occur due to the participation in the reaction of recombination products of acetylenyl radicals, for example, with methyl or ethynyl radicals. The subsequent formation of nitro-substituted alkylbenzene, according to the authors, could occur as a result of interactions with highly active $NO_2^+$ cations that appeared as a result of ionizing radiation. During the next step, the nitro group with hydrogen is converted to an amino group in the presence of a metal catalyst, and an alkyl aminobenzene is subsequently formed. Finally, oxidation of the alkyl radical to a carboxyl group led to the formation of the corresponding amino acid.

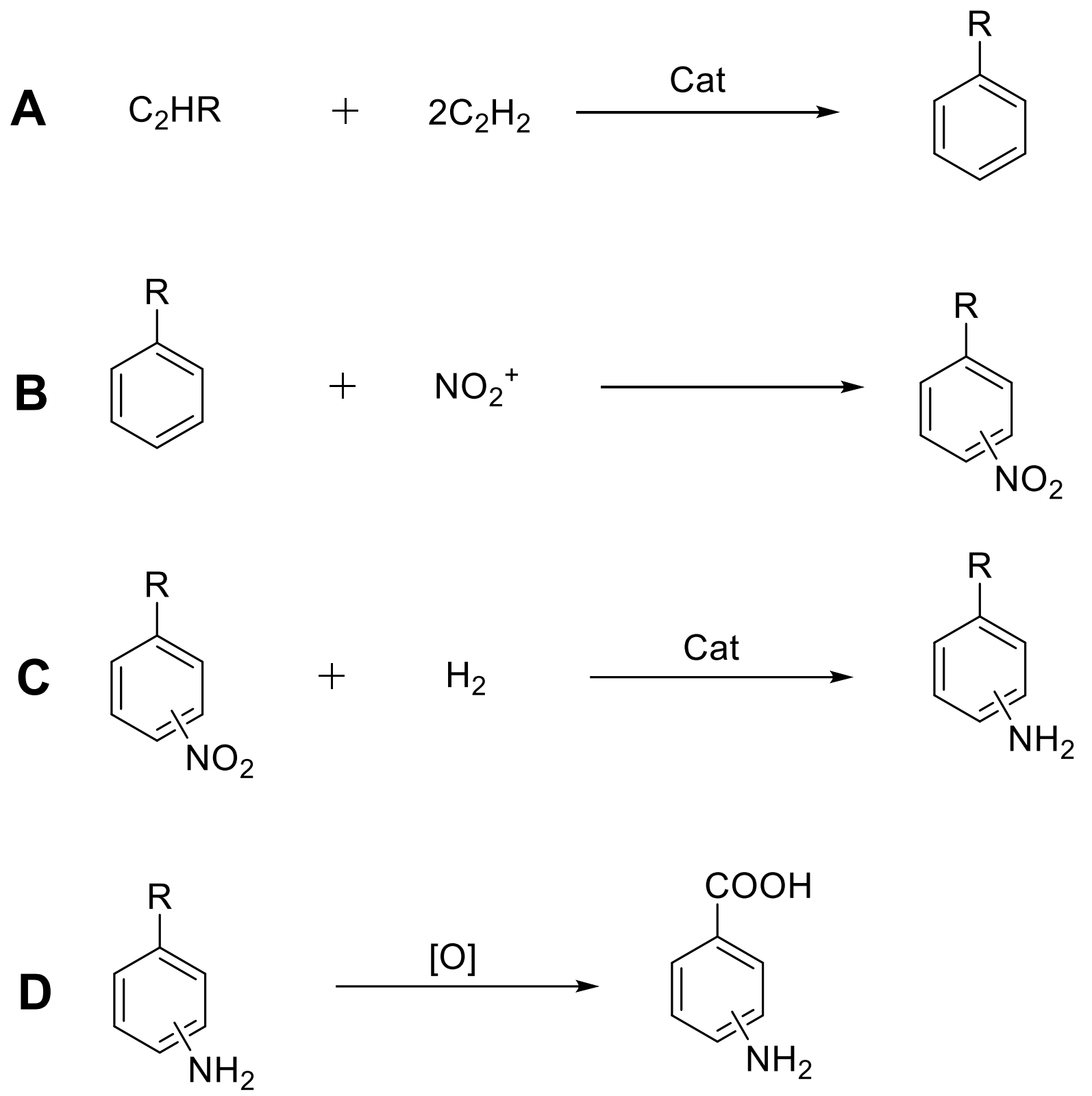

**Scheme 21.** Formation of amino acids from acetylene according to model I by Abbas et al.

In the second model, acetaldehyde is formed from acetylene in the presence of peroxide and hindered borane, which is illustrated in Scheme 22. The authors admit that the presence of hydrogen peroxide on the surface of Titan in significant quantities is unlikely. However, these findings indicate the possibility of peroxide formation from water under the influence of UV radiation. The aldehyde that is formed from acetylene or its derivatives must further react with ammonia and hydrogen cyanide to form the α-amino nitrile. Acid hydrolysis of this compound under hot spring conditions can lead to the formation of an amino acid such as alanine.

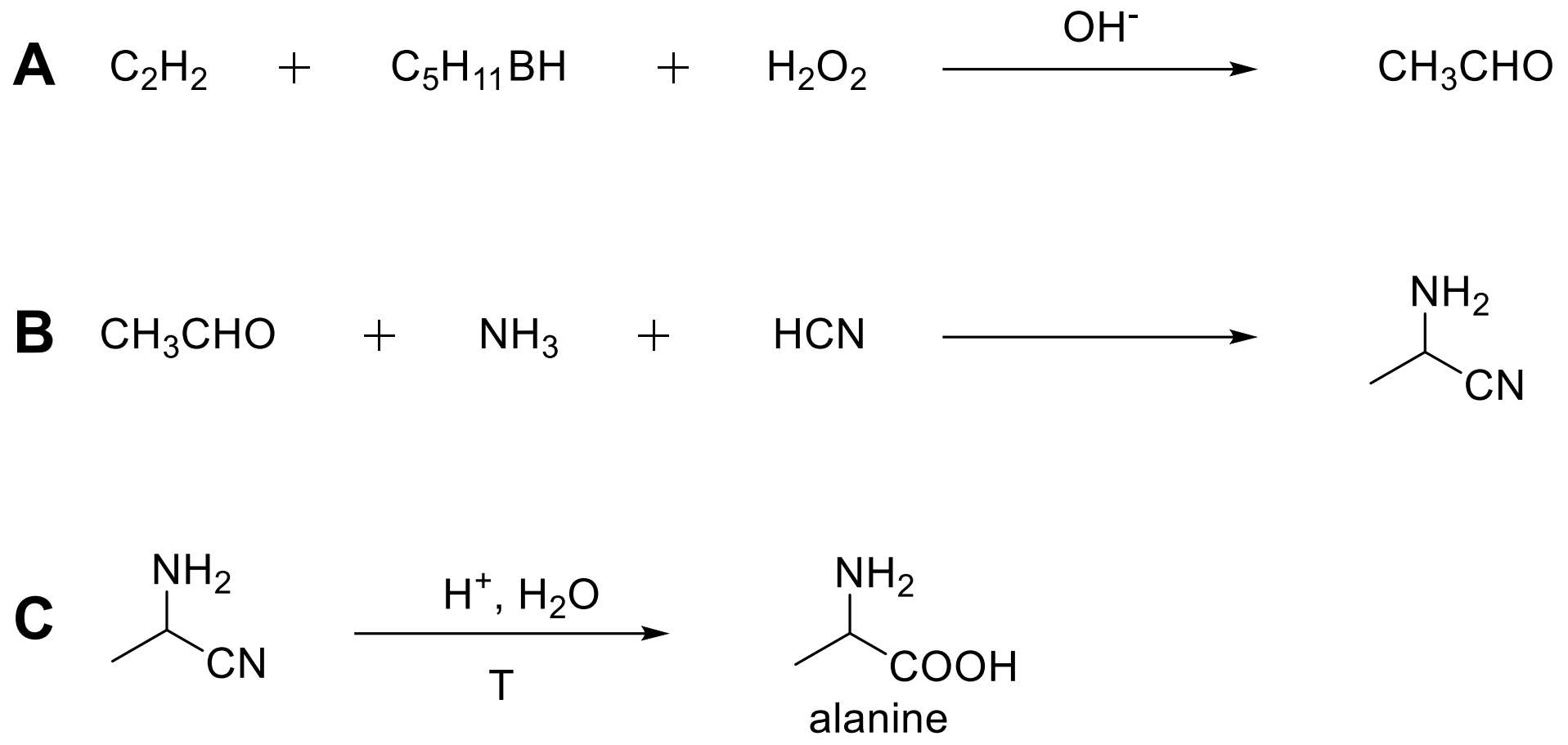

**Scheme 22.** Formation of alanine from acetylene according to the model II by Abbas et al.

The third model (shown in Scheme 23) is associated with the formation of an alkynyl radical and a halogen radical under the influence of UV radiation. These interactions will lead to the appearance of alkynyl halides. In turn, the reaction of an alkynyl halide as a result of a substitution reaction can give the corresponding nitrile. The corresponding carboxylic acid will result from hydrolysis of the nitrile in an acidic environment. The addition of ammonia to a double bond results in the corresponding amino acid. The authors suggested that the first route of amino acid synthesis is the most relevant to the conditions of Titan due to the lack of liquid water.

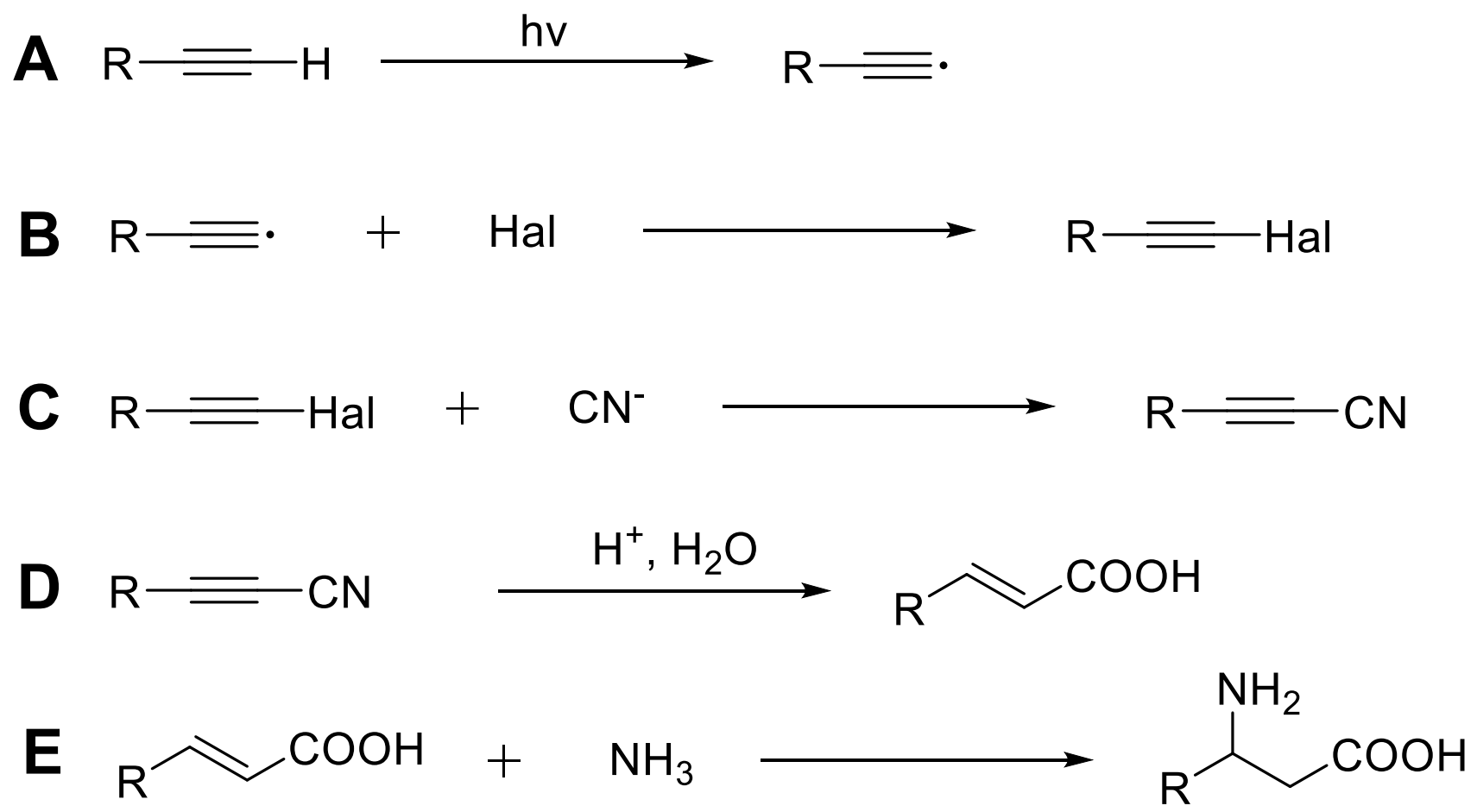

**Scheme 23.** Formation of alanine from acetylene according to the model III by Abbas et al.

Considering the role of acetylene in the formation of key prebiotic organics on premordial Earth, acetylene could be expected to be effective in reducing the $CO_2$- and $H_2$-rich atmosphere of primitive Earth. In this case, UV photolysis and volcanic lightning produce $CH_4$ and other light hydrocarbons, which, in turn, can form HCN and other N-species via reactions with dissociated atmospheric $N_2$. These simple organic molecules can be converted into prebiotic species (nucleobases, fatty acids, amino acids, alcohols, etc.) via aqueous synthesis. In a reducing atmosphere, acetylene could appear in significant quantities. For example, Segura et al.[488] modeled the chemical processes that occur under the influence of volcanic lightning under conditions of explosive volcanism. Volcanic lightning was simulated by focusing a high-energy infrared laser beam on a volcanic gas mixture consisting of 64% $CH_4$, 24% $H_2$, 10% $H_2O$ and 2% $N_2$. The authors relied on an accretion model and measured the nitrogen content in Martian meteorites. Among the products, 11 hydrocarbons were identified, and the most common compound was acetylene. Rimmer et al.[126] studied potential processes that could occur in the atmosphere of a young Earth-like exoplanet under conditions of protoplanetary disk collapse. A mixture of methane ($CH_4$), carbon monoxide (CO), nitrogen ($N_2$) and water was used as a model atmosphere. In the simulation of the high-speed impact of an extraterrestrial body on the early atmosphere of a planet, significant quantities of substances such as acetylene (8%), hydrogen cyanide (5%), and cyanoacetylene (5%) form inside the reactor as does ammonia (1%). The authors concluded that acetylene can be considered one of the signatures of the impacts of bodies from protoplanetary disk on planets with reducing the atmospheres.

Similar chemical processes could have occurred on the early Earth. However, according to modern geophysical studies, after the solidification of the magma ocean after the Moon-forming impact, the atmosphere of Hadean Earth should have been rather neutral, $CO_2$-rich and $H_2$-poor.[489,490,491] Under these conditions, organic syntheses from feedstock atmospheric gases became much less effective[492], and the delivery of organics via giant impacts or localized synthesis during volcanic eruptions likely became a key process.

Acetylene and other key feedstock molecules for prebiotic chemistry can be produced in hydrothermal and volcanic springs. Currently, hydrocarbon gases can be found in the fumaroles of geothermal areas.[493] Acetylene has also been detected in Hawaiian volcanic glasses.[494] Moreover, modern volcanic activity is very different from the volcanic activity of the primordial Earth. Early experimental simulations of volcanic processes on primordial Earth revealed the formation of unsaturated compounds such as acetylene, propylene, vinyl acetylene, methyl acetylene, and butadiene.[495] Rimmer et al.[496] calculated the chemical composition of volcanic gases feeding shallow and superficial vents over a range of pressures and C/N/O ratios in basalt. In the case of an ultrareduced and carbon- and nitrogen-rich gas phase of magma, interactions with groundwater in volcanic vents lead to $10^{-3}$–1 mol $l^{-1}$ concentrations of diacetylene ($C_4H_2$), acetylene ($C_2H_2$), cyanoacetylene ($HC_3N$), hydrogen cyanide (HCN), bisulfite ($HSO^{3-}$), hydrogen sulfide ($SH^-$) and soluble iron in wastewater (Figure 23).

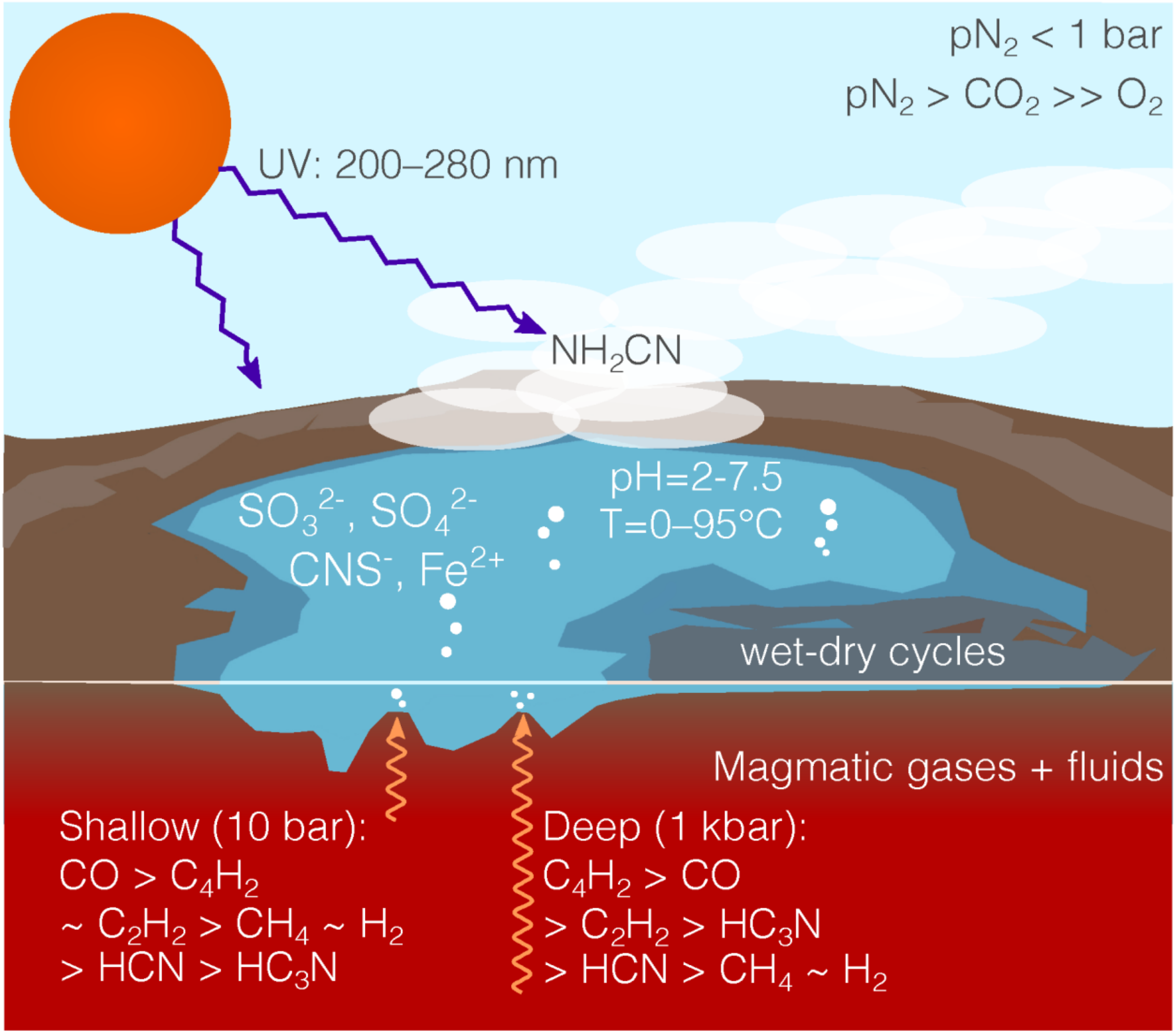

**Figure 23.** Chemistry of a surface hydrothermal vent fed by carbon- and nitrogen-rich magma. The yield of diacetylene ($C_4H_2$), acetylene ($C_2H_2$), cyanoacetylene ($HC_3N$), hydrogen cyanide (HCN), hydrogen ($H_2$) and methane ($CH_4$) from the pool is shown in ratios determined by the depth at which the gas is located. The interaction of species under the influence of UV leads to the formation of a small amount of cyanamide ($NH_2CN$). Reproduced with permission from ref 494. Copyright 2019 MDPI.

Acetylene could have been an energy source for early microorganisms on Earth and could also be an energy source for hypothetical life forms on other planets. *Pelobacter acetylenicus* was discovered in 1985 using acetylene as the sole source of carbon and energy.[497,498] *Pelobacter acetylenicus* can only grow by acetylene. The key enzyme of these bacteria is acetylene hydratase.[499] Acetylene hydratase contains transition metals, one [4Fe-4S] cluster and one tungsten atom, coordinated by two molybdopterin-guanosine dinucleotide ligands.[500] During the initial stage of acetylene fermentation, this enzyme catalyzes the hydration of acetylene to acetaldehyde through an enol intermediate. Acetaldehyde is transformed into acetyl-coenzyme CoA. Acetyl-CoA is subsequently used to construct metabolites with $C_2$-units. At subsequent stages, the formation of ethanol, acetate and hydrogen occurs (Scheme 24). Therefore, acetylene has the potential to fuel complete metabolism on its own.

$$C_2H_2 + H_2O \xrightarrow[\text{enzyme system}]{\text{acetylene hydrating}} CH_3CHO$$

$$2CH_3CHO + H_2O \longrightarrow C_2H_5OH + CH_3COO^- + H^+ + 1/2H_2$$

**Scheme 24.** Acetate formation from acetylene by bacterial metabolism.

The biochemistry of such organisms is much simpler than that of methanogens, which may indicate that acetylene may be the dominant food for primitive organisms early in evolution.[501] Considering these findings, Oremland and Voytek called acetylene a ‘fast food’ for early life and suggested that such microorganisms that consume acetylene could play a key role in the evolution of the Earth's early biosphere.[42] In Table 24, we provide the results of prebiotic chemistry discussed in this section.

**Table 24.** Summary of considered reactions, kinetics (if available) and conditions where the reactions are relevant.

| **Reactions** | **Kinetics** | **Objects & conditions** |
|---|---|---|
| 1. Mixture $C_2H_2/NH_3$ + hν → [a chain of reactions] → HCN + others[41] | Rate constants can be found or calculated for the most of reactions in the chain | Atmospheres of Jovian planets and comets.<br>Conditions:<br>$n_H = 10^{10}$-$10^{16}$ $cm^{-3}$<br>T = 50-300 K<br>$G_0$ = up to $10^2$<br>Galactic CR field |
| 2. $C_2H_2$ + $NH_3$ → co-crystal $C_2H_2$:$NH_3$[477]<br>3. 2c-$C_6H_6$ + $C_2H_2$ + HCN → co-crystal 2c-$C_6H_6$:$C_2H_2$:HCN[478]<br>4. $C_2H_2$ + $CH_3CN$ → co-crystal $C_2H_2$:$CH_3CN$[479] | n.d. | Titan’s atmosphere.<br>Conditions:<br>$n_H = 10^{10}$-$10^{16}$ $cm^{-3}$<br>T = 60-300 K<br>$G_0$ = up to 10<br>Galactic CR field |
| 5. $C_2H_2 + 2H_2 \rightarrow C_2H_6$<br>$C_2H_2 + 3H_2 \rightarrow 2CH_4$[485] | n.d. | Titan’s atmosphere, planetary atmospheres, probably early Earth. |

**Table 24.** Summary of considered reactions, kinetics (if available) and conditions where the reactions are relevant.

| **Reactions** | **Kinetics** | **Objects & conditions** |
|---|---|---|
| 6. $C_2H_2$ + $H_2O$ [acetylene hydrating enzyme system] → $CH_3CHO$<br>$2CH_3CHO$ + $H_2O$ → $C_2H_5OH$ + $CH_3COO^-$ + $H^+$ +1/2$H_2$[499] | | Conditions:<br>$n_H = 10^{10}$-$10^{16}$ cm$^{-3}$<br>T = 50-300 K<br>$G_0$ = up to $10^2$<br>Galactic CR field. |

## 8.2. From acetylene to nitrogen-bearing PAHs and nucleobases

Previously, we described the mechanisms of PAH formation from acetylene. In this section, specific PAHs that are connected with life precursors are considered. Many COMs, including prebiotic molecules, are formed in astrophysical ices; consequently, the ices and chemical reactions inside them are the subject of interest. Most PAHs are depleted in ices in dense and cold molecular clouds. A fraction of PAHs can attach different functional groups (O, CN, $NH_2$, $CH_3$, COOH, etc.) instead of peripheral H atoms.[502,503,504] Taking into account that the ices mostly consist of water and that there are also other oxygen-bearing molecules such as methanol, oxygen-bearing functional groups, -OH and =OH, are highly expected. Consequently, quinones, ethers and aromatic alcohols are also expected.[505] For example, the formation of quionone (1,4-napthaquinone) in water ice was traced experimentally[503], and this quinone performs important biochemical functions, such as electron transport and oxidative phosphorylation.

Apart from functional groups, carbon atoms in PAHs can be replaced by nitrogen or oxygen. Nitrogen (N-) heterocycles, pyrimidine ($C_4H_4N_2$) and purine ($C_5H_4N_4$), are the main components of DNA and RNA nucleobases. N-heterocycles, associated nucleobases and other biorelevant molecules have been found in meteorites.[506,507,508,509,510,511,512,513] The detection of $^{15}N/^{14}N$ isotope enrichment[514] as well as the terrestrially rare nucleobases 2,6-diaminopurine and 6,8-diaminopurine in Murchison[508,509] points to an extraterrestrial origin. N-heterocycles have not yet been detected in the ISM, although some observations can be attributed to their presence. Specifically, Hudgins et al. showed that the commonly observed PAH band at 6.2 μm can be associated with traces nitrogen-substituted PAHs.[515]

Previously, it was shown that nitrogenous bases are unlikely to be formed and survive in interstellar and circumstellar environments, as these compounds are highly sensitive to UV radiation. It was estimated that gas phase N-heterocycles are destroyed within 10–100 years in diffuse ISM[516]. Their lifetime does not exceed a few hours at a distance of 1 AU from the Sun either. The presence of N-heterocycles at least in meteorites means that there must be mechanisms and conditions that make possible their formation and survival outside the Earth.

Research by Parker et al. showed that the formation of (iso)quinoline ($C_9H_7N$) probably proceeds with the participation of acetylene molecules in the circumstellar envelopes of evolved C-rich stars. Experiments in a pyrolytic reactor in combination with DFT calculations showed that the formation of (iso)quinoline can occur as a result of the reaction of the *m*-pyridyl radical in series with two molecules of acetylene.[517] In turn, as shown by Kaiser et al., pyridine can be obtained through barrierless elemental reactions: 1) by the reaction of vinyl cyanide ($C_2H_3CN$) with cyanovinyl (HC=CH–CN)[518] and 2) by the reaction of cyano radical (CN) with 1,3-butadiene ($C_4H_6$)[519,520].

Under low-temperature conditions in molecular clouds (for example, TMC-1 and OMC-1), planetary atmospheres and their satellites (for example, Titan and Pluto), quinoline and isoquinoline ($C_9H_7N$) can be synthesized via fast, de facto barrierless reactions involving *o*-, *m*-, and *p*-pyridinyl radicals ($C_5H_4N$) with vinylacetylene ($C_4H_4$), similar to the HAVA mechanism of naphthalene formation. Molecular beam experiments combined with electronic structure calculations of the reaction (Figure 24) conducted by Mebel et al. point to this mechanism.[521]

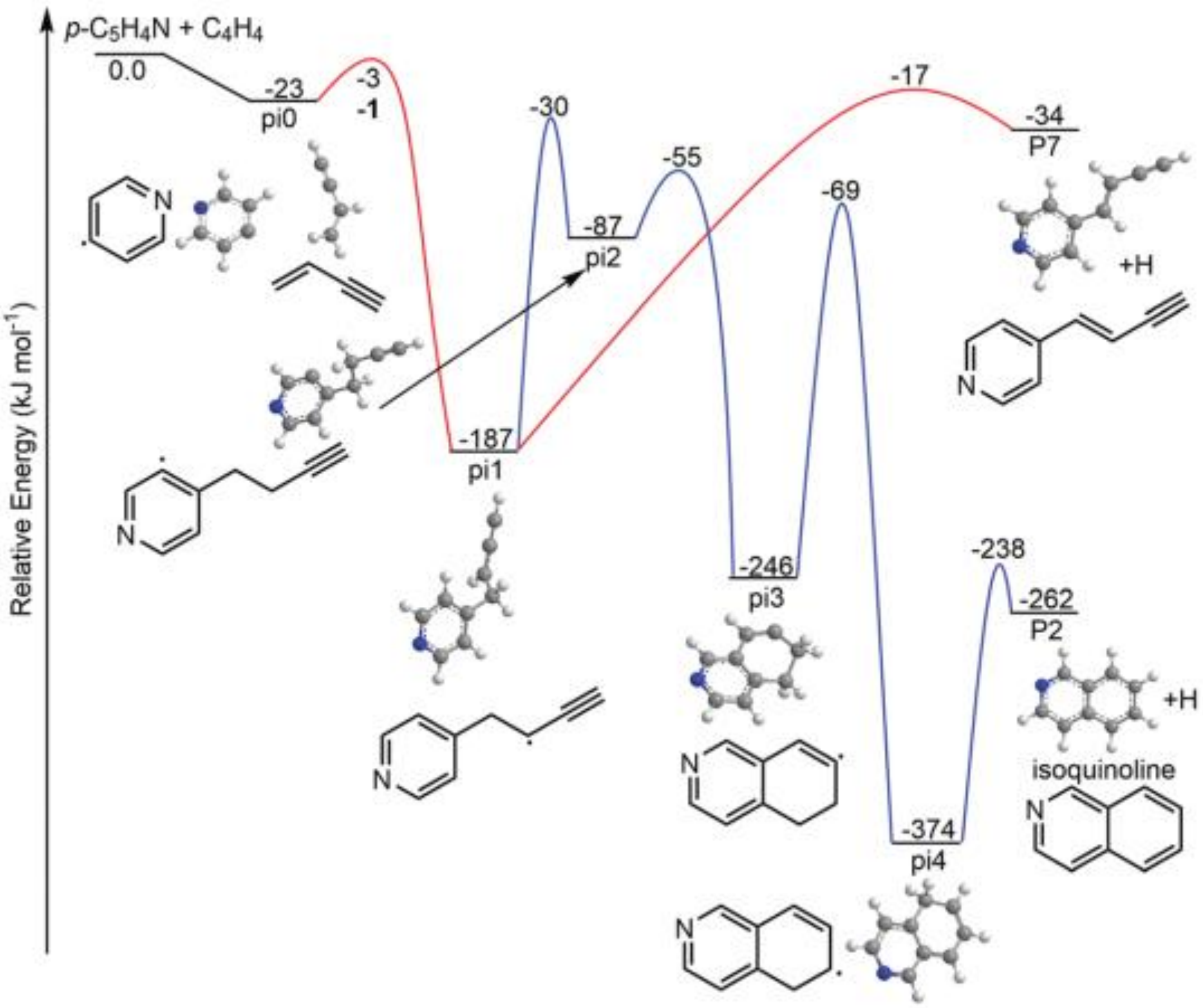

**Figure 24.** Computed potential energy surface for the reaction of p-pyridinyl ($C_5H_4N$) radical with vinylacetylene ($C_4H_4$); the energies calculated at the G3(MP2,CC)//B3LYP/6-311G(d,p)+ZPE level of theory. Reproduced with permission from ref 519. Copyright 2021 Royal Society of Chemistry.

Ricca et al.[522] carried out a computational study of possible mechanisms for the formation of PAHs containing nitrogen atoms in Titan's stratosphere. According to their model, the synthesis of pyridine ($C_5H_5N$) is driven by radical reactions of HCN with $C_2H_2$. The authors noted that mechanisms involving the addition of HCN and $C_2H_2$ to a pure aromatic ring to form an additional ring have barriers lower than 15 kcal/mol. Such barriers are probably too high for Titan, where the temperature is approximately 70-200 K[523], to allow reactions to occur. However, aromatic molecules are strong UV absorbers and can efficiently convert this energy into vibrational energy.[524] Therefore, the authors suggested that the reactions are possible even at 170 K due to the vibrational energy. The reducing nature of Titan's atmosphere could help reconstruct abiogenetic processes on the early Earth, although chemical evolution on Titan is limited by surface temperatures that are too low, which hinders the development of biochemical systems.

Rap et al.[525] in 2022 demonstrated another possible low-temperature pathway for the formation of nitrogen-containing PAHs, highlighting the underestimation of ion–molecule reactions (Scheme 25). A study of the ion–molecule reaction of pyridinium radical ($C_6H_5N^{\cdot+}$) with acetylene showed branched pathways for the formation of ionic species such as quinolizinium and H-quinolinium. The intermediate structures of the 2-ethynyl-N-pyridine cation, 2-vinyl-β-distone-N-pyridine cation, quinolysinium cation and H-quinoline cation were identified by mass spectrometry.

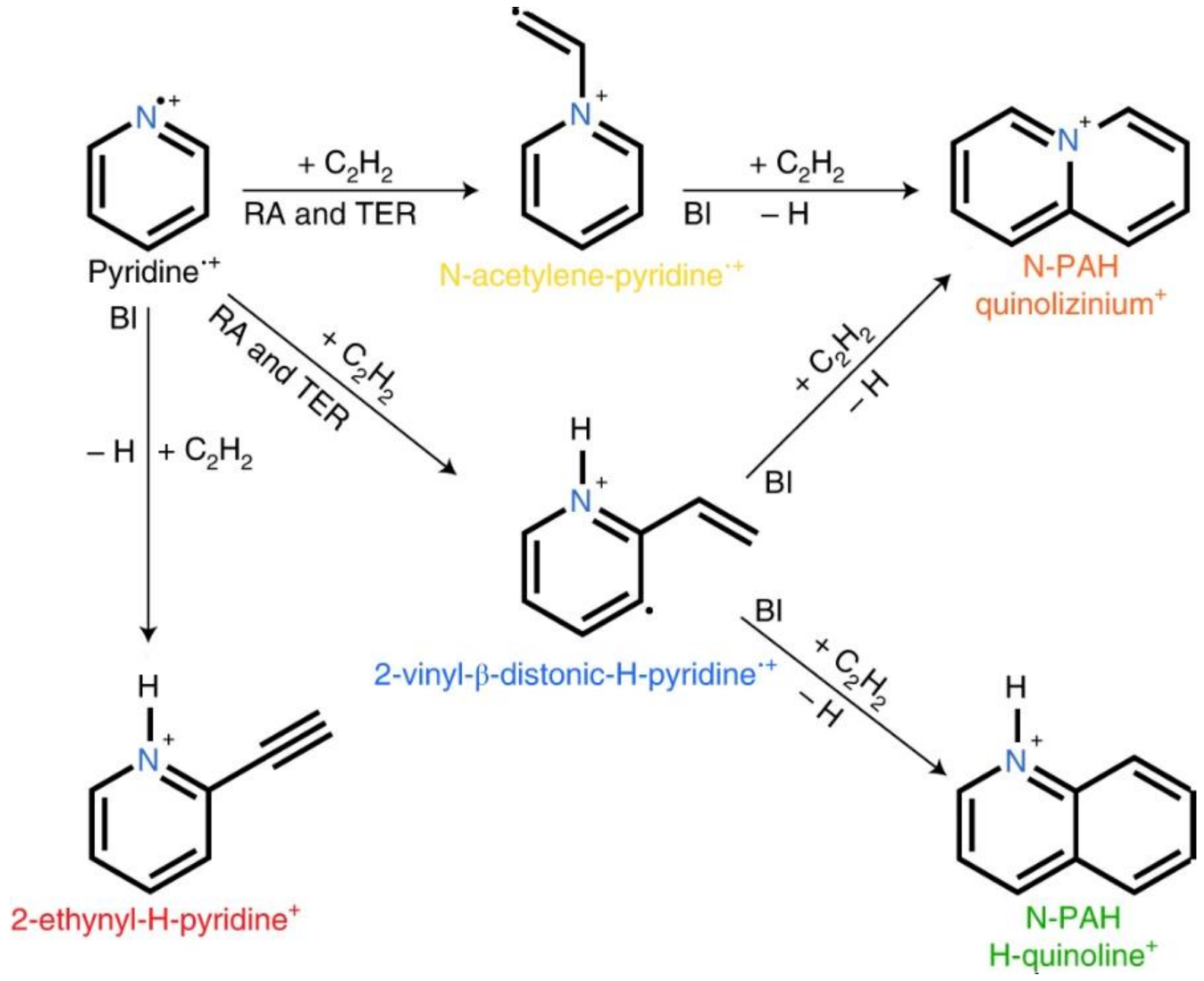

**Scheme 25.** Mechanisms of reaction pathways to N-PAHs via ion–molecule reaction of pyridinium radical and acetylene. The different reaction types are designated: radiative association (RA), bimolecular reaction (BI) and termolecular association (TER). Reproduced with permission from ref 523. Copyright 2022 Springer Nature.

Menor-Salvá et al.[526] showed that UV irradiation (in the range of 185–254 nm) of a urea solution subjected to freeze-thaw cycles (from -21 °C to +5 °C) in $C_2H_2$ leads to the formation of purines, pyrimidines and hydantoin. Guanine, cytosine, and uracil were detected using GC-MS. The authors explained the absence of adenine by the possible derivatization of adenine and guanine with the formation of 8-hydroxyadenine, the peak of which was detected. Moreover, Jeilani et al.[29] investigated possible pathways for the formation of pyrimidine bases such as cytosine and uracil under cold conditions relevant to Titan chemistry. The mechanisms for the formation of pyrimidine bases from urea, acetylene and ethynyl radicals were studied by DFT. The authors showed that the precursor of uracil and cytosine may be the biuret molecule. In addition, the pathways for the formation of melamine, ammeline, ammelide, cyanuric acid, uric acid and 8-oxoguanine are shown in Figure 25.

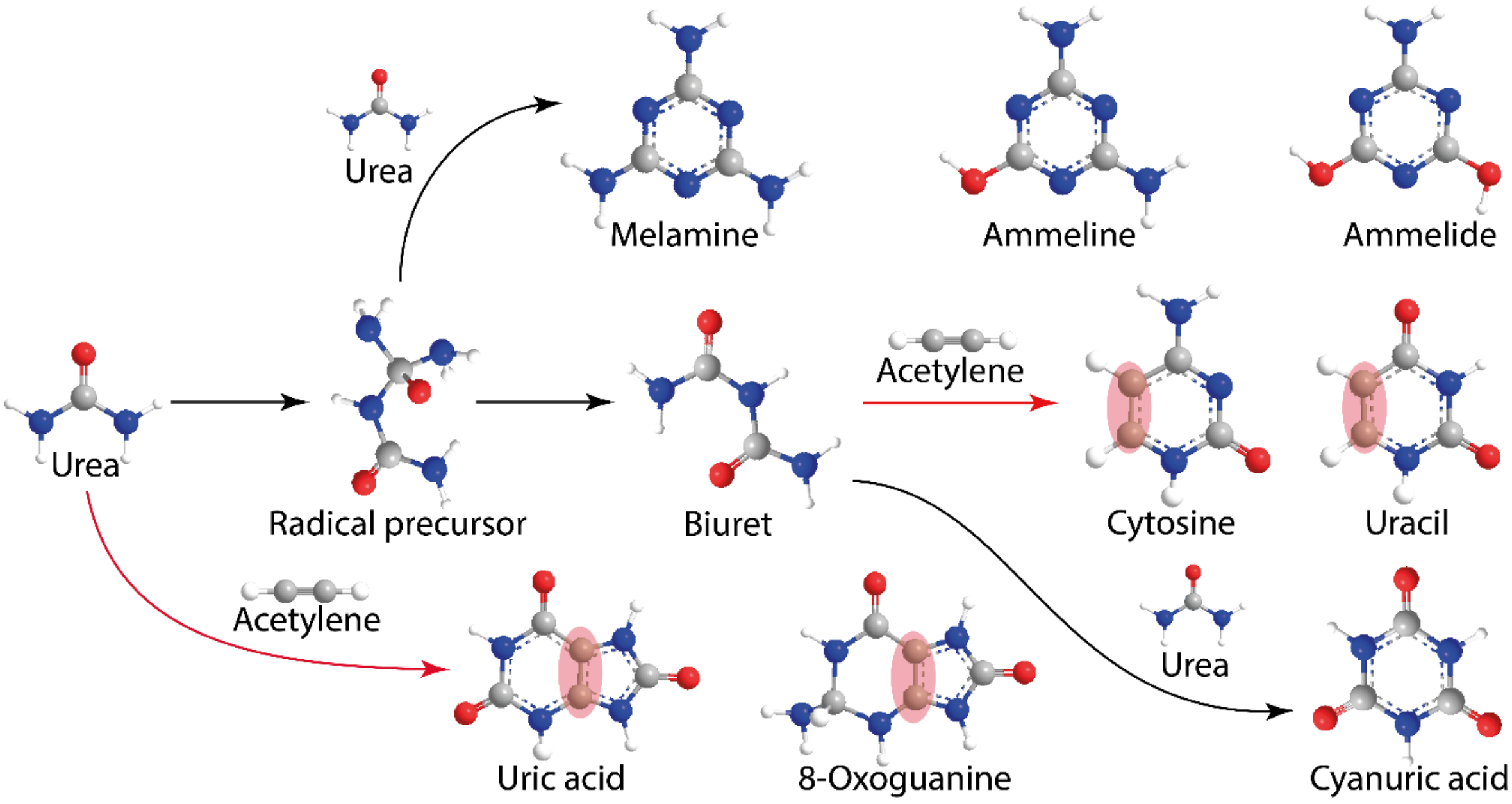

**Figure 25.** Formation of pyrimidine bases and triazines from urea and acetylene. Red ovals indicate atoms from acetylene.[29]

The synthesis of cytosine from urea and cyanoacetaldehyde was also examined.[527] Cyanoaldehyde may be a product of the hydrolysis of cyanoacetylene, which, in turn, is a derivative of acetylene[528,529] (cyanoacetylene can be produced by spark discharge in a $CH_4/N_2$ mixture).[530,531] Moreover, purine bases can be the product of the reaction of cyanoacetylene with cyanamide through barrierless cyclization through a small number of stages.[532]

Another important class of nitrogenous organic compounds is pyrolles. Pyrol structures underlie important biological molecules, such as heme, chlorophyll, cobalamin and others. In his recent work, Seitz et al.[533] demonstrated that pyrrole and dimethylpyrrole can be formed under the simulated volcanic hydrothermal conditions of the Early Earth. As was shown, the starting materials for such important molecules could be acetylene, propine and ammonium salts, the transformations of which were achieved by NiS or CoS as catalysts (Scheme 26).

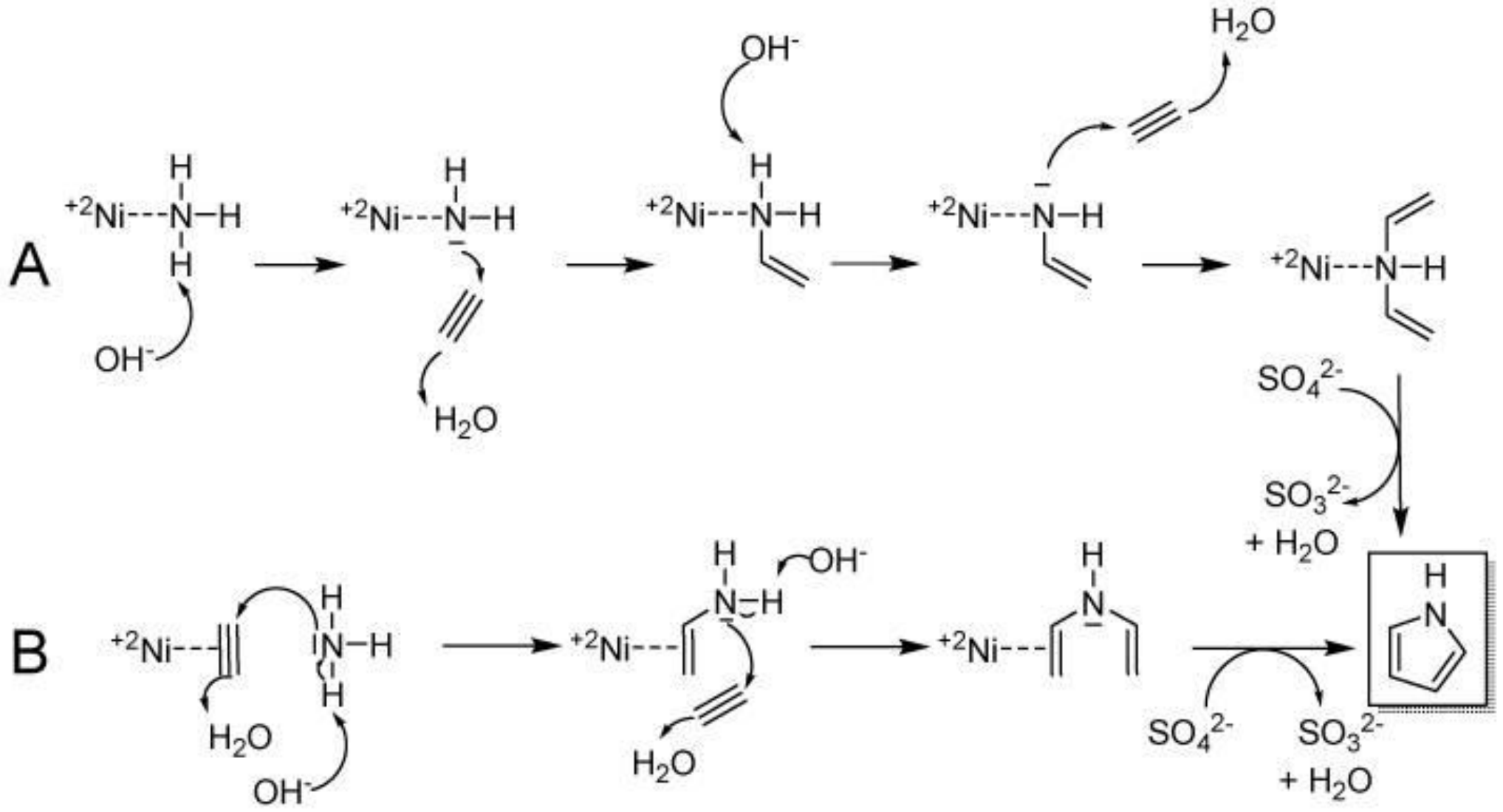

**Scheme 26.** Proposed mechanism for the formation of pyrrole from acetylene in the presence of Ni. Begins with the binding of ammonia to the Ni ion (A), begins with the coordination of acetylene on nickel (B). Reproduced with permission from ref 531. Copyright 2021 MDPI.

The important prebiotic role of acetylene and nitrile interactions was noted by Kaiser.[217] In addition, the mechanisms of the formation of polyaromatic compounds discussed in this review also indicate the important role of acetylene in biochemical evolution from the standpoint of the PAH-world hypothesis. Platts[534] and Chen et al. suggested[43] that certain types of PAHs can specifically interact with each other. Ionic and π-stacking interactions, as well as hydrogen bonds, can promote the formation of highly ordered structures. Thus, a matrix block polymer with a certain sequence of PAH fragments can be formed. In turn, various neighboring complementary blocks of PAHs, such as polynucleotide sequences, can provide block-specific replication. The distance between self-ordering PAH species and between nucleotides of RNA and DNA in both cases is approximately 0.34 nm. UV irradiation of astrophysical ice analogs ($H_2O$, $NH_3$, and $CH_4$) with pyrimidine or purine led to the formation of the main nucleases, such as uracil, cytosine, thymine, adenine, guanine and their isomers.[535,536,537,538,539] In addition, pyrimidine and purine-based nucleases can be formed in a simple astrophysical ice analog consisting of a mixture of $H_2O$, CO, $NH_3$, and $CH_3OH$, i.e., initially without pyridine or purine[540]; thus, the formation of pyrimidines and purines and nucleobases containing these compounds seems to be inevitable. Finally, according to astrochemical modeling, PAHs substantially increase the abundance of several COMs and sulfur-bearing molecules[541], while the latter are considered to be important for several amino acids.[542,543] We summarize the main prebiotic reactions discussed in this section in Table 25.

**Table 25.** The reactions of prebiotic chemistry with PAHs.

| **Reactions** | **Kinetics** | **Objects & conditions** |
|---|---|---|
| Ice reactions<br>1. $C_{10}H_8$ + $H_2O$ ice → $C_{10}H_6O_2$ (1,4-naphthoquinone)[503] | n.d. | Molecular clouds, prestellar cores.<br>Conditions:<br>T = 10 K<br>$G_0$ = very low<br>Galactic CR field.<br>+ Comets, meteorites. |
| 2. $C_5H_4N$ + $2C_2H_2$ → $C_9H_7N$ (isoquinoline)[517] | n.d. | Stellar envelopes (outer edge).<br>Conditions:<br>$n_H = 10^8$-$10^{12}$ cm$^{-3}$<br>T = 500-1500 K |
| 3. $C_2H_2CN$ + $C_2H_3CN$ → $C_5H_5N$ (pyridine) + CN[518] | n.d. | Stellar envelopes (outer edge).<br>Conditions:<br>$n_H = 10^8$-$10^{12}$ cm$^{-3}$<br>T = 500-1500 K<br>Molecular clouds, prestellar cores.<br>Conditions:<br>$n_H = 10^2$-$10^7$ cm$^{-3}$<br>T = 10 K<br>$G_0$ = very low<br>Galactic CR field. |
| 4. CN + $C_4H_6$ → $C_5H_5N$ + H (pyridine)[520] | $k_4 = 10$-$10^4$ s$^{-1}$<br>~10 - $10^4$ s$^{-1}$<br>(depending on collision energy) | Molecular clouds, prestellar cores.<br>Conditions:<br>T = 10 K<br>$G_0$ = very low<br>Galactic CR field.<br>+ Comets, meteorites. |
| 5. $C_5H_4N\cdot$ + $C_4H_4$ → $C_9H_7N$ + H (isoquinoline)[521] | n.d. | |
| 6. HCN + 2 $C_2H_2$ → $C_5H_5N$ (pyridine)[522] | n.d. | Titan's atmosphere.<br>Conditions:<br>$n_H$: $10^{10}$-$10^{16}$ cm$^{-3}$<br>T = 50-300 K<br>$G_0$ = up to 10<br>Galactic CR field. |
| 7. Ice reactions<br>$H_2O/CH_4N_2O$(urea)/$C_2H_2$ + hν → guanine, adenine, xanthine, uric acid, uracil, cytosine[526] | n.d. | Planetary atmospheres, comets, meteorites<br>Conditions:<br>$n_H = 10^{10}$-$10^{16}$ cm$^{-3}$<br>T = 50-300 K<br>$G_0$ = up to $10^2$<br>Galactic CR field. |

## 8.3. Role of acetylene in monocarboxylic acid formation

In addition to amino acids and nucleobases, fatty acids are considered important components in the formation of the first protocellular structures.[544,545] A number of studies[546,547,548] suggest that the first membranes encapsulating information polymers consist of fatty acids. Like amino acids and nitrogenous bases, fatty acids are found in meteorites. Moreover, fatty acids are the most common water-soluble organic compounds in carbonaceous chondrites.[549] Fatty acids with chain lengths of 2–12 carbon atoms have been found in meteorites. In water, these molecules spontaneously assemble into bilayers that form vesicles.[550]

There are many possible noncatalytic and catalytic mechanisms for the natural formation of fatty acids in meteorites and under primordial Earth conditions. One such mechanism is based on the reaction of acetylene ($C_2H_2$) and carbon monoxide (CO) upon contact with nickel sulfide (NiS) in a hot aqueous environment.[544,545,551] An experimental study showed that this reaction can lead to the formation of C3,5,7,9-monocarboxylic acids. Scheidler et al.[552] suggested that such reactions could occur in Hadean volcanic-hydrothermal vents saturated with $C_2H_2$ and CO. These currents could bring gases into contact with NiS under conditions conducive to the formation of monocarboxylic acids. In this case, a large length of the alkyl radical could be achieved due to the high pressure.

Scheidler et al.[552] have made some interesting observations linking the proposed pathway with abiogenetic processes. These pathways are illustrated in Scheme 27. For example, for higher C7–9 monocarboxylic acids, the process of formation of the saturated form is not complete. The resulting products are a mixture of unsaturated monocarboxylic acids with one or more double bonds. Double bonds are predominantly in the form of trans isomers. In the case of several double bonds, a conjugated system is generally formed. The authors also noted that the modern organisms Bacteria and Eukarya contain unsaturated fatty acyl lipids within the membranes of saturated lipids. Moreover, the arrangement of double bonds in the middle of the hydrophobic tail is a common feature of natural unsaturated monocarboxylic acids and unsaturated lipids formed during the interaction of $C_2H_2$ and CO.

The evolution of a complex prebiotic mixture containing $C_2H_2$, CO and NiS was also shown in recent works by Sobotta et al.[553] and Diederich et al.[554] A suspension of precipitated NiS was shown to convert acetylene and carbon monoxide into a set of C2–4 products. Moreover, the authors demonstrated the formation of a homologous series of thioacids upon repeated addition of $C_2$ to the resulting compounds via acetylene. Over 7 days, molecules with alkyl radical sizes ranging from 15 to 27 carbon atoms and a mass range of 230–500 m/z were formed. In this case, only up to four carbon tags from $^{13}CO$ were present. Thus, we note the role of acetylene as the main driver of mass growth and as a building material for the synthesis of high-molecular weight compounds. The general reactions that are part of this synthesis are provided in Table 26.

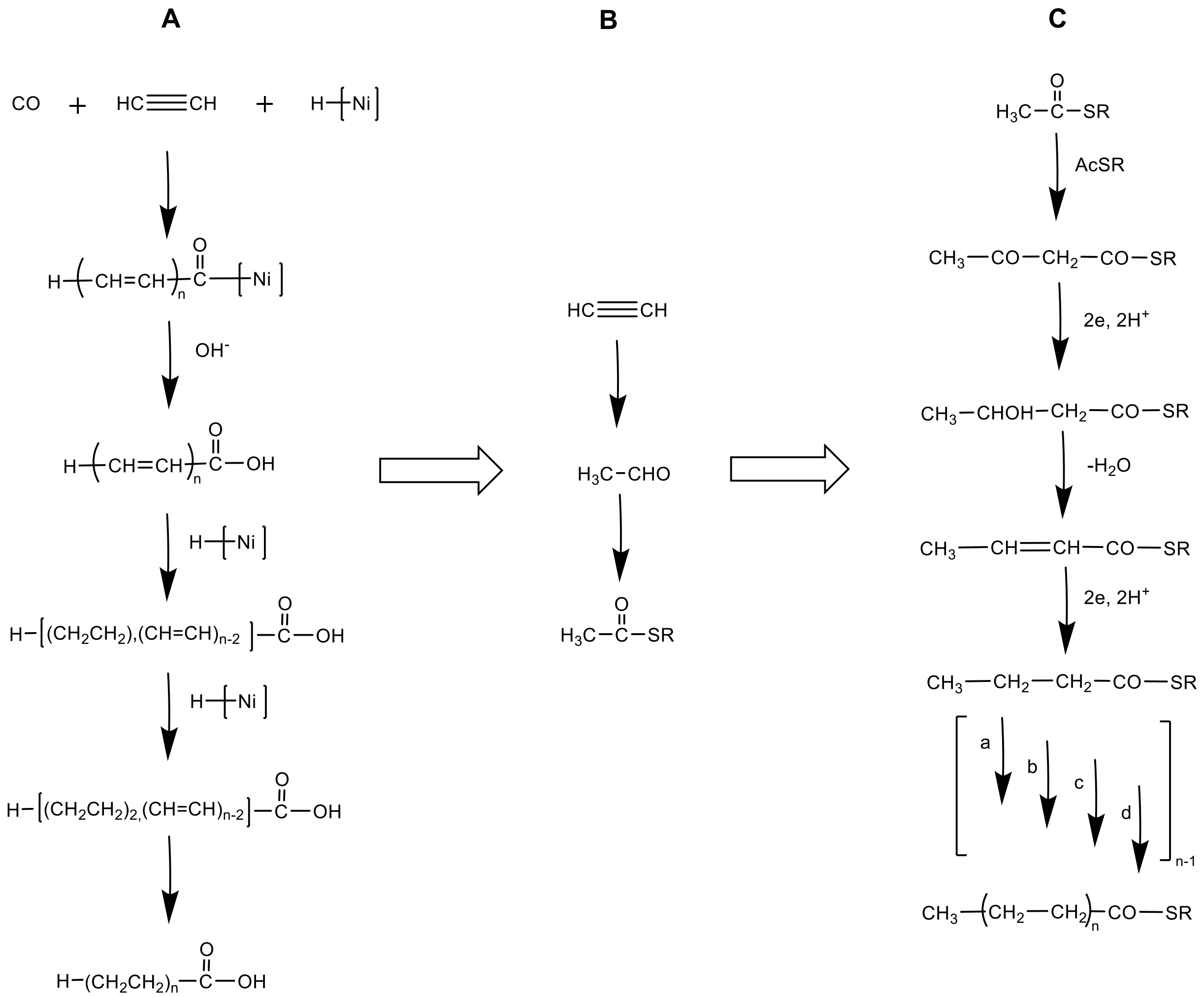

**Scheme 27.** Evolution of monocarboxylic acid biosynthesis. Formation of carboxylic acids from acetylene on a nickel catalyst under prebiotic conditions (A); Conversion of acetylene to acetyl-thioester (B); Extant biosynthesis based on acetyl-thioester condensation (C).[550]

The described processes may be relevant for carbonaceous chondrites. Chondrites contain carbon grains stuck together, with water ice in the layers between them. Under the influence of solar radiation, the surface can be heated, and water ice can melt, initiating the synthesis of nitrogen bases, amino acids and fatty acids in the resulting solution.

We conclude that acetylene is a universal, stable, and flexible building block and could be the basis for the formation of a variety of precursors of biochemical molecules. Studying the chemistry of acetylene in space environments can provide important insights into markers indicating various stages of biochemical evolution, from nitrogenous bases to putative PAH templates for the first informational molecules. Moreover, the presence of metal and carbon catalysts can significantly diversify synthetic pathways and increase molecular complexity. Currently, we are still at the origin of knowledge about the role of catalytic processes in the formation of prebiotic molecular systems. Moreover, acetylene has high complementarity with respect to its ability to participate in catalytic cycles involving metals, sulfos, and carbon species, suggesting that this area of research is promising.

**Table 26.** The reactions of formation monocarboxylic acids from acetylene and other reactants.

| Reactions | Kinetics | Objects & conditions |
|---|---|---|
| 1. $C_2H_2 + CO + NiS + H_2O \rightarrow$ C3,5,7,9-Monocarboxylic Acids[552]<br>2. $C_2H_2 + CO + NiS + H_2O \rightarrow$ products + $C_2H_2 \rightarrow C_5H_5OS$ (thio acids)[554] | n.d. | early Earth |

# 9. Conclusions

Acetylene consists of two sp-hybridized carbon atoms connected by one σ and two π bonds. This structure predefines acetylene with a high exothermicity for addition reactions at the triple bond. Additionally, a pronounced shift in the electron density from a proton to a carbon atom results in the high acidity of the acetylene molecules. Due to its structure, the acetylene molecule is a reactive yet sufficiently robust and stable $C_2$ species. This position makes it an almost ideal building block for a myriad of organic compounds. Acetylene serves as a versatile chemical currency, with the potential to link molecules into complex structures possessing infinite carbon skeletons.

The significant role of acetylene in chemical evolution correlates with the widespread distribution of this molecule across various cosmic environments, spanning from hot circumstellar spaces to cold molecular clouds, planetary atmospheres and icy comets. Acetylene molecules are synthesized under very different conditions in these media. While acetylene is merely one component of the chemical cocktail of the universe, the array of processes in which it participates commands special attention. Despite not being the most prevalent carbon molecule in space, acetylene acts as a critical bridge to the chemistry of complex macromolecular compounds.

Most mechanisms of benzene synthesis across diverse space environments involve acetylene, thereby paving the way to a world of aromatic compounds. The primary mechanism for PAH synthesis, HACA, is rooted in the chemistry surrounding the acetylene molecule. Our understanding of the formation of high-molecular-weight aromatic compounds has undergone a seismic shift over the past decade. The emergence of new mechanisms based on barrierless reactions has radically altered our perception of space-carbon chemistry, shifting focus from high-temperature combustion-based reactions to low-temperature neutral-neutral reactions. This revolutionary concept suggests that PAHs can be formed not only at high temperatures but also at extremely low temperatures, leading to the creation of complex macromolecular compounds in cold the ISM. This includes not only "pure" carbon PAHs but also nitrogen-containing compounds. These mechanisms provide fresh perspectives on the chemistry of celestial bodies such as Titan. Further studies may aid in pinpointing the conditions conducive to determining the origin of PAHs based on their structural characteristics.

Despite advancements in the field, many factors remain elusive. Traditionally, studies of the underlying mechanisms involved the formation of bulk organic matter in rarefied interstellar or circumstellar media. However, critical chemical processes might transpire on the surface of solids at both low and high temperatures. Such processes could be catalytic in nature. For instance, catalytic pathways for the formation of aromatic compounds in space environments are likely underexplored. Catalytic processes, encompassing crucial events such as the formation of aromatic compounds, could occur on the surfaces of comets, asteroids, planets, and interstellar dust grains, as corroborated by recent findings. We can anticipate more exciting discoveries in this area in the coming years.

Our understanding of cosmic chemical processes will continue to broaden in the future. There is no doubt that missions such as the James Webb Space Telescope[36, 555], the forthcoming Dragonfly spacecraft[556, 557] and the Jupiter Icy Moons Explorer[558] will revolutionize our comprehension of the evolution of organic compounds in space. This approach will facilitate the precise quantification of how physical conditions influence organic transformations, thereby allowing us to track the evolution of acetylene and other organic molecules across various space environments.

ASSOCIATED CONTENT

**Supporting Information**

Table S1 contains data on the occurrence of acetylene in different space regions and objects; data on the main mechanisms of PAH formation and the role of acetylene in this pathway; Table S2 contains data on several important PAHs and their formation pathways based on crossed molecular beam experiments and pyrolytic reactors combined with quantum chemical calculations (PDF).

AUTHOR INFORMATION

**Corresponding Author**

* **Valentine P. Ananikov** − Zelinsky Institute of Organic Chemistry, Russian Academy of Sciences, Moscow 119991, Russia; orcid.org/0000-0002-6447-557X;
Email: val@ioc.ac.ru

**Authors**

**Evgeniy O. Pentsak** - Zelinsky Institute of Organic Chemistry, Russian Academy of Sciences, Moscow 119991, Russia; orcid.org/0000-0001-9040-0013

**Maria S. Murga** - Institute of Astronomy, Russian Academy of Sciences, Pyatnitskaya str. 48, Moscow 119017, Russia; orcid.org/0000-0002-7910-6662

**Author Contributions**

The manuscript was written through contributions of all the authors. All the authors have given approval to the final version of the manuscript.

**Notes**

The authors declare no competing financial interest.

[4] Pässler, P; Hefner, W; Buckl, K; Meinass, H; Meiswinkel, A; Wernicke, H.-J.; Ebersberg, G; Müller, R; Bässler, J; Behringer, H; Mayer, D. Acetylene, in Ullmann's Encyclopedia of Industrial Chemistry, Wiley-VCH, Weinheim, **2011** [DOI: 10.1002/14356007.a01_097.pub4]

[5] Schobert, H., Production of acetylene and acetylene-based chemicals from coal. *Chem. Rev.* **2014,** *114* (3), 1743-60. [DOI: 10.1021/cr400276u]

[6] Trotuş, I.-T.; Zimmermann, T.; Schüth, F. Catalytic reactions of acetylene: A feedstock for the chemical industry revisited. *Chem. Rev.* **2014,** *114,* 1761–1782. [DOI: 10.1021/cr400357r]

[7] Voronin, V. V.; Ledovskaya, M. S.; Bogachenkov, A. S.; Rodygin, K. S.; Ananikov, V. P., Acetylene in Organic Synthesis: Recent Progress and New Uses. *Molecules* **2018,** *23* (10). [DOI: 10.3390/molecules23102442]

[8] Rodygin, K. S.; Ananikov, V. P., An efficient metal-free pathway to vinyl thioesters with calcium carbide as the acetylene source. *Green Chem.* **2016,** *18* (2), 482-486. [DOI: 10.1039/C5GC01552A]

[9] Rodygin, K. S.; Vikenteva, Y. A.; Ananikov, V. P., Calcium-Based Sustainable Chemical Technologies for Total Carbon Recycling. *ChemSusChem* **2019,** *12* (8), 1483-1516. [DOI: 10.1002/cssc.201802412]

[10] Tracy, J.; Zietz, O.; Olson, S.; Jiao, J., Plasma-Enhanced Chemical Vapor Deposition of Acetylene on Codeposited Bimetal Catalysts Increasing Graphene Sheet Continuity Under Low-Temperature Growth Conditions. *Nanoscale Res. Lett.* **2019,** *14* (1), 335. [DOI: 10.1186/s11671-019-3156-y]

[11] Osikoya, A. O.; Parlak, O.; Murugan, N. A.; Dikio, E. D.; Moloto, H.; Uzun, L.; Turner, A. P.; Tiwari, A., Acetylene-sourced CVD-synthesised catalytically active graphene for electrochemical biosensing. *Biosens. Bioelectron.* **2017,** *89* (Pt 1), 496-504. [DOI: 10.1016/j.bios.2016.03.063]

[12] Bulyarskiy, S. V.; Gorelik, V. S.; Ryazanov, R. M., Nitrogen Doping of Carbon Nanotubes Synthesized in Flowing Acetylene and Ammonia. *Inorganic Materials* **2020,** *56* (10), 1006-1010. [DOI: 10.1134/S0020168520100027]

[13] Afrin, R.; Abbas, S. M.; Shah, N. A.; Mustafa, M. F.; Ali, Z.; Ahmad, N., Effect of Varying Inert Gas and Acetylene Concentration on the Synthesis of Carbon Nanotubes. *J. Nanosci. Nanotechnol.* **2016,** *16* (3), 2956-9. [DOI: 10.1166/jnn.2016.12469]

[14] Magrez, A.; Seo, J. W.; Kuznetsov, V. L.; Forro, L., Evidence of an equimolar C2H2-CO2 reaction in the synthesis of carbon nanotubes. *Angew. Chem. Int. Ed. Engl.* **2007,** *46* (3), 441-4. [DOI: 10.1002/anie.200603764]

[15] Zhong, G.; Hofmann, S.; Yan, F.; Telg, H.; Warner, J. H.; Eder, D.; Thomsen, C.; Milne, W. I.; Robertson, J., Acetylene: A Key Growth Precursor for Single-Walled Carbon Nanotube Forests. *J. Phys. Chem. C.* **2009,** *113* (40), 17321-17325. [DOI: 10.1021/jp905134b]

[16] Wang, H.; Zhou, Y.; Qi, M.; Liu, C.; Xu, X.; Ren, Z., Direct Growth of Graphene on Fused Quartz by Atmospheric Pressure Chemical Vapor Deposition with Acetylene. *J. Phys. Chem. C.* **2018,** *123* (4), 2370-2377. [DOI: 10.1021/acs.jpcc.8b08806]

[17] Ananikov, V. P.; Ledovskaya, M. S.; Voronin, V. V.; Rodygin, K. S., Acetylene and Ethylene: Universal $C_2$ Molecular Units in Cycloaddition Reactions. *Synthesis* **2021**. [DOI: 10.1055/a-1654-2318]

[18] Zhao, L.; Kaiser, R. I.; Xu, B.; Ablikim, U.; Ahmed, M.; Joshi, D.; Veber, G.; Fischer, F. R.; Mebel, A. M., Pyrene synthesis in circumstellar envelopes and its role in the formation of 2D nanostructures. *Nat. Astron.* **2018,** *2* (5), 413-419. [DOI: 10.1038/s41550-018-0399-y]

[19] Zhao, L.; Kaiser, R. I.; Lu, W.; Xu, B. et al. Molecular mass growth through ring expansion in polycyclic aromatic hydrocarbons via radical-radical reactions. *Nat. Commun.* **2019,** *10,* 3689 [DOI: 10.1038/s41467-019-11652-5]

[20] Contreras, C. S.; Salama, F. Laboratory Investigations of Polycyclic Aromatic Hydrocarbon Formation and Destruction in the Circumstellar Outflows of Carbon Stars. *Astrophys. J., Suppl. Ser.* **2013,** *208,* 6. [DOI: 10.1088/0067-0049/208/1/6]

[21] Jones, B. M.; Zhang, F.; Kaiser, R. I.; Jamal, A. et al. Formation of benzene in the interstellar medium. *Proc. Natl. Acad. Sci. U. S. A.* **2011,** *108,* 452. [DOI: 10.1073/pnas.1012468108]

[22] Chiar, J. E.; Tielens, A. G. G. M.; Adamson, A. J.; Ricca, A. The Structure, Origin, and Evolution of Interstellar Hydrocarbon Grains. *Astrophys. J.* **2013,** *770,* 78. [DOI: 10.1088/0004-637X/770/1/78]

[23] Chen, T.; Li, A. Synthesizing carbon nanotubes in space. *Astron. Astrophys.* **2019,** *631,* A54. [DOI: 10.1051/0004-6361/201935789]

[24] García-Hernández, D. A.; Villaver, E.; García-Lario, P.; Acosta-Pulido, J. A. et al. Infrared Study of Fullerene Planetary Nebulae. *Astrophys. J.* **2012,** *760,* 107. [DOI: 10.1088/0004-637X/760/2/107]

[25] García-Hernández, D. A.; Manchado, A.; García-Lario, P.; Stanghellini, L. et al. Formation of Fullerenes in H-Containing Planetary Nebulae. *Astrophys. J.* **2010,** *724,* L39. [DOI: 10.1088/2041-8205/724/1/L39]

[26] Tennyson, J.; Yurchenko, S. N. Laboratory spectra of hot molecules: Data needs for hot super-Earth exoplanets. *Mol. Astrophys.* **2017,** 8, 1. [DOI: 10.1016/j.molap.2017.05.002]

[27] Chubb, K. L.; Tennyson, J.; Yurchenko, S. N. ExoMol molecular line lists – XXXVII. Spectra of acetylene. *Mon. Not. R. Astron. Soc.* **2020,** *493,* 1531. [DOI: 10.1093/mnras/staa229]

[28] Abplanalp, M. J.; Frigge, R.; Kaiser, R. I. Low-temperature synthesis of polycyclic aromatic hydrocarbons in Titan's surface ices and on airless bodies. *Sci. Adv.* **2019,** *5,* eaaw5841. [DOI: 10.1126/sciadv.aaw5841]

[29] Jeilani, Y. A.; Fearce, C.; Nguyen, M. T. Acetylene as an essential building block for prebiotic formation of pyrimidine bases on Titan. *Phys. Chem. Chem. Phys.* **2015,** *17,* 24294. [DOI: 10.1039/c5cp03247d]

[30] Frankland, V. L.; James, A. D.; Sánchez, J. D. C.; Mangan, T. P. et al. Uptake of acetylene on cosmic dust and production of benzene in Titan's atmosphere. *Icarus* **2016,** *278,* 88. [DOI: 10.1016/j.icarus.2016.06.007]

[31] Postberg, F.; Khawaja, N.; Abel, B.; Choblet, G. et al. Macromolecular organic compounds from the depths of Enceladus. *Nature* **2018,** *558,* 564. [DOI: 10.1038/s41586-018-0246-4]

[32] Waite, J. H., Jr.; Combi, M. R.; Ip, W. H.; Cravens, T. E. et al. Cassini ion and neutral mass spectrometer: Enceladus plume composition and structure. *Science* **2006**, *311*, 1419. [DOI: 10.1126/science.1121290]

[33] Kaiser, R. I.; Hansen, N., An Aromatic Universe-A Physical Chemistry Perspective. *J. Phys. Chem. A.* **2021,** *125* (18), 3826-3840. [DOI: 10.1021/acs.jpca.1c00606]

[34] Mebel, A. M.; Landera, A.; Kaiser, R. I. Formation Mechanisms of Naphthalene and Indene: From the Interstellar Medium to Combustion Flames. *J. Phys. Chem. A* **2017**, *121*, 901. [DOI: 10.1021/acs.jpca.6b09735]

[35] Frenklach, M.; Mebel, A. M. On the mechanism of soot nucleation. *Phys. Chem. Chem. Phys.* **2020**, *22*, 5314. [DOI: 10.1039/D0CP00116C]

[70] Courtin, R.; Gautier, D.; Marten, A.; Bezard, B.; Hanel, R., The composition of Saturn's atmosphere at northern temperate latitudes from Voyager IRIS spectra - $NH_3$, $PH_3$, $C_2H_2$, $C_2H_6$, $CH_3D$, $CH_4$, and the Saturnian D/H isotopic ratio. *Astrophys. J.* **1984,** *287*. [DOI: 10.1086/162748]

[71] Voyager. Mission Overview https://voyager.jpl.nasa.gov/mission/ (accessed 2024-03-04).

[72] Hanel, R.; Conrath, B.; Flasar, F. M.; Kunde, V.; Maguire, W.; Pearl, J.; Pirraglia, J.; Samuelson, R.; Herath, L.; Allison, M.; Cruikshank, D.; Gautier, D.; Gierasch, P.; Horn, L.; Koppany, R.; Ponnamperuma, C., Infrared observations of the saturnian system from voyager 1. *Science* **1981,** *212* (4491), 192-200. [DOI: 10.1126/science.212.4491.192]

[73] Noll, K. S.; Knacke, R. F.; Tokunaga, A. T.; Lacy, J. H.; Beck, S.; Serabyn, E., The abundances of ethane and acetylene in the atmospheres of Jupiter and Saturn. *Icarus* **1986,** *65* (2-3), 257-263. [DOI: 10.1016/0019-1035(86)90138-7]

[74] Esposito, L. W.; Barth, C. A.; Colwell, J. E.; Lawrence, G. M.; McClintock, W. E.; Stewart, A. I. F.; Keller, H. U.; Korth, A.; Lauche, H.; Festou, M. C.; Lane, A. L.; Hansen, C. J.; Maki, J. N.; West, R. A.; Jahn, H.; Reulke, R.; Warlich, K.; Shemansky, D. E.; Yung, Y. L., The Cassini Ultraviolet Imaging Spectrograph Investigation. *Space Sci. Rev.* **2004,** *115 (1-4),* 299-361. [DOI: 10.1007/s11214-004-1455-8]

[75] PIA13699: Acetylene around Jupiter's Poles https://photojournal.jpl.nasa.gov/catalog/PIA13699 (accessed 2024-03-01).

[76] Jennings, D. E.; Sada, P. V.; Romani, P. N.; Bjoraker, G. L.; Simon-Miller, A. A., Brightening of Acetylene and Ethane in the South Pole of Saturn from Ground-Based and CIRS Observations. American Astronomical Society, DPS meeting #37, id.60.09; Bulletin of the American Astronomical Society, **2005**, *37*, 757. [https://ui.adsabs.harvard.edu/abs/2005DPS....37.6009J/abstract]

[77] Fletcher, L. N.; Cavalie, T.; Grassi, D.; Hueso, R.; Lara, L. M.; Kaspi, Y.; Galanti, E.; Greathouse, T. K.; Molyneux, P. M.; Galand, M.; Vallat, C.; Witasse, O.; Lorente, R.; Hartogh, P.; Poulet, F.; Langevin, Y.; Palumbo, P.; Gladstone, G. R.; Retherford, K. D.; Dougherty, M. K.; Wahlund, J. E.; Barabash, S.; Iess, L.; Bruzzone, L.; Hussmann, H.; Gurvits, L. I.; Santolik, O.; Kolmasova, I.; Fischer, G.; Muller-Wodarg, I.; Piccioni, G.; Fouchet, T.; Gerard, J. C.; Sanchez-Lavega, A.; Irwin, P. G. J.; Grodent, D.; Altieri, F.; Mura, A.; Drossart, P.; Kammer, J.; Giles, R.; Cazaux, S.; Jones, G.; Smirnova, M.; Lellouch, E.; Medvedev, A. S.; Moreno, R.; Rezac, L.; Coustenis, A.; Costa, M., Jupiter Science Enabled by ESA's Jupiter Icy Moons Explorer. *Space Sci. Rev.* **2023,** *219* (7), 53. [DOI: 10.1007/s11214-023-00996-6]

[78] Sinclair, J. A.; Moses, J. I.; Hue, V.; Greathouse, T. K.; Orton, G. S.; Fletcher, L. N.; Irwin, P. G. J., Jupiter's auroral-related stratospheric heating and chemistry III: Abundances of $C_2H_4$, $CH_3C_2H$, $C_4H_2$ and $C_6H_6$ from Voyager-IRIS and Cassini-CIRS. *Icarus* **2019,** *328*, 176-193. [DOI: 10.1016/j.icarus.2019.03.012]

[79] Summers, M. E.; Strobel, D. F., Photochemistry of the atmosphere of Uranus. *Astrophys. J.* **1989,** *346*. [DOI: 10.1086/168031]

[80] Orton, G. S.; Aitken, D. K.; Smith, C.; Roche, P. F.; Caldwell, J.; Snyder, R., The spectra of Uranus and Neptune at 8–14 and 17–23 μm. *Icarus* **1987,** *70* (1), 1-12. [DOI: 10.1016/0019-1035(87)90070-4]

[81] Caldwell, J.; Wagener, R.; Fricke, K.-H., Observations of Neptune and Uranus below 2000 Å with the IUE. *Icarus* **1988,** *74* (1), 133-140. [DOI: 10.1016/0019-1035(88)90035-8]

[82] Stone, E. C.; Miner, E. D., The voyager 2 encounter with the neptunian system. *Science* **1989,** *246* (4936), 1417-21. [DOI: 10.1126/science.246.4936.1417]

[83] Moses, J. I.; Poppe, A. R., Dust Ablation on the Giant Planets: Consequences for Stratospheric Photochemistry. *Icarus* **2017,** *297*, 33-58. [DOI: 10.1016/j.icarus.2017.06.002]

[84] Bar-nun, A., Acetylene formation on Jupiter: Photolysis or thunderstorms? *Icarus* **1979,** *38* (2), 180-191. [DOI: 10.1016/0019-1035(79)90177-5]

[85] Moses, J. I.; Cavalie, T.; Fletcher, L. N.; Roman, M. T., Atmospheric chemistry on Uranus and Neptune. *Philos. Trans. A. Math. Phys. Eng. Sci.* **2020,** *378* (2187), 20190477. [DOI: 10.1098/rsta.2019.0477]

[86] Moses, J., Photochemistry of Saturn's Atmosphere I. Hydrocarbon Chemistry and Comparisons with ISO Observations. *Icarus* **2000,** *143* (2), 244-298. [DOI: 10.1006/icar.1999.6270]

[87] Burgdorf, M.; Orton, G.; Vancleve, J.; Meadows, V.; Houck, J., Detection of new hydrocarbons in Uranus' atmosphere by infrared spectroscopy. *Icarus* **2006,** *184* (2), 634-637. [DOI: 10.1016/j.icarus.2006.06.006]

[88] Lo, J.-I.; Lin, M.-Y.; Peng, Y.-C.; Chou, S.-L.; Lu, H.-C.; Cheng, B.-M.; Ogilvie, J. F., Far-ultraviolet photolysis of solid methane. *Mon. Not. R. Astron. Soc.* **2015,** *451* (1), 159-166. [DOI: 10.1093/mnras/stv935]

[89] Romani, P. N.; Atreya, S. K., Methane photochemistry and haze production on Neptune. *Icarus* **1988,** *74* (3), 424-445. [DOI: 10.1016/0019-1035(88)90113-3]

[90]. Atreya, S. K. and Romani P. N. Photochemistry and clouds of Jupiter, Saturn, and Uranus. In Recent Advances in Planetary Meterology (G. E. Hint, Ed.), pp. 17-68. Cambridge University Press, Cambridge. **1985**. [https://ui.adsabs.harvard.edu/abs/1985rapm.book...17A/abstract]

[91] Gans, B.; Boye-Peronne, S.; Broquier, M.; Delsaut, M.; Douin, S.; Fellows, C. E.; Halvick, P.; Loison, J. C.; Lucchese, R. R.; Gauyacq, D., Photolysis of methane revisited at 121.6 nm and at 118.2 nm: quantum yields of the primary products, measured by mass spectrometry. *Phys. Chem. Chem. Phys.* **2011,** *13* (18), 8140-52. [DOI: 10.1039/C0CP02627A]

[92] Blitz, M. A.; Seakins, P. W., Laboratory studies of photochemistry and gas phase radical reaction kinetics relevant to planetary atmospheres. *Chem. Soc. Rev.* **2012,** *41* (19), 6318-47. [DOI: 10.1039/C2CS35204D]

[93] Mordaunt, D. H.; Lambert, I. R.; Morley, G. P.; Ashfold, M. N. R.; Dixon, R. N.; Western, C. M.; Schnieder, L.; Welge, K. H., Primary product channels in the photodissociation of methane at 121.6 nm. *J. Chem. Phys.* **1993,** *98* (3), 2054-2065. [DOI: 10.1063/1.464237]

[94] Bishop, J.; Romani, P. N.; Atreya, S. K., Voyager 2 ultraviolet spectrometer solar occultations at neptune: photochemical modeling of the 125–;165 nm lightcurves. *Planet. Space Sci.* **1998,** *46 (1),* 1-20. [DOI: 10.1016/S0032-0633(97)00126-8]

[95] Gibbard, S.; Levy, E. H.; Lunine, J. I., Generation of lightning in Jupiter's water cloud. *Nature* **1995,** *378* (6557), 592-5. [DOI: 10.1038/378592a0]

[96] Podolak, M.; Bar-Nun, A., Moist convection and the abundances of lightning-produced CO, $C_2H_2$, and HCN on Jupiter. *Icarus* **1988,** *75* (3), 566-570. [DOI: 10.1016/0019-1035(88)90165-0]

[97] Orville, R. E., Spectrum of the lightning stepped leader. *J. Geophys. Res.* **1968,** *73* (22), 6999-7008. [DOI: 10.1029/JB073i022p06999]

[98] Bar-Nun, A., Thunderstorms on Jupiter. *Icarus* **1975,** *24* (1), 86-94. [DOI: 10.1016/0019-1035(75)90162-1]

[99] Hurley, J.; Irwin, P. G. J.; Fletcher, L. N.; Moses, J. I.; Hesman, B.; Sinclair, J.; Merlet, C., Observations of upper tropospheric acetylene on Saturn: No apparent correlation with 2000km-sized thunderstorms. *Planet. Space Sci.* **2012,** *65* (1), 21-37. [DOI: 10.1016/j.pss.2011.12.026]

[118] Naraoka, H.; Takano, Y.; Dworkin, J. P.; Oba, Y.; Hamase, K.; Furusho, A.; Ogawa, N. O.; Hashiguchi, M.; Fukushima, K.; Aoki, D.; Schmitt-Kopplin, P.; Aponte, J. C.; Parker, E. T.; Glavin, D. P.; McLain, H. L.; Elsila, J. E.; Graham, H. V.; Eiler, J. M.; Orthous-Daunay, F. R.; Wolters, C.; Isa, J.; Vuitton, V.; Thissen, R.; Sakai, S.; Yoshimura, T.; Koga, T.; Ohkouchi, N.; Chikaraishi, Y.; Sugahara, H.; Mita, H.; Furukawa, Y.; Hertkorn, N.; Ruf, A.; Yurimoto, H.; Nakamura, T.; Noguchi, T.; Okazaki, R.; Yabuta, H.; Sakamoto, K.; Tachibana, S.; Connolly, H. C., Jr.; Lauretta, D. S.; Abe, M.; Yada, T.; Nishimura, M.; Yogata, K.; Nakato, A.; Yoshitake, M.; Suzuki, A.; Miyazaki, A.; Furuya, S.; Hatakeda, K.; Soejima, H.; Hitomi, Y.; Kumagai, K.; Usui, T.; Hayashi, T.; Yamamoto, D.; Fukai, R.; Kitazato, K.; Sugita, S.; Namiki, N.; Arakawa, M.; Ikeda, H.; Ishiguro, M.; Hirata, N.; Wada, K.; Ishihara, Y.; Noguchi, R.; Morota, T.; Sakatani, N.; Matsumoto, K.; Senshu, H.; Honda, R.; Tatsumi, E.; Yokota, Y.; Honda, C.; Michikami, T.; Matsuoka, M.; Miura, A.; Noda, H.; Yamada, T.; Yoshihara, K.; Kawahara, K.; Ozaki, M.; Iijima, Y. I.; Yano, H.; Hayakawa, M.; Iwata, T.; Tsukizaki, R.; Sawada, H.; Hosoda, S.; Ogawa, K.; Okamoto, C.; Hirata, N.; Shirai, K.; Shimaki, Y.; Yamada, M.; Okada, T.; Yamamoto, Y.; Takeuchi, H.; Fujii, A.; Takei, Y.; Yoshikawa, K.; Mimasu, Y.; Ono, G.; Ogawa, N.; Kikuchi, S.; Nakazawa, S.; Terui, F.; Tanaka, S.; Saiki, T.; Yoshikawa, M.; Watanabe, S. I.; Tsuda, Y., Soluble organic molecules in samples of the carbonaceous asteroid (162173) Ryugu. *Science* **2023,** *379* (6634), eabn9033. [DOI: 10.1126/science.abn9033]

[119] Okazaki, R.; Miura, Y. N.; Takano, Y.; Sawada, H.; Sakamoto, K.; Yada, T.; Yamada, K.; Kawagucci, S.; Matsui, Y.; Hashizume, K.; Ishida, A.; Broadley, M. W.; Marty, B.; Byrne, D.; Furi, E.; Meshik, A.; Pravdivtseva, O.; Busemann, H.; Riebe, M. E. I.; Gilmour, J.; Park, J.; Bajo, K. I.; Righter, K.; Sakai, S.; Sekimoto, S.; Kitajima, F.; Crowther, S. A.; Iwata, N.; Shirai, N.; Ebihara, M.; Yokochi, R.; Nishiizumi, K.; Nagao, K.; Lee, J. I.; Clay, P.; Kano, A.; Caffee, M. W.; Uemura, R.; Inagaki, M.; Krietsch, D.; Maden, C.; Yamamoto, M.; Fawcett, L.; Lawton, T.; Nakamura, T.; Naraoka, H.; Noguchi, T.; Yabuta, H.; Yurimoto, H.; Tsuda, Y.; Watanabe, S. I.; Abe, M.; Arakawa, M.; Fujii, A.; Hayakawa, M.; Hirata, N.; Hirata, N.; Honda, R.; Honda, C.; Hosoda, S.; Iijima, Y. I.; Ikeda, H.; Ishiguro, M.; Ishihara, Y.; Iwata, T.; Kawahara, K.; Kikuchi, S.; Kitazato, K.; Matsumoto, K.; Matsuoka, M.; Michikami, T.; Mimasu, Y.; Miura, A.; Morota, T.; Nakazawa, S.; Namiki, N.; Noda, H.; Noguchi, R.; Ogawa, N.; Ogawa, K.; Okada, T.; Okamoto, C.; Ono, G.; Ozaki, M.; Saiki, T.; Sakatani, N.; Senshu, H.; Shimaki, Y.; Shirai, K.; Sugita, S.; Takei, Y.; Takeuchi, H.; Tanaka, S.; Tatsumi, E.; Terui, F.; Tsukizaki, R.; Wada, K.; Yamada, M.; Yamada, T.; Yamamoto, Y.; Yano, H.; Yokota, Y.; Yoshihara, K.; Yoshikawa, M.; Yoshikawa, K.; Furuya, S.; Hatakeda, K.; Hayashi, T.; Hitomi, Y.; Kumagai, K.; Miyazaki, A.; Nakato, A.; Nishimura, M.; Soejima, H.; Iwamae, A.; Yamamoto, D.; Yogata, K.; Yoshitake, M.; Fukai, R.; Usui, T.; Ireland, T.; Connolly, H. C., Jr.; Lauretta, D. S.; Tachibana, S., First asteroid gas sample delivered by the Hayabusa2 mission: A treasure box from Ryugu. *Sci Adv* **2022,** *8* (46), eabo7239. [DOI: 10.1126/sciadv.abo7239]

[120] Brooke, T. Y.; Tokunaga, A. T.; Weaver, H. A.; Crovisier, J.; Bockelee-Morvan, D.; Crisp, D., Detection of acetylene in the infrared spectrum of comet Hyakutake. *Nature* **1996,** *383* (6601), 606-8. [DOI: 10.1038/383606a0]

[121] Mumma, M. J.; DiSanti, M. A.; Dello Russo, N.; Magee-Sauer, K.; Gibb, E.; Novak, R., Remote infrared observations of parent volatiles in comets: A window on the early solar system. *Adv. Space Res.* **2003,** *31* (12), 2563-2575. [DOI: 10.1016/S0273-1177(03)00578-7]

[122] W.M. Keck Observatory https://www.keckobservatory.org/ .

[123] Eberhardt, P. Comet Halley's Gas Composition and Extended Sources: Results from the Neutral Mass Spectrometer on Giotto. *Space Sci. Rev.* **1999,** *90* (1/2), 45-52. [DOI: 10.1023/A:1005221309219]

[124] Le Roy, L.; Altwegg, K.; Balsiger, H.; Berthelier, J.-J.; Bieler, A.; Briois, C.; Calmonte, U.; Combi, M. R.; De Keyser, J.; Dhooghe, F.; Fiethe, B.; Fuselier, S. A.; Gasc, S.; Gombosi, T. I.; Hässig, M.; Jäckel, A.; Rubin, M.; Tzou, C.-Y., Inventory of the volatiles on comet

[171] Draine, B. T.; Li, A., Infrared Emission from Interstellar Dust. I. Stochastic Heating of Small Grains. *Astrophys. J.* **2001,** *551* (2), 807-824. [DOI: 10.1086/320227]

[172] Ling, Y.; Lifshitz, C., Time-Dependent Mass Spectra and Breakdown Graphs. 21. $C_{14}H_{10}$ Isomers. *The J. Phys. Chem. A* **1998,** *102* (4), 708-716. [DOI: 10.1021/jp973167w]

[173] Le Page, V.; Snow, T. P.; Bierbaum, V. M., Hydrogenation and Charge States of Polycyclic Aromatic Hydrocarbons in Diffuse Clouds. II. Results. *Astrophys. J.* **2003,** *584* (1), 316-330. [DOI: 10.1086/345595]

[174] Murga, M. S.; Wiebe, D. S., Uncertainty in Estimates of Dissociation Rates of Polycyclic Aromatic Hydrocarbons in the Interstellar Medium. *Bull. Lebedev Phys. Inst.* **2023,** *49* (12), 416-421. [DOI: 10.3103/s1068335622120077]

[175] Murga, M. S.; Kirsanova, M. S.; Vasyunin, A. I.; Pavlyuchenkov, Y. N., Impact of PAH photodissociation on the formation of small hydrocarbons in the Orion Bar and the horsehead PDRs. *Mon. Not. R. Astron. Soc.* **2020,** *497* (2), 2327-2339. [DOI: 10.1093/mnras/staa2026]

[176] Pilleri, P.; Montillaud, J.; Berné, O.; Joblin, C., Evaporating very small grains as tracers of the UV radiation field in photo-dissociation regions. *Astron. Astrophys.* **2012,** *542*. [DOI: 10.1051/0004-6361/201015915]

[177] Alata, I.; Cruz-Diaz, G. A.; Muñoz Caro, G. M.; Dartois, E., Vacuum ultraviolet photolysis of hydrogenated amorphous carbons. *Astron. Astrophys.* **2014,** *569,* A119. [DOI: 10.1051/0004-6361/201323118]

[178] Alata, I.; Jallat, A.; Gavilan, L.; Chabot, M.; Cruz-Diaz, G. A.; Munoz Caro, G. M.; Béroff, K.; Dartois, E., Vacuum ultraviolet of hydrogenated amorphous carbons. *Astron. Astrophys.* **2015,** *584*. [DOI: 10.1051/0004-6361/201526368]

[179] Scott, A.; Duley, W. W.; Pinho, G. P., Polycyclic Aromatic Hydrocarbons and Fullerenes as Decomposition Products of Hydrogenated Amorphous Carbon. *Astrophys. J.* **1997,** *489* (2), L193-L195. [DOI: 10.1086/316789]

[180] Li, A.; Draine, B. T., The Carriers of the Interstellar Unidentified Infrared Emission Features: Aromatic or Aliphatic? *Astrophys. J.* **2012,** *760* (2). [DOI: 10.1088/2041-8205/760/2/L35]

[181] Jones, A. P.; Fanciullo, L.; Köhler, M.; Verstraete, L.; Guillet, V.; Bocchio, M.; Ysard, N., The evolution of amorphous hydrocarbons in the ISM: dust modelling from a new vantage point. *Astron. Astrophys.* **2013,** *558*. [DOI: 10.1051/0004-6361/201321686]

[182] Murga, M. S.; Vasyunin, A. I.; Kirsanova, M. S., Impact of HAC evolution on the formation of small hydrocarbons in the Orion Bar and the Horsehead PDRs. *Mon. Not. R. Astron. Soc.* **2023,** *519* (2), 2466-2474. [DOI: 10.1093/mnras/stac3656]

[183] Cuadrado, S.; Goicoechea, J. R.; Pilleri, P.; Cernicharo, J.; Fuente, A.; Joblin, C., The chemistry and spatial distribution of small hydrocarbons in UV-irradiated molecular clouds: the Orion Bar PDR. *Astron. Astrophys.* **2015,** *575*. [DOI: 10.1051/0004-6361/201424568]

hν [184] Godard, B.; Falgarone, E.; Pineau des Forêts, G., Models of turbulent dissipation regions in the diffuse interstellar medium. *Astron. Astrophys.* **2009,** *495* (3), 847-867. [DOI: 10.1051/0004-6361:200810803]

[185] Murga, M. S.; Kirsanova, M. S.; Wiebe, D. S.; Boley, P. A., Orion Bar as a window to the evolution of PAHs. *Mon. Not. R. Astron. Soc.* **2022,** *509* (1), 800-817. [DOI: 10.1093/mnras/stab3061]

[186] Carr, J. S.; Najita, J. R., Organic molecules and water in the planet formation region of young circumstellar disks. *Science* **2008,** *319* (5869), 1504-6. [DOI: 10.1126/science.1153807]

[198] Morris, M.; Turner, B. E.; Palmer, P.; Zuckerman, B., Cyanoacetylene in dense interstellar clouds. *Astrophys. J.* **1976,** *205*. [DOI: 10.1086/154252]

[199] Gupta, H.; Gottlieb, C. A.; McCarthy, M. C.; Thaddeus, P., A Survey of C4h, C6h, and C6h– with the Green Bank Telescope. *Astrophys. J.* **2009,** *691* (2), 1494-1500. [DOI: 10.1088/0004-637X/691/2/1494]

[200] Agundez, M.; Cernicharo, J.; Quintana-Lacaci, G.; Castro-Carrizo, A.; Velilla Prieto, L.; Marcelino, N.; Guelin, M.; Joblin, C.; Martin-Gago, J. A.; Gottlieb, C. A.; Patel, N. A.; McCarthy, M. C., The growth of carbon chains in IRC +10216 mapped with ALMA. *Astron. Astrophys.* **2017,** *601*. [DOI: 10.1051/0004-6361/201630274]

[201] Pardo, J. R.; Cernicharo, J.; Goicoechea, J. R.; Guelin, M.; Asensio Ramos, A., Molecular Line Survey of CRL 618 from 80 to 276 GHz and Complete Model. *Astrophys. J.* **2007,** *661* (1), 250-261. [DOI: 10.1086/513734]

[202] Winnewisser, G.; Walmsley, C. M., Carbon chain molecules in interstellar clouds. *Astrophys. Space Sci.* **1979,** *65* (1), 83-93. [DOI: 10.1007/BF00643491]

[203] Cernicharo, J. The Polymerization of Acetylene, Hydrogen Cyanide, and Carbon Chains in the Neutral Layers of Carbon-rich Protoplanetary Nebulae. *Astrophys. J.* **2004**, *608*, L41. [DOI: 10.1086/422170]

[204] Fukuzawa, K.; Osamura, Y.; Schaefer Iii, H. F., Are Neutral-Neutral Reactions Effective for the Carbon-Chain Growth of Cyanopolyynes and Polyacetylenes in Interstellar Space? *Astrophys. J.* **1998,** *505* (1), 278-285. [DOI: 10.1086/306168]

[205] Chapman, J. F.; Millar, T. J.; Wardle, M.; Burton, M. G.; Walsh, A. J., Cyanopolyynes in hot cores: modelling G305.2+0.2. *Mon. Not. R. Astron. Soc.* **2009,** *394* (1), 221-230. [DOI: 10.1111/j.1365-2966.2008.14144.x]

[206] Balucani, N.; Asvany, O.; Huang, L. C. L.; Lee, Y. T.; Kaiser, R. I.; Osamura, Y.; Bettinger, H. F., Formation of Nitriles in the Interstellar Medium via Reactions of Cyano Radicals, CN(X 2Σ+), with Unsaturated Hydrocarbons. *Astrophys. J.* **2000,** *545* (2), 892-906. [DOI: 10.1086/317848]

[207] Santoro, G.; Martinez, L.; Lauwaet, K.; Accolla, M.; Tajuelo-Castilla, G.; Merino, P.; Sobrado, J. M.; Pelaez, R. J.; Herrero, V. J.; Tanarro, I.; Mayoral, A. L.; Agundez, M.; Sabbah, H.; Joblin, C.; Cernicharo, J.; Martin-Gago, J. A., The Chemistry of Cosmic Dust Analogues from C, C(2), and C(2)H(2) in C-Rich Circumstellar Envelopes. *Astrophys J* **2020,** *895* (2). [DOI: 10.3847/1538-4357/ab9086]

[208] Gu, X.; Kim, Y. S.; Kaiser, R. I.; Mebel, A. M.; Liang, M. C.; Yung, Y. L., Chemical dynamics of triacetylene formation and implications to the synthesis of polyynes in Titan's atmosphere. *Proc Natl Acad Sci U S A* **2009,** *106* (38), 16078-83. [DOI: 10.1073/pnas.0900525106]

[209] He, C.; Thomas, A. M.; Galimova, G. R.; Mebel, A. M.; Kaiser, R. I., Gas Phase Formation of the Interstellar Molecule Methyltriacetylene. *Chem. Phys. Chem.* **2019,** *20* (15), 1912-1917. [DOI: 10.1002/cphc.201900305]

[210] Remijan, A. J.; Hollis, J. M.; Snyder, L. E.; Jewell, P. R.; Lovas, F. J., Methyltriacetylene ($CH_3C_6H$) toward TMC-1: The Largest Detected Symmetric Top. *Astrophys. J.* **2006,** *643* (1), L37-L40. [DOI: 10.1086/504918]

[211] Cuylle, S. H.; Zhao, D.; Strazzulla, G.; Linnartz, H., Vacuum ultraviolet photochemistry of solid acetylene: a multispectral approach. *Astron. Astrophys.* **2014,** *570*. [DOI: 10.1051/0004-6361/201424379]

[212] Compagnini, G.; D'Urso, L.; Puglisi, O.; Baratta, G. A.; Strazzulla, G., The irradiation of solid hydrocarbons and the formation of linear carbon chains. *Carbon* **2009,** *47* (6), 1605-1607. [DOI: 10.1016/j.carbon.2009.02.009]

[213] Cernicharo, J.; Guélin, M.; Agúndez, M.; Kawaguchi, K.; McCarthy, M.; Thaddeus, P., Astronomical detection of $C_4H^-$, the second interstellar anion. *Astron. Astrophys.* **2007,** *467* (2), L37-L40. [DOI: 10.1051/0004-6361:20077415]

[214] Remijan, A. J.; Hollis, J. M.; Lovas, F. J.; Cordiner, M. A.; Millar, T. J.; Markwick-Kemper, A. J.; Jewell, P. R., Detection of $C_8H^-$ and Comparison with $C_8H$ toward IRC +10216. *Astrophys. J.* **2007,** *664* (1), L47-L50. [DOI: 10.1086/520704]

[215] Millar, T. J.; Walsh, C.; Field, T. A., Negative Ions in Space. *Chem. Rev.* **2017,** *117* (3), 1765-1795. [DOI: 10.1021/acs.chemrev.6b00480]

[216] Bastian, B.; Michaelsen, T.; Meyer, J.; Wester, R., Anionic Carbon Chain Growth in Reactions of $C_2^-$, $C_4^-$, $C_6^-$, $C_2H^-$, $C_4H^-$, and $C_6H^-$ with $C_2H_2$. *Astrophys. J.* **2019,** *878* (2). [ref: 10.3847/1538-4357/ab2042]

[217] Kaiser, R. I.; Mebel, A. M. On the formation of polyacetylenes and cyanopolyacetylenes in Titan's atmosphere and their role in astrobiology. *Chem. Soc. Rev.* **2012,** *41* (16), 5490-501. [DOI: 10.1039/c2cs35068h]

[218] McGuire, B. A., 2021 Census of Interstellar, Circumstellar, Extragalactic, Protoplanetary Disk, and Exoplanetary Molecules. *Astrophys. J., Suppl. Ser.* **2022,** *259* (2). [DOI: 10.3847/1538-4365/ac2a48]

[219] Herbst, E.; Garrod, R. T., Synthetic Approaches to Complex Organic Molecules in the Cold Interstellar Medium. *Front. Astron. Space Sci.* **2022,** *8*. [DOI: 10.3389/fspas.2021.789428]

[220] Burke, D. J.; Brown, W. A., Ice in space: surface science investigations of the thermal desorption of model interstellar ices on dust grain analogue surfaces. *Phys. Chem. Chem. Phys.* **2010,** *12* (23), 5947-69. [DOI: 10.1039/B917005G]

[221] Oberg, K. I., Photochemistry and Astrochemistry: Photochemical Pathways to Interstellar Complex Organic Molecules. *Chem. Rev.* **2016,** *116* (17), 9631-63. [DOI: 10.1021/acs.chemrev.5b00694]

[222] Wakelam, V.; Loison, J. C.; Mereau, R.; Ruaud, M., Binding energies: New values and impact on the efficiency of chemical desorption. *Mol. Astrophys.* **2017,** *6*, 22-35. [DOI: 10.1016/j.molap.2017.01.002]

[223] Garrod, R. T.; Weaver, S. L. W.; Herbst, E., Complex Chemistry in Star-forming Regions: An Expanded Gas-Grain Warm-up Chemical Model. *Astrophys. J.* **2008,** *682* (1), 283-302. [DOI: 10.1086/588035]

[224] Jin, M.; Garrod, R. T., Formation of Complex Organic Molecules in Cold Interstellar Environments through Nondiffusive Grain-surface and Ice-mantle Chemistry. *Astrophys. J., Suppl. Ser.* **2020,** *249* (2). [DOI: 10.3847/1538-4365/ab9ec8]

[225] Chuang, K. J.; Fedoseev, G.; Scirè, C.; Baratta, G. A.; Jäger, C.; Henning, T.; Linnartz, H.; Palumbo, M. E., Formation of complex organic molecules in molecular clouds: acetaldehyde, vinyl alcohol, ketene, and ethanol via the "energetic" processing of $C_2H_2$ ice. *Astron. Astrophys.* **2021,** *650*. [DOI: 10.1051/0004-6361/202140780]

[226] Hudson, R. L.; Loeffler, M. J., Ketene Formation in Interstellar Ices: A Laboratory Study. Astrophys. J. 2013, 773 (2). [DOI: 10.1088/0004-637x/773/2/109]

[227] Bergner, J. B.; Öberg, K. I.; Rajappan, M., Oxygen Atom Reactions with $C_2H_6$, $C_2H_4$, and $C_2H_2$ in Ices. *Astrophys. J.* **2019,** *874* (2). [DOI: 10.3847/1538-4357/ab07b2]

[228] Bergner, J. B.; Öberg, K. I.; Garrod, R. T.; Graninger, D. M., Complex Organic Molecules toward Embedded Low-mass Protostars. *Astrophys. J.* **2017,** *841* (2). [DOI: 10.3847/1538-4357/aa72f6]

[229] Zasimov, P. V.; Ryazantsev, S. V.; Tyurin, D. A.; Feldman, V. I., Radiation-induced chemistry in the $C_2H_2$–$H_2O$ system at cryogenic temperatures: a matrix isolation study. *Mon. Not. R. Astron. Soc.* **2019**. [DOI: 10.1093/mnras/stz3228]

[230] Hudson, R.; Moore, M., Hydrocarbon Radiation Chemistry in Ices of Cometary Relevance☆. *Icarus* **1997,** *126* (1), 233-235. [DOI: 10.1006/icar.1997.5678]

[231] Robert Wu, C., Extreme Ultraviolet Photon-Induced Chemical Reactions in the C2H2–H2O Mixed Ices at 10 K. *Icarus* **2002,** *156* (2), 456-473. [DOI: 10.1006/icar.2001.6801]

[232] Hudson, R. L.; Moore, M. H., Solid-Phase Formation of Interstellar Vinyl Alcohol. *Astrophys. J.* **2003,** *586* (1), L107-L110. [DOI: 10.1086/374580]

[233] Tielens, A. G.G.M. Grain Surface Chemistry. In *Chemistry and Spectroscopy of Interstellar Molecules*. Bohme, K.K., Ed.; Tokyo: University of Tokyo Press, p. 237, 1992. [https://ui.adsabs.harvard.edu/abs/1992csim.conf..237T/abstract]

[234] Charnley, S. B., Acetaldehyde in star-forming regions. *Adv. Space Res.* **2004,** *33* (1), 23-30. [DOI: 10.1016/j.asr.2003.08.005]

[235] Chuang, K. J.; Fedoseev, G.; Qasim, D.; Ioppolo, S.; Jäger, C.; Henning, T.; Palumbo, M. E.; van Dishoeck, E. F.; Linnartz, H., Formation of complex molecules in translucent clouds: acetaldehyde, vinyl alcohol, ketene, and ethanol via “nonenergetic” processing of $C_2H_2$ice. *Astron. Astrophys.* **2020,** *635*. [DOI: 10.1051/0004-6361/201937302]

[236] Perrero, J.; Enrique-Romero, J.; Martinez-Bachs, B.; Ceccarelli, C.; Balucani, N.; Ugliengo, P.; Rimola, A., Non-energetic Formation of Ethanol via CCH Reaction with Interstellar H(2)O Ices. A Computational Chemistry Study. *ACS Earth Space Chem.* **2022,** *6* (3), 496-511. [DOI: 10.1021/acsearthspacechem.1c00369]

[237] Zhou, L.; Kaiser, R. I.; Gao, L. G.; Chang, A. H. H.; Liang, M. C.; Yung, Y. L., Pathways to Oxygen-Bearing Molecules in the Interstellar Medium and in Planetary Atmospheres: Cyclopropenone (c-$C_3H_2O$) and Propynal (HCCCHO). *Astrophys. J.* **2008,** *686* (2), 1493-1502. [DOI: 10.1086/591072]

[238] Wu, Y.-J.; Chuang, S.-J.; Chen, S.-C.; Huang, T.-P., Infrared Spectra of Acetylene Diluted in Solid Nitrogen Upon Irradiation with Vacuum Ultraviolet Light and Electrons. *Astrophys. J., Suppl. Ser.* **2014,** *212* (1). [DOI: 10.1088/0067-0049/212/1/7]

[239] Canta, A.; Öberg, K. I.; Rajappan, M., The Formation of Imines and Nitriles during VUV Photoirradiation of $NH_3$:$C_2H$ x Ice Mixtures. *Astrophys. J.* **2023,** *953* (1). [DOI: 10.3847/1538-4357/acda99]

[240] Zhang, C.; Wang, J.; Turner, A. M.; Marks, J. H.; Chandra, S.; Fortenberry, R. C.; Kaiser, R. I., On the Formation of Vinylamine ($C_2H_3NH_2$) in Interstellar Ice Analogs. *Astrophys. J.* **2023,** *952* (2). [DOI: 10.3847/1538-4357/acd451]

[241] Molpeceres, G.; Rivilla, V. M., Radical addition and H abstraction reactions in $C_2H_2$, $C_2H_4$, and $C_2H_6$: A gateway for ethyl- and vinyl-bearing molecules in the interstellar medium. *Astron. Astrophys.* **2022,** *665*. [DOI: 10.1051/0004-6361/202243892]

[242] Baines, K. H.; Sromovsky, L. A.; Carlson, R. W.; Momary, T. W.; Fry, P. M., The visual spectrum of Jupiter's Great Red Spot accurately modeled with aerosols produced by photolyzed ammonia reacting with acetylene. *Icarus* **2019,** *330*, 217-229. [DOI: 10.1016/j.icarus.2019.04.008]

[243] Carlson, R. W.; Baines, K. H.; Anderson, M. S.; Filacchione, G.; Simon, A. A., Chromophores from photolyzed ammonia reacting with acetylene: Application to Jupiter's Great Red Spot. *Icarus* **2016,** *274*, 106-115. [DOI: 10.1016/j.icarus.2016.03.008]

[244] Bernard-Salas, J.; Peeters, E.; Sloan, G. C.; Cami, J.; Guiles, S.; Houck, J. R., The Spitzer IRS spectrum of SMP LMC 11. *Astrophys. J.* **2006,** *652* (1), L29-L32. [DOI: 10.1086/509777]

[245] McGuire, B. A.; Burkhardt, A. M.; Kalenskii, S.; Shingledecker, C. N.; Remijan, A. J.; Herbst, E.; McCarthy, M. C., Detection of the aromatic molecule benzonitrile (c-C(6)H(5)CN) in the interstellar medium. *Science* **2018,** *359* (6372), 202-205. [DOI: 10.1126/science.aao4890]

[246] Cherchneff, I. The formation of Polycyclic Aromatic Hydrocarbons in evolved circumstellar environments. *EAS Publ. Ser.* **2011**, *46*, 177. [DOI: 10.1051/eas/1146019]

[247] Allamandola L.J., Tielens A.G.G.M., Barker J.R. Infrared Emission from Interstellar PAHs. In: Morfill G.E., Scholer M. (eds) Physical Processes in Interstellar Clouds. NATO ASI Series (Series C: Mathematical and Physical Sciences), vol 210. Springer, Dordrecht, 1987, pp. 305-331. [DOI: 10.1007/978-94-009-3945-5_20]

[248] Ehrenfreund, P.; Cami, J., Cosmic carbon chemistry: from the interstellar medium to the early Earth. *Cold Spring Harbor Perspect. Biol.* **2010,** *2* (12), a002097. [DOI: 10.1101/cshperspect.a002097]

[249] Kwok, S., The synthesis of organic and inorganic compounds in evolved stars. *Nature* **2004,** *430* (7003), 985-91. [DOI: 10.1038/nature02862]

[250] Kwok, S., Delivery of Complex Organic Compounds from Planetary Nebulae to the Solar System. *Int. J. Astrobiol.* **2009,** *8* (3), 161-167. [DOI: 10.1017/S1473550409004492]

[251] Cherchneff, I.; Barker, J. R.; Tielens, A. G. G. M. Polycyclic aromatic hydrocarbon formation in carbon-rich stellar envelopes. *Astrophys. J.* **1992**, *401*, 269. [DOI: 10.1086/172059]

[252] Cole, J. A.; Bittner, J. D.; Longwell, J. P.; Howard, J. B. Formation mechanisms of aromatic compounds in aliphatic flames. *Combust. Flame* **1984**, *56*, 51. [DOI: 10.1016/0010-2180(84)90005-1]

[253] Frenklach, M.; Warnatz, J. Detailed Modeling of PAH Profiles in a Sooting Low-Pressure Acetylene Flame. *Combust. Sci. Technol.* **1987**, *51*, 265. [DOI: 10.1080/00102208708960325]

[254] Frenklach, M.; Feigelson, E. D. Formation of polycyclic aromatic hydrocarbons in circumstellar envelopes. *Astrophys. J.* **1989**, *341*, 372. [DOI: 10.1086/167501]

[255] Miller, J. A.; Melius, C. F. Kinetic and thermodynamic issues in the formation of aromatic compounds in flames of aliphatic fuels. *Combust. Flame* **1992**, *91*, 21. [DOI: 10.1016/0010-2180(92)90124-8]

[256] Zhao, L.; Lu, W.; Ahmed, M.; Zagidullin, M. V.; Azyazov, V. N.; Morozov, A. N.; Mebel, A. M.; Kaiser, R. I., Gas-phase synthesis of benzene via the propargyl radical self-reaction. *Sci. Adv.* **2021**, *7* (21). [DOI: 10.1126/sciadv.abf0360]

[257] Richter, H.; Howard, J. B. Formation of polycyclic aromatic hydrocarbons and their growth to soot—a review of chemical reaction pathways. *Prog. Energy Combust. Sci.* **2000**, *26*, 565. [DOI: 10.1016/S0360-1285(00)00009-5]

[258] Richter, H.; Howard, J. B. Formation and consumption of single-ring aromatic hydrocarbons and their precursors in premixed acetylene, ethylene and benzene flames. *Phys. Chem. Chem. Phys.* **2002**, *4*, 2038. [DOI: 10.1039/B110089K]

[259] Lindstedt, R. P.; Skevis, G., Benzene formation chemistry in premixed, 1,3-butadiene flames. *Twenty-Sixth Symposium (International) on Combustion, Vols 1 and 2* **1996**, 703-709. [DOI: 10.1016/S0082-0784(96)80278-5]

[260] Ikeda, E.; Tranter, R. S.; Kiefer, J. H.; Kern, R. D.; Singh, H. J.; Zhang, Q. The pyrolysis of methylcyclopentadiene: Isomerization and formation of aromatics. *Proc. Combust. Inst.* **2000**, *28*, 1725. [DOI: 10.1016/S0082-0784(00)80573-1]

[261] Moskaleva, L. V.; Mebel, A. M.; Lin, M. C. The CH3+C5H5 reaction: A potential source of benene at high temperatures. *Symposium (International) on Combustion* **1996**, *26*, 521. [DOI: 10.1016/S0082-0784(96)80255-4]

[262] Kaiser, R. I.; Zhao, L.; Lu, W.; Ahmed, M.; Zagidullin, M. V.; Azyazov, V. N.; Mebel, A. M., Formation of Benzene and Naphthalene through Cyclopentadienyl-Mediated Radical-Radical Reactions. *J. Phys. Chem. Lett.* **2022**, 13 (1), 208-213. [DOI: 10.1021/acs.jpclett.1c03733]

[263] Kalvāns, J., Temperature Spectra of Interstellar Dust Grains Heated by Cosmic Rays. II. Dark Cloud Cores. *Astrophys. J., Suppl. Ser.* **2018,** *239* (1). [DOI: 10.3847/1538-4365/aae527]

[264] Mitchell, J. L.; Lora, J. M., The Climate of Titan. *Annu. Rev. Earth Planet. Sci.* **2016,** *44* (1), 353-380. [DOI: 10.1146/annurev-earth-060115-012428]

[265] Alkemade, U.; Homann, K. H., Formation of $C_6H_6$ Isomers by Recombination of Propynyl in the System Sodium Vapour/Propynylhalide. *Z. Phys. Chem.* **1989,** *161* (1-2), 19-34. [DOI: 10.1524/zpch.1989.161.Part_1_2.019]

[266] Cernicharo, J.; Heras, A. M.; Pardo, J. R.; Tielens, A. G. G. M.; Guélin, M.; Dartois, E.; Neri, R.; Waters, L. B. F. M. Methylpolyynes and Small Hydrocarbons in CRL 618. *Astrophys. J.* **2001**, *546*, L127. [DOI: 10.1086/318872]

[267] Kaiser, R. I. Experimental investigation on the formation of carbon-bearing molecules in the interstellar medium via neutral-neutral reactions. *Chem. Rev.* **2002**, *102*, 1309. [DOI: 10.1021/cr970004v]

[268] Hiraoka, K.; Takayama, T.; Euchi, A.; Handa, H.; Sato, T. Study of the Reactions of H and D Atoms with Solid $C_2H_2$, $C_2H_4$, and $C_2H_6$ at Cryogenic Temperatures. *Astrophys. J.* **2000**, *532*, 1029. [DOI: 10.1086/308612]

[269] Flower, D. R.; Forets, G. P. d., Grain-mantle erosion in magnetohydrodynamic shocks. *Mon. Not. R. Astron. Soc.* **1994,** *268* (3), 724-732. [DOI: 10.1093/mnras/268.3.724]

[270] Markwick, A. J.; Millar, T. J.; Charnley, S. B., On the Abundance Gradients of Organic Molecules along the TMC-1 Ridge. *Astrophys. J.* **2000,** *535* (1), 256-265. [DOI: 10.1086/308814]

[271] Lukianova, M. A.; Feldman, V. I., Direct evidence for a single-step radiation-induced assembling of benzene ring from acetylene trimer at cryogenic temperatures. *Radiat. Phys. Chem.* **2021,** *183*. [DOI: 10.1016/j.radphyschem.2021.109417]

[272] Dangi, B. B.; Parker, D. S. N.; Yang, T.; Kaiser, R. I.; Mebel, A. M., Gas-Phase Synthesis of the Benzyl Radical ($C_6H_5CH_2$). *Angew. Chem.* **2014,** *126* (18), 4696-4701. [DOI: 10.1002/ange.201310612]

[273] Zhang, F.; Parker, D.; Kim, Y. S.; Kaiser, R. I.; Mebel, A. M., On the formation of ortho-benzyne (o-$C_6H_4$) under single collision conditions and its role in interstellar chemistry. *Astrophys. J.* **2011,** *728* (2). [DOI: 10.1088/0004-637X/728/2/141]

[274] Bera, P. P.; Peverati, R.; Head-Gordon, M.; Lee, T. J., Hydrocarbon growth via ion-molecule reactions: computational studies of the isomers of $C_4H_2$+, $C_6H_2$+and $C_6H_4$+and their formation paths from acetylene and its fragments. *Phys Chem Chem Phys* **2015,** *17* (3), 1859-69. [DOI: 10.1039/C4CP04480K]

[275] Peverati, R.; Bera, P. P.; Lee, T. J.; Head-Gordon, M., Formation and stability of $C_6H_3$+ isomers. *J Phys Chem A* **2014,** *118* (43), 10109-16. [DOI: 10.1021/jp5081862]

[276] Peverati, R.; Bera, P. P.; Lee, T. J.; Head-Gordon, M., Insights into hydrocarbon chain and aromatic ring formation in the interstellar medium: computational study of the isomers of $C_4H_3$+, $C_6H_3$+ and C6H5+ and their formation pathways. *Astrophys. J.* **2016,** *830* (2). [DOI: 10.3847/0004-637X/830/2/128]

[277] Coustenis, A.; Salama, A.; Schulz, B.; Ott, S.; Lellouch, E.; Encrenaz, T. h.; Gautier, D.; Feuchtgruber, H. Titan’s atmosphere from ISO mid-infrared spectroscopy. *Icarus* **2003**, *161*, 383. [DOI: 10.1016/S0019-1035(02)00028-3]

[278] Wong, A.-S.; Yung, Y. L.; Friedson, A. J. Benzene and Haze Formation in the Polar Atmosphere of Jupiter. *Geophys. Res. Lett.* **2003**, *30* [DOI: 10.1029/2002GL016661]

[279] Wilson, E. H.; Atreya, S. K.; Coustenis, A. Mechanisms for the formation of benzene in the atmosphere of Titan. *J. Geophys. Res.: Planets* **2003**, *108*, n/a. [DOI: 10.1029/2002JE001896]

[280] Lebonnois, S. Benzene and aerosol production in Titan and Jupiter's atmospheres: a sensitivity study. *Planet. Space Sci.* **2005**, *53*, 486. [DOI: 10.1016/j.pss.2004.11.004]

[281] Lavvas, P. P.; Coustenis, A.; Vardavas, I. M. Coupling photochemistry with haze formation in Titan's atmosphere, Part II: Results and validation with Cassini/Huygens data. *Planet. Space Sci.* **2008**, *56*, 67. [DOI: 10.1016/j.pss.2007.05.027]

[282] Waite, J. H., Jr.; Niemann, H.; Yelle, R. V.; Kasprzak, W. T.; Cravens, T. E.; Luhmann, J. G.; McNutt, R. L.; Ip, W. H.; Gell, D.; De La Haye, V.; Muller-Wordag, I.; Magee, B.; Borggren, N.; Ledvina, S.; Fletcher, G.; Walter, E.; Miller, R.; Scherer, S.; Thorpe, R.; Xu, J.; Block, B.; Arnett, K., Ion neutral mass spectrometer results from the first flyby of Titan. *Science* **2005,** *308* (5724), 982-6. [DOI: 10.1126/science.1110652]

[283] Waite, J. H., Jr.; Young, D. T.; Cravens, T. E.; Coates, A. J.; Crary, F. J.; Magee, B.; Westlake, J., The process of tholin formation in Titan's upper atmosphere. *Science* **2007,** *316* (5826), 870-5. [DOI: 10.1126/science.1139727]

[284] Vuitton, V.; Yelle, R. V.; Cui, J. Formation and distribution of benzene on Titan. *J. Geophys. Res.* **2008**, *113*. [DOI: 10.1029/2007JE002997]

[285] Anicich, V. G.; McEwan, M. J. Ion-molecule chemistry in Titan's ionosphere. *Planet. Space Sci.***1997**, *45*, 897. [DOI: 10.1016/S0032-0633(97)00053-6]

[286] Zhou, L.; Zheng, W.; Kaiser, R. I.; Landera, A.; Mebel, A. M.; Liang, M.-C.; Yung, Y. L. Cosmic-Ray-Mediated Formation of Benzene on the Surface of Saturn’s Moon Titan. *Astrophys. J.* **2010**, *718*, 1243. [DOI: 10.1088/0004-637X/718/2/1243]

[287] Kleimeier, N. F.; Liu, Y.; Turner, A. M.; Young, L. A.; Chin, C. H.; Yang, T.; He, X.; Lo, J. I.; Cheng, B. M.; Kaiser, R. I., Excited state photochemically driven surface formation of benzene from acetylene ices on Pluto and in the outer solar system. *Phys. Chem. Chem. Phys.* **2022**, 24 (3), 1424-1436. [DOI: 10.1039/d1cp04959c]

[288] Tsuge, M.; Bahou, M.; Wu, Y.-J.; Allamandola, L.; Lee, Y.-P. The Infrared Spectrum of Protonated Ovalene in Solid Para-Hydrogen and Its Possible Contribution to Interstellar Unidentified Infrared Emission. *Astrophys. J.* **2016**, *825*, 96. [DOI: 10.3847/0004-637X/825/2/96]

[289] Kim, H. S.; Wagner, D. R.; Saykally, R. J. Single photon infrared emission spectroscopy of the gas phase pyrene cation: support for a polycyclic aromatic hydrocarbon origin of the unidentified infrared emission bands. *Phys. Rev. Lett.* **2001**, *86*, 5691. [DOI: 10.1103/PhysRevLett.86.5691]

[290] Doddipatla, S.; Galimova, G. R.; Wei, H.; Thomas, A. M.; He, C.; Yang, Z.; Morozov, A. N.; Shingledecker, C. N.; Mebel, A. M.; Kaiser, R. I. Low-temperature gas-phase formation of indene in the interstellar medium. *Sci. Adv.* **2021**, *7*. [DOI: 10.1126/sciadv.abd4044]

[291] Tielens, A. G. G. M. Interstellar Polycyclic Aromatic Hydrocarbon Molecules. *Annu. Rev. Astron. Astrophys.* **2008**, *46*, 289. [DOI: 10.1146/annurev.astro.46.060407.145211]

[292] Rawlings, M. G.; Adamson, A. J.; Marshall, C. C. M.; Sarre, P. J. Near-infrared diffuse interstellar bands towards Her 36. *Mon. Not. R. Astron. Soc.* **2019**, *485*, 3398. [DOI: 10.1093/mnras/stz663]

[293] Cox, N. L. J.; Cami, J.; Farhang, A.; Smoker, J.; Monreal-Ibero, A.; Lallement, R.; Sarre, P. J.; Marshall, C. C. M.; Smith, K. T.; Evans, C. J.; Royer, P.; Linnartz, H.; Cordiner, M. A.; Joblin, C.; van Loon, J. T.; Foing, B. H.; Bhatt, N. H.; Bron, E.; Elyajouri, M.; de Koter, A.; Ehrenfreund, P.; Javadi, A.; Kaper, L.; Khosroshadi, H. G.; Laverick, M.; Le Petit, F.; Mulas, G.; Roueff, E.; Salama, F.; Spaans, M. The ESO Diffuse Interstellar Bands Large Exploration Survey: EDIBLES I. Project description, survey sample and quality assessment. *Astron. Astrophys.* **2017**, *606*. [DOI: 10.1051/0004-6361/201730912]

[294] Ruiterkamp, R.; Halasinski, T.; Salama, F.; Foing, B. H.; Allamandola, L. J.; Schmidt, W.; Ehrenfreund, P. Spectroscopy of large PAHs. *Astron. Astrophys.* **2002**, *390*, 1153. [DOI: 10.1051/0004-6361:20020478]

[295] Tsuge, M.; Tseng, C. Y.; Lee, Y. P. Spectroscopy of prospective interstellar ions and radicals isolated in para-hydrogen matrices. *Phys. Chem. Chem. Phys.* **2018**, *20*, 5344. [DOI: 10.1039/C7CP05680J]

[296] Ruiterkamp, R.; Cox, N. L. J.; Spaans, M.; Kaper, L.; Foing, B. H.; Salama, F.; Ehrenfreund, P. PAH charge state distribution and DIB carriers: Implications from the line of sight toward HD 147889. *Astron. Astrophys.* **2005**, *432*, 515. [DOI: 10.1051/0004-6361:20040331]

[297] Salama, F.; Galazutdinov, G. A.; Krełowski, J.; Biennier, L.; Beletsky, Y.; Song, I.-O., Polycyclic Aromatic Hydrocarbons and the Diffuse Interstellar Bands: A Survey. *Astrophys. J.* **2011,** *728* (2). [DOI: 10.1088/0004-637X/728/2/154]

[298] Omont, A.; Bettinger, H. F.; Tönshoff, C., Polyacenes and diffuse interstellar bands. *Astron. Astrophys.* **2019,** *625*. [DOI: 10.1051/0004-6361/201834953]

[299] Hensley, B. S.; Draine, B. T., The Astrodust+PAH Model: A Unified Description of the Extinction, Emission, and Polarization from Dust in the Diffuse Interstellar Medium. *Astrophys. J.* **2023,** *948* (1). [DOI: 10.3847/1538-4357/acc4c2]

[300] McGuire, B. A.; Loomis, R. A.; Burkhardt, A. M.; Lee, K. L. K.; Shingledecker, C. N.; Charnley, S. B.; Cooke, I. R.; Cordiner, M. A.; Herbst, E.; Kalenskii, S.; Siebert, M. A.; Willis, E. R.; Xue, C.; Remijan, A. J.; McCarthy, M. C., Detection of two interstellar polycyclic aromatic hydrocarbons via spectral matched filtering. *Science* **2021,** *371* (6535), 1265-1269. [DOI: 10.1126/science.abb7535]

[301] Cernicharo, J.; Agundez, M.; Cabezas, C.; Tercero, B.; Marcelino, N.; Pardo, J. R.; de Vicente, P., Pure hydrocarbon cycles in TMC-1: Discovery of ethynyl cyclopropenylidene, cyclopentadiene and indene. *Astron. Astrophys.* **2021,** *649*. [DOI: 10.1051/0004-6361/202141156]

[302] Zenobi, R.; Philippoz, J. M.; Zare, R. N.; Buseck, P. R. Spatially resolved organic analysis of the allende meteorite. *Science* **1989**, *246*, 1026. [DOI: 10.1126/science.246.4933.1026]

[303] Wang, Y.; Huang, Y.; Alexander, C. M. O. D.; Fogel, M.; Cody, G. Molecular and compound-specific hydrogen isotope analyses of insoluble organic matter from different carbonaceous chondrite groups. *Geochim. Cosmochim. Acta* **2005**, *69*, 3711. [DOI: 10.1016/j.gca.2005.03.008]

[304] Messenger, S.; Amari, S.; Gao, X.; Walker, R. M.; Clemett, S. J.; Chillier, X. D. F.; Zare, R. N.; Lewis, R. S. Indigenous Polycyclic Aromatic Hydrocarbons in Circumstellar Graphite Grains from Primitive Meteorites. *Astrophys. J.* **1998**, *502*, 284. [DOI: 10.1086/305874]

[305] Hahn, J. H.; Zenobi, R.; Bada, J. L.; Zare, R. N. Application of two-step laser mass spectrometry to cosmogeochemistry: direct analysis of meteorites. *Science* **1988**, *239*, 1523. [DOI: 10.1126/science.239.4847.1523]

[306] Plows, F. L.; Elsila, J. E.; Zare, R. N.; Buseck, P. R. Evidence that polycyclic aromatic hydrocarbons in two carbonaceous chondrites predate parent-body formation. *Geochim. Cosmochim. Acta* **2003**, *67*, 1429. [DOI: 10.1016/S0016-7037(02)01277-2]

[307] Elsila, J. E.; de Leon, N. P.; Buseck, P. R.; Zare, R. N. Alkylation of polycyclic aromatic hydrocarbons in carbonaceous chondrites. *Geochim. Cosmochim. Acta* **2005**, *69*, 1349. [DOI: 10.1016/j.gca.2004.09.009]

[308] Kaiser, K.; Schulz, F.; Maillard, J. F.; Hermann, F.; Pozo, I.; Peña, D.; Cleaves, H. J.; Burton, A. S.; Danger, G.; Afonso, C.; Sandford, S.; Gross, L., Visualization and identification of single meteoritic organic molecules by atomic force microscopy. *Meteorit. Planet. Sci.* **2022**. [DOI: 10.1111/maps.13784]

[309] Boersma, C.; Bregman, J.; Allamandola, L. J. The Charge State of Polycyclic Aromatic Hydrocarbons across a Reflection Nebula, an H ii Region, and a Planetary Nebula. *Astrophys. J.* **2018**, *858*, 67. [DOI: 10.3847/1538-4357/aabcbe]

[310] Ehrenfreund, P., Irvine, W.M., Owen, T., Becker, L., Blank, J., Brucato, J.R., Colangeli, L., Derenne, S., Dutrey, A., Despois, D., Lazcano, A., Robert, F. (Eds.) Astrobiology. Future Perspectives, Springer Netherlands, **2004**. [DOI: 10.1007/1-4020-2305-7]

[311] Duley, W. W. Polycyclic aromatic hydrocarbons, carbon nanoparticles and the diffuse interstellar bands. *Faraday Discuss.* **2006**, *133*, 415. [DOI: 10.1039/B516323D]

[312] d'Hendecourt, L.; Ehrenfreund, P. Spectroscopic properties of polycyclic aromatic hydrocarbons (PAHs) and astrophysical implications. *Adv. Space Res.* **1997**, *19*, 1023. [DOI: 10.1016/S0273-1177(97)00349-9]

[313] Rhee, Y. M.; Lee, T. J.; Gudipati, M. S.; Allamandola, L. J.; Head-Gordon, M. Charged polycyclic aromatic hydrocarbon clusters and the galactic extended red emission. *Proc. Natl. Acad. Sci. U. S. A.* **2007**, *104*, 5274. [DOI: 10.1073/pnas.0609396104]

[314] Lemmens, A. K.; Rap, D. B.; Thunnissen, J. M. M.; Willemsen, B.; Rijs, A. M. Polycyclic aromatic hydrocarbon formation chemistry in a plasma jet revealed by IR-UV action spectroscopy. *Nat. Commun.* **2020**, *11*, 269. [DOI: 10.1038/s41467-019-14092-3]

[315] Cook, D. J.; Schlemmer, S.; Balucani, N.; Wagner, D. R.; Steiner, B.; Saykally, R. J. Infrared emission spectra of candidate interstellar aromatic molecules. *Nature* **1996**, *380*, 227. [DOI: 10.1038/380227a0]

[316] Herbst, E. Astrochemistry and Star Formation: Successes and Challenges, WILEY-VCH, Weinheim, 2006.

[317] Habart, E.; Boulanger, F.; Verstraete, L.; Walmsley, C. M.; Forêts, G. P. d., Some empirical estimates of the H2formation rate in photon-dominated regions. *Astron. Astrophys.* **2004,** *414* (2), 531-544. [DOI: 10.1051/0004-6361:20031659]

[318] Zhang, F.; Jones, B.; Maksyutenko, P.; Kaiser, R. I.; Chin, C.; Kislov, V. V.; Mebel, A. M., Formation of the phenyl radical [$C_6H_5$(X(2)A1)] under single collision conditions: a crossed molecular beam and ab initio study. *J. Am. Chem. Soc.* **2010,** *132* (8), 2672-83. [DOI: 10.1021/ja908559v]

[319] Sagan, C.; Khare, B. N.; Thompson, W. R.; McDonald, G. D.; Wing, M. R.; Bada, J. L.; Vo-Dinh, T.; Arakawa, E. T. Polycyclic aromatic hydrocarbons in the atmospheres of Titan and Jupiter. *Astrophys. J.* **1993**, *414*, 399. [DOI: 10.1086/173086]

[320] Kahl, G. PAH world hypothesis (polycyclic aromatic hydrocarbons world). In *The Dictionary of Genomics, Transcriptomics and Proteomics*, Kahl, G., Ed.; 2015. [DOI: 10.1002/9783527678679.dg09094]

[321] d'Ischia, M.; Manini, P.; Martins, Z.; Remusat, L.; Alexander, C. M. O'D.; Puzzarini, C.; Barone, V.; Saladino, R., Insoluble organic matter in chondrites: Archetypal melanin-like PAH-based multifunctionality at the origin of life? *Phys. Life Rev.* **2021,** *37,* 65-93. [DOI: 10.1016/j.plrev.2021.03.002]

[322] Micelotta, E. R.; Jones, A. P.; Tielens, A. G. G. M. Polycyclic aromatic hydrocarbon processing by cosmic rays. *Astron. Astrophys.* **2010**, *526*, A52. [DOI: 10.1051/0004-6361/201015741]

[323] Micelotta, E. R.; Jones, A. P.; Tielens, A. G. G. M., Polycyclic aromatic hydrocarbon processing in interstellar shocks. *Astron. Astrophys.* **2010,** *510*, A36. [DOI: 10.1051/0004-6361/200911682]

[324] Micelotta, E. R.; Jones, A. P.; Tielens, A. G. G. M. Polycyclic aromatic hydrocarbon processing in a hot gas. *Astron. Astrophys.* **2010**, *510*, A37. [DOI: 10.1051/0004-6361/200911683]

[325] Murga, M. S.; Wiebe, D. S.; Sivkova, E. E.; Akimkin, V. V. SHIVA: a dust destruction model. *Mon. Not. R. Astron. Soc.* **2019,** *488* (1), 965-977. [DOI: 10.1093/mnras/stz1724]

[326] Parker, D. S.; Kaiser, R. I. On the formation of nitrogen-substituted polycyclic aromatic hydrocarbons (NPAHs) in circumstellar and interstellar environments. *Chem. Soc. Rev.* **2017,** *46,* 452. [DOI: 10.1039/C6CS00714G]

[327] Hirashita, H., Shattering by turbulence as a production source of very small grains. *Mon. Not. R. Astron. Soc.* **2010,** *407* (1), L49-L53. [DOI: 10.1111/j.1745-3933.2010.00902.x]

[328] Matsuura, M.; De Buizer, J. M.; Arendt, R. G.; Dwek, E.; Barlow, M. J.; Bevan, A.; Cigan, P.; Gomez, H. L.; Rho, J.; Wesson, R.; Bouchet, P.; Danziger, J.; Meixner, M., SOFIA mid-infrared observations of Supernova 1987A in 2016 – forward shocks and possible dust re-formation in the post-shocked region. *Mon. Not. R. Astron. Soc.* **2019,** *482* (2), 1715-1723. [DOI: 10.1093/mnras/sty2734]

[329] Sandstrom, K. M.; Bolatto, A. D.; Draine, B. T.; Bot, C.; Stanimirović, S., Thespitzersurvey of the Small Magellanic Cloud (S3mc): Insights into the Life Cycle of Polycyclic Aromatic Hydrocarbons. *Astrophys. J.* **2010,** *715* (2), 701-723. [DOI: 10.1088/0004-637X/715/2/701]

[330] He, C.; Kaiser, R. I.; Lu, W.; Ahmed, M.; Krasnoukhov, V. S.; Pivovarov, P. S.; Zagidullin, M. V.; Azyazov, V. N.; Morozov, A. N.; Mebel, A. M., Unconventional gas-phase preparation of the prototype polycyclic aromatic hydrocarbon naphthalene (C(10)H(8)) via the reaction of benzyl (C(7)H(7)) and propargyl (C(3)H(3)) radicals coupled with hydrogen-atom assisted isomerization. *Chem. Sci.* **2023,** *14* (20), 5369-5378. [DOI: 10.1039/D3SC00911D]

[331] Hirashita, H.; Murga, M. S., Self-consistent modelling of aromatic dust species and extinction curves in galaxy evolution. *Mon. Not. R. Astron. Soc.* **2020,** *492* (3), 3779-3793. [DOI: 10.1093/mnras/stz3640]

[332] Lemmens, A. K.; Rap, D. B.; Brunken, S.; Buma, W. J.; Rijs, A. M., Polycyclic aromatic hydrocarbon growth in a benzene discharge explored by IR-UV action spectroscopy. *Phys. Chem. Chem. Phys*. **2022**, 24 (24), 14816-14824. [DOI: 10.1039/d2cp01631a]

[333] Bockhorn, H.; Fetting, F.; Wenz, H. W. *Berichte der Bunsengesellschaft für physikalische Chemie* **1983**, *87*, 1067. [DOI: 10.1002/bbpc.19830871121]

[334] Frenklach, M.; Wang, H. Detailed modeling of soot particle nucleation and growth. *Symposium (International) on Combustion* **1991**, *23*, 1559. [DOI: 10.1016/S0082-0784(06)80426-1]

[335] Frenklach, M., Reaction mechanism of soot formation in flames. *Phys. Chem. Chem. Phys.* **2002,** *4* (11), 2028-2037. [DOI: 10.1039/B110045A]

[336] Wang, H.; Frenklach, M. A detailed kinetic modeling study of aromatics formation in laminar premixed acetylene and ethylene flames. *Combust. Flame* **1997**, *110*, 173. [DOI: 10.1016/S0010-2180(97)00068-0]

[337] Chu, T. C.; Buras, Z. J.; Smith, M. C.; Uwagwu, A. B.; Green, W. H., From benzene to naphthalene: direct measurement of reactions and intermediates of phenyl radicals and acetylene. *Phys. Chem. Chem. Phys.* **2019,** *21* (40), 22248-22258. [DOI: 10.1039/C9CP04554F]

[338] Yang, T.; Troy, T. P.; Xu, B.; Kostko, O.; Ahmed, M.; Mebel, A. M.; Kaiser, R. I. Hydrogen-Abstraction/Acetylene-Addition Exposed. *Angew. Chem. Int. Ed. Engl.* **2016**, *55*, 14983. [DOI: 10.1002/anie.201607509]

[339] Cherchneff, I., The inner wind of IRC +10216 revisited: new exotic chemistry and diagnostic for dust condensation in carbon stars. *Astron. Astrophys.* **2012,** *545*. [DOI: 10.1051/0004-6361/201118542]

[340] Mebel, A. M.; Kislov, V. V., Can the C(5)H(5) + C(5)H(5) --> C(10)H(10) --> C(10)H(9) + H/C(10)H(8) + H(2) reaction produce naphthalene? An Ab initio/RRKM study. *J. Phys. Chem. A* **2009,** *113* (36), 9825-33. [DOI: 10.1021/jp905931j]

[341] Kislov, V. V.; Sadovnikov, A. I.; Mebel, A. M. Formation mechanism of polycyclic aromatic hydrocarbons beyond the second aromatic ring. *J. Phys. Chem. A* **2013**, *117*, 4794. [DOI: 10.1021/jp402481y]

[342] Parker, D. S.; Kaiser, R. I.; Bandyopadhyay, B.; Kostko, O.; Troy, T. P.; Ahmed, M. Unexpected chemistry from the reaction of naphthyl and acetylene at combustion-like temperatures. *Angew. Chem. Int. Ed. Engl.* **2015**, *54*, 5421. [DOI: 10.1002/anie.201411987]

[343] Kislov, V. V.; Islamova, N. I.; Kolker, A. M.; Lin, S. H.; Mebel, A. M., Hydrogen Abstraction Acetylene Addition and Diels-Alder Mechanisms of PAH Formation: A Detailed Study Using First Principles Calculations. *J. Chem. Theory Comput.* **2005,** *1* (5), 908-24. [DOI: 10.1021/ct0500491]

[344] Badger, G. M.; Donnelly, J. K.; Spotswood, T. M., The formation of aromatic hydrocarbons at high temperatures. XXIV. The pyrolysis of some tobacco constituents. *Aust. J. Chem.* **1965,** *18* (8). [DOI: 10.1071/CH9651249]

[345] Sarofim A.F., Longwell J.P., Wornat M.J., Mukherjee J. The Role of Biaryl Reactions in PAH and Soot Formation. In: Bockhorn H. (eds) Soot Formation in Combustion. Springer Series in Chemical Physics, vol 59. Springer, Berlin, Heidelberg, 1994. [DOI: 10.1007/978-3-642-85167-4_27]

[346] Griesheimer, J.; Homann, K. H. Large molecules, radicals ions, and small soot particles in fuel-rich hydrocarbon flames. *Symposium (International) on Combustion* **1998**, *27*, 1753. [DOI: 10.1016/S0082-0784(98)80016-7]

[347] Böhm, H.; Jander, H.; Tanke, D. PAH growth and soot formation in the pyrolysis of acetylene and benzene at high temperatures and pressures: Modeling and experiment. *Symp. (Int.) Combust.* **1998**, *27 (1)*, 1605-1612. [DOI: 10.1016/S0082-0784(98)80570-5]

[348] Dong, G. L.; Hüttinger, K. J. Consideration of reaction mechanisms leading to pyrolytic carbon of different textures. *Carbon* **2002**, *40*, 2515. [DOI: 10.1016/S0008-6223(02)00174-4]

[349] Zhang, F.; Gu, X.; Kaiser, R. I. Formation of the diphenyl molecule in the crossed beam reaction of phenyl radicals with benzene. *J. Chem. Phys.* **2008**, *128*, 084315. [DOI: 10.1063/1.2834221]

[350] Schuetz, C. A.; Frenklach, M. Nucleation of soot: Molecular dynamics simulations of pyrene dimerization. *Proc. Combust. Inst.* **2002**, *29*, 2307. [DOI: 10.1016/S1540-7489(02)80281-4]

[351] Sabbah, H.; Biennier, L.; Klippenstein, S. J.; Sims, I. R.; Rowe, B. R. Exploring the Role of PAHs in the Formation of Soot: Pyrene Dimerization. *J. Phys. Chem. Lett.* **2010**, *1*, 2962. [DOI: 10.1021/jz101033t]

[352] He, C.; Fujioka, K.; Nikolayev, A. A.; Zhao, L.; Doddipatla, S.; Azyazov, V. N.; Mebel, A. M.; Sun, R.; Kaiser, R. I., A chemical dynamics study of the reaction of the methylidyne radical (CH, X(2)Pi) with dimethylacetylene (CH(3)CCCH(3), X(1)A(1g)). *Phys. Chem. Chem. Phys.* **2021**, 24 (1), 578-593. [DOI: 10.1039/D1CP04443E]

[353] Zhao, L.; Kaiser, R. I.; Lu, W.; Ahmed, M.; Oleinikov, A. D.; Azyazov, V. N.; Mebel, A. M.; Howlader, A. H.; Wnuk, S. F., Gas phase formation of phenalene via 10pi-aromatic, resonantly stabilized free radical intermediates. *Phys. Chem. Chem. Phys.* **2020**, 22 (27), 15381-15388. [DOI: 10.1039/D0CP02216K]

[354] Zhao, L.; Prendergast, M. B.; Kaiser, R. I.; Xu, B.; Ablikim, U.; Ahmed, M.; Sun, B. J.; Chen, Y. L.; Chang, A. H. H.; Mohamed, R. K.; Fischer, F. R., Synthesis of Polycyclic Aromatic Hydrocarbons by Phenyl Addition-Dehydrocyclization: The Third Way. *Angew. Chem. Int. Ed. Engl.* **2019,** *58* (48), 17442-17450. [DOI: 10.1002/anie.201909876]

[355] Yang, J.; Smith, M. C.; Prendergast, M. B.; Chu, T. C.; Green, W. H., C(14)H(10) polycyclic aromatic hydrocarbon formation by acetylene addition to naphthalenyl radicals observed. *Phys. Chem. Chem. Phys.* **2021**, 23 (26), 14325-14339. [DOI: 10.1039/d1cp01565f]

[356] Parker, D. S.; Kaiser, R. I.; Kostko, O.; Ahmed, M. Selective Formation of Indene through the Reaction of Benzyl Radicals with Acetylene. *ChemPhysChem* **2015**, *16*, 2091. [DOI: 10.1002/cphc.201500313]

[357] Bittner, J. D.; Howard, J. B. Composition profiles and reaction mechanisms in a near-sooting premixed benzene/oxygen/argon flame. *Symposium (International) on Combustion* **1981**, *18*, 1105. [DOI: 10.1016/S0082-0784(81)80115-4]

[358] He, C.; Kaiser, R. I.; Lu, W.; Ahmed, M.; Pivovarov, P. S.; Kuznetsov, O. V.; Zagidullin, M. V.; Mebel, A. M., Unconventional Pathway in the Gas-Phase Synthesis of 9H-Fluorene ($C_{13}H_{10}$) via the Radical–Radical Reaction of Benzyl ($C_7H_7$) with Phenyl ($C_6H_5$). *Angew. Chem*. **2023**, 135 (6). [DOI: 10.1002/ange.202216972]

[359] Li, W.; Zhao, L.; Kaiser, R. I., A unified reaction network on the formation of five-membered ringed polycyclic aromatic hydrocarbons (PAHs) and their role in ring expansion processes through radical-radical reactions. *Phys. Chem. Chem. Phys*. **2023**, 25 (5), 4141-4150. [DOI: 10.1039/D2CP05305E]

[360] Zhao, L.; Kaiser, R. I.; Lu, W.; Kostko, O.; Ahmed, M.; Evseev, M. M.; Bashkirov, E. K.; Oleinikov, A. D.; Azyazov, V. N.; Mebel, A. M.; Howlader, A. H.; Wnuk, S. F., Gas phase formation of cyclopentanaphthalene (benzindene) isomers via reactions of 5- and 6-indenyl radicals with vinylacetylene. *Phys. Chem. Chem. Phys*. **2020**, 22 (39), 22493-22500. [DOI: 10.1039/d0cp03846f]

[361] Necula, A.; Scott, L. T., Trapping Aryl Radicals with Acetylene: Evidence for C2-Accretion as a Mechanism for Polycyclic Aromatic Hydrocarbon Growth1. *J. Am. Chem. Soc.* **2000,** *122* (7), 1548-1549. [DOI: 10.1021/ja992115c]

[362] Zhao, L.; Doddipatla, S.; Kaiser, R. I.; Lu, W.; Kostko, O.; Ahmed, M.; Tuli, L. B.; Morozov, A. N.; Howlader, A. H.; Wnuk, S. F.; Mebel, A. M.; Azyazov, V. N.; Mohamed, R. K.; Fischer, F. R., Gas-phase synthesis of corannulene - a molecular building block of fullerenes. *Phys. Chem. Chem. Phys*. **2021**, 23 (10), 5740-5749. [DOI: 10.1039/D0CP06537D]

[392] Lukianova, M. A.; Feldman, V. I., Radiation-induced closure of the second aromatic ring: Possible way to PAH starting from a styrene-acetylene complex. *Radiat. Phys. Chem.* **2022,** *191*. [DOI: 10.1016/j.radphyschem.2021.109847]

[393] Protopapa, S.; Olkin, C. B.; Grundy, W. M.; Li, J.-Y.; Verbiscer, A.; Cruikshank, D. P.; Gautier, T.; Quirico, E.; Cook, J. C.; Reuter, D.; Howett, C. J. A.; Stern, A.; Beyer, R. A.; Porter, S.; Young, L. A.; Weaver, H. A.; Ennico, K.; Dalle Ore, C. M.; Scipioni, F.; Singer, K., Disk-resolved Photometric Properties of Pluto and the Coloring Materials across its Surface. *Astron. J.* **2020,** *159* (2). [DOI: 10.3847/1538-3881/ab5e82]

[394] Brown, M. E.; Schaller, E. L.; Blake, G. A. Irradiation Products on Dwarf Planet Makemake. *Astron. J.* **2015**, *149*, 105. [DOI: 10.1088/0004-6256/149/3/105]

[395] Combe, J.-P.; Singh, S.; Johnson, K. E.; McCord, T. B.; De Sanctis, M. C.; Ammannito, E.; Carrozzo, F. G.; Ciarniello, M.; Frigeri, A.; Raponi, A.; Tosi, F.; Zambon, F.; Scully, J. E. C.; Raymond, C. A.; Russell, C. T. The surface composition of Ceres' Ezinu quadrangle analyzed by the Dawn mission. *Icarus* **2019**, *318*, 124. [DOI: 10.1016/j.icarus.2017.12.039]

[396] Barucci, M. A.; Merlin, F.; Guilbert, A.; de Bergh, C.; Alvarez-Candal, A.; Hainaut, O.; Doressoundiram, A.; Dumas, C.; Owen, T.; Coradini, A., Surface composition and temperature of the TNO Orcus. *Astron. Astrophys.* **2008,** *479* (1), L13-L16. [DOI: 10.1051/0004-6361:20079079]

[397] Brown, M. E., The Compositions of Kuiper Belt Objects. *Annu. Rev. Earth Planet. Sci.* **2012,** *40* (1), 467-494. [DOI: 10.1146/annurev-earth-042711-105352]

[398] Brown, R. H.; Baines, K. H.; Bellucci, G.; Buratti, B. J.; Capaccioni, F.; Cerroni, P.; Clark, R. N.; Coradini, A.; Cruikshank, D. P.; Drossart, P.; Formisano, V.; Jaumann, R.; Langevin, Y.; Matson, D. L.; McCord, T. B.; Mennella, V.; Nelson, R. M.; Nicholson, P. D.; Sicardy, B.; Sotin, C.; Baugh, N.; Griffith, C. A.; Hansen, G. B.; Hibbitts, C. A.; Momary, T. W.; Showalter, M. R. Observations in the Saturn system during approach and orbital insertion, with Cassini's visual and infrared mapping spectrometer (VIMS). *Astron. Astrophys.* **2006**, *446*, 707. [DOI: 10.1051/0004-6361:20053054]

[399] Buratti, B. J.; Soderlund, K.; Bauer, J.; Mosher, J. A.; Hicks, M. D.; Simonelli, D. P.; Jaumann, R.; Clark, R. N.; Brown, R. H.; Cruikshank, D. P.; Momary, T. Infrared (0.83–5.1 μm) photometry of Phoebe from the Cassini Visual Infrared Mapping Spectrometer. *Icarus* **2008**, *193*, 309. [DOI: 10.1016/j.icarus.2007.09.014]

[400] Cruikshank, D. P.; Dalle Ore, C. M.; Clark, R. N.; Pendleton, Y. J. Aromatic and aliphatic organic materials on Iapetus: Analysis of Cassini VIMS data. *Icarus* **2014**, *233*, 306. [DOI: 10.1016/j.icarus.2014.02.011]

[401] Dalle Ore, C. M.; Protopapa, S.; Cook, J. C.; Grundy, W. M.; Cruikshank, D. P.; Verbiscer, A. J.; Ennico, K.; Olkin, C. B.; Stern, S. A.; Weaver, H. A.; Young, L. A. Ices on Charon: Distribution of H2O and NH3 from New Horizons LEISA observations. *Icarus* **2018**, *300*, 21. [DOI: 10.1016/j.icarus.2017.08.026]

[402] Protopapa, S.; Grundy, W. M.; Reuter, D. C.; Hamilton, D. P.; Dalle Ore, C. M.; Cook, J. C.; Cruikshank, D. P.; Schmitt, B.; Philippe, S.; Quirico, E.; Binzel, R. P.; Earle, A. M.; Ennico, K.; Howett, C. J. A.; Lunsford, A. W.; Olkin, C. B.; Parker, A.; Singer, K. N.; Stern, A.; Verbiscer, A. J.; Weaver, H. A.; Young, L. A. Pluto's global surface composition through pixel-by-pixel Hapke modeling of New Horizons Ralph/LEISA data. *Icarus* **2017**, *287*, 218. [DOI: 10.1016/j.icarus.2016.11.028]

[403] Abplanalp, M. J.; Kaiser, R. I., Implications for Extraterrestrial Hydrocarbon Chemistry: Analysis of Acetylene ($C_2H_2$) and D2-acetylene ($C_2D_2$) Ices Exposed to Ionizing Radiation via Ultraviolet–Visible Spectroscopy, Infrared Spectroscopy, and Reflectron Time-of-flight Mass Spectrometry. *Astrophys. J.* **2020,** *889* (1). [DOI: 10.3847/1538-4357/ab616c]

[404] Zhang, C.; Zhu, C.; Turner, A. M.; Antonov, I. O.; Garcia, A. D.; Meinert, C.; Young, L. A.; Jewitt, D. C.; Kaiser, R. I., Processing of methane and acetylene ices by galactic cosmic rays and implications to the color diversity of Kuiper Belt objects. *Sci. Adv.* **2023,** *9* (22), eadg6936. [DOI: 10.1126/sciadv.adg6936]

[405] Dalle Ore, C. M.; Barucci, M. A.; Emery, J. P.; Cruikshank, D. P.; Dalle Ore, L. V.; Merlin, F.; Alvarez-Candal, A.; de Bergh, C.; Trilling, D. E.; Perna, D.; Fornasier, S.; Mastrapa, R. M. E.; Dotto, E. *Astron. Astrophys.* **2009**, *501*, 349. [DOI: 10.1051/0004-6361/200911752]

[406] Krätschmer, W.; Lamb, L. D.; Fostiropoulos, K.; Huffman, D. R., Solid C60: a new form of carbon. *Nature* **1990,** *347* (6291), 354-358. [DOI: 10.1038/347354a0]

[407] Smalley, R. E., Self-assembly of the fullerenes. *Acc. Chem. Res.* **2002,** *25* (3), 98-105. [DOI: 10.1021/ar00015a001],

[408] Dunk, P. W.; Kaiser, N. K.; Hendrickson, C. L.; Quinn, J. P.; Ewels, C. P.; Nakanishi, Y.; Sasaki, Y.; Shinohara, H.; Marshall, A. G.; Kroto, H. W., Closed network growth of fullerenes. *Nat Commun* **2012,** *3*, 855. [DOI: 10.1038/ncomms1853]

[409] Micelotta, E. R.; Jones, A. P.; Cami, J.; Peeters, E.; Bernard-Salas, J.; Fanchini, G., The Formation of Cosmic Fullerenes from Arophatic Clusters. *Astrophys. J.* **2012,** *761* (1). [DOI: 10.1088/0004-637X/761/1/35]

[410] Berne, O.; Tielens, A. G., Formation of buckminsterfullerene (C60) in interstellar space. *Proc Natl Acad Sci U S A* **2012,** *109* (2), 401-6. [DOI: 10.1073/pnas.1114207108]

[411] Otero, G.; Biddau, G.; Sanchez-Sanchez, C.; Caillard, R.; Lopez, M. F.; Rogero, C.; Palomares, F. J.; Cabello, N.; Basanta, M. A.; Ortega, J.; Mendez, J.; Echavarren, A. M.; Perez, R.; Gomez-Lor, B.; Martin-Gago, J. A., Fullerenes from aromatic precursors by surface-catalysed cyclodehydrogenation. *Nature* **2008,** *454* (7206), 865-8. [DOI: 10.1038/nature07193]

[412] Chuvilin, A.; Kaiser, U.; Bichoutskaia, E.; Besley, N. A.; Khlobystov, A. N., Direct transformation of graphene to fullerene. *Nat. Chem.* **2010,** *2* (6), 450-3. [DOI: 10.1038/nchem.644]

[413] Berné, O.; Montillaud, J.; Joblin, C., Top-down formation of fullerenes in the interstellar medium (Corrigendum). *Astron. Astrophys.* **2016,** *588*. [DOI: 10.1051/0004-6361/201425338e]

[414] Tuli, L. B.; Goettl, S. J.; Turner, A. M.; Howlader, A. H.; Hemberger, P.; Wnuk, S. F.; Guo, T.; Mebel, A. M.; Kaiser, R. I., Gas phase synthesis of the C40 nano bowl C(40)H(10). *Nat Commun* **2023,** *14* (1), 1527. [DOI: 10.1038/s41467-023-37058-y]

[415] Shaitanov, A. G.; Surovikin, Y. V.; Rezanov, I. V.; Shtertser, A. A.; Ul'yanitskii, V. Y.; Vasil'ev, A. A.; Likholobov, V. A., Synthesis and Study of Nanodispersed Carbon in the Combustion of Acetylene in a Flow Detonation Tube. *Russ. J. Appl. Chem.* **2019,** *91* (12), 2003-2011. [DOI: 10.1134/S1070427218120121]

[416] Martin, J. W.; Salamanca, M.; Kraft, M., Soot inception: Carbonaceous nanoparticle formation in flames. *Prog. Energy Combust. Sci.* **2022,** *88*. [DOI: 10.1016/j.pecs.2021.100956]

[417] Gail HP., Sedlmayr E. Dust Formation in Stellar Winds. In: Morfill G.E., Scholer M. (eds) Physical Processes in Interstellar Clouds. NATO ASI Series (Series C: Mathematical and Physical Sciences), vol 210. Springer, Dordrecht, 1987, pp. 275-303 [DOI: 10.1007/978-94-009-3945-5_19]

[418] Mironenko, R. M.; Likholobov, V. A.; Belskaya, O. B., Nanoglobular carbon and palladium - carbon catalysts for liquid-phase hydrogenation of organic compounds. *Russ. Chem. Rev.* **2022**, *91 (1),* RCR5017. [DOI: 10.1070/RCR5017]

[434] Galan, B. R.; Rovis, T., Beyond Reppe: building substituted arenes by [2+2+2] cycloadditions of alkynes. *Angew. Chem. Int. Ed. Engl.* **2009,** *48* (16), 2830-4. [DOI: 10.1002/anie.200804651]

[435] Roglans, A.; Pla-Quintana, A.; Sola, M., Mechanistic Studies of Transition-Metal-Catalyzed [2+2+2] Cycloaddition Reactions. *Chem. Rev.* **2021,** *121* (3), 1894-1979. [DOI: 10.1021/acs.chemrev.0c00062]

[436] Takahashi, T.; Xi, Z.; Yamazaki, A.; Liu, Y.; Nakajima, K.; Kotora, M., Cycloaddition Reaction of Zirconacyclopentadienes to Alkynes: Highly Selective Formation of Benzene Derivatives from Three Different Alkynes. *J. Am. Chem. Soc.* **1998,** *120* (8), 1672-1680. [DOI: 10.1021/ja970869q]

[437] Tanaka, R.; Yuza, A.; Watai, Y.; Suzuki, D.; Takayama, Y.; Sato, F.; Urabe, H., One-pot synthesis of metalated pyridines from two acetylenes, a nitrile, and a titanium(II) alkoxide. *J. Am. Chem. Soc.* **2005,** *127* (21), 7774-80. [DOI: 10.1021/ja050261e]

[438] Yamamoto, Y.; Ishii, J.; Nishiyama, H.; Itoh, K., Cp(*)RuCl-catalyzed formal intermolecular cyclotrimerization of three unsymmetrical alkynes through a boron temporary tether: regioselective four-component coupling synthesis of phthalides. *J. Am. Chem. Soc.* **2005,** *127* (26), 9625-31. [DOI: 10.1021/ja051377d]

[439] Hilt, G.; Hengst, C.; Hess, W., Solvent-Dependent Regiochemical Cyclotrimerisation of Phenylacetylene with Cobalt Catalysts Containing Disulfide Ligands: A Case Study. *Eur. J. Org. Chem.* **2008,** *2008* (13), 2293-2297. [DOI: 10.1002/ejoc.200800106]

[440] Tanaka, K.; Shirasaka, K., Highly chemo- and regioselective intermolecular cyclotrimerization of alkynes catalyzed by cationic rhodium(I)/modified BINAP complexes. *Org. Lett.* **2003,** *5* (24), 4697-9. [DOI: 10.1021/ol035963s]

[441] Geetharani, K.; Tussupbayev, S.; Borowka, J.; Holthausen, M. C.; Ghosh, S., A mechanistic study of the utilization of arachno-diruthenaborane [(Cp*RuCO)2B2H6] as an active alkyne-cyclotrimerization catalyst. *Chem. - Eur. J.* **2012,** *18* (27), 8482-9. [DOI: 10.1002/chem.201200291]

[442] Rassadin, V. A.; Nicolas, E.; Six, Y., Ti(OiPr)4/nBuLi: an attractive reagent system for [2+2+2] cyclotrimerisation reactions. *Chem. Commun. (Camb)* **2014,** *50* (57), 7666-9. [DOI: 10.1039/C4CC02698E]

[443] Orsino, A. F.; Gutierrez Del Campo, M.; Lutz, M.; Moret, M. E., Enhanced Catalytic Activity of Nickel Complexes of an Adaptive Diphosphine-Benzophenone Ligand in Alkyne Cyclotrimerization. *ACS Catal.* **2019,** *9* (3), 2458-2481. [DOI: 10.1021/acscatal.8b05025]

[444] Tsuji, H.; Yamagata, K.; Fujimoto, T.; Nakamura, E., Manganese-catalyzed benzene synthesis by [2+2+2] coupling of 1,3-dicarbonyl compound and terminal acetylene. *J. Am. Chem. Soc.* **2008,** *130* (25), 7792-3. [DOI: 10.1021/ja8015186]

[445] Kuninobu, Y.; Takata, H.; Kawata, A.; Takai, K., Rhenium-catalyzed synthesis of multisubstituted aromatic compounds via C-C single-bond cleavage. *Org. Lett.* **2008,** *10* (14), 3133-5. [DOI: 10.1021/ol801226t]

[446] Heller, B.; Sundermann, B.; Buschmann, H.; Drexler, H. J.; You, J.; Holzgrabe, U.; Heller, E.; Oehme, G., Photocatalyzed [2 + 2 + 2]-cycloaddition of nitriles with acetylene: an effective method for the synthesis of 2-pyridines under mild conditions. *J. Org. Chem.* **2002,** *67* (13), 4414-22. [DOI: 10.1021/jo011032n]

[447] More, A. A.; Ramana, C. V., Total synthesis of the putative structure of xylarinol B. *Chem. Asian. J.* **2014,** *9* (6), 1557-62. [DOI: 10.1002/asia.201301647]

[477] Cable, M. L.; Vu, T. H.; Maynard-Casely, H. E.; Choukroun, M.; Hodyss, R., The Acetylene-Ammonia Co-crystal on Titan. *ACS Earth Space Chem.* **2018,** *2* (4), 366-375. [DOI: 10.1021/acsearthspacechem.7b00135]

[478] Ennis, C.; Cable, M. L.; Hodyss, R.; Maynard-Casely, H. E., Mixed Hydrocarbon and Cyanide Ice Compositions for Titan's Atmospheric Aerosols: A Ternary-Phase Co-crystal Predicted by Density Functional Theory. *ACS Earth Space Chem.* **2020,** *4* (7), 1195-1200. [DOI: 10.1021/acsearthspacechem.0c00130]

[479] Cable, M. L.; Vu, T. H.; Malaska, M. J.; Maynard-Casely, H. E.; Choukroun, M.; Hodyss, R., Properties and Behavior of the Acetonitrile–Acetylene Co-Crystal under Titan Surface Conditions. *ACS Earth Space Chem.* **2020,** *4* (8), 1375-1385. [DOI: 10.1021/acsearthspacechem.0c00129]

[480] Czaplinski, E. C.; Vu, T. H.; Cable, M. L.; Choukroun, M.; Malaska, M. J.; Hodyss, R., Experimental Characterization of the Pyridine:Acetylene Co-crystal and Implications for Titan's Surface. *ACS Earth Space Chem.* **2023,** *7* (3), 597-608. [DOI: 10.1021/acsearthspacechem.2c00377]

[481] Norman, L. H., Is there life on … Titan? *Astron. Geophys.* **2011,** *52* (1), 1.39-1.42. [DOI: 10.1111/j.1468-4004.2011.52139.x]

[482] Seager, S.; Schrenk, M.; Bains, W., An astrophysical view of Earth-based metabolic biosignature gases. *Astrobiology* **2012,** *12* (1), 61-82. [DOI: 10.1089/ast.2010.0489]

[483] Seager, S.; Bains, W.; Hu, R., A Biomass-Based Model to Estimate the Plausibility of Exoplanet Biosignature Gases. *Astrophys. J.* **2013,** *775* (2). [DOI: 10.1088/0004-637X/775/2/104]

[484] Bains, W., Many chemistries could be used to build living systems. *Astrobiology* **2004,** *4* (2),

137-67. [DOI: 10.1089/153110704323175124]

[485] McKay, C. P.; Smith, H. D., Possibilities for methanogenic life in liquid methane on the surface of Titan. *Icarus* **2005,** *178* (1), 274-276. [DOI: 10.1016/j.icarus.2005.05.018]

[486] Catling D.C. Astrobiology: A Very Short Introduction, OUP Oxford, Oxford, UK, 2013.

[487] Abbas, O.; Schulze-Makuch, D., Acetylene-based pathways for prebiotic evolution on Titan. *Esa. Sp. Publ.* **2002,** *518,* 345-348. [https://ui.adsabs.harvard.edu/abs/2002ESASP.518..345A/abstract]

[488] Segura, A.; Navarro-Gonzalez, R., Production of low molecular weight hydrocarbons by volcanic eruptions on early Mars. *Orig. Life Evol. Biosph.* **2005,** *35* (5), 477-87. [DOI: 10.1007/s11084-005-6420-3]

[489] Kasting, J. F., Earth's early atmosphere. *Science* **1993,** *259* (5097), 920-6. [DOI: 10.1126/science.11536547]

[490] Sossi, P. A.; Burnham, A. D.; Badro, J.; Lanzirotti, A.; Newville, M.; O'Neill, H. S. C., Redox state of Earth's magma ocean and its Venus-like early atmosphere. *Sci. Adv.* **2020,** *6* (48). [DOI: 10.1126/sciadv.abd1387]

[491] Lupu, R. E.; Zahnle, K.; Marley, M. S.; Schaefer, L.; Fegley, B.; Morley, C.; Cahoy, K.; Freedman, R.; Fortney, J. J., The Atmospheres of Earthlike Planets after Giant Impact Events. *Astrophys. J.* **2014,** *784* (1). [DOI: 10.1088/0004-637X/784/1/27]

[492] Tian, F.; Kasting, J. F.; Zahnle, K., Revisiting HCN formation in Earth's early atmosphere. *Earth Planet. Sci. Lett.* **2011,** *308* (3-4), 417-423. [DOI: 10.1016/j.epsl.2011.06.011]

[493] Capaccioni, B.; Martini, M.; Mangani, F.; Giannini, L.; Nappii, G.; Prati, F., Light hydrocarbons in gas-emissions from volcanic areas and geothermal fields. *Geochem. J.* **1993,** *27* (1), 7-17. [DOI: 10.2343/geochemj.27.7]

[494] Muenow, D. W., High temperature mass spectrometric gas-release studies of Hawaiian volcanic glass: Pele's tears. *Geochim. Cosmochim. Acta* **1973,** *37* (6), 1551-1561. [DOI: 10.1016/0016-7037(73)90090-2]

[495] Mukhin, L. M., Volcanic processes and synthesis of simple organic compounds on primitive earth. *Orig. Life* **1976,** *7* (4), 355-68. [DOI: 10.1007/BF00927931]

[496] Rimmer, P. B.; Shorttle, O., Origin of Life's Building Blocks in Carbon- and Nitrogen-Rich Surface Hydrothermal Vents. *Life (Basel)* **2019,** *9* (1). [DOI: 10.3390/life9010012]

[497] Schink, B., Fermentation of acetylene by an obligate anaerobe, Pelobacter acetylenicus sp. nov. *Arch. Microbiol.* **1985,** *142* (3), 295-301. [DOI: 10.1007/bf00693407]

[498] Sutton, J. M.; Baesman, S. M.; Fierst, J. L.; Poret-Peterson, A. T.; Oremland, R. S.; Dunlap, D. S.; Akob, D. M., Complete Genome Sequences of Two Acetylene-Fermenting Pelobacter acetylenicus Strains. *Genome Announc.* **2017,** *5* (6). [DOI: 10.1128/genomeA.01572-16]

[499] Boll, M.; Einsle, O.; Ermler, U.; Kroneck, P. M.; Ullmann, G. M., Structure and Function of the Unusual Tungsten Enzymes Acetylene Hydratase and Class II Benzoyl-Coenzyme A Reductase. *J. Mol. Microbiol. Biotechnol.* **2016,** *26* (1-3), 119-37. [DOI: 10.1159/000440805]

[500] Tenbrink, F.; Schink, B.; Kroneck, P. M., Exploring the active site of the tungsten, iron-sulfur enzyme acetylene hydratase. *J. Bacteriol.* **2011,** *193* (5), 1229-36. [DOI: 10.1128/jb.01057-10]

[501] Lovett, R. A., Enceladus named sweetest spot for alien life. *Nature* **2011**. [DOI: 10.1038/news.2011.337]

[502] Bernstein, M. P.; Moore, M. H.; Elsila, J. E.; Sandford, S. A.; Allamandola, L. J.; Zare, R. N., Side Group Addition to the Polycyclic Aromatic Hydrocarbon Coronene by Proton Irradiation in Cosmic Ice Analogs. *Astrophys. J.* **2003,** *582* (1), L25-L29. [DOI: 10.1086/345941]

[503] Bernstein, M. P.; Dworkin, J. P.; Sandford, S. A.; Allamandola, L. J., Ultraviolet irradiation of naphthalene in H2O ice: Implications for meteorites and biogenesis. *Meteorit. Planet. Sci.* **2001,** *36* (3), 351-358. [DOI: 10.1111/j.1945-5100.2001.tb01878.x]

[504] Ashbourn, S. F. M.; Elsila, J. E.; Dworkin, J. P.; Bernstein, M. P.; Sandford, S. A.; Allamandola, L. J., Ultraviolet photolysis of anthracene in H2O interstellar ice analogs: Potential connection to meteoritic organics. *Meteorit. Planet. Sci.* **2007,** *42* (12), 2035-2041. [DOI: 10.1111/j.1945-5100.2007.tb01005.x]

[505] Bernstein, M. P.; Sandford, S. A.; Allamandola, L. J.; Gillette, J. S.; Clemett, S. J.; Zare, R. N., UV irradiation of polycyclic aromatic hydrocarbons in ices: production of alcohols, quinones, and ethers. *Science* **1999,** *283* (5405), 1135-8. [DOI: 10.1126/science.283.5405.1135]

[506] Peeters, Z.; Botta, O.; Charnley, S. B.; Ruiterkamp, R.; Ehrenfreund, P. The Astrobiology of Nucleobases. *Astrophys. J.* **2003**, *593*, L129. [DOI: 10.1086/378346]

[507] Smith, K. E.; Callahan, M. P.; Gerakines, P. A.; Dworkin, J. P.; House, C. H. Investigation of pyridine carboxylic acids in CM2 carbonaceous chondrites: Potential precursor molecules for ancient coenzymes. *Geochim. Cosmochim. Acta* **2014**, *136*, 1. [DOI: 10.1016/j.gca.2014.04.001]

[508] Callahan, M. P.; Smith, K. E.; Cleaves, H. J., 2nd; Ruzicka, J.; Stern, J. C.; Glavin, D. P.; House, C. H.; Dworkin, J. P. Carbonaceous meteorites contain a wide range of extraterrestrial nucleobases. *Proc. Natl. Acad. Sci. U. S. A.* **2011**, *108*, 13995. [DOI: 10.1073/pnas.1106493108]

[509] Martins, Z.; Botta, O.; Fogel, M. L.; Sephton, M. A.; Glavin, D. P.; Watson, J. S.; Dworkin, J. P.; Schwartz, A. W.; Ehrenfreund, P. Extraterrestrial nucleobases in the Murchison meteorite. *Earth Planet. Sci. Lett.* **2008**, *270*, 130. [DOI: 10.1016/j.epsl.2008.03.026]

[510] Folsome, C. E.; Lawless, J.; Romiez, M.; Ponnamperuma, C., Heterocyclic compounds indigenous to the Murchison meteorite. *Nature* **1971,** *232* (5306), 108-9. [DOI: 10.1038/232108a0]

[511] Hayatsu, R.; Studier, M. H.; Moore, L. P.; Anders, E., Purines and triazines in the Murchison meteorite. *Geochim. Cosmochim. Acta* **1975,** *39* (4), 471-488. [DOI: 10.1016/0016-7037(75)90101-5]

[512] Stoks, P. G.; Schwartz, A. W., Uracil in carbonaceous meteorites. *Nature* **1979,** *282* (5740), 709-710. [DOI: 10.1038/282709a0]

[513] Martins, Z., The Nitrogen Heterocycle Content of Meteorites and Their Significance for the Origin of Life. *Life (Basel)* **2018,** *8* (3). [DOI: 10.3390/life8030028]

[514] Pizzarello, S.; Holmes, W. Nitrogen-containing compounds in two CR2 meteorites: 15N composition, molecular distribution and precursor molecules. *Geochim. Cosmochim. Acta* **2009**, *73*, 2150. [DOI: 10.1016/j.gca.2009.01.022]

[515] Hudgins, D. M.; Bauschlicher, J. C. W.; Allamandola, L. J., Variations in the Peak Position of the 6.2 μm Interstellar Emission Feature: A Tracer of N in the Interstellar Polycyclic Aromatic Hydrocarbon Population. *Astrophys. J.* **2005,** *632* (1), 316-332. [DOI: 10.1086/432495]

[516] Peeters, Z.; Botta, O.; Charnley, S. B.; Kisiel, Z.; Kuan, Y. J.; Ehrenfreund, P. Formation and photostability of N-heterocycles in space. *Astron. Astrophys.* **2005**, *433*, 583. [DOI: 10.1051/0004-6361:20042443]

[517] Parker, D. S. N.; Kaiser, R. I.; Kostko, O.; Troy, T. P.; Ahmed, M.; Mebel, A. M.; Tielens, A. G. G. M. Gas Phase Synthesis of (Iso)Quinoline and Its Role in the Formation of Nucleobases in the Interstellar Medium. *Astrophys. J.* **2015**, *803*, 53. [DOI: 10.1088/0004-637X/803/2/53]

[518] Parker, D. S.; Kaiser, R. I.; Kostko, O.; Troy, T. P.; Ahmed, M.; Sun, B. J.; Chen, S. H.; Chang, A. H. On the formation of pyridine in the interstellar medium. *Phys. Chem. Chem. Phys.* **2015**, *17*, 32000. [DOI: 10.1039/C5CP02960K]

[519] Morales, S. B.; Bennett, C. J.; Le Picard, S. D.; Canosa, A.; Sims, I. R.; Sun, B. J.; Chen, P. H.; Chang, A. H. H.; Kislov, V. V.; Mebel, A. M.; Gu, X.; Zhang, F.; Maksyutenko, P.; Kaiser, R. I. A Crossed Molecular Beam, Low-Temperature Kinetics, and Theoretical Investigation of the Reaction of the Cyano Radical (Cn) with 1,3-Butadiene (C4h6). A Route to Complex Nitrogen-Bearing Molecules in Low-Temperature Extraterrestrial Environments. *Astrophys. J.* **2011**, *742*, 26. [DOI: 10.1088/0004-637X/742/1/26]

[520] Sun, B. J.; Huang, C. H.; Chen, S. Y.; Chen, S. H.; Kaiser, R. I.; Chang, A. H. Theoretical study on reaction mechanism of ground-state cyano radical with 1,3-butadiene: prospect of pyridine formation. *J. Phys. Chem.* **2014**, *118*, 7715. [DOI: 10.1021/jp5056864]

[521] Zhao, L.; Prendergast, M.; Kaiser, R. I.; Xu, B.; Lu, W.; Ahmed, M.; Hasan Howlader, A.; Wnuk, S. F.; Korotchenko, A. S.; Evseev, M. M.; Bashkirov, E. K.; Azyazov, V. N.; Mebel, A. M., A molecular beam and computational study on the barrierless gas phase formation of (iso)quinoline in low temperature extraterrestrial environments. *Phys. Chem. Chem. Phys.* **2021**, 23 (34), 18495-18505. [DOI: 10.1039/d1cp02169a]

[522] Ricca, A., A Computational Study of the Mechanisms for the Incorporation of a Nitrogen Atom into Polycyclic Aromatic Hydrocarbons in the Titan Haze. *Icarus* **2001,** *154* (2), 516-521. [DOI: 10.1006/icar.2001.6694]

[523] Lorenz, R. D.; Lunine, J. I. Titan's surface reviewed: the nature of bright and dark terrain. *Planet. Space Sci.* **1997**, *45*, 981. [DOI: 10.1016/S0032-0633(97)00087-1]

[524] Allamandola, L. J.; Tielens, A. G.; Barker, J. R. Interstellar polycyclic aromatic hydrocarbons: the infrared emission bands, the excitation/emission mechanism, and the astrophysical implications. *Astrophys. J., Suppl. Ser.* **1989**, *71*, 733. [DOI: 10.1086/191396]

[525] Rap, D. B.; Schrauwen, J. G. M.; Marimuthu, A. N.; Redlich, B.; Brünken, S., Low-temperature nitrogen-bearing polycyclic aromatic hydrocarbon formation routes validated by infrared spectroscopy. *Nat. Astron* **2022,** *6* (9), 1059-1067. [DOI: 10.1038/s41550-022-01713-z]

[526] Menor-Salvan, C.; Marin-Yaseli, M. R., A new route for the prebiotic synthesis of nucleobases and hydantoins in water/ice solutions involving the photochemistry of acetylene. *Chem. - Eur. J.* **2013,** *19* (20), 6488-97. [DOI: 10.1002/chem.201204313]

[527] Shapiro, R., Prebiotic cytosine synthesis: a critical analysis and implications for the origin of life. *Proc. Natl. Acad. Sci. U S A* **1999,** *96* (8), 4396-401. [DOI: 10.1073/pnas.96.8.4396]

[528] Albernaz, A. F.; da Silva, W. B.; Correa, E., Isocyanoacetylene and cyanoacetylene formation study from $C_2H_2$ + CN reaction. *Int. J. Chem. Kinet.* **2022,** *54* (5), 309-316. [DOI: 10.1002/kin.21560]

[529] Hopf, H.; Mlynek, C.; McMahon, R. J.; Menke, J. L.; Lesarri, A.; Rosemeyer, M.; Grabow, J. U., On the trimerization of cyanoacetylene: mechanism of formation of tricyanobenzene isomers and laboratory detection of their radio spectra. *Chem. - Eur. J.* **2010,** *16* (47), 14115-23. [DOI: 10.1002/chem.201001648]

[530] Miller, S. L.; Urey, H. C., Organic compound synthesis on the primitive earth. *Science* **1959,** *130* (3370), 245-51. [DOI: 10.1126/science.130.3370.245]

[531] Sanchez, R. A.; Ferris, J. P.; Orgel, L. E., Cyanoacetylene in prebiotic synthesis. *Science* **1966,** *154* (3750), 784-5. [DOI: 10.1126/science.154.3750.784]

[532] Kaur, S.; Ohri, A.; Sharma, P., Could Purines Be Formed from Cyanamide and Cyanoacetylene in a Prebiotic Earth Environment? *ACS Omega* **2019,** *4* (7), 12771-12781. [DOI: 10.1021/acsomega.9b01169]

[533] Seitz, C.; Eisenreich, W.; Huber, C., The Abiotic Formation of Pyrrole under Volcanic, Hydrothermal Conditions-An Initial Step towards Life's First Breath? *Life (Basel)* **2021,** *11* (9). [DOI: 10.3390/life11090980]

[534] Platts S. N. The PAH World: Discotic polynuclear aromatic compounds as a mesophase scaffolding at the origin of life, http://www.pahworld.com/ (accessed 2004-11-11).

[535] Nuevo, M.; Milam, S. N.; Sandford, S. A.; Elsila, J. E.; Dworkin, J. P., Formation of uracil from the ultraviolet photo-irradiation of pyrimidine in pure H2O ices. *Astrobiology* **2009,** *9* (7), 683-95. [DOI: 10.1089/ast.2008.0324]

[536] Nuevo, M.; Milam, S. N.; Sandford, S. A., Nucleobases and prebiotic molecules in organic residues produced from the ultraviolet photo-irradiation of pyrimidine in NH(3) and H(2)O+NH(3) ices. *Astrobiology* **2012,** *12* (4), 295-314. [DOI: 10.1089/ast.2011.0726]

[537] Nuevo, M.; Materese, C. K.; Sandford, S. A., The Photochemistry of Pyrimidine in Realistic Astrophysical Ices and the Production of Nucleobases. *Astrophys. J.* **2014,** *793* (2). [DOI: 10.1088/0004-637X/793/2/125]

[538] Materese, C. K.; Nuevo, M.; Sandford, S. A., The Formation of Nucleobases from the Ultraviolet Photoirradiation of Purine in Simple Astrophysical Ice Analogues. *Astrobiology* **2017,** *17* (8), 761-770. [DOI: 10.1089/ast.2016.1613]

prebiotic hydrothermal conditions. *Commun. Chem.* **2023,** *6* (1), 220. [DOI: 10.1038/s42004-023-01021-1]

[555] James Webb Space Telescope https://webb.nasa.gov (accessed 2024-03-01).

[556] Lorenz, R. D.; MacKenzie, S. M.; Neish, C. D.; Gall, A. L.; Turtle, E. P.; Barnes, J. W.; Trainer, M. G.; Werynski, A.; Hedgepeth, J.; Karkoschka, E. Selection and Characteristics of the Dragonfly Landing Site near Selk Crater, Titan. *Planet. Sci. J.* **2021**, *2*, 24. [DOI: 10.3847/PSJ/abd08f]

[557] Dragonfly News https://dragonfly.jhuapl.edu/ (accessed 2024-03-01).

[558] JUICE https://sci.esa.int/web/juice (accessed 2024-03-01).